\documentclass{aa}  
\usepackage{graphicx}
\usepackage{float}
\usepackage{indentfirst}
\usepackage{lscape}
\usepackage{longtable}
\usepackage{amsmath}
\usepackage{amssymb}
\usepackage{graphicx}
\usepackage{lastpage}
\usepackage{natbib}
\usepackage{txfonts}
\begin{document}
   \title{Gas phase Elemental abundances in Molecular cloudS (GEMS)}

   \subtitle{VI. A sulphur journey across star-forming regions: study of thioformaldehyde emission}

   \author{G. Esplugues
          \inst{1},
          A. Fuente
          \inst{1},
          D. Navarro-Almaida  
          \inst{1},    
          M. Rodr\'iguez-Baras
          \inst{1},
          L. Majumdar
          \inst{2, 3}, 
          P. Caselli
          \inst{4},
          V. Wakelam  
          \inst{5},    
          E. Roueff
          \inst{6},
          R. Bachiller  
          \inst{1},
          S. Spezzano  
          \inst{4}, 
          P. Rivi\`ere-Marichalar
          \inst{1},    
          R. Mart\'in-Dom\'enech
          \inst{7},         
          \and
          G. M. Mu\~noz Caro
          \inst{8}       
         }

   \institute{Observatorio Astron\'omico Nacional (OAN), Alfonso XII, 3, 28014, Madrid, Spain\\
              \email{g.esplugues@oan.es}
        \and
School of Earth and Planetary Sciences, National Institute of Science Education and Research, Jatni 752050, Odisha, India
        \and
Homi Bhabha National Institute, Training School Complex, Anushaktinagar, Mumbai 400094, India
   		\and
   		Max-Planck-Institut für extraterrestrische Physik, 85748 Garching, Germany
   		\and
   		Laboratoire d’astrophysique de Bordeaux, Univ. Bordeaux, CNRS, B18N, all\'ee Geoffroy Saint-Hilaire, 33615 Pessac, France
   		\and
   		LERMA, Observatoire de Paris, PSL Research University, CNRS, Sorbonne Universit\'e, 92190 Meudon, France
   		\and
   		Center for Astrophysics | Harvard \& Smithsonian, 60 Garden St., Cambridge, MA 02138, USA
   		\and
   		Centro de Astrobiología, INTA-CSIC, Torrejón de Ardoz, E-28850 Madrid, Spain
             }


 
  \abstract   
   {In the context of the IRAM 30m Large Program GEMS, we present a study of (deuterated) thioformaldehyde in several starless cores located in star-forming filaments of Taurus, Perseus, and Orion.  
   } 
   {We investigate the influence of the environmental conditions on the abundances of these molecules in the cores, and the effect of time evolution. 
   }
   {We have modelled the observed lines of H$_2$CS, HDCS, and D$_2$CS using the radiative transfer code RADEX. We have also used the chemical code Nautilus to model the evolution of these species depending on the characteristics of the starless cores.    
   }
   {We derive column densities and abundances for all the cores. We also derive deuterium fractionation ratios, $D$$_{\mathrm{frac}}$, which allow us to determine and compare the evolutionary stage between different parts of each star-forming region. Our results indicate that the north region of the B\,213 filament in Taurus is more evolved than the south, while the north-eastern part of Perseus presents an earlier evolutionary stage than the south-western zone. Model results also show that $D$$_{\mathrm{frac}}$ decreases with the cosmic-ray ionisation rate, while it increases with density and with the degree of sulphur depletion. In particular, we can only reproduce the observations when the initial sulphur abundance in the starless cores is at least one order of magnitude lower than the solar elemental sulphur abundance.  
}   
   {The progressive increase in HDCS/H$_2$CS and D$_2$CS/H$_2$CS with time makes these ratios powerful tools for deriving the chemical evolutionary stage of starless cores. However, they cannot be used to derive the temperature of these regions, since both ratios present a similar evolution at two different temperature ranges ($\sim$7-11 K and $\sim$15-19 K). Regarding chemistry, (deuterated) thioformaldehyde is mainly formed through gas-phase reactions (double-replacement and neutral–neutral displacement reactions), while surface chemistry plays an important role as a destruction mechanism.   
}

   \titlerunning{A sulphur journey across star-forming regions: study of thioformaldehyde emission}
   \authorrunning{G. Esplugues et al.}
   \maketitle
%

\section{\textbf{Introduction}}

Molecular clouds are known to be the birthplaces of stars. Observations of different star-forming regions at several wavelengths suggest molecular clouds have complex morphologies, with the dust and gas arranged mostly along elongated and filamentary structures \citep[e.g.,][]{Hartmann2002, Myers2009}. These filaments appear to be key structures that are required to reach the densities necessary for star formation according to Spitzer and Herschel observations, from sub-parsec scales in nearby star-forming regions \citep{Andre2010} to tens-of-parsecs scales along spiral arms \citep{Molinari2010}. In particular, filaments are where the initial conditions for star formation are set since they funnel interstellar gas and dust into increasingly denser concentrations that will contract and fragment, leading to gravitationally bound starless cores that will eventually form stars.

Dust and gas properties are key to determining the evolution of starless cores within filaments. In dense regions, grains are covered by icy mantles, which may affect grain coagulation \citep[e.g.,][]{Kimura2020}, and the presence of a dust size distribution, which in turn influences the grain emissivity. Moreover, grain coagulation may affect the charge balance in the gas phase, hence modifying the coupling of the gas with the magnetic field. In addition, grain surface chemistry plays an important role in the formation of the molecules that are key to the chemical network (e.g. H$_{{2}}$ and H$_{{2}}$O) and in the freeze-out process of molecules in the cold interiors of starless cores (e.g. CO).

One way to understand the dynamics of starless cores is through the study of molecule deuteration since the deuterium fraction, $D$$_{\mathrm{frac}}$, defined as the abundance ratio between a deuterated molecule and its hydrogenated counterpart, changes with their evolution \citep[e.g.,][]{Sakai2012}. In particular, $D$$_{\mathrm{frac}}$ is predicted to increase when a core evolves towards the onset of gravitational collapse as the core density profile becomes more and more centrally peaked \citep{Crapsi2005}. Then, $D$$_{\mathrm{frac}}$ drops when the young stellar object (YSO) formed at the core centre begins to heat its surroundings \citep[e.g.,][]{Emprechtinger2009}. 

In molecular clouds ($T$$\sim$10 K), as neutral-neutral reactions often have activation barriers, the dominant reactions are those involving ions, {H$^{+}_{3}$} being the first molecular ion formed as a product of the cosmic-ray (CR) ionisation of H$_2$ and H \citep{Ceccarelli2014}. This ion is the main one that initiates the deuterium enrichment process through an exothermic reaction with HD \citep[e.g.,][]{Millar1989}, and it forms H$_2$D$^+$, D$_2$H$^+$, and D$^{+}_{3}$. The collision of all these multi-deuterated forms of {H$^{+}_{3}$} with neutral species produces deuterated molecules, such as N$_2$D$^+$ and DCO$^+$. On the other hand, their dissociative electronic recombination increases the D/H atomic ratio by several orders of magnitude with respect to the D cosmic abundance \citep{Caselli2019}, thus allowing the deuteration of molecules (e.g. methanol) on the surface of dust grains.  
In cold environments, HD can also react with CH$^{+}_{3}$ and C$_2$H$^{+}_{2}$, leading to large enhancements of deuterated ions as well, such as CH$_2$D$^+$ and C$_2$HD$^+$, since the back reactions are inhibited \citep{Millar2005}. All these newly formed deuterated ions react in turn with other molecules and atoms, transferring the D atoms to all the other species. 

In pre-stellar cores \citep[characterised by low temperatures, $T$$\lesssim$10 K;][]{Ceccarelli2014}, CO and O freeze-out onto dust grains, which implies fewer destruction events for H$^{+}_{3}$ and its deuterated forms. This leads to an increase in H$_2$D$^+$, D$_2$H$^+$, and D$^{+}_{3}$ \citep{Dalgarno1984, Roberts2003, Walmsley2004}, favouring deuteration processes in the presence of high-density gas \citep[e.g.,][]{Roberts2000, Caselli2002, Bacmann2003, Crapsi2005}. Therefore, in dense and cold cores, the deuterated fraction is expected to be much higher than the average [D/H] interstellar abundance ratio \citep[of the order of 10$^{-5}$;][]{Oliveira2003, Linsky2006}.

Evolution of starless cores can also be studied through molecule depletion. At temperatures of roughly 10 K and densities above 10$^{4}$ cm$^{-3}$, several molecules, such as CO and CS, condense out onto dust grain surfaces \citep[e.g.,][]{Caselli1999, Crapsi2005}, and the amount of depletion increases with time \citep[e.g.,][]{Bergin1997, Aikawa2003}. Therefore, depletion can be used as a time marker since evolved cores should be more depleted of certain species than younger cores. Apart from core evolution, the study of depletion is of paramount importance because of its effects on molecular clouds; depletion causes variations in the deuterium fractionation and the degree of ionisation \citep{Dalgarno1984, Caselli1998}, which is one of the fundamental parameters regulating the star formation rate \citep[e.g.,][]{Shu1987}. It also affects the gas-phase chemical composition, leading to chemical variations between similar types of clouds. In addition, the thermal balance of clouds may also be affected by the depletion of major gas coolants, such as CO \citep[e.g.,][]{Goldsmith1978}.     

Although sulphur is one of the most abundant species in the Universe \citep[S/H$\sim$1.5$\times$10$^{-5}$;][]{Asplund2009} and plays a crucial role in biological systems on Earth \citep[e.g.,][]{Leustek2002, Francioso2020}, S-bearing molecules are not as abundant as expected in the interstellar medium (ISM). In fact, one needs to assume a significant sulphur depletion to reproduce observations not only in cold starless cores, but also in hot corinos and hot cores \citep{Esplugues2014, Crockett2014}. In particular, it is thought that sulphur is depleted by a factor of up to 1000 compared to its estimated cosmic abundance \citep{Ruffle1999, Wakelam2004}. Several studies have been carried out to shed light on the sulphur reservoir in molecular clouds and the sulphur depletion issue \citep[e.g.,][]{Martin-Domenech2016, Navarro-Almaida2020}. Chemical models predict that, in the dense ISM, atomic sulphur would stick on grains and be mostly hydrogenated to form H$_{{2}}$S \citep{Hatchell1998, Garrod2007, Esplugues2014}, which is thought to be the main sulphur reservoir in the ice \citep{Vidal2017, Navarro-Almaida2020}. In fact, according to \citet{Druard2012}, the lower the (gas and dust) temperature (<20 K), the greater the H$_{{2}}$S abundance on the grain surfaces. However, H$_{{2}}$S has never been detected in interstellar ices, and its abundance has been estimated to be smaller than 5$\times$10$^{-8}$ (/H) \citep{Jimenez-Escobar2011}. Other solid species have been proposed as possible sulphur reservoirs, such as OCS, SO$_{{2}}$, H$_{{2}}$S$_{{2}}$, CS$_{{2}}$, and S$_8$ \citep[e.g.,][]{Palumbo1997, Druard2012, Laas2019, Shingledecker2020, Cazaux2022}, OCS being the only S-bearing molecule unambiguously detected in ice mantles in the infrared \citep{Geballe1985, Palumbo1995} along with, tentatively, SO$_{{2}}$ \citep{Boogert1997, Zasowski2009}.    

Gas phase Elemental abundances in Molecular CloudS (GEMS) is an Institut de Radioastronomie Millimétric (IRAM) 30m Large Program, which aims to estimate S, C, N, and O depletions and the gas ionisation fraction, $X$(e$^{-}$), as a function of visual extinction in a selected set of prototypical star-forming filaments \citep{Fuente2019}. To achieve this goal, it is first necessary to determine the abundances of the main reservoirs of the elements in the gas phase by selecting a sample of filaments located in several clouds, covering different types of star formation activity. Here we focus on a sample of starless cores of the nearby star-forming regions Taurus, Perseus, and Orion. These molecular cloud complexes were previously observed with $\it{Herschel}$ and SCUBA as part of the Gould Belt Survey \citep{Andre2010}, and accurate visual extinction $(A_{\mathrm{V}})$ and dust temperature $(T_{\mathrm{d}})$ maps are available \citep{Malinen2012, Hatchell2005, Lombardi2014, Zari2016}. These regions each have different types of star formation activity and therefore different levels of external illumination, allowing us to investigate the influence of UV radiation on the gas composition. In particular, observing several starless cores within each filament will allow us to investigate the effect of time evolution on the chemistry of these cores, while comparing cores in different regions will let us explore the effect of the environment on the chemistry therein. 
To carry out the present study, we selected the species H$_{{2}}$CS and its deuterated counterparts (HDCS and D$_{{2}}$CS). Several studies \citep[e.g.,][]{Drozdovskaya2018} confirm that H$_{{2}}$CS is expected to be involved in many grain-surface reactions and that it plays a key role in the synthesis of larger sulphur-bearing species, such as CH$_3$SH (in an analogous process to the sequential hydrogenation of CO that leads to CH$_3$OH through H$_2$CO). H$_2$CS can be directly formed from the hydrogenation of HCS, which, in turn, is formed through a neutral-neutral reaction between atomic carbon and H$_2$S 
Recent results also demonstrate that H$_2$CS can be formed via the reaction between CH and H$_2$S \citep{Doddipatla2020}, as well as via the irradiation of CO:H$_2$S ice \citep{Jimenez-Escobar2011}. In any case, these reactions are limited by the available amount of H$_2$S within the ices. Nevertheless, as previously mentioned, H$_2$S is considered the main sulphur reservoir in ices, which would facilitate the formation of H$_2$CS. In this way, theoretical studies \citep[e.g.,][]{Laas2019} also conclude that, at high densities, H$_2$CS, in addition to being present at a significant level in the gas phase, might also be an abundant C-bearing sulphur species in the ice. All this makes H$_{{2}}$CS and its deuterated counterparts good candidates for characterising starless cores and studying their evolution.  

In this paper the observations of H$_2$CS, HDCS, and D$_2$CS are described in Sect. \ref{section:observations}, and the considered source sample in Sect. \ref{section:sample}. In Sect. \ref{section:results} we present the data and use the non-local thermodynamic equilibrium radiative transfer code RADEX to derive column densities and abundances. A discussion about fractional abundance differences between the sources of the sample is presented in Sect. \ref{Discussion}. In that section, we also use the Nautilus time-dependent chemical code to analyse the deuterium fraction evolution and to study the CR impact on this fraction. Section \ref{Discussion} also provides a comparison between theoretical and observational results, as well as an analysis of the main chemical formation and destruction routes of (deuterated) thioformaldehyde. Finally, we summarise our conclusions in Sect. \ref{section:summary}.

\begin{table*}
\caption{Cores included in the GEMS sample and the observation cuts associated with them and shown in Figs. \ref{figure:B213_map}-\ref{figure:Orion_map}.}
\label{Table: GEMS sample}
\centering
\begin{tabular}{l|llllll}
\hline\hline
Region & Cloud & Core & \multicolumn{2}{c}{Coordinates}  & Cut  \\
       &       & ID   &   RA (J2000)      & Dec (J2000)        &  \\
\hline
\noalign{\smallskip}                          
%
& B\,213$^1$  &  \#1  &  04:17:41.8   &     $+$28:08:47.0        &  C1     \\   
&            &  \#2  &   04:17:50.6   &     $+$27:56:01.0      &  C2     \\
&            &  \#5  &  04:18:03.8   &     $+$28:23:06.0       &   C5    \\
{Taurus} & &  \#6  &  04:18:08.4   &     $+$28:05:12.0       &  C6     \\    
&            &  \#7  &  04:18:11.5   &     $+$27:35:15.0      &  C7       \\   
&            &  \#10 &  04:19:37.6   &     $+$27:15:31.0      &  C10      \\  
&            &  \#12 &  04:19:51.7   &     $+$27:11:33.0      &   C12      \\  
&            &  \#16 &  04:21:21.0   &     $+$27:00:09.0      &  C16      \\
\hline
& L1448$^2$   &  \#32 &  03:25:49.0   &     $+$30:42:24.6      &       C1   \\
& NGC\,1333$^2$           &  \#46 &  03:29:11.0   &      $+$31:18:27.4       &   {C3-1}   \\
 &           &  \#60 & 03:28:39.4    &      $+$31:18:27.4       &   C3-14   \\
&            &  \#51 &  03:29:08.8   &      $+$31:15:18.1         &    C4    \\
{Perseus} &   &  \#53 & 03:29:04.5  &      $+$31:20:59.1        &   C5    \\  
&            &  \#57 &  03:29:18.2     &      $+$31:25:10.8       &   C6    \\ 
&            &  \#64 &  03:29:25.5     &      $+$31:28:18.1       &   C7        \\ 
& Barnard 5$^2$   &  \#79 &  03:47:38.9    &      $+$32:52:15.0       &    C1      \\ 
& IC\,348$^2$     &  \#1  &    03:44:01.0    &   $+$32:01:54.8     &  C1   \\
&            &  \#10 &  03:44:05.7   &       $+$32:01:53.5         &  C10    \\                   
\hline
& Orion A     &       &   05:35:19.5   &      $-$05:00:41.5         &   C1          \\
{Orion} &            &       &  05:35:08.1    &      $-$05:35:41.5         &   C2           \\
&            &       & 05:35:23.6  &     $-$05:12:31.8     &      C3          \\
\noalign{\smallskip}                 
\hline 
\label{table:Cores_considered}
\end{tabular}

$^1$Taurus core IDs are from \citep{Hacar2013}. \\
$^2$Perseus core IDs are from \citep{Hatchell2007}. \\
\end{table*}

\section{Observations}
\label{section:observations}

The  data  used  in  this  work  are  taken  from  the  GEMS project. A detailed description of the observations is given by \citet{Fuente2019} and \citet{Rodriguez-Baras2021}. For clarity, the main observational parameters are summarised below. The 3 mm and 2 mm observations (covering the frequency range 85-172 GHz) were carried out using the IRAM 30m telescope at Pico Veleta (Spain) during three observing periods in July 2017, August 2017, and February 2018. The observational parameters are listed in Table \ref{table:observational_parameters} (see the appendix) with the beam size varying with the frequency as HPBW($\arcsec$)=2460/$\nu$ where $\nu$ is in GHz. The observing mode was frequency switching with a frequency throw of 6 MHz well adapted to removing standing waves between the secondary mirror and the receivers. The Eight MIxer Receivers (EMIR) and the fast Fourier transform spectrometers with a spectral resolution of 49 kHz (equivalent to 0.165 km s$^{-1}$ and 0.087 km s$^{-1}$ for frequencies at 89 GHz and 168 GHz, respectively) were used for these observations. The achieved rms was $\sim$10-20 mK for $\nu$<150 GHz and $\sim$20-30 mK for $\nu$>150 GHz. 

The intensity scale is $T$$_{\mathrm{MB}}$, which is related to $T$$^{\star}_{\mathrm{A}}$ by

\begin{equation}
T_{\mathrm{MB}}=\left(F_{\mathrm{eff}}/B_{\mathrm{eff}}\right)\times{T^{\star}_{\mathrm{A}}}
\end{equation}

\noindent where $F$$_{\mathrm{eff}}$ is the telescope forward efficiency and $B$$_{\mathrm{eff}}$ is the main beam efficiency\footnote{http://www.iram.es/IRAMES/mainWiki/Iram30mEfficiencies}. The difference between $T$$^{\star}_{\mathrm{A}}$ and $T$$_{\mathrm{MB}}$ is $\sim$17$\%$ at 86 GHz and $\sim$27$\%$ at 145 GHz.

Calibration errors are estimated to be $\sim$10$\%$. The data were reduced and processed using the CLASS and GREG packages from IRAM GILDAS software\footnote{http://www.iram.fr/IRAMFR/GILDAS}, developed by IRAM.


\section{Source sample}
\label{section:sample}

\begin{figure*}
\centering
\includegraphics[scale=0.30, angle=0]{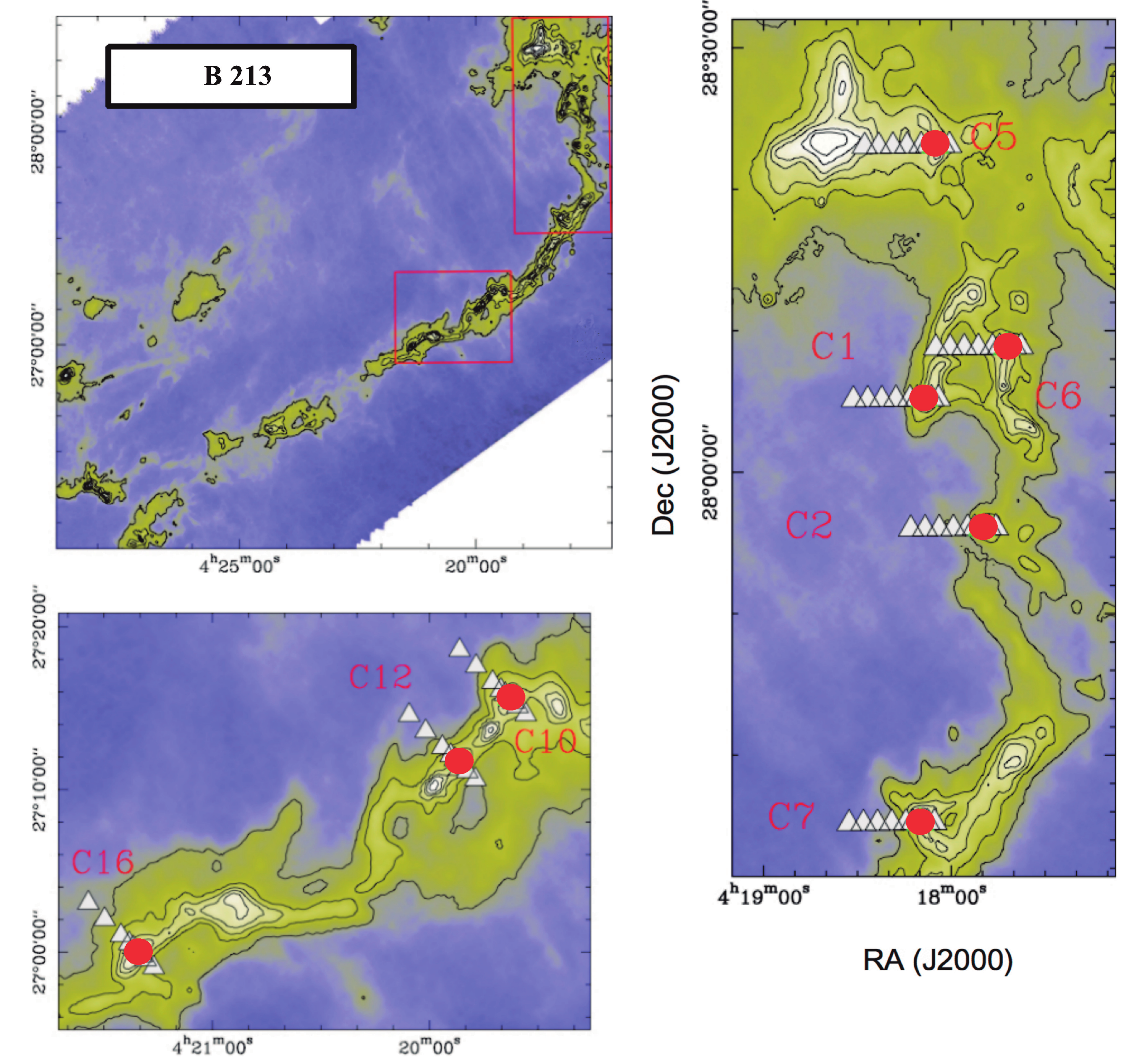} 
\\
\caption{B\,213 molecular hydrogen column density maps as derived by \citet{Palmeirim2013}, reconstructed at an angular resolution of 18.2$\arcsec$.
A general view of the region is shown in the top-right panel, and main regions of interest are enlarged. Contours are (3, 6, 9,12, 15, 20, and 25)$\times$10$^{21}$ cm$^{-2}$. Positions observed by GEMS with the 30m telescope are indicated with white triangles. Red circles represent the position of the starless cores. Labels in red indicate the cut IDs. See Table \ref{table:Cores_considered} for further details.}
\label{figure:B213_map}
\end{figure*}

\begin{figure*}
\centering
\includegraphics[scale=0.34, angle=0]{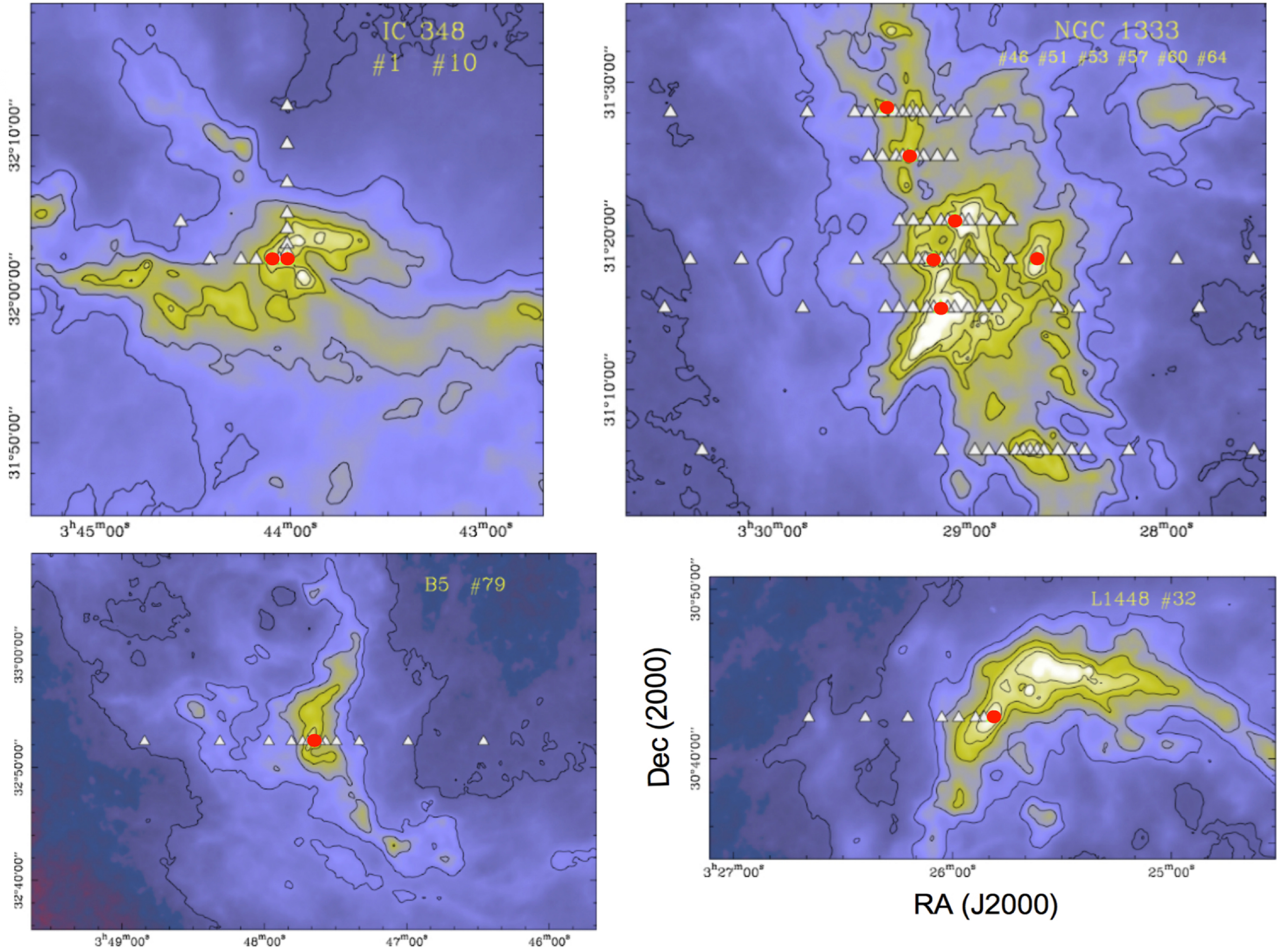} 
\\
\caption{Perseus filament (from right to left and top to bottom: NGC\,1333, IC\,348, L\,1448, and B5) dust opacity maps at 850 µm by \citet{Zari2016}, convolved at an angular resolution of 36$\arcsec$. Contours are (0.056, 0.13, 0.24, 0.56, 1.01, and 1.6) $\times$10$^{21}$ cm$^{-2}$, which, according to expression (7) from \citet{Zari2016}, corresponds to visual extinctions of $\sim$5, 7.5, 10, 15, 20, and 25 mag, respectively. Positions observed with the 30m telescope are indicated with white triangles. Red circles represent the positions of the starless cores.}
\label{figure:Perseus_map}
\end{figure*}

\begin{figure}
\centering
\includegraphics[scale=0.25, angle=0]{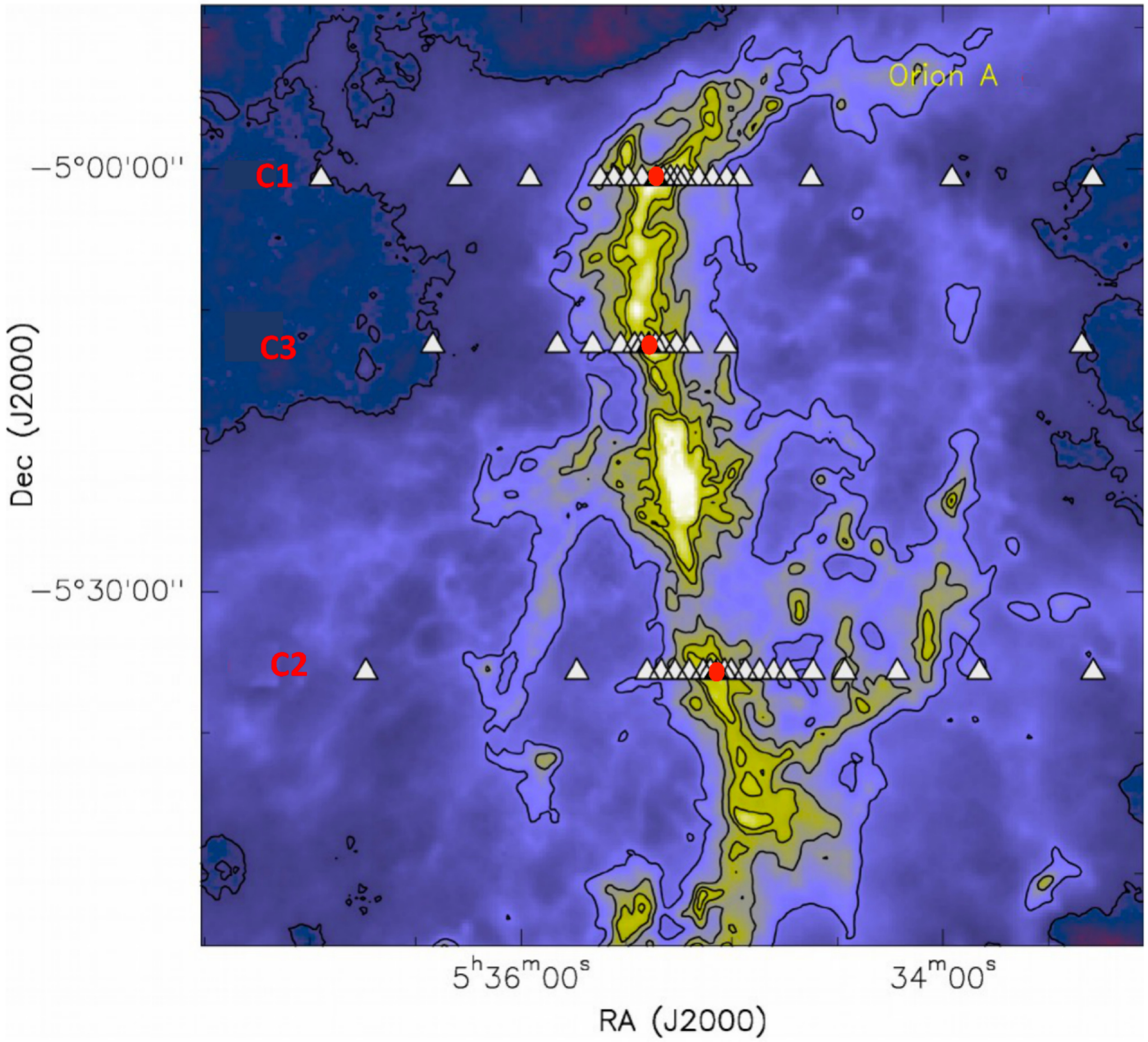}  
\\
\caption{Orion dust opacity map at 850 µm by Lombardi et al. (2014), convolved at an angular resolution of 36$\arcsec$. Contours are (0.056, 0.24, 0.56, 1.36, and 1.61)$\times$10$^{21}$ cm$^{-2}$, which, according to \citet{Lombardi2014}, correspond to visual extinctions of $\sim$1.3, 5.6, 13.2, 23.8, and 38 mag. Positions observed with the 30m telescope are indicated with white triangles. Red circles represent the positions of the starless cores. Labels in red indicate the cut IDs (see Table \ref{table:Cores_considered}).}
\label{figure:Orion_map}
\end{figure}

GEMS considers the molecular cloud complexes Taurus, Perseus, and Orion. In particular, observations include several cuts roughly perpendicular to selected filaments, where the separation between one position and another in a given cut is selected
to sample the visual extinction range in regular intervals of $A_{\mathrm{V}}$. In this project, we only focus on the starless cores within filaments of these regions (see Figs. \ref{figure:B213_map}, \ref{figure:Perseus_map}, and \ref{figure:Orion_map}). In the following, we describe the observed positions
in more detail.

The Taurus (L\,1495/B\,213) molecular cloud, with a total mass of about 1.5$\times$10$^4$ M$_{\sun}$ derived from CO data analysis \citep{Pineda2010}, is one of the closest regions of star formation, at a distance of about 145 pc \citep{Qian2015, Yan2019}, and is known to contain more than 250 YSOs. Taurus is considered an archetypal low-mass star-forming region. Multiple studies, using different ground-based and space telescopes, have been carried out to analyse its evolution and structure. For instance, \citet{Palla2002} found a young population inside the filaments and a more dispersed and older population outside them, concluding the presence of an age spread in the region. \citet{Goldsmith2008} showed a very complex and highly structured cloud morphology including filaments, cavities, and rings, while \citet{Goodman1992} studied the presence and orientation of magnetic fields with respect to several filaments within the Taurus cloud. We also highlight the $\it{Herschel}$ observations of Taurus obtained in the context of the Gould Belt survey \citep{Kirk2013, Palmeirim2013}, where a large-scale continuum map of this complex was obtained \citep{Schmalz2010}, and the recent astrometric studies carried out with Gaia \citep[e.g.,][]{Zari2018, Roccatagliata2020}.

One of the main filaments in Taurus is known as the Lynds Dark Nebula 1495 (L1495). L1495 contains several Barnard Dark Nebulae, which are dust-filled regions. Dark nebulae are extremely dense regions of dust that obscure visible light. The central region is known as B10, with B211 and B213 stretching out from the centre. The L\,1495/B\,213 nebula is a clear example of a star-forming region where the magnetic field lines are perpendicular to the main filament \citep{Soler2019}. It has been extensively studied by \citet{Hacar2013} in high-density tracers, such as C$^{18}$O, N$_2$H$^+$, and SO, and by \citet{Hacar2016} in the three main isotopologues of $^{12}$CO, $^{13}$CO, and C$^{18}$O, deducing the presence of several dense cores embedded in B213 \citep{Benson1989, Onishi2002, Tatematsu2004, Punanova2018}. Some of these dense cores are starless, while others are associated with YSOs of different ages, with a density of stars decreasing from north to south \citep{Davis2010}. This suggests a different dynamical and chemical age along the filament.
Here, we consider the eight (Table \ref{table:Cores_considered}, Fig. \ref{figure:B213_map}) starless cores \#1, \#2, \#5, \#6, \#7, \#10, \#12, and \#16 (core numbers from the catalogue of Hacar et al. 2013).

The Perseus molecular cloud (L\,1448, NGC\,1333, Barnard 5, IC\,348) is one of the most active nearby star-forming regions, extending $\sim$10 pc on the sky (\citep{Bachiller1986}). According to recent Gaia parallaxes and photometric data, it is located at a distance ranging from 234 to 331 pc (\citep{Zucker2020}). The infrared survey of \citep{Ladd1993} suggests that Perseus is intermediate in its star-forming properties between the Taurus complex, with stars predominantly forming in relative isolation (YSO surface density $\sim$10 pc$^{-2}$), and the Orion complex, with large clusters with densities of 100 pc$^{-2}$ as the main stage for the star formation. 

The Perseus complex contains six regions with star-formation activity -L1448, L1455, NGC\,1333, Barnard 1 (B1), IC\,348, and Barnard 5 (B5)- \citep{Bachiller1986, Ladd1994} as well as more than 400 YSOs, and $\sim$100 dense cores \citep{Hatchell2005, Dunham2013}. Most of the protostars in Perseus are associated with IC\,348 and NGC\,1333 \citep{Yang2021}, with the latter having many active outflows that may regulate the ongoing star formation \citep{Bachiller1990, Knee2000, Davis2008, Curtis2010, Plunkett2013}. B1, with a high deuterium enrichment \citep{Marcelino2005}, has been studied in different wavelengths \citep[e.g.,][]{Walawender2005}, revealing the presence of 10 protostars, multiple molecular outflows, and Herbig-Haro objects. All this suggests that the surrounding region of B1 is very actively forming stars. An energetic outflow can also be found in B5 \citep{Langer1996} driving into the surrounding cloud material. B5 presents a morphology, with multiple filaments within the velocity coherent region of the core \citep{Pineda2011}. The starless cores from Perseus considered in this paper are listed in Table \ref{table:Cores_considered} (see also Fig. \ref{figure:Perseus_map}). All the cores were classified as starless cores by \citet{Hatchell2007}, except \#60 (1333-C3-14) that is a young Class 0 object. 

The Orion molecular cloud (Orion A), located at $\sim$428 pc \citep{Zucker2019}, is the nearest region with a presence of recent OB star formation \citep[e.g.,][]{Brown1995, Lombardi2014}. From CO maps \citep{Maddalena1986}, two big molecular regions are distinguished, the north molecular complex (Orion B) and the south molecular complex (Orion A). Orion B is an extensive CO emission region associated with the dark cloud L\,1630 \citep{Lynds1962}. This emission is spread over a region of 4 degrees in the north-south direction from the Horsehead nebula, NGC\,2023, and NGC\,2024 to the reflection nebulae NGC\,2068 and NGC\,2071. On the other hand, Orion A is associated with the dark clouds L\,1640, L\,1641, and L\,1647, with emission extending 6 degrees south from the Orion Nebula. The molecular emission of this southern complex is subdivided into three clouds of neutral material: OMC\,1, OMC\,2, and OMC\,3. OMC\,1, which is located behind the HII M\,42 region, is identified as a dense gas directly associated with Orion KL \citep{Wilson1970}. OMC\,2 is related to the HII M\,43 region \citep{Gatley1974}, and OMC\,3 is located approximately 16$\arcmin$ north of OMC\,2 \citep{Kutner1976}. These clouds structured as an integral-shaped filament of molecular gas \citep{Bally1987} present concentrations of sub-millimetre continuum emission in the southern part of the filament, which are referred to as OMC\,4 \citep{Johnstone1999} and OMC\,5 \citep{Johnstone2006}.

In GEMS, we have considered three cuts, along OMC\,2 (ORI-C3), OMC\,3 (ORI-C1), and
OMC\,4 (ORI-C2) (see Table \ref{table:Cores_considered} and Fig. \ref{figure:Orion_map}). The cuts avoid protostars and stars in these active star-forming regions and represent distinct environments because of their different distances from the Orion nebula. We selected the visual extinction peaks along each cut for this study.


\section{Data analysis and results}
\label{section:results}

Interstellar thioformaldehyde (H$_2$CS) is a slightly asymmetric rotor with two interchangeable hydrogen nuclei; therefore, its rotational levels are grouped into ortho ($K_{\mathrm{a}}$ odd) and para ($K_{\mathrm{a}}$ even), with statistical weights 3:1. The ortho ground state lies 14.9 K above the para ground state, and its dipole moment is µ$_{\mathrm{a}}$=1.647 D \citep{Fabricant1977}. An advantage of the spectrum of a slightly asymmetric rotor molecule is that lines arising between levels of different energies lie nearby in the spectrum.

H$_2$CS was first detected in the observation of the $K$-doublet 2$_{1}$$_{,}$$_{1}$-2$_{1}$$_{,}$$_{2}$ transition at 3 GHz in absorption towards Sgr\,B2 \citep{Sinclair1973}. Subsequently, this molecule has been observed towards several interstellar sources, such as interstellar clouds \citep[e.g.,][]{Cummins1986, Blake1987, Minh1991, Wootten2009}, including cold dark clouds (e.g. \citep{Irvine1989, Vastel2018}), in circumstellar envelopes \citep[e.g.,][]{Agundez2008}, in presence of shocked gas \citep[e.g.,][]{Bachiller1997}, and in regions dominated by UV photons \citep[e.g.,][]{Cuadrado2017, Riviere-Marichalar2019}.

Regarding the deuterated versions of thioformaldehyde, HDCS was first detected in TMC-1 through a spectral survey \citep{Ohishi1998, Kaifu2004}, using the frequencies calculated in the laboratory by \citet{Minowa1997}.  
Twice deuterated thioformaldehyde (D$_2$CS) was observed for the first time in a radio astronomical source (the dark cloud Barnard 1) by \citet{Marcelino2005}.

\subsection{Line profiles}
\label{Line_profiles}

Figures \ref{figure:H2CS_B213}-\ref{figure:D2CS_NGC1333} (see the appendix) show the lines of H$_2$CS, HDCS, and D$_2$CS observed with the IRAM 30m telescope. We have detected two transitions (3$_{1}$$_{,}$$_{3}$-2$_{1}$$_{,}$$_{2}$ and 4$_{1}$$_{,}$$_{4}$-3$_{1}$$_{,}$$_{3}$) of o-H$_2$CS, two transitions (4$_{0}$$_{,}$$_{4}$-3$_{0}$$_{,}$$_{3}$ and 5$_{0}$$_{,}$$_{5}$-4$_{0}$$_{,}$$_{4}$) of p-H$_2$CS, two transitions (3$_{1}$$_{,}$$_{3}$-2$_{1}$$_{,}$$_{2}$ and 3$_{0}$$_{,}$$_{3}$-2$_{0}$$_{,}$$_{2}$) of HDCS, one transition (3$_{0}$$_{,}$$_{3}$-2$_{0}$$_{,}$$_{2}$) of o-D$_2$CS, and one transition (5$_{1}$$_{,}$$_{5}$-4$_{1}$$_{,}$$_{4}$) of p-D$_2$CS (Tables \ref{table:H2CS_parameters_gaussian}-\ref{table:D2CS_parameters_gaussian}). These transitions span an energy range of $E_{\mathrm{up}}$=8.1-24.7 K.

We first fitted the observed lines with Gaussian profiles using the CLASS software to derive the radial velocity ($V_{\mathrm{LSR}}$), the line-width, and the intensity for each line. Results are shown in Tables \ref{table:H2CS_parameters_gaussian}-\ref{table:D2CS_parameters_gaussian}. The contribution to the intensity arises from one narrow velocity component in all the cases. Line widths for H$_{2}$CS, HDCS, and D$_{2}$CS vary between 0.18 and $\sim$1.20 km s$^{-1}$, with the widest line profiles found in the starless core 1333-C4-1 (Perseus). 
All the detected lines present $T_{\mathrm{MB}}$$\geq$35 mK, with H$_2$CS having $T_{\mathrm{MB}}$$<$0.8 K, $T_{\mathrm{MB}}$$<$0.5 K for HDCS, and D$_2$CS with $T_{\mathrm{MB}}$$<$0.3 K.

\subsection{Radiative transfer code}
\label{Radiative transfer code}

In order to derive column densities and abundances, we carried out a more advanced analysis of the emission of H$_2$CS, HDCS, and D$_2$CS using the molecular excitation and radiative transfer code RADEX (van der Tak et al. 2007). RADEX is a one-dimensional non-local thermodynamic equilibrium radiative transfer code, that uses the escape probability formulation assuming an isothermal and homogeneous medium without large-scale velocity fields. 

The collisional rates have been scaled from the (ortho- and para-) H$_2$CO rates derived by \citet{Wiesenfeld2013}, which were calculated including energy levels up to about 180 cm$^{-1}$ for collisions with H$_2$. For deuterated thioformaldehyde, we also used the same collision rates as for H$_2$CS, with a correction for the reduced mass of
HDCS. To obtain the fit that better reproduces the observed line profiles, we let temperature ($T_{\mathrm{k}}$) and gas density ($n_{\mathrm{H_2}}$) vary as free parameters. However, several good fits can be obtained for different combinations of $n_{\mathrm{H_2}}$ and $T_{\mathrm{k}}$. In order to avoid this degeneracy, we let the gas temperature to vary in a small range around the dust temperature\footnote{Obtained by 
\citet{Palmeirim2013}, \citet{Lombardi2014}, and \citet{Zari2016} on the basis of the Herschel Gould Belt Survey \citep{Andre2010} and Planck data \citep{Bernard2010}.}, $T$$_{\mathrm{K}}$ = $T$$_{\mathrm{d}}$$\pm$5 K, and let the density and the H$_2$CS column density to vary to constrain accurately the physical conditions and reproduce the observed intensities. The best fit model is obtained by finding the minimum root mean square (rms) value of log$_{10}$($I_{\mathrm{obs}}$/$I_{\mathrm{mod}}$), following \citet{Neufeld2014}). This is defined as 

\textbf{
\begin{equation}
{\mathrm{rms}}= \displaystyle{ \sqrt { {\frac{1}{n}} \displaystyle\sum_{i=1}^{n} (log_{10} {\frac{ I^{i}_{\mathrm{obs}}}{ I^{i}_{\mathrm{mod}} }}}})^2\,, 
\end{equation}
}

\noindent where $n$ is the number of observed lines, $I$$^{i}_{\mathrm{obs}}$ is the observed line intensity calculated from Gaussian fits (Sect. \ref{Line_profiles} and Tables \ref{table:H2CS_parameters_gaussian}-\ref{table:D2CS_parameters_gaussian}), and $I$$^{i}_{\mathrm{mod}}$ is the model line intensity using RADEX. The values obtained for the H$_2$ volume density of the cores in the best fit models are in the same order of magnitude that the ones obtained by \citet{Rodriguez-Baras2021} using the species CS to derive the densities. In particular, the differences between both cases are within a factor of $\sim$3. Column densities for HDCS and D$_2$CS were derived assuming physical conditions obtained for H$_2$CS in each core. 

Apart from calibration, given that lines are optically thin, one of the main sources of uncertainty to consider is the low angular resolution of the telescope (between $\sim$14$\arcsec$ and $\sim$29$\arcsec$ depending on frequency), which implies that the emission from the inner region of the pre-stellar core is blended with the outer cold envelope. This mainly affects Orion and Perseus cores at low frequencies. Also, the possible volume density gradients along the line of sight may influence the results as well. Taking this into account, the uncertainties in the obtained column densities are estimated to be 20$\%$ for H$_2$CS, and 25$\%$ for HDCS and D$_2$CS (higher uncertainty for the deuterated versions of H$_2$CS because their lines are weaker). The derived column densities for H$_2$CS, HDCS, and D$_2$CS are shown in Table \ref{table:column_densities_Radex}.

We also calculated the fractional abundances, $X$, of H$_2$CS, HDCS, and D$_2$CS with respect to total H nuclei using visual extinction data for the cores \citep{Rodriguez-Baras2021} and the relation between extinction and hydrogen column density in the Galaxy \citep{Guver2009}\footnote{$N$$_{\mathrm{H}}$ (cm$^{-2}$) = (2.21$\pm$0.09)$\times$10$^{21}$ $A$$_{\mathrm{V}}$ (mag)}. 
The visual extinction data were calculated for B\,213 from the $N$$_{\mathrm{H_2}}$ values provided by \citet{Palmeirim2013}. In the Orion case, we used dust opacity maps at 850 $\mu$m \citep{Lombardi2014} and the expression $A$$_{\mathrm{V}}$=$A$$_{\mathrm{K}}$/0.112, where the $K$-band extinction $A$$_{\mathrm{K}}$=2640$\times$$\tau$$_{850}$+0.012 (see \citet{Rodriguez-Baras2021} for more details). In Perseus, we also used the dust opacity at 850$\mu$m ($\tau$$_{850}$) and expression (7) from \citet{Zari2016}. 
We have therefore assumed an uncertainty of 30$\%$ for $X$(H$_2$CS) and of 35$\%$ for $X$(HDCS) and $X$(D$_2$CS) due to the uncertainty introduced by the visual extinction values. Fractional abundance results are listed in Table \ref{table:abundances}.

\subsubsection{Thioformaldehyde}
\label{Thioformaldehyde}

Figures \ref{figure:H2CS_B213}-\ref{figure:H2CS_ORI} (in the appendix) show our best fit models for the detected rotational lines of o-H$_2$CS and p-H$_2$CS in B\,213, L\,1448, NGC\,1333, Barnard\,5, IC\,348, and Orion\,A, respectively.   

For the cores located in the Taurus complex, we find the highest H$_2$CS column density, $N$(H$_{2}$CS)=(7$\pm$1)$\times$10$^{12}$ cm$^{-2}$ (Table \ref{table:column_densities_Radex}), in the core C5. This core can be found in a region located to the north of the filament L\,1495/B\,213 (Fig. \ref{figure:B213_map}), which is characterised by the presence of Class I/Flat objects, but whose stellar population is dominated by more evolved objects. In fact, the region hosting C5 has the largest number of Class III objects \citep[7 out of 25 YSOs;][]{Hacar2013}. The starless core C16, located towards the south-west of B\,213, presents  the second highest H$_2$CS column density (Table \ref{table:column_densities_Radex}). This core is in a region that has recently started to form stars and represents a younger active region compared to C5. Other cores of the sample, such as C1, C2, and C6, which are located north of B\,213, although slightly more south from C5 (see maps shown in Figs. 1 and 2 in Hacar et al. 2013), present H$_2$CS column densities up to three times lower than C5.

Regarding Perseus, the lowest H$_2$CS column densities ($<$2$\times$10$^{12}$ cm$^{-2}$) are found in the cores C5, C6, and C7 of NGC\,1333, and in IC\,348, which is located in the eastern part of the Perseus molecular cloud and associated with a cluster containing a pre-main-sequence star with an age of 0.5-3.5 Myr \citep{Luhman2003}. By contrast, the highest value of $N$(H$_{2}$CS), (11$\pm$2)$\times$10$^{12}$ cm$^{-2}$, is found in L\,1448 situated to the west of the complex.  
The core 79-C1-1 (in Barnard\,5, north-east of Perseus) presents an intermediate H$_2$CS column density (5$\pm$1$\times$10$^{12}$ cm$^{-2}$). 

\begin{table*}
\centering
\caption{Fractional abundances for H$_{2}$CS, HDCS, and D$_{2}$CS.}
\begin{tabular}{l|llll}
\hline 
\hline
Region &  Core & $N$(H$_{2}$CS)/$N$$_{\mathrm{H}}$ & $N$(HDCS)/$N$$_{\mathrm{H}}$ & $N$(D$_{2}$CS)/$N$$_{\mathrm{H}}$  \\ & &  $\times$10$^{-11}$ & $\times$10$^{-11}$ & $\times$10$^{-11}$  \\
\hline
\hline 
& B\,213-C1-1           & 7$\pm$2            & 3.0$\pm$1.0         & 1.8$\pm$0.6      \\ 
& B\,213-C2-1           & 5$\pm$2            & 1.6$\pm$0.6         & 1.3$\pm$0.4      \\ 
& B\,213-C5-1           & 14$\pm$4           & 1.6$\pm$0.6         & 0.7$\pm$0.3        \\
{Taurus} & B\,213-C6-1  & 9$\pm$3            & 2.4$\pm$0.9         & 2.3$\pm$0.8      \\   
& B\,213-C7-1           & 10$\pm$3           & 2.2$\pm$0.8         & 1.2$\pm$0.4      \\  
& B\,213-C10-1          & 4$\pm$1            & 0.5$\pm$0.2         & $<$0.6$\pm$0.2                 \\ 
& B\,213-C12-1          & 1.3$\pm$0.4        & $<$0.4$\pm$0.2      & 0.6$\pm$0.2     \\ 
& B\,213-C16-1          & 10$\pm$3           & 1.4$\pm$0.5         & 1.1$\pm$0.4         \\
\hline
& L\,1448-1             & 18$\pm$5           & 6$\pm$2             & 4$\pm$1        \\ 
& 1333-C3-1             & 4$\pm$1            & 0.4$\pm$0.1         & $<$0.10$\pm$0.04   \\  
& 1333-C4-1             & 13$\pm$4           & 4$\pm$1             & 4$\pm$1        \\ 
& 1333-C5-1             & 1.6$\pm$0.5        & $<$0.12$\pm$0.04    & $<$0.25$\pm$0.09                  \\ 
{Perseus} & 1333-C6-1   & 1.4$\pm$0.4 & 1.1$\pm$0.4                & $<$0.11$\pm$0.04       \\
& 1333-C3-14            & 12$\pm$4           & 1.8$\pm$0.6         & 1.7$\pm$0.6         \\ 
& 1333-C7-1             & 5$\pm$1            & 2.5$\pm$0.9         & 1.0$\pm$0.4      \\  
& 79-C1-1               & 12$\pm$4           & 3$\pm$1             & 1.9$\pm$0.7         \\  
& IC\,348-1             & 2.4$\pm$0.7        & $<$0.12$\pm$0.04    & $<$0.4$\pm$0.1     \\ 
& IC\,348-10            & 2.6$\pm$0.8      & $<$0.10$\pm$0.04    & $<$0.3$\pm$0.1    \\ 
\hline
& ORI-C1-2              & 1.3$\pm$0.4        & $<$0.03$\pm$0.01    & $<$0.02$\pm$0.01   \\
{Orion} & ORI-C2-3      & 3.0$\pm$0.9        & 0.3$\pm$0.1         & $<$0.08$\pm$0.03   \\ 
& ORI-C3-1              & 0.8$\pm$0.3        & $<$0.05$\pm$0.02    & $<$0.04$\pm$0.01    \\ 
\hline
\end{tabular}

\label{table:abundances}
\end{table*}

In Orion A, we obtain $N$(H$_2$CS)=(2.5-5)$\times$10$^{12}$ cm$^{-2}$ for the three cores (Table \ref{table:column_densities_Radex}), which are very similar column densities considering their uncertainty values (Table \ref{table:column_densities_Radex}). Nevertheless, the highest value is found in ORI-C1 \citep[located in OMC-3, one of the most active regions of the sample, with the presence of several 3.6 cm free–free emission sources, nine embedded mid-infrared sources, and a molecular outflow,][]{Johnstone1999, Shimajiri2015}.  

In our analysis, we treated o-H$_2$CS and p-H$_2$CS separately, which allows us to derive the ortho-to-para ratio (OPR) for the different regions. We found that the values of the OPR(H$_2$CS) range from 1.8 to 5.0 with an average value of 2.4$\pm$0.9, without any clear trend with the gas temperature and/or the environment when taking the uncertainties into account.

\subsubsection{Deuterated thioformaldehyde}
\label{Deuterated_thioformaldehyde}

Figures \ref{figure:HDCS_B213}-\ref{figure:D2CS_NGC1333} (in the ppendix) show our best fit models for the detected rotational lines of HDCS, o-D$_2$CS, and p-D$_2$CS in the sample of starless cores. As for H$_2$CS, we observe that, although most of the line profiles are very well reproduced by the model, there are a few cases where the best fits underestimate the observed spectra. The presence of residuals may be mainly due to the noise added to the data, which is especially important in the cases where the line intensities are very weak ($<$0.1 K). This is, for instance, the case for the transition 3$_{1}$$_{,}$$_{3}$-2$_{1}$$_{,}$$_{2}$ of HDCS in B213-C2-1, B213-C5-1, and B213-C7-1. The underestimation of the observed spectra by the fits can be also due to the presence of blended species, or simply due to a limited model, since, depending on the physical structure of each source, emission can arise from different possible source components that are not considered in the model.  

In the Taurus complex, we detected the deuterated species HDCS in all the starless cores of B\,213, except in B\,213-C12 (located in the south part of L\,1495/B\,213, Fig. \ref{figure:B213_map}) for which we provide an upper limit of $N$(HDCS)$<$(2.2$\pm$0.5)$\times$10$^{11}$ cm$^{-2}$. In fact, the lowest column densities of HDCS, (2.1-8)$\times$10$^{11}$ cm$^{-2}$ (Table \ref{table:column_densities_Radex}), are also found in this central part of the B\,213 filament, in particular in B\,213-C10 and B\,213-C16. By contrast, the highest values are found in B\,213-C1, C6, and C7, located all of them in the north of the filament, suggesting a higher deuterium fraction in the north of this complex than in the south region.

Similar results are found for D$_2$CS, with the lowest column density values (3.0$\times$10$^{11}$ cm$^{-2}$) in the south part of the B\,213 filament (core C-12), and the highest ones in cores (C1 and C6) of the north. In particular, we detect o-D$_2$CS in all the cores, except in one (B\,213-C10) for which we provide an upper limit. On the other hand, p-D$_2$CS is only detected in B\,213-C1 with a column density of (4$\pm$1)$\times$10$^{11}$ cm$^{-2}$. For the rest of the cores, we calculated the column density of p-D$_2$CS from o-D$_2$CS and assuming OPR=2.0$\pm$1.0. This OPR(D$_2$CS) is obtained by averaging the OPRs from the cores B\,213-C1-1 and L\,1448-1, since only in these two cores we observe o-D$_2$CS and p-D$_2$CS. If there is no detection of o-D$_2$CS, we just provide an upper limit for p-D$_2$CS (as for B\,213-C10) in Table \ref{table:column_densities_Radex}.

In the Perseus complex, we find the highest HDCS column density, (3.5$\pm$0.9)$\times$10$^{12}$ cm$^{-2}$, in L\,1448 and NGC\,1333-C4 like in the case of H$_2$CS. Non-detection of HDCS nor D$_2$CS is, however, found in IC\,348, which is immersed in a more active environment compared to the one where L\,1448 is located \citep{Knee2000, Plunkett2013}. 
Values for $N$(HDCS) in 79-C1 and in most of the starless cores of NGC\,1333 are found to be lower than in L\,1448, with $N$(HDCS)$<$3$\times$10$^{12}$ cm$^{-2}$. 
Regarding D$_2$CS, this double deuterated species is only detected in its two versions (o-D$_2$CS and p-D$_2$CS) in L\,1448, while in the rest of the cores we provide upper limits or calculate the column density of p-D$_2$CS where o-D$_2$CS has been detected by assuming an OPR=2.0$\pm$1.0 as previously mentioned.

In the three starless cores observed in Orion A, we only detect HDCS, with $N$=(3.7$\pm$0.9)$\times$10$^{11}$ cm$^{-2}$, in ORI-C2, which is located in OMC-4. The OMC-4 region is less luminous and turbulent than OMC-2 and OMC-3 (where the cores ORI-C3 and ORI-C1 are located, respectively), and it does not show evidence of outflow activity \citep{Johnstone1999}. It suggests that OMC-4 might be in a pre-collapse phase of protostar evolution with colder temperatures, explaining thus the presence of deuterated thioformaldehyde. Regarding doubly deuterated thioformaldehyde, we do not detect its presence in any of the sampled starless cores, but we also provide upper limits in Table \ref{table:column_densities_Radex}.

\begin{figure*}[h!]
\centering
\includegraphics[scale=0.39, angle=0]{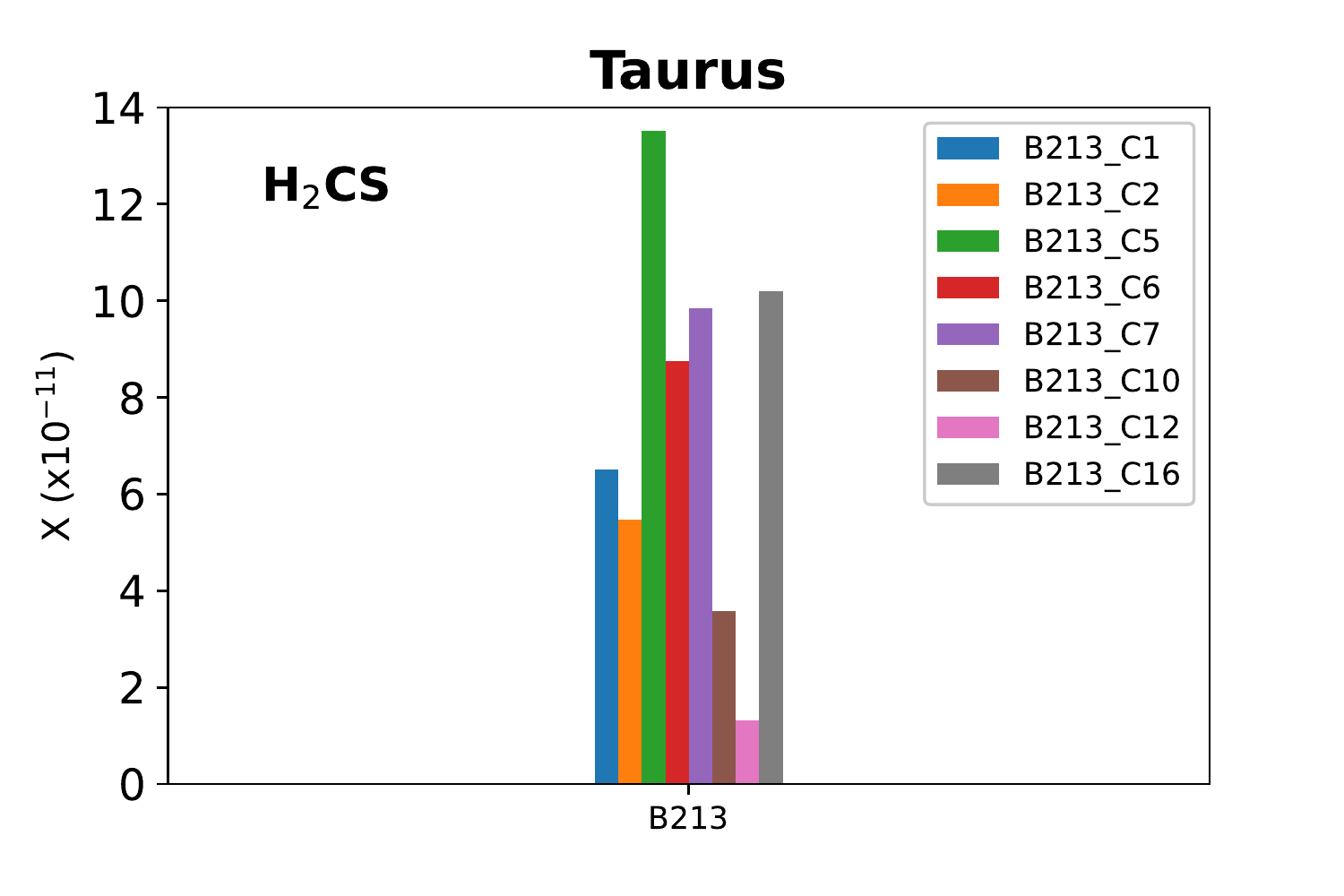}  \hspace{0.0cm}
\includegraphics[scale=0.39, angle=0]{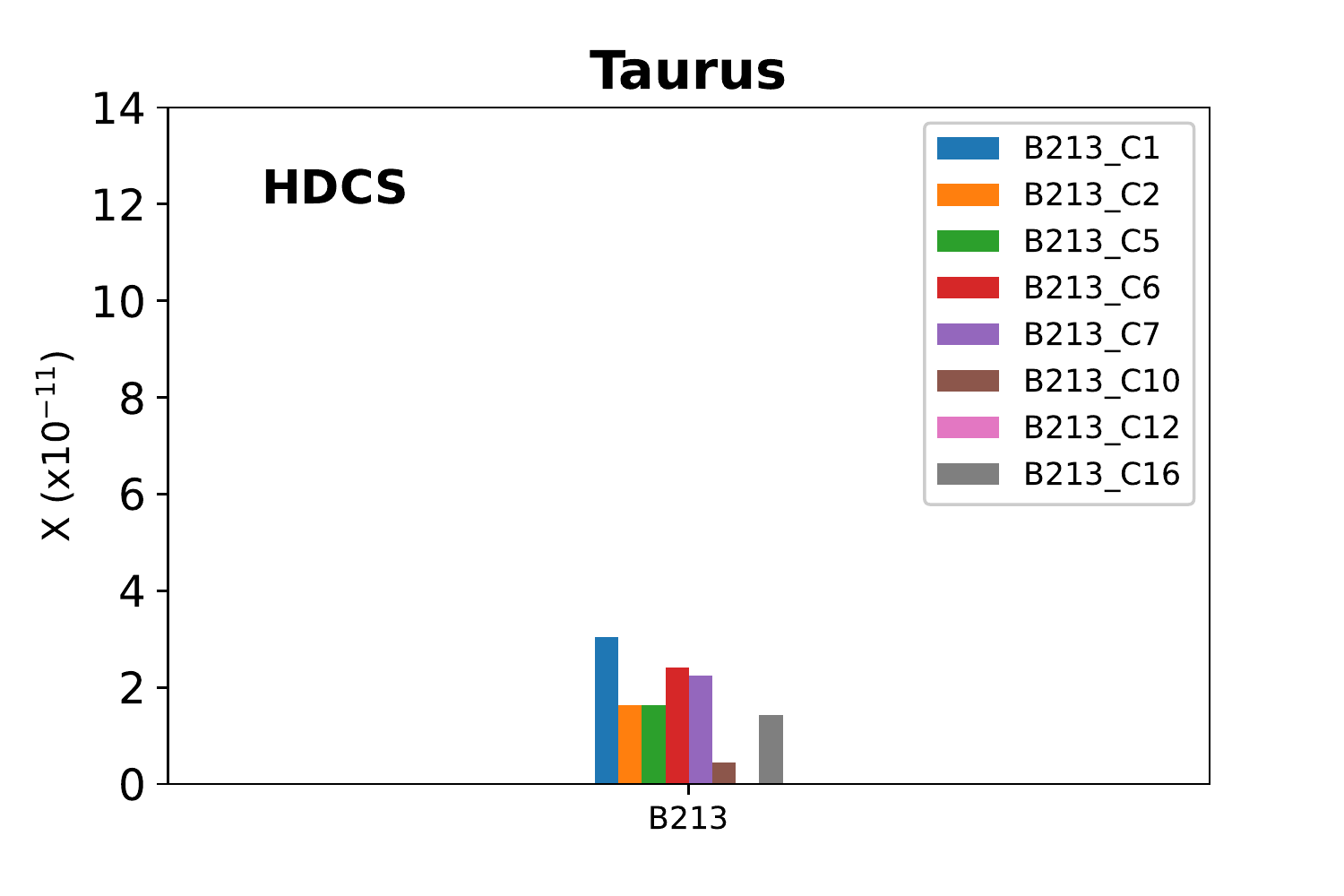}  \hspace{0.0cm}
\includegraphics[scale=0.39, angle=0]{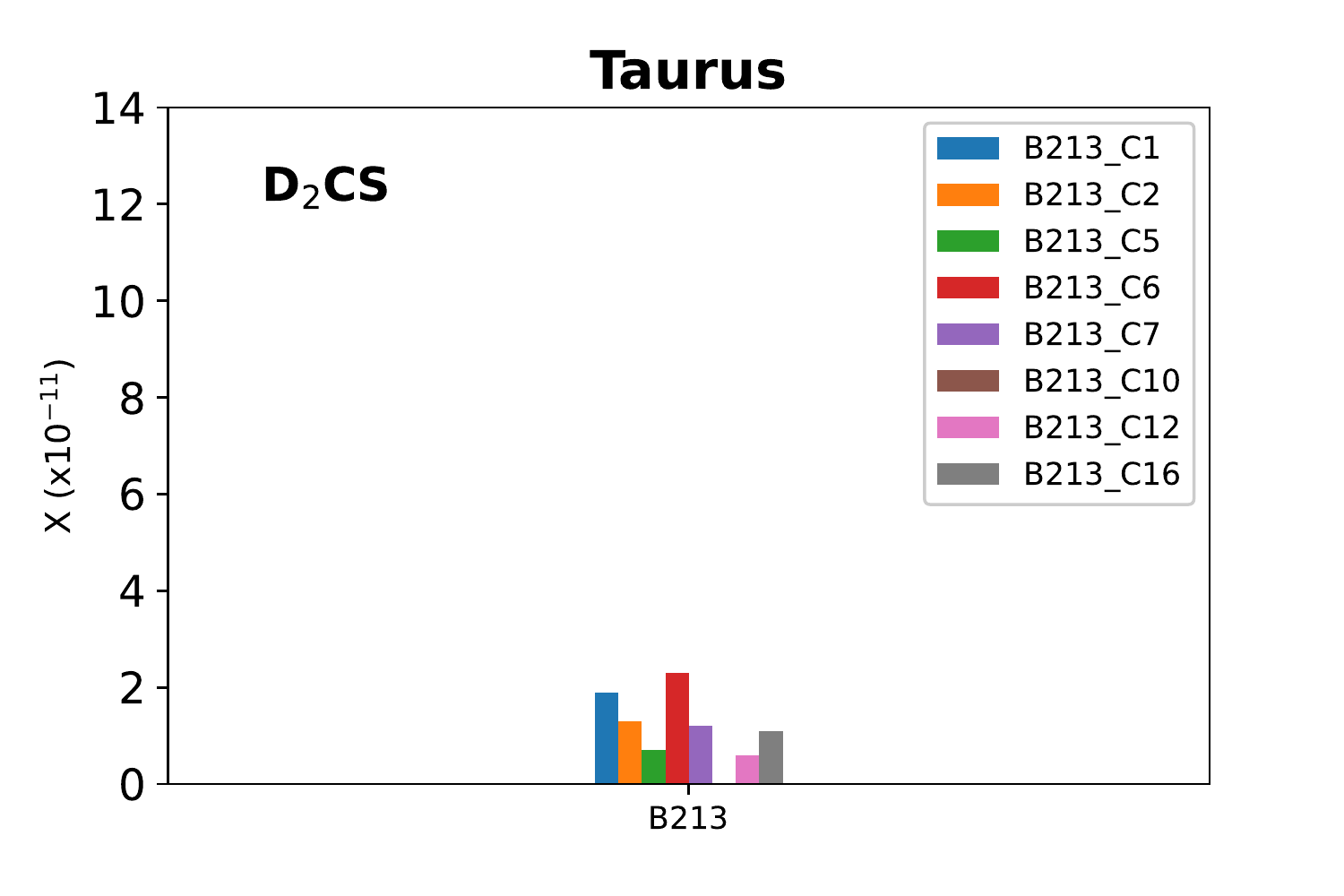}  \hspace{0.0cm}
\includegraphics[scale=0.39, angle=0]{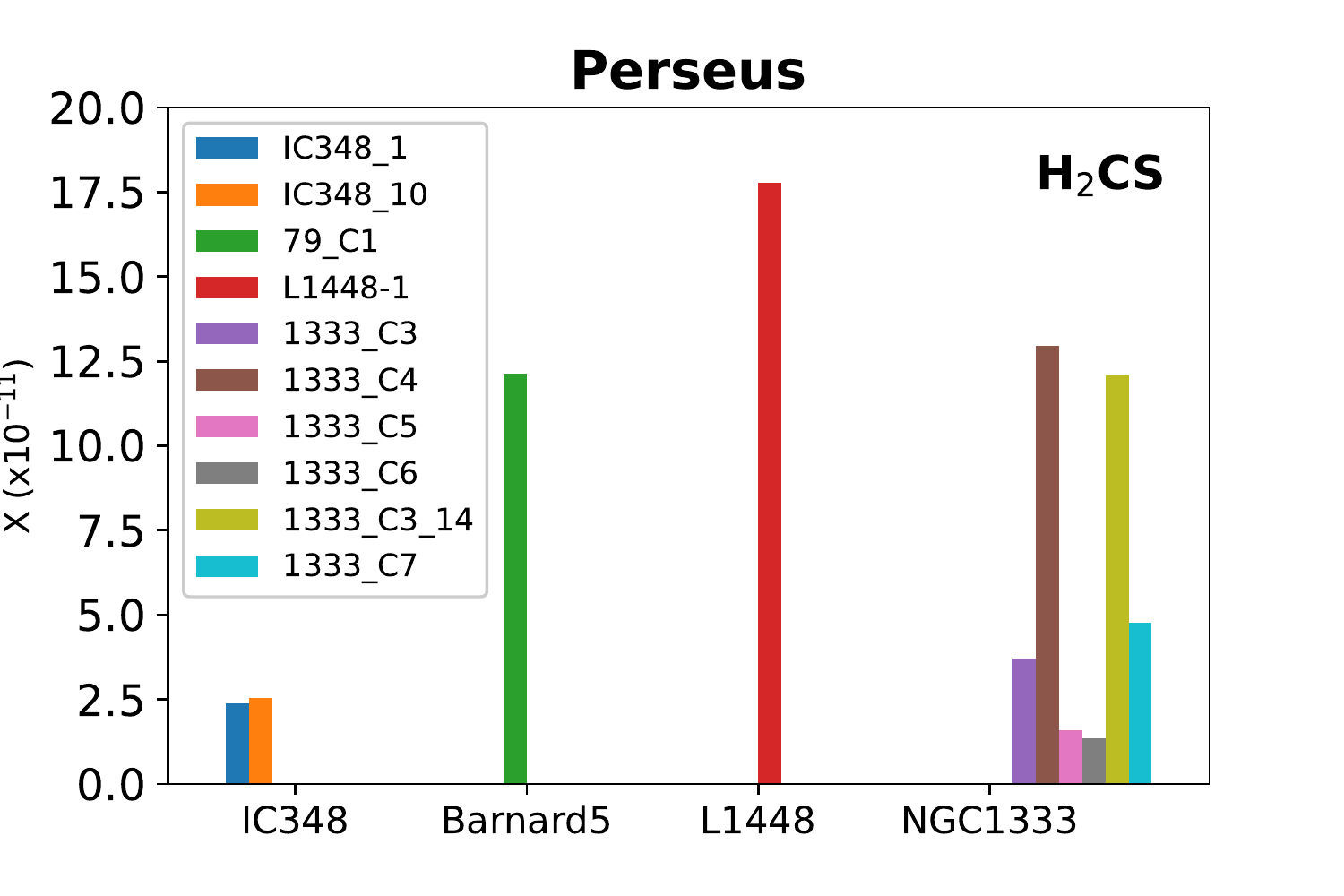}  \hspace{0.00cm}
\includegraphics[scale=0.39, angle=0]{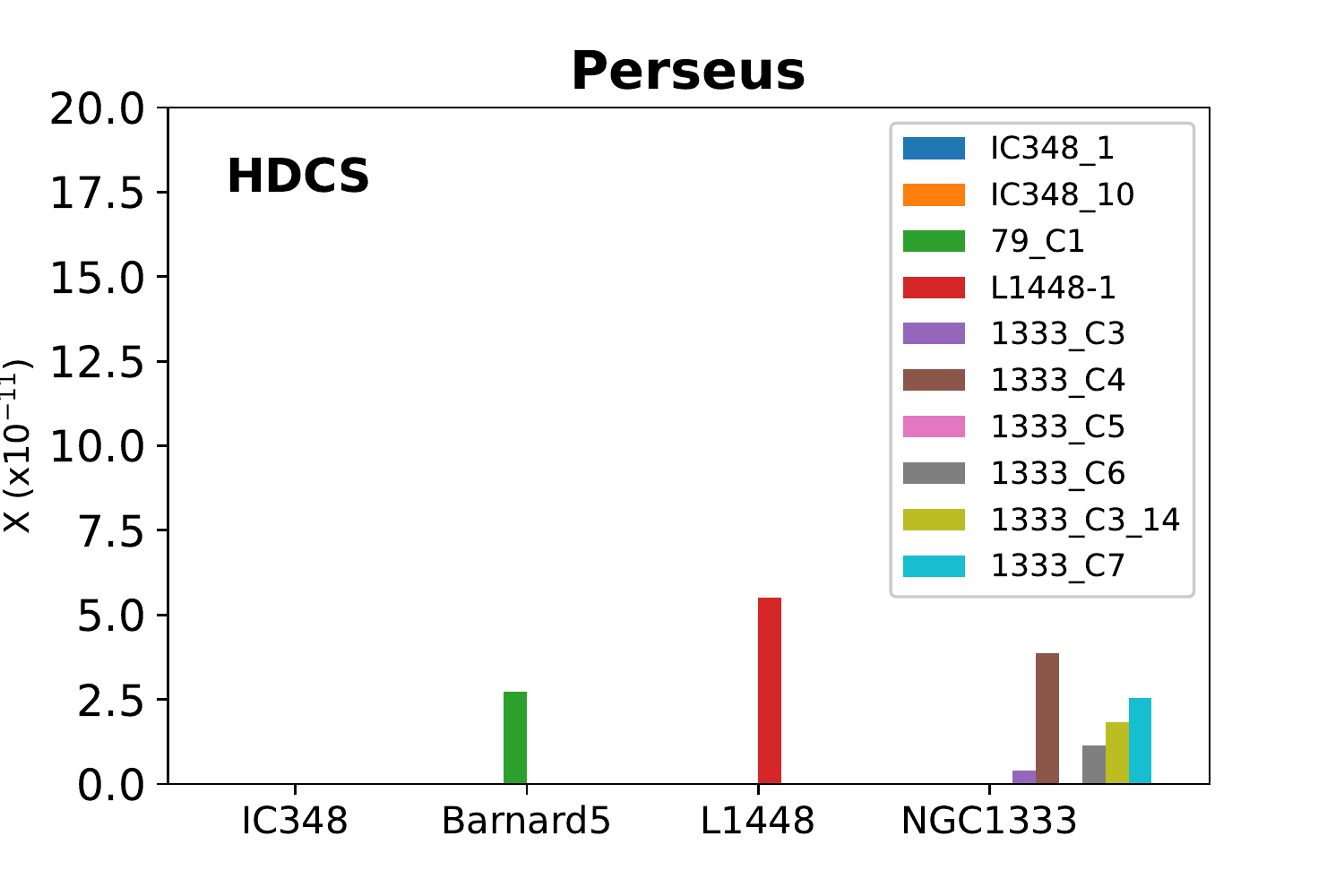}  \hspace{0.00cm}
\includegraphics[scale=0.39, angle=0]{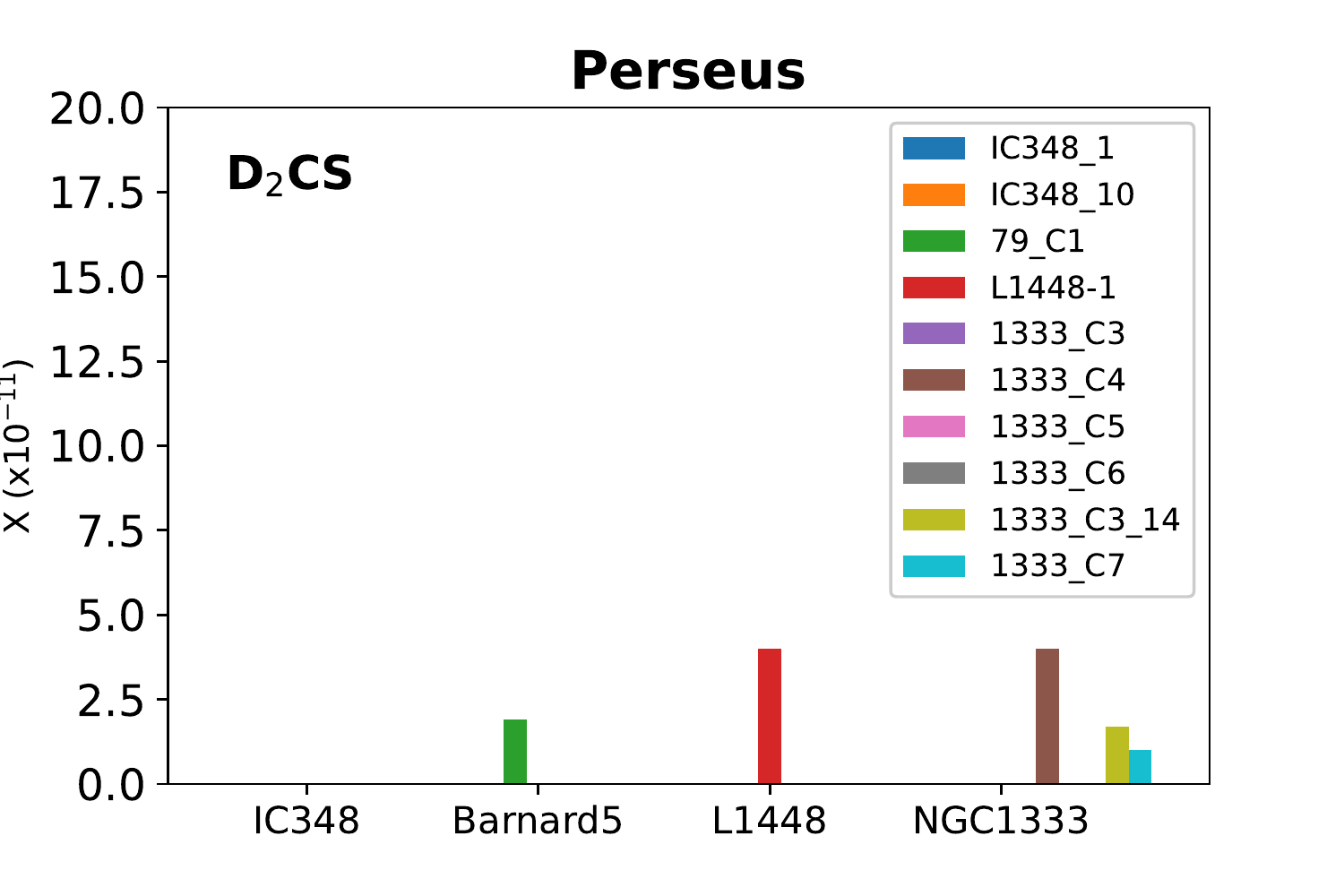}  \hspace{0.00cm}
\vspace{0.cm}

\hspace{-6.3cm}
\includegraphics[scale=0.39, angle=0]{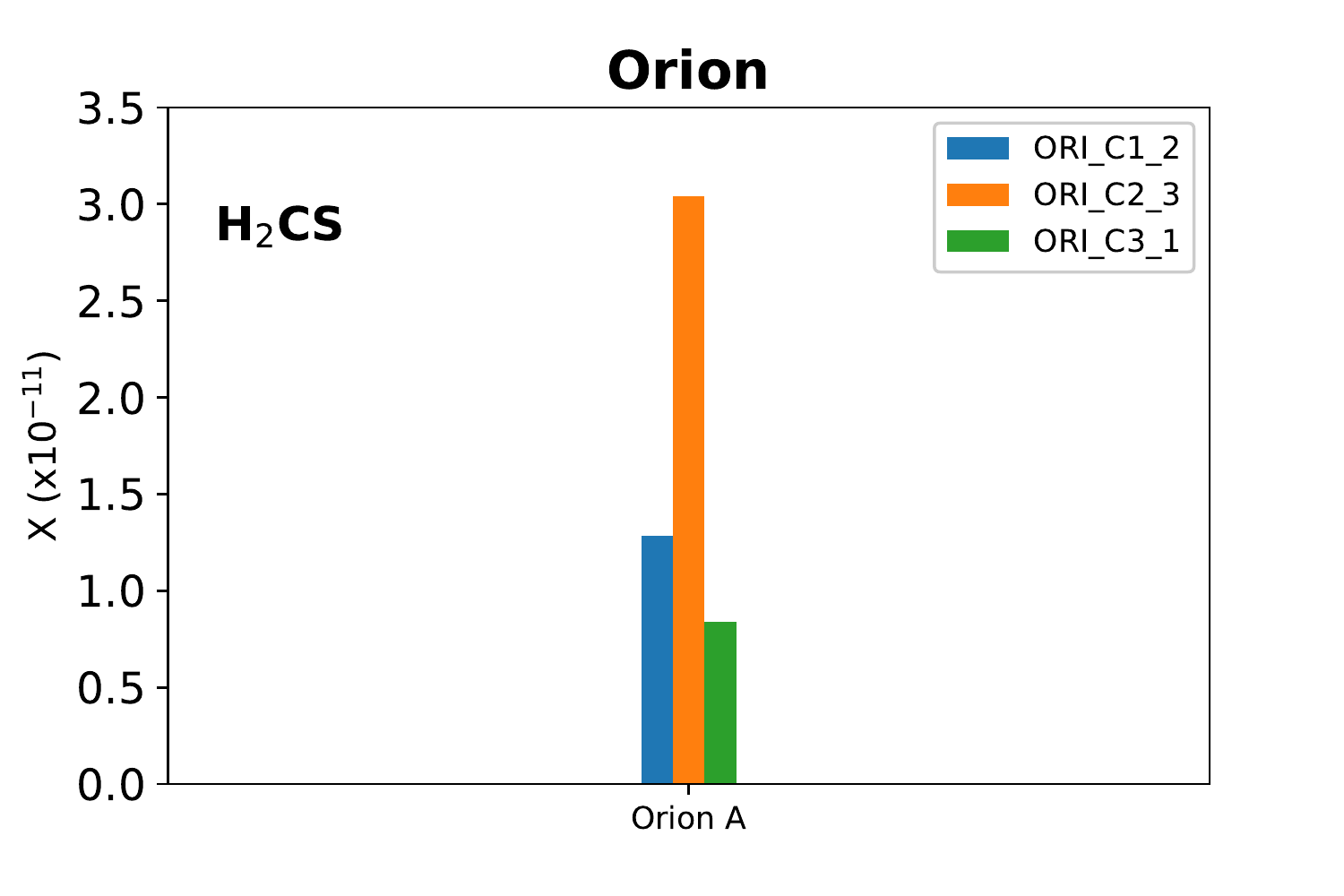}  \hspace{0.00cm}
\includegraphics[scale=0.39, angle=0]{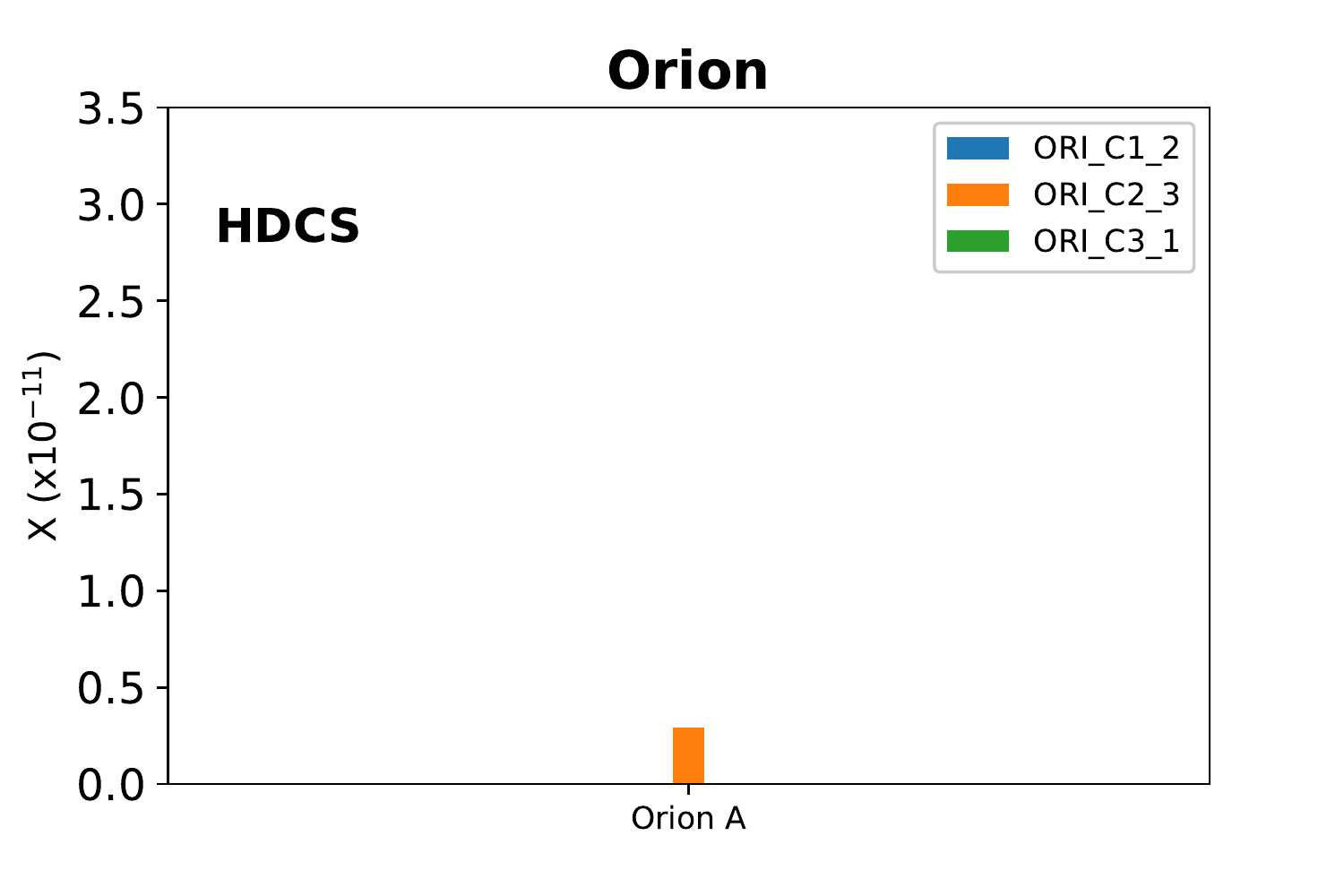}  \hspace{9.00cm}
\\
\caption{Fractional abundances of H$_2$CS, HDCS, and D$_2$CS with respect to total H nuclei for the core sample.}
\label{figure:abundances_histograms}
\end{figure*}

\section{Discussion}
\label{Discussion}

\subsection{Fractional abundances}
\label{fractional_abundances}

Figure \ref{figure:abundances_histograms} shows the H$_2$CS, HDCS, and D$_2$CS fractional abundances, $X$, with respect to total H nuclei in the core sample (see also Table \ref{table:abundances}).

In the Taurus cloud, we find the highest H$_2$CS fractional abundance (1.4$\times$10$^{-10}$) in the starless core B\,213-C5. The rest of cores present fractional abundances up to ten times lower than C5. As previously mentioned, the region hosting C5 is characterised by the presence of several YSOs, some of them being Class III \citep{Hacar2013}. This implies that the core B\,213-C5 may be affected by nearby star formation activity and, therefore, by the presence of winds and outflows that could excavate molecular material, leading to a more efficient penetration of the interstellar radiation field into the molecular cloud. The presence of low-mass stars nearby can also accelerate energetic particles and increase the local CR flux and ionisation rate \citep[see discussion in][]{Spezzano2022}, leading to an increase in the H$_2$CS abundance as also found in our model results (see Fig. \ref{figure:H2CS_evolution_vs_CR}).  
For deuterated thioformaldehyde, we find fractional abundances for HDCS and D$_2$CS in Taurus in the range (0.5-3)$\times$10$^{-11}$ (Table \ref{table:abundances}). In Fig. \ref{figure:abundances_histograms}, we observe that the highest deuterium fractional abundances are found in the cores located in the north of B\,213 (e.g. C1, C6, and C7), while in cores located in the south part of the filament (C10, C12, and C16), $X$(HDCS) and $X$(D$_2$CS) are lower than in the north or simply not detected.  

In Perseus complex, we find $X$(H$_2$CS) in the range (1.4-18)$\times$10$^{-11}$, similar to the range found in Taurus, with the highest value in L\,1448, which hosts a few young stars associated with extremely long outflows \citep[of up to $\sim$240 arcsec;][]{Curtis2010}. By contrast, some of the lowest $X$(H$_2$CS) values are found in the starless cores of IC\,348, a young region where only $<$0.05$\%$ of its mass is contained in outflows \citep{Curtis2010}. 
The highest $X$(HDCS) value in the Perseus sample is also found in L\,1448, while we do not detect it in IC\,348, and the values obtained for most of the cores of NGC\,1333 \citep[which is the most active star-forming region site in the whole Perseus cloud,][]{Pineda2008} are low (0.4-2.5$\times$10$^{-11}$). Regarding D$_2$CS, its obtained fractional abundances are between 1-4$\times$10$^{-11}$, with one of the highest values also found in L\,1448.  

In Orion A, the highest $X$(H$_2$CS) is (3.0$\pm$0.9)$\times$10$^{-11}$, and it is found in ORI-C2-3. This value is more than double those found in the other cores of the Orion sample. The fractional abundance of HDCS in ORI-C2-3 is 0.3$\pm$0.1)$\times$10$^{-11}$, which is ten times lower than $X$(H$_2$CS).

\subsection{Deuterium fractionation} 

Deuterium was formed at the birth of the Universe with an abundance D/H estimated to be 1.6$\times$10$^{-5}$ and is destroyed in the interiors of stars at $T$$\sim$0.5$\times$10$^6$ K \citep{Molinari1986, Tsujimoto2010, Moscoso2021}. The deuterium fractionation ratio, defined as the ratio of the column density of a deuterated molecule to its hydrogen counterpart, is found to increase greatly in some sources. This is, for instance, the case of cold ($T$$_{\mathrm{g}}$$\sim$10 K) dark clouds, where the increase is a few orders of magnitude compared to the cosmic D/H ratio \citep[e.g.][]{Roueff2003, Ceccarelli2007, Herbst2009}.

\begin{table}
\centering
\caption{Deuterium fractionation ratios.}
\begin{center}
\begin{tabular}{l|lll}
\hline 
\hline
Region &  Core & [HDCS]/[H$_2$CS] & [D$_2$CS]/[H$_2$CS]  \\
\hline
\hline 
& B\,213-C1-1           & 0.5$\pm$0.2        & 0.3$\pm$0.1              \\ 
& B\,213-C2-1           & 0.3$\pm$0.1        & 0.2$\pm$0.1               \\ 
& B\,213-C5-1           & 0.12$\pm$0.05      & 0.05$\pm$0.02                    \\
{Taurus} & B\,213-C6-1  & 0.3$\pm$0.1        & 0.3$\pm$0.1               \\   
& B\,213-C7-1           & 0.2$\pm$0.1        & 0.11$\pm$0.05               \\  
& B\,213-C10-1          & 0.13$\pm$0.06      & $<$0.16$\pm$0.07                   \\ 
& B\,213-C12-1          & $<$0.3$\pm$0.2     & 0.5$\pm$0.2           \\ 
& B\,213-C16-1          & 0.14$\pm$0.06      & 0.11$\pm$0.05                  \\
\hline
& L\,1448-1             & 0.3$\pm$0.1        & 0.2$\pm$0.1                     \\ 
& 1333-C3-1             & 0.10$\pm$0.05      & $<$0.03$\pm$0.01            \\  
& 1333-C4-1             & 0.3$\pm$0.1        & 0.3$\pm$0.1                      \\ 
& 1333-C5-1             & $<$0.08$\pm$0.04   & $<$0.15$\pm$0.07                  \\ 
{Perseus} & 1333-C6-1   & 0.8$\pm$0.4        & $<$0.08$\pm$0.04               \\
& 1333-C3-14            & 0.15$\pm$0.07      & 0.14$\pm$0.06                 \\ 
& 1333-C7-1             & 0.5$\pm$0.2        & 0.2$\pm$0.1              \\  
& 79-C1-1               & 0.2$\pm$0.1        & 0.16$\pm$0.07                    \\  
& IC\,348-1             & $<$0.05$\pm$0.02   & $<$0.18$\pm$0.08       \\ 
& IC\,348-10            & $<$0.04$\pm$0.02   & $<$0.12$\pm$0.05        \\ 
\hline
& ORI-C1-2              & $<$0.02$\pm$0.01   & $<$0.013$\pm$0.006       \\
{Orion} & ORI-C2-3      & 0.10$\pm$0.04      & $<$0.03$\pm$0.01             \\ 
& ORI-C3-1              & $<$0.07$\pm$0.03   & $<$0.05$\pm$0.02        \\ 
\hline
\end{tabular}
\end{center}
\label{table:deuterated_ratios}
\end{table}

\begin{figure*}
\centering
\includegraphics[scale=0.4, angle=0]{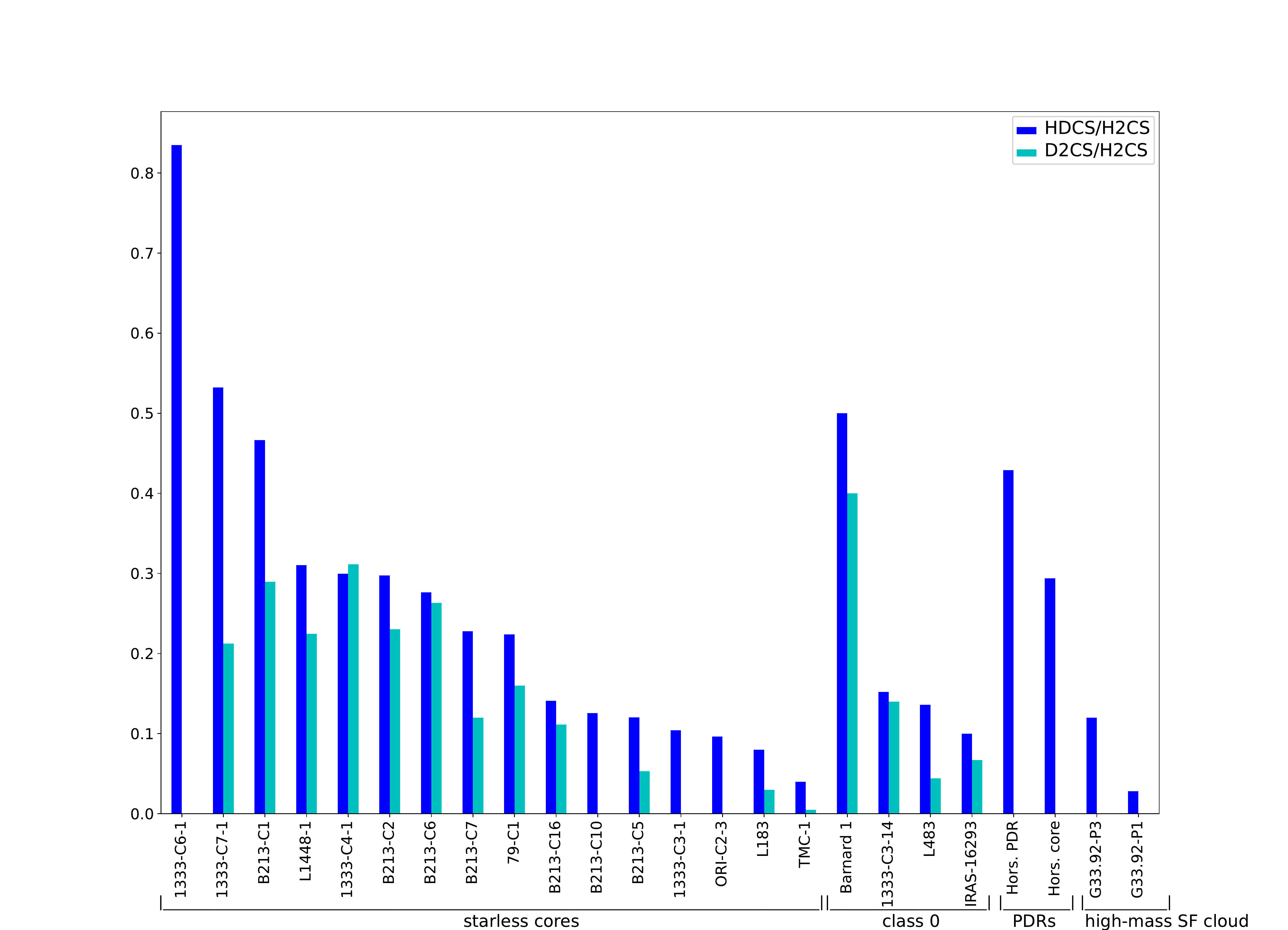}  
\\
\caption{Deuterium fractionation ratios in different interstellar sources. References are
 \citep{Minowa1997} for TMC-1, \citep{Irvine1989} for L\,183, \citep{Marcelino2005} for Barnard 1, \citep{Agundez2019} for L\,483, \citep{Drozdovskaya2018} for IRAS 16293-2422, \citep{Riviere-Marichalar2019} for the Horeshead PDR, and \citep{Minh2018} for G33.92+0.11.}
\label{figure:deuterated_ratios}
\end{figure*}

Table \ref{table:deuterated_ratios} shows the deuterium fractionation ratios obtained in our starless core sample. Figure \ref{figure:deuterated_ratios} shows a comparison between these results and the ones obtained for other starless cores from the literature, as well as a comparison with other type of interstellar sources, such as Class 0 objects and photodissociation regions (PDRs). The HDCS/H$_2$CS ratios obtained in our sample of starless cores are in the range $\sim$0.1-0.8, with Orion showing the lowest value (HDCS/H$_2$CS=0.10$\pm$0.04) compared to those obtained in Taurus and Perseus. The obtained HDCS/H$_2$CS ratios are up to $\sim$4 orders of magnitude higher than the cosmic D/H ratio, but they are similar to those found in other starless cores (see Fig. \ref{figure:deuterated_ratios}), such as TMC-1 and L\,183 \citep[e.g.][]{Irvine1989, Minowa1997, Marcelino2005}, where the low ($\sim$7-13 K) temperatures of these regions, as well as the large fraction of neutral heavy species freeze-out onto dust grains, favour the formation and enhancement of deuterated species \citep{Millar1989, Ceccarelli2014}. In particular, the deuterium enrichment in cold regions occurs through the reaction between the molecular ion H$^{+}_{3}$ with HD to form H$_2$D$^+$, which reacts with other molecules and atoms transferring deuterium to other species \citep[e.g.][]{Howe1993}. Nevertheless, in Fig. \ref{figure:deuterated_ratios}, we also observe that similar HDCS/H$_2$CS ratios ($\sim$0.1-0.5) are found as well in interstellar regions characterised by higher temperatures than those found in starless cores, such as Class 0 protostar \citep[e.g. L\,483 and IRAS\,16293-2422:][]{Agundez2019, Drozdovskaya2018}, high-mass star-forming clouds \citep[e.g. G33.92+0.11:][]{Minh2018}, and PDRs \citep[e.g. Horsehead PDR:][]{Riviere-Marichalar2019}. A rich deuterium chemistry is observed in PDRs, such as the Orion Bar \citep{Parise2009} and Monoceros R2 \citep{Trevino-Morales2014}. In the regions characterised by warm gas temperatures (30$\lesssim$$T$$\lesssim$100 K), the deuterium fractionation is driven through a chemistry that is different to that from cold regions. In particular, the D atoms are transferred to molecules by CH$_2$D$^+$, whose activation barrier when reacting with H$_2$ is larger than that of H$_2$D$^+$ \citep{Roberts2000, Roueff2007b}. In  the  case  of  single-dish  observations of dense cores around Class 0 protostars, the origin of the deuterated compounds emission could be  in  the  cold  and dense region of the core that still remains unaffected by the heating produced by the recently born star \citep{Agundez2019}. In hot corinos, with temperatures $T$$>$100 K, the ice mantles evaporate releasing deuterated compounds formed in the gas and/or the ice during the cold phase \citep{Drozdovskaya2018}.

The derived fractionation ratio for the double deuterated thioformaldehyde, D$_2$CS/H$_2$CS, is in the range $\sim$0.05-0.4 (Table \ref{table:deuterated_ratios} and Fig. \ref{figure:deuterated_ratios}) for the starless cores in Taurus and Perseus, while only upper limits are reported for those in Orion, where D$_2$CS has not been detected. For the cases where D$_2$CS is detected, the double deuterated ratio is on average a factor $\sim$1.9 lower than the single deuterated ratio (HDCS/H$_2$CS). This is a similar range to that found for Class 0 objects, such as Barnard 1 and L\,483, where D$_2$CS/H$_2$CS is $\sim$1.3 and $\sim$3 times lower than HDCS/H$_2$CS, respectively \citep{Marcelino2005, Agundez2019}. For the case of Orion where only upper limits are provided (i.e. cores ORI-C1-2 and ORI-C3-1), we obtain HDCS/H$_2$CS values of $<$0.02 and $<$0.07, respectively, which are about $\sim$10 times lower than those found in Taurus and Perseus. In the case of D$_2$CS/H$_2$CS, we also observe this difference between the Orion cores and the cores in Taurus and Perseus. Only in 1333-C3-1 and B\,213-C5-1 (affected by the presence of nearby star formation activity as ORI-C3-1; see Sects. \ref{Thioformaldehyde} and \ref{fractional_abundances}), we derive a deuterium fractionation ratio similar to the ones obtained in Orion. In regions characterised by more extreme environments and higher temperatures, such as PDRs, D$_2$CS has not been detected either.

Our results, therefore, make it evident that the deuterium fractionation is higher in the pre-stellar cores of Taurus and Perseus than those of Orion. Nevertheless, we should also take the different distances of the studied regions (Taurus, Perseus, and Orion) into account. In particular, typical pre-stellar core sizes are of about $\sim$10,000 AU \citep[][and references therein]{Ceccarelli2014}, which it is equivalent to $\sim$0.05 pc or $\sim$24$\arcsec$ at a distance of $d$$\sim$428 pc (average distance to Orion). In the case of Perseus ($d$$\sim$300 pc) and Taurus ($d$$\sim$145 pc), the angular sizes would be $\sim$30$\arcsec$ and $\sim$71$\arcsec$, respectively. These angular sizes are significantly larger than the telescope beam (up to $\sim$29”). However, in the Orion case, the angular size is smaller than the beam for the lowest frequencies. This implies that in the nearest regions (in particular in Taurus), the observed emission may be limited to the high-density central region of the cores, while, in the Orion case, we are not able to distinguish the emission coming from the densest zone from that coming from the envelope. This could cause the observed trend in the deuterium fractionation, which shows the lowest values in the Orion cores and the highest values in Taurus. In particular, the D/H values obtained in Orion are a factor of $\sim$10 lower than those obtained in Taurus. This suggests that the D/H ratio in the envelope of the Orion cores should be (much) lower than a factor of 10 with respect to the D/H in the high-density central regions. We would need observations at higher frequencies or interferometry to resolve the emission from the pre-stellar cores in Orion.

\subsection{Deuterium fractionation evolution}
\label{Deuterium fractionation evolution}

Apart from temperature, deuterium fractionation is also thought to vary along the dynamical evolutionary stages of starless cores. In particular, when starless cores evolve towards star formation, the deuterium fraction of molecules formed in the gas phase increases, reaching their maximum at the onset of star formation \citep{Crapsi2005, Hirota2006, Emprechtinger2009, Feng2019}, while after the star birth, the deuterium fraction decreases \citep{Fontani2011, Sakai2012, Gerner2015}. Deuterium fractionation ratios can therefore be used as evolutionary tracers.

\begin{table}
\caption{Abundances with respect to total hydrogen nuclei (the initial H$_2$ OPR is 3).
}             
\centering 
\begin{tabular}{l l l l l}     
\hline\hline       
Species &  Abundance & \vline  & Species & Abundance                       \\ 
\hline 
HD      & 1.6$\times$10$^{-5}$    & \vline   &  S$^+$   & 1.5$\times$10$^{-5}$/$\times$10$^{-6}$/$\times$10$^{-7}$       \\
He      & 9.0$\times$10$^{-2}$    & \vline   &  Na$^+$  & 2.0$\times$10$^{-9}$       \\
N       & 6.2$\times$10$^{-5}$    & \vline   &  Mg$^+$  & 7.0$\times$10$^{-9}$       \\
O       & 2.4$\times$10$^{-4}$    & \vline   &  P$^+$   & 2.0$\times$10$^{-10}$      \\ 
C$^+$   & 1.7$\times$10$^{-4}$    & \vline   &  Cl$^+$  & 1.0$\times$10$^{-9}$        \\
Si$^+$  & 8.0$\times$10$^{-9}$    & \vline   &  F$^+$   & 6.7$\times$10$^{-9}$       \\
Fe$^+$  & 3.0$\times$10$^{-9}$    & \vline   &          &   \\
\hline 
\label{table:abundances_Nautilus}                 
\end{tabular}
\end{table}

In order to study the evolution of the HDCS/H$_2$CS and D$_2$CS/H$_2$CS ratios in starless cores, we used the Nautilus gas-grain chemical code \citep{Ruaud2016}. Nautilus is a three-phase model in which gas, grain surface, grain mantle phases, and their interactions are considered. Nautilus solves the kinetic equations for both the gas phase and the surface of interstellar dust grains and computes the evolution with time of chemical abundances for a given physical structure. We use the chemical network presented by \citet{Majumdar2017}, which considers multiple deuterated molecules and includes the spin chemistry in Nautilus. Moreover, we used this state-of-the-art chemical model to explore the influence that poorly known parameters, such as the sulphur elemental abundance and the CR ionisation rate for H$_2$ molecules, would have on the H$_2$CS deuterium fractionation. 

\begin{figure*}
\centering
\includegraphics[scale=0.38, angle=0]{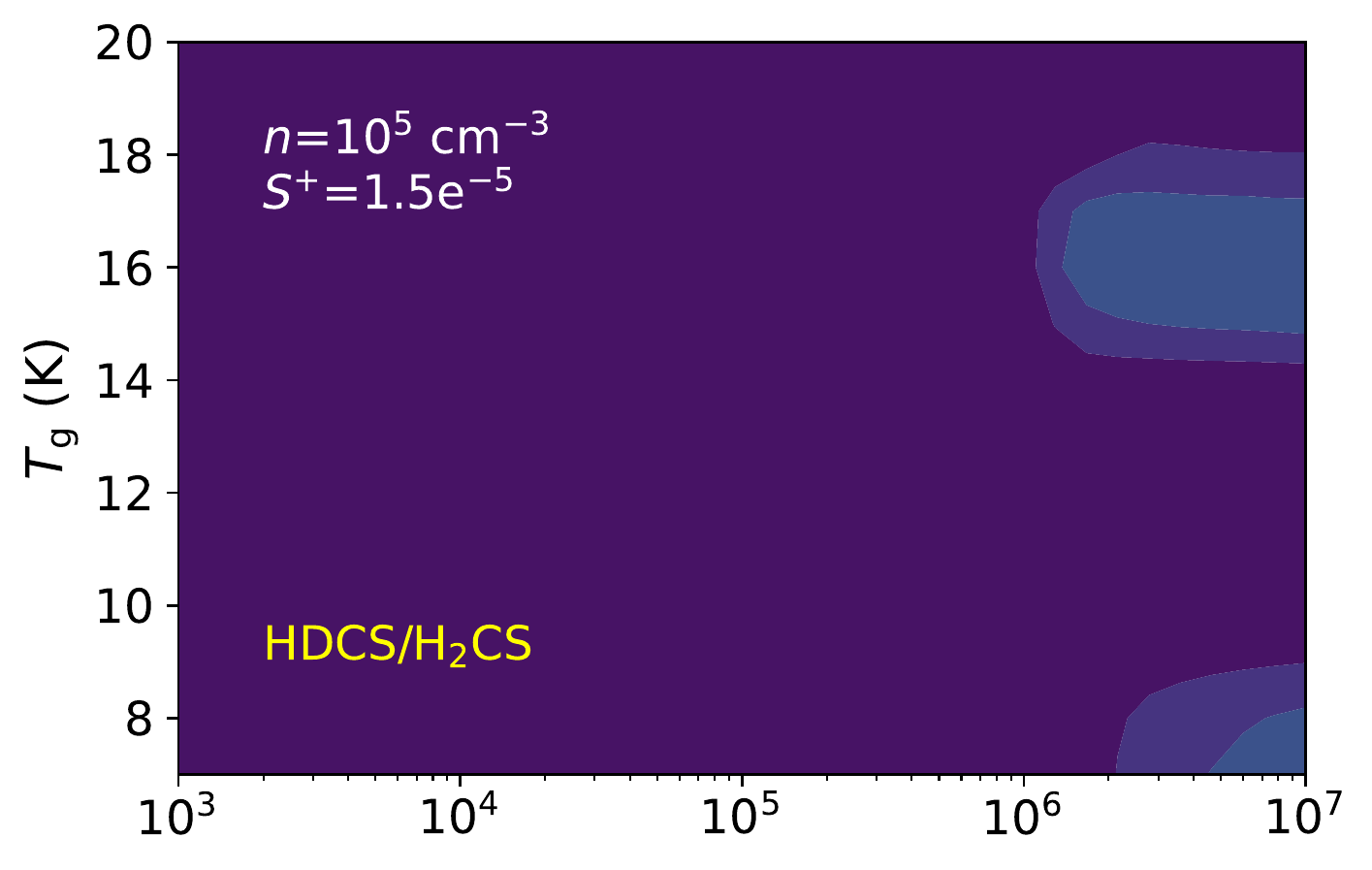}  
\hspace{-0.2cm}
\includegraphics[scale=0.38, angle=0]{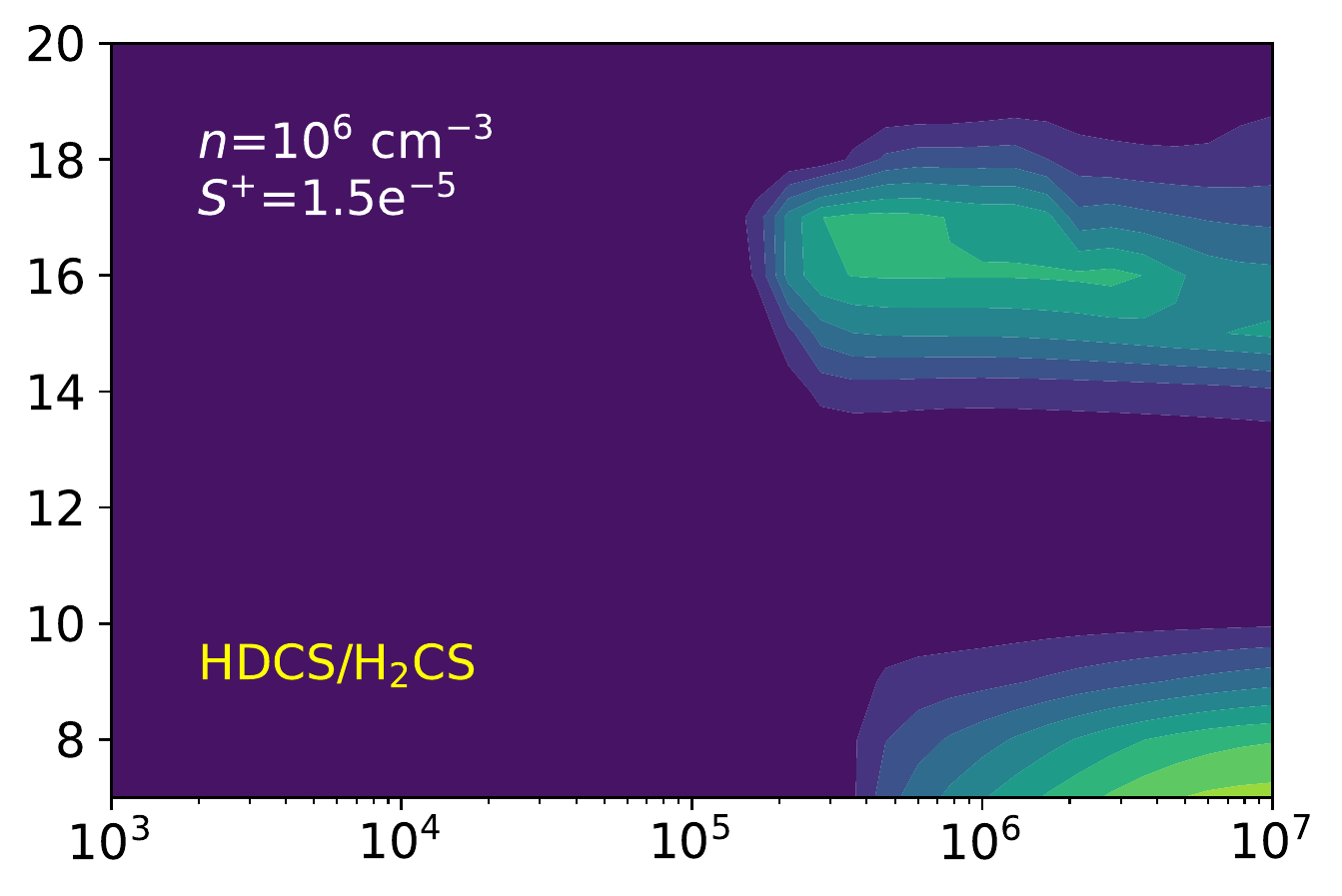} 
\hspace{-0.2cm}
\includegraphics[scale=0.38, angle=0]{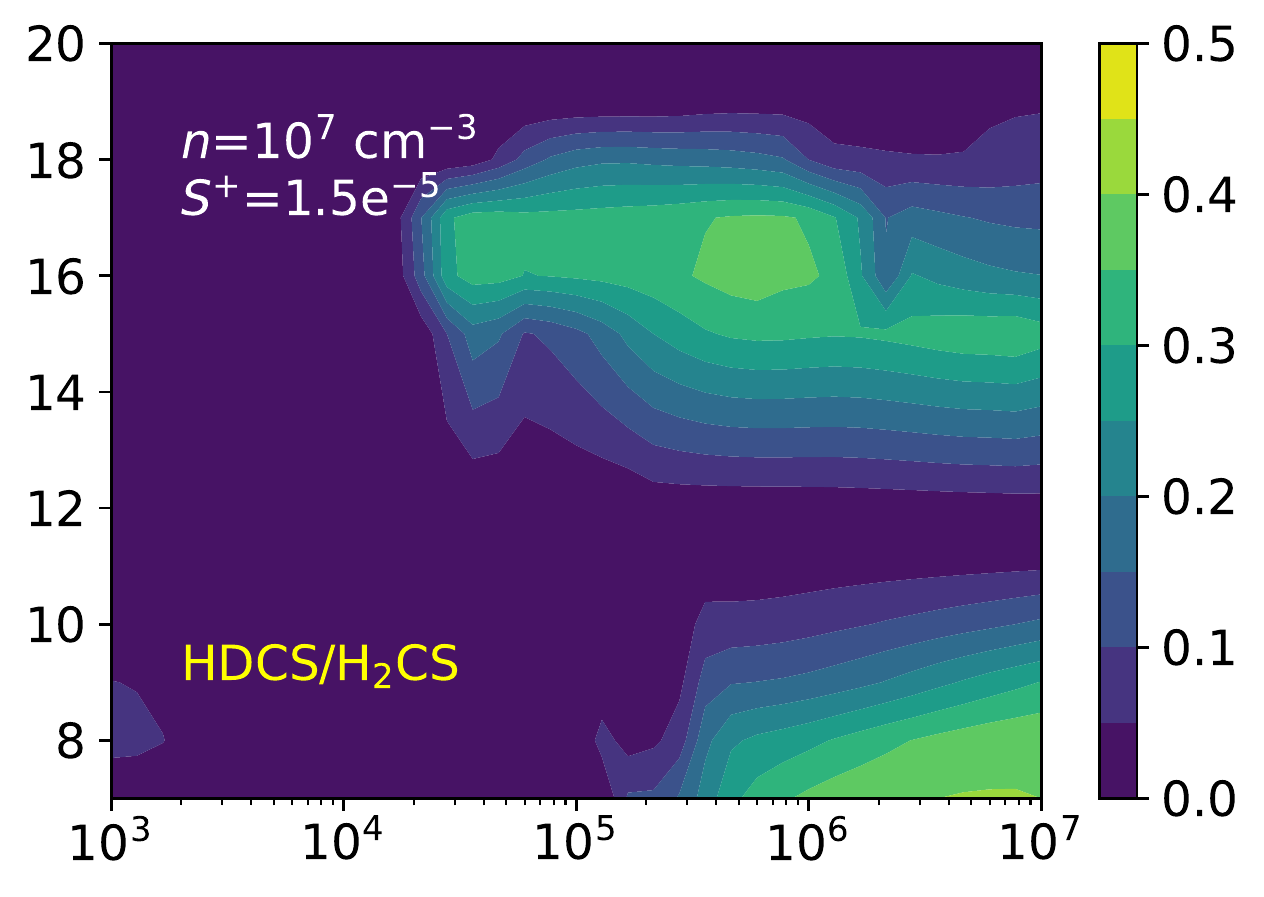}  

\includegraphics[scale=0.38, angle=0]{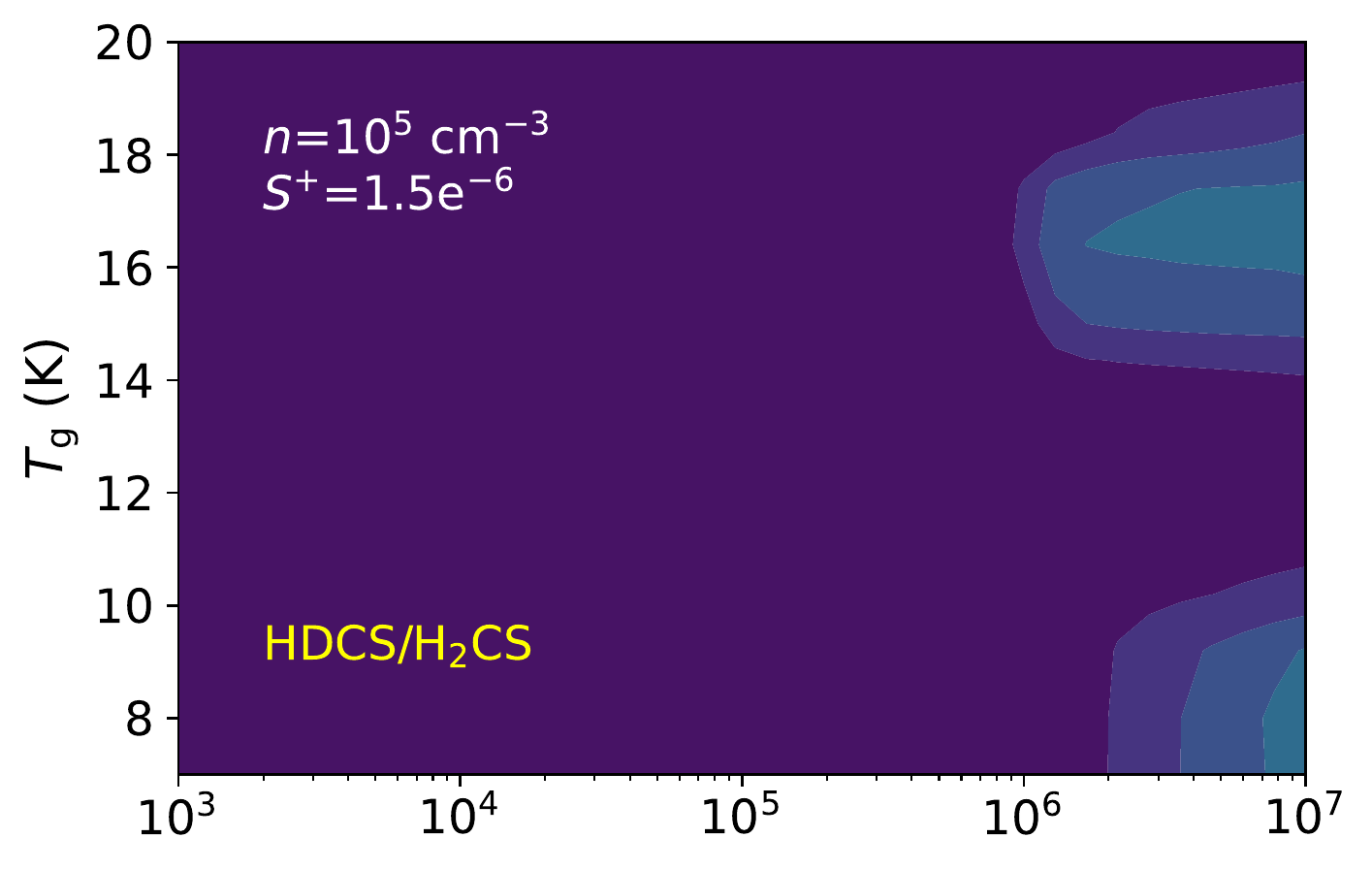}  
\hspace{-0.2cm}
\includegraphics[scale=0.38, angle=0]{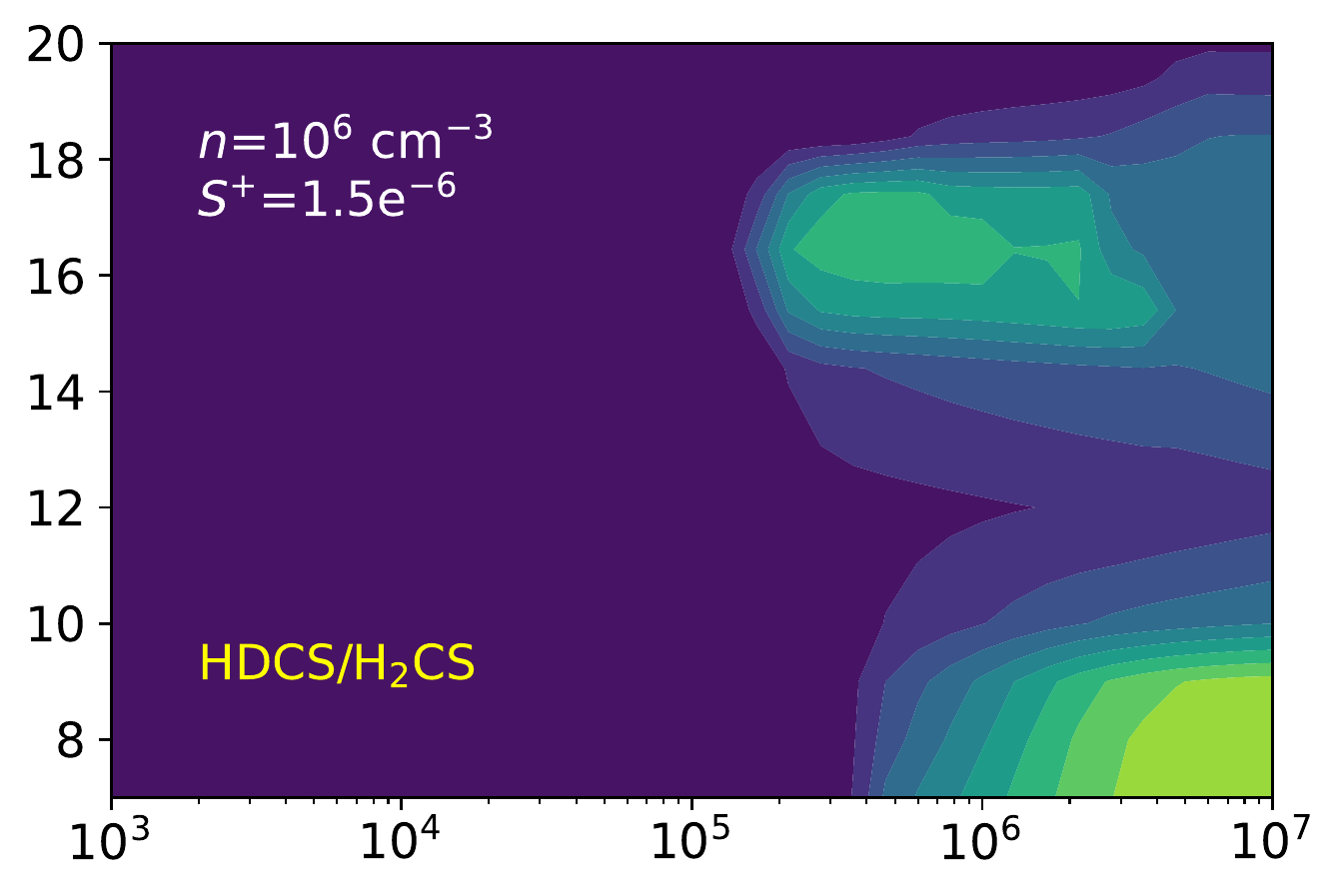}  
\hspace{-0.2cm}
\includegraphics[scale=0.38, angle=0]{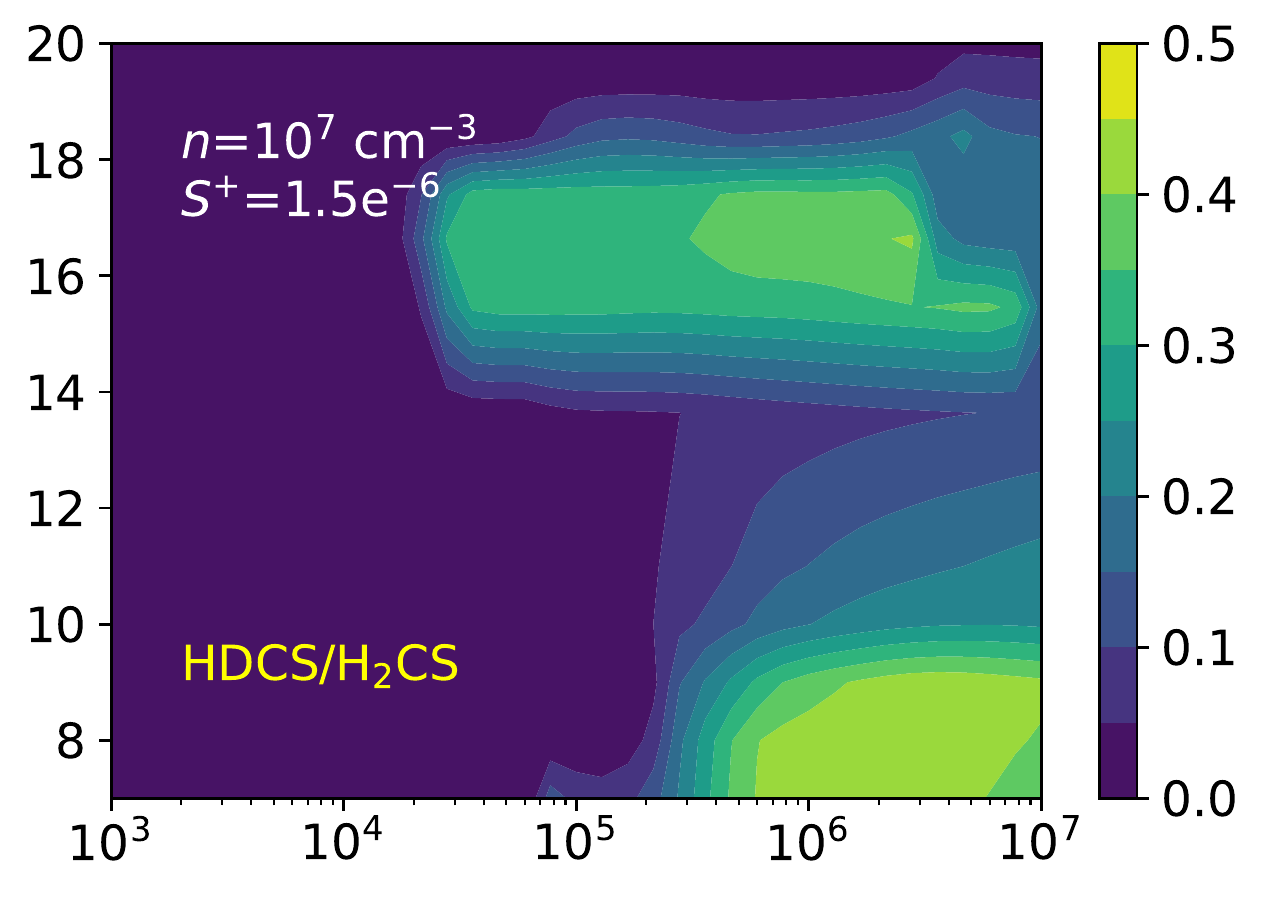}  

\includegraphics[scale=0.38, angle=0]{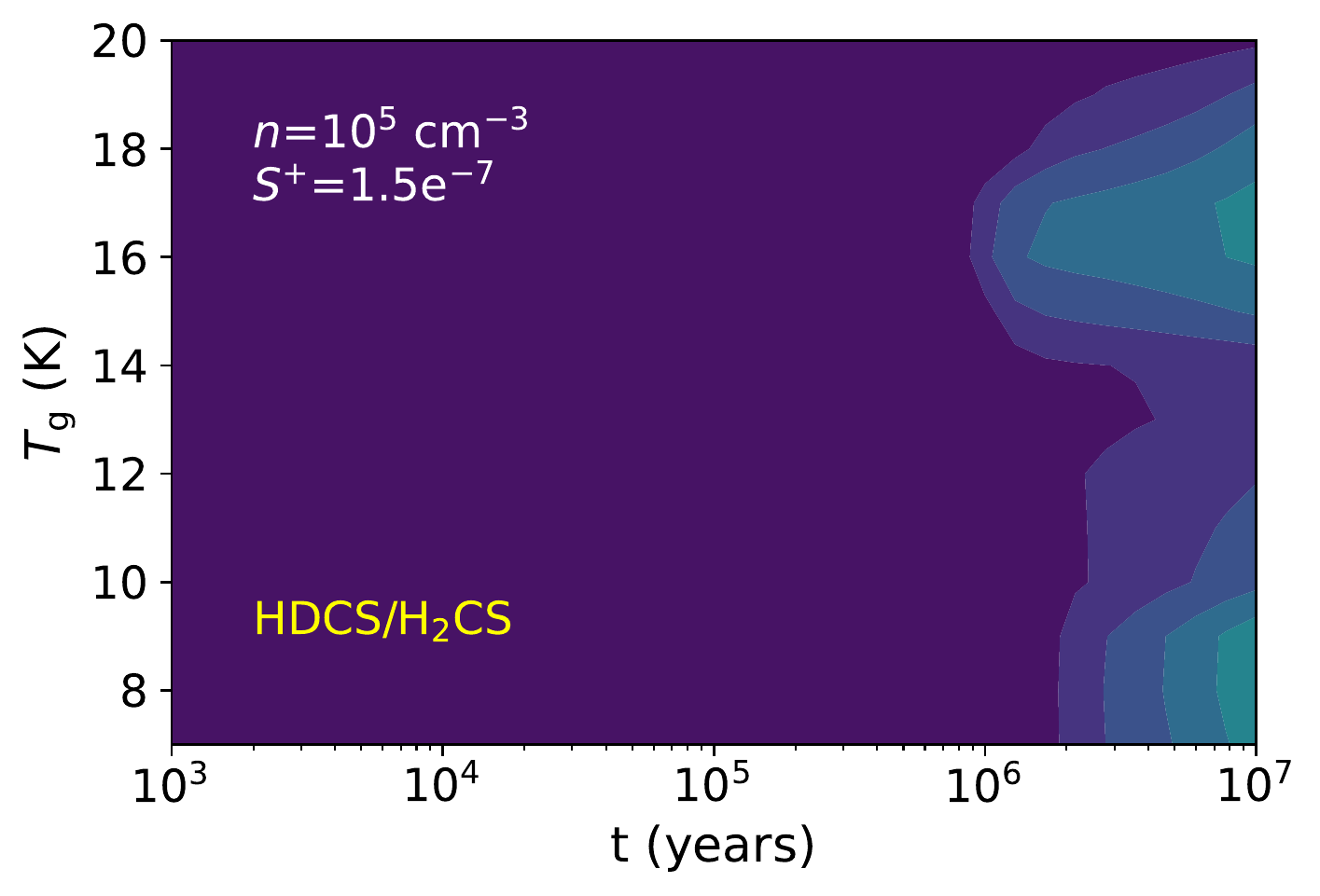}  
\hspace{-0.2cm}
\includegraphics[scale=0.38, angle=0]{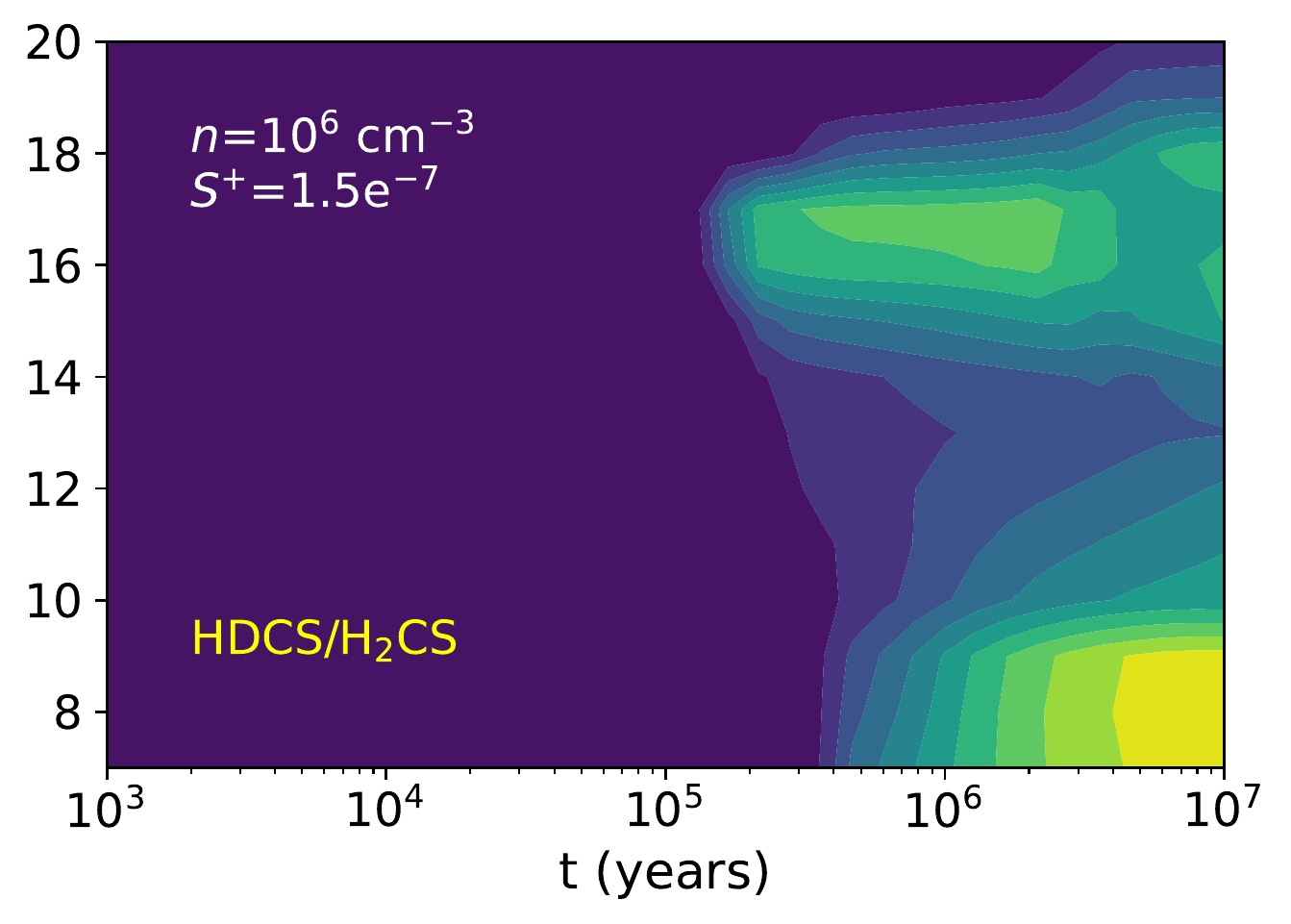}  
\hspace{-0.2cm}
\includegraphics[scale=0.38, angle=0]{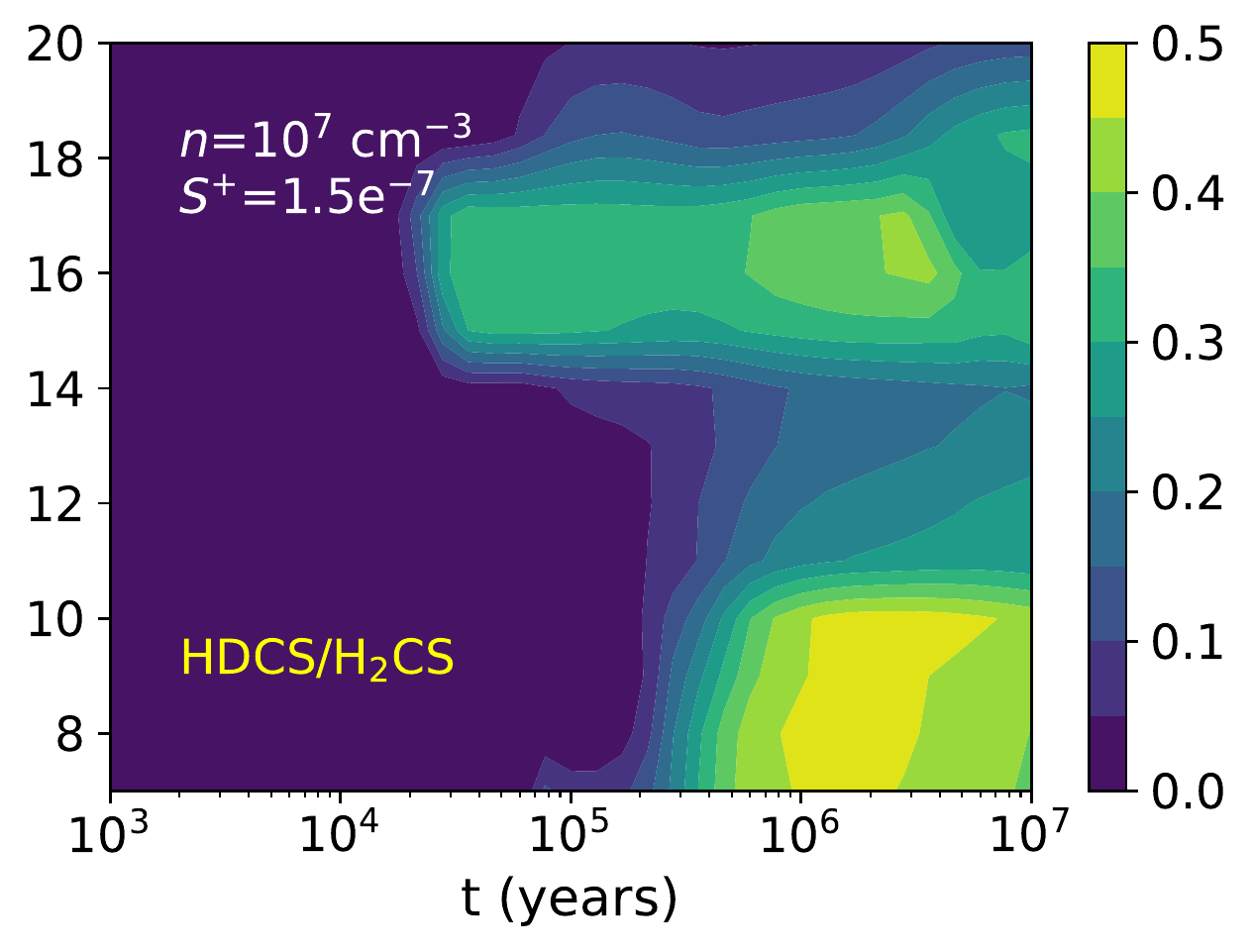}  

\hspace{-0.5cm}
\\
\caption{Evolution of the HDCS/H$_2$CS ratio (colour bar) as a function of time and temperature for a CR ionisation rate $\xi$=1.3$\times$10$^{-17}$ s$^{-1}$, a hydrogen number density $n_{\mathrm{H}}$=10$^5$, 10$^6$, and 10$^7$ cm$^{-3}$, and an initial sulphur abundance $S$$^+$=1.5$\times$10$^{-5}$, 1.5$\times$10$^{-6}$, and 1.5$\times$10$^{-7}$.}
\label{figure:HDCS/H2CS_evolution}
\end{figure*}

\begin{figure*}
\centering
\includegraphics[scale=0.38, angle=0]{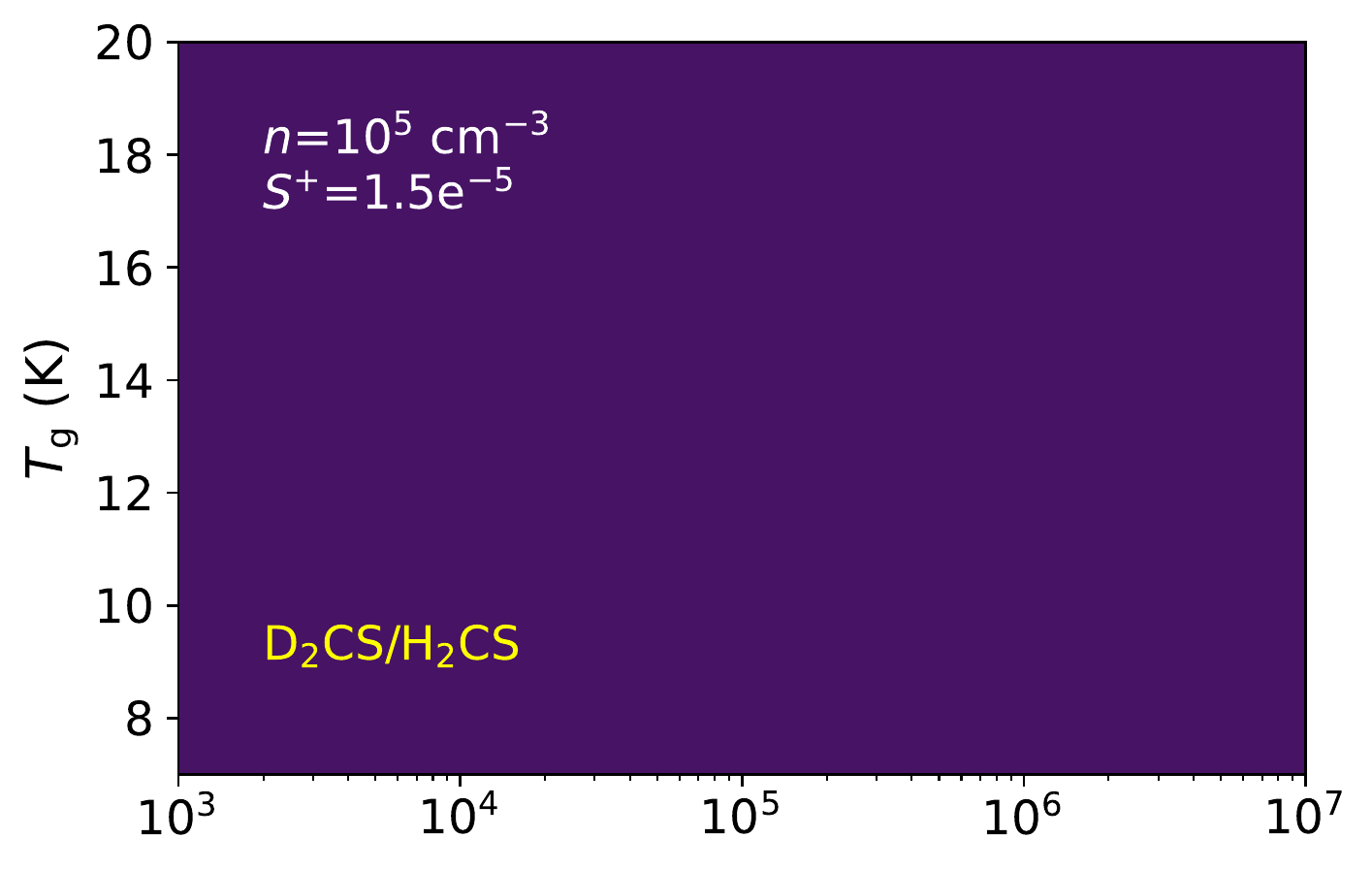}  
\hspace{-0.2cm}
\includegraphics[scale=0.38, angle=0]{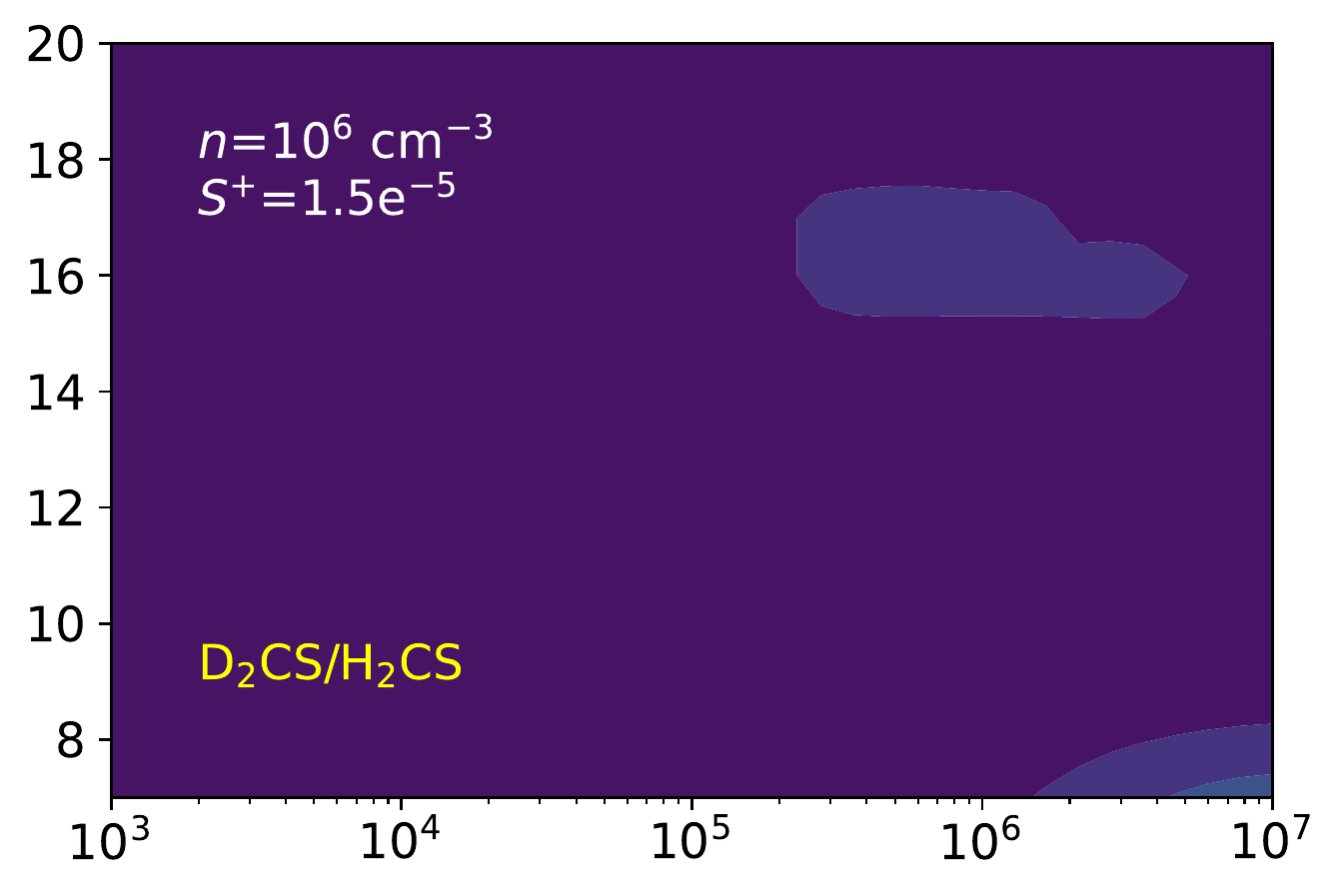} 
\hspace{-0.2cm}
\includegraphics[scale=0.38, angle=0]{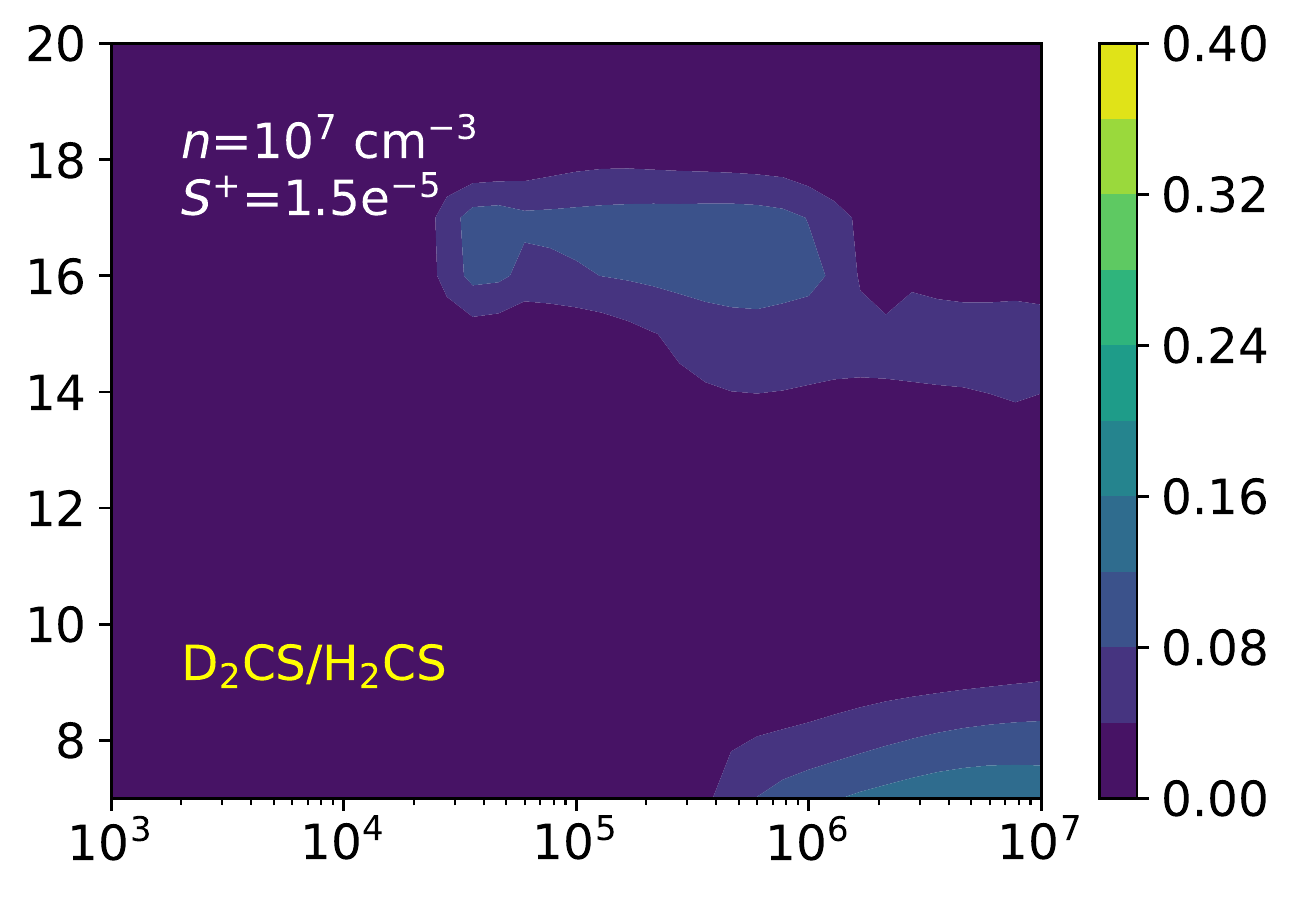}  

\includegraphics[scale=0.38, angle=0]{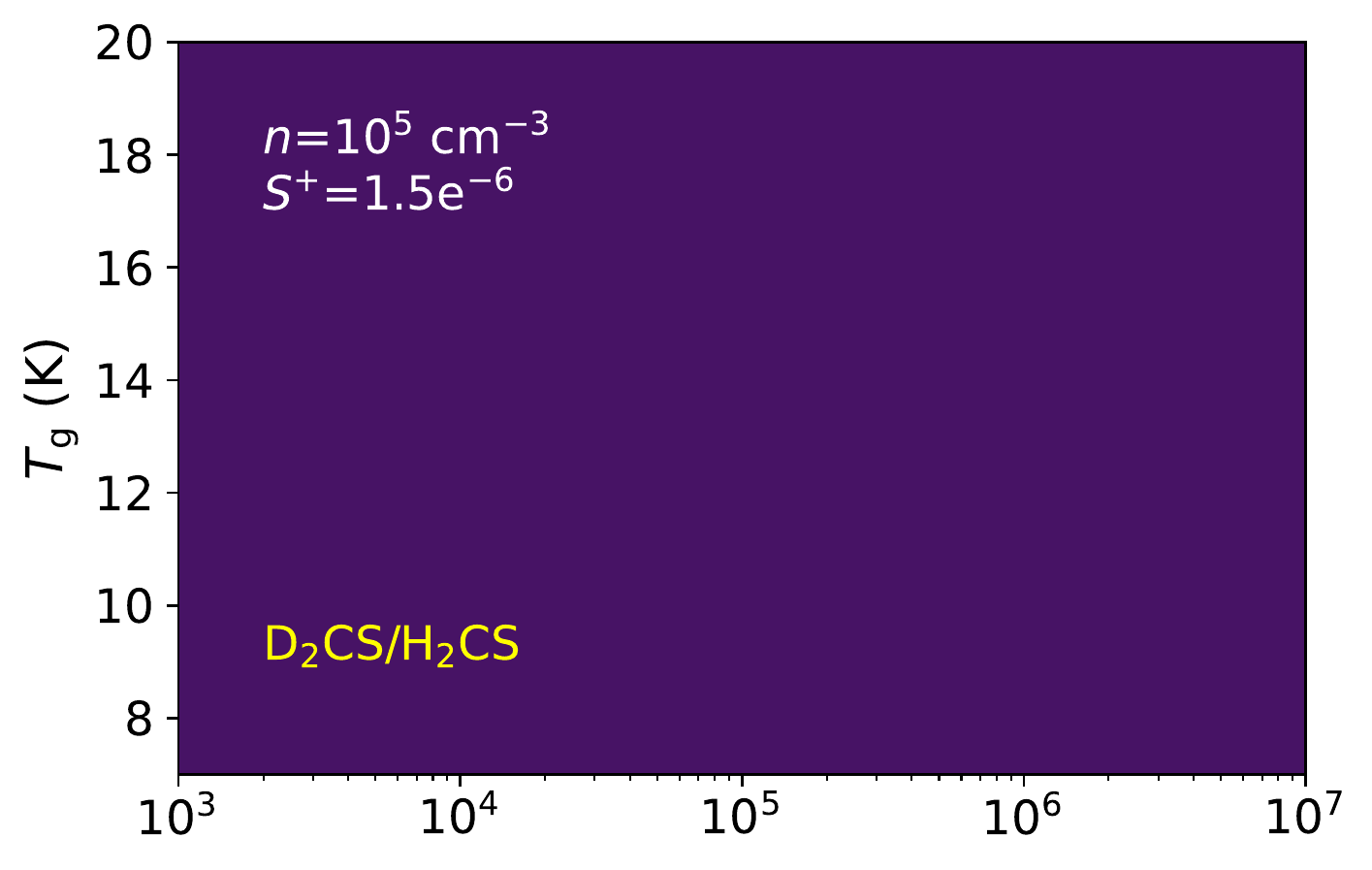}  
\hspace{-0.2cm}
\includegraphics[scale=0.38, angle=0]{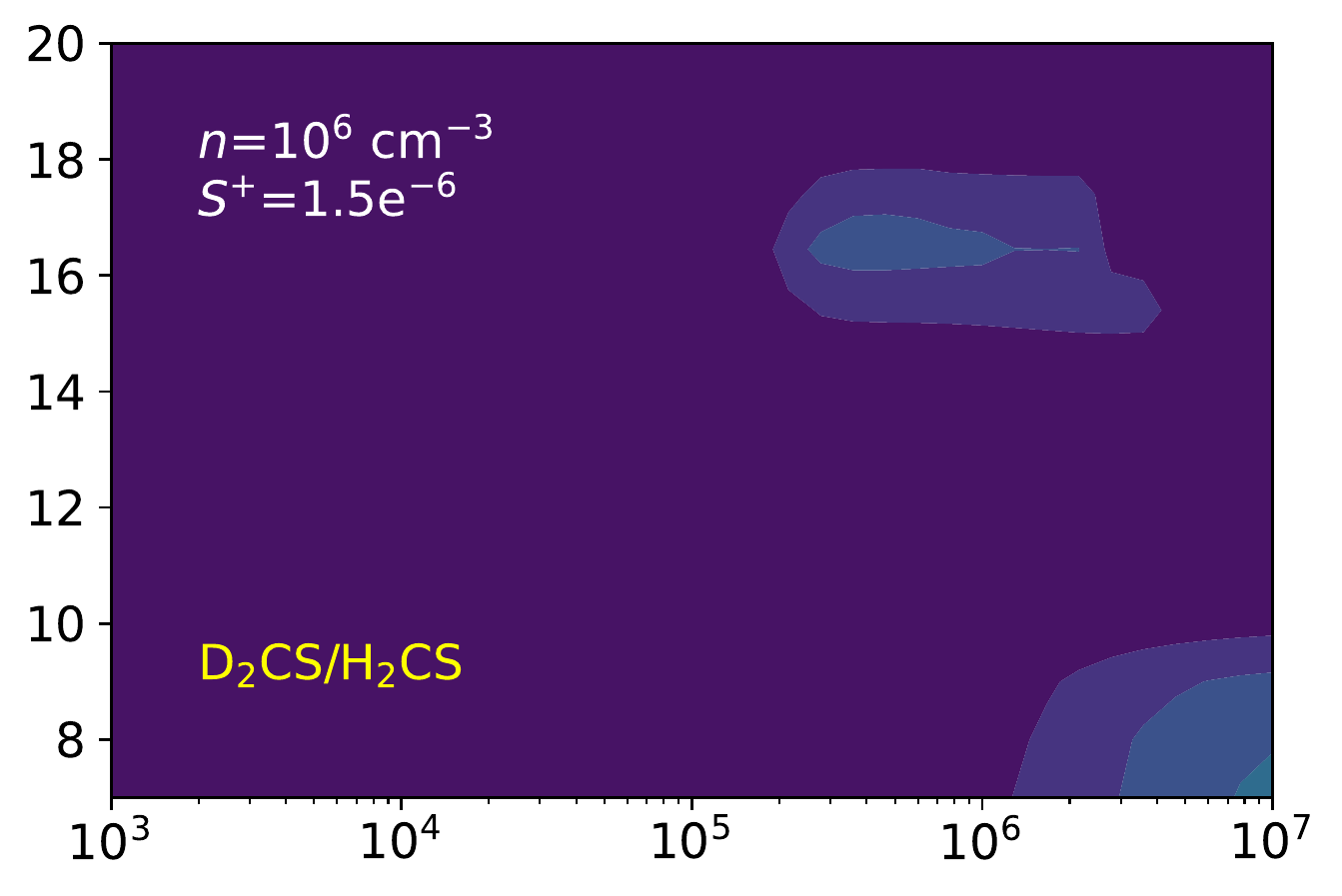}  
\hspace{-0.2cm}
\includegraphics[scale=0.38, angle=0]{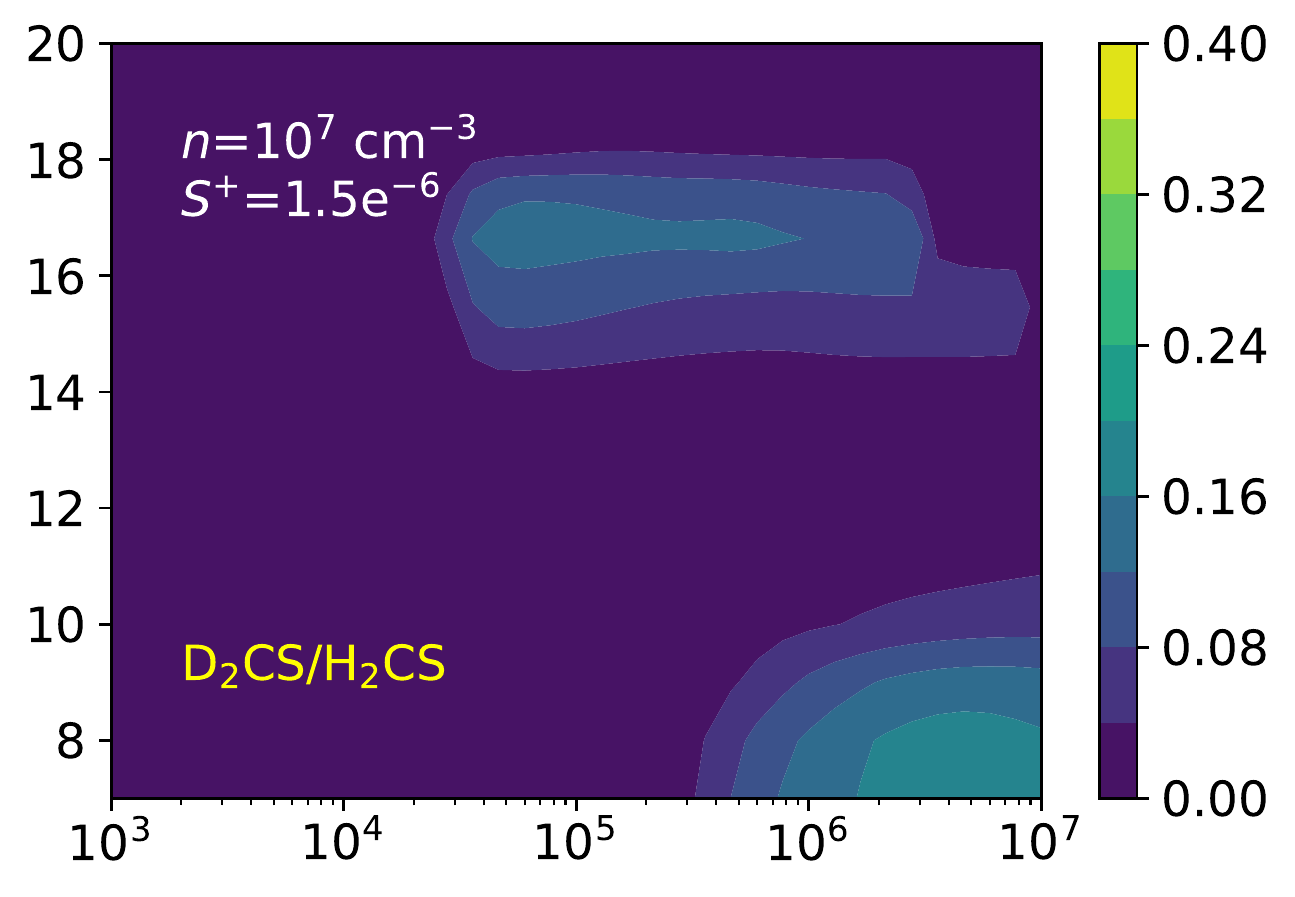}  

\includegraphics[scale=0.38, angle=0]{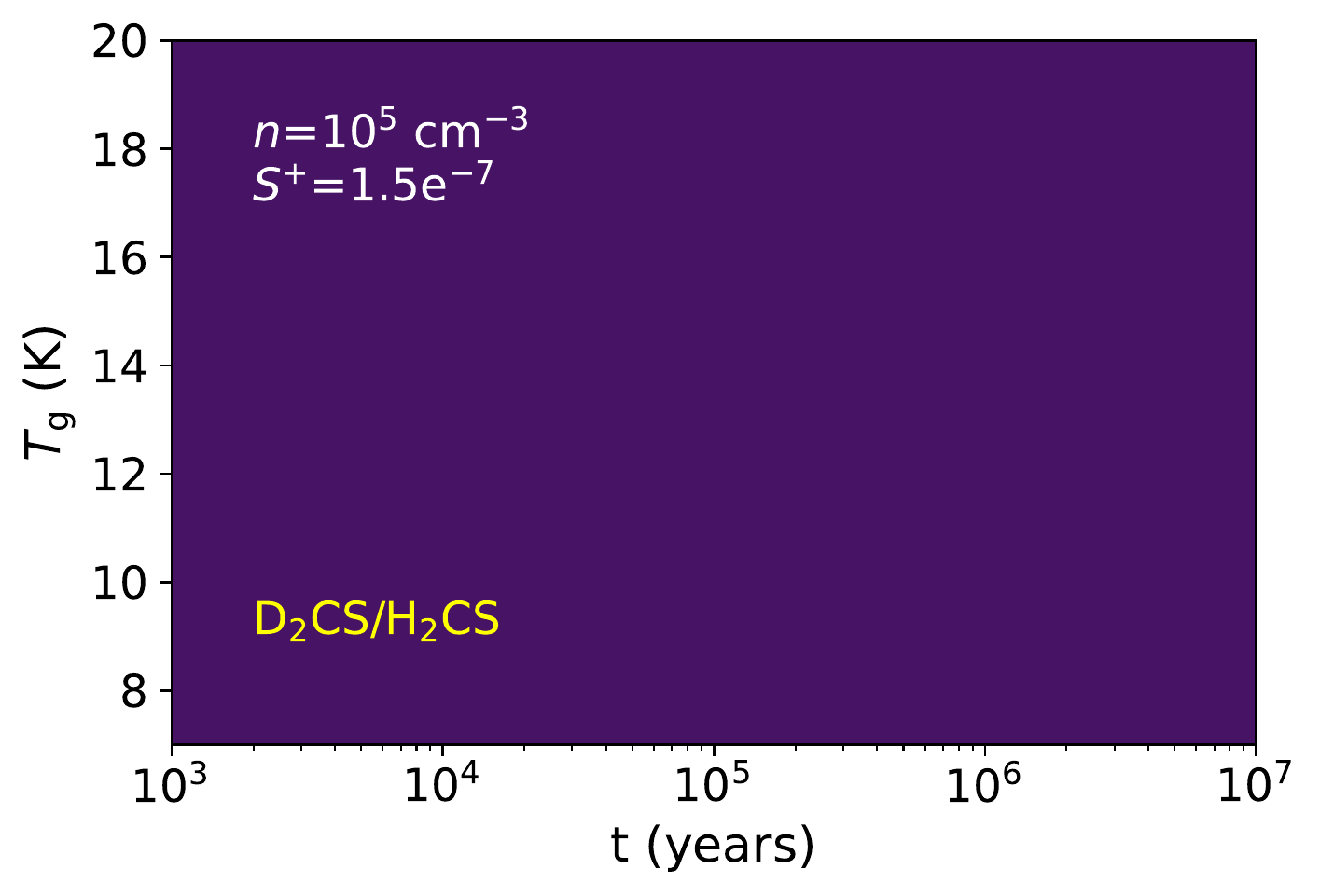}  
\hspace{-0.2cm}
\includegraphics[scale=0.38, angle=0]{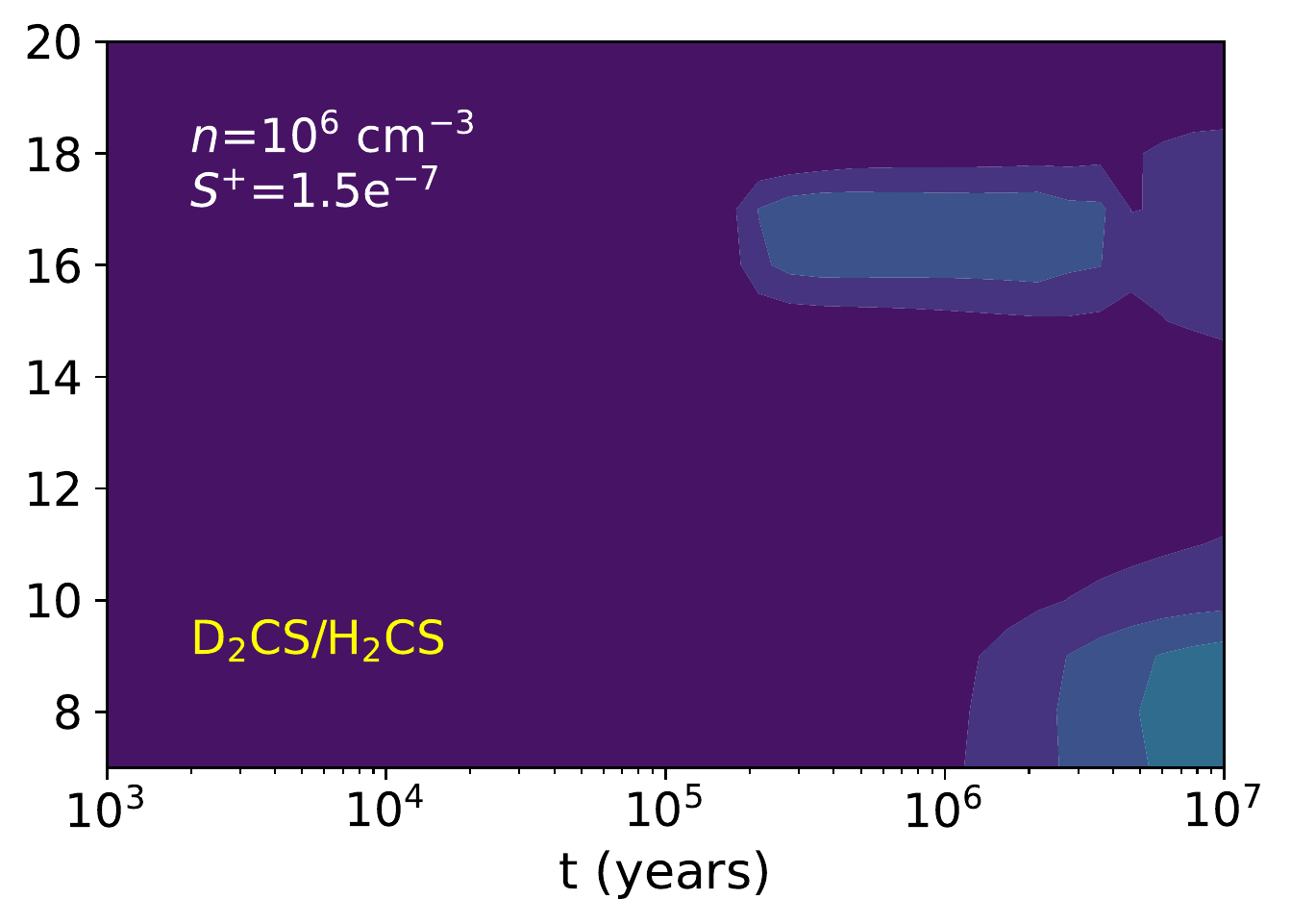}  
\hspace{-0.2cm}
\includegraphics[scale=0.38, angle=0]{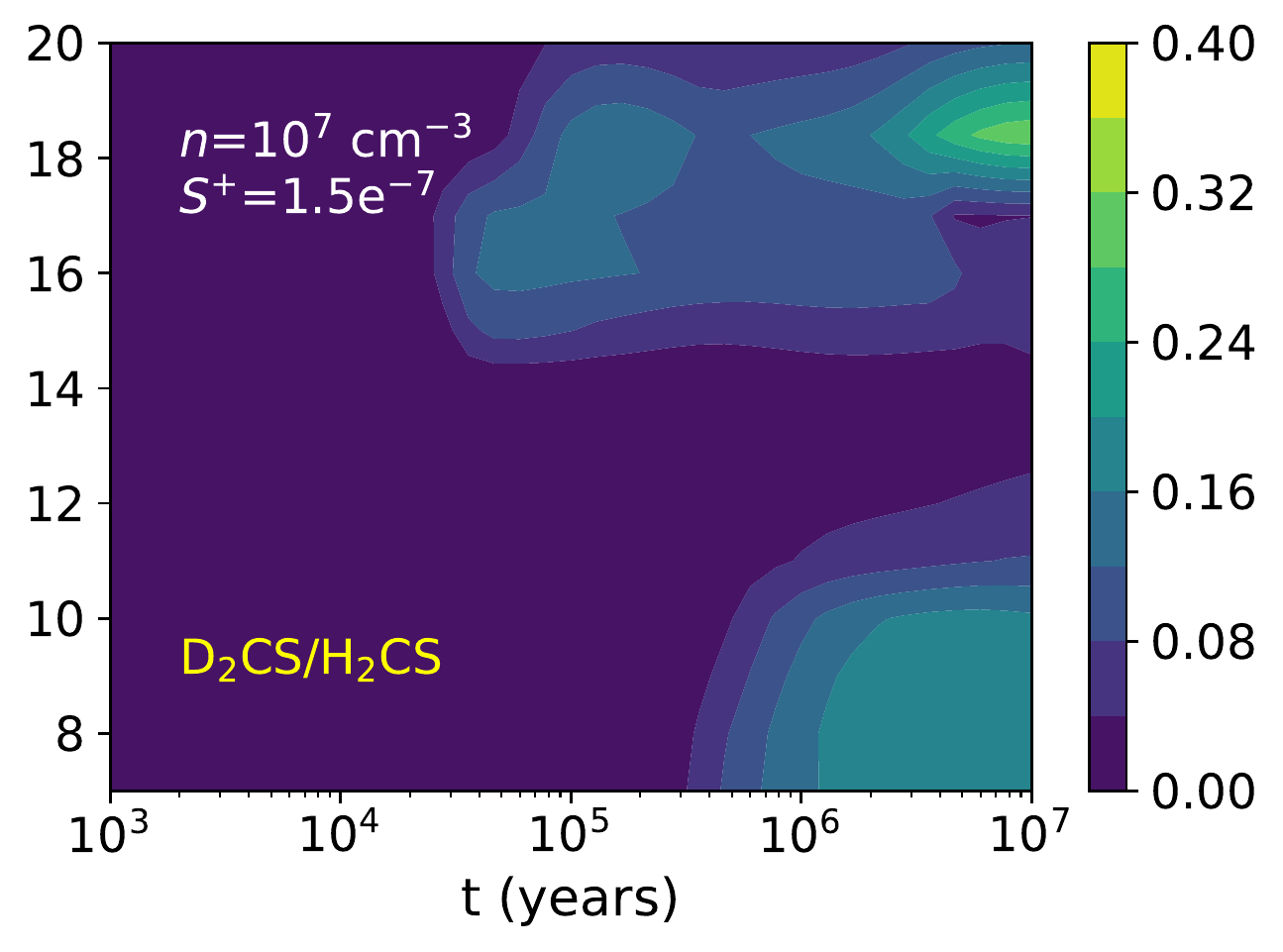}  

\hspace{-0.5cm}
\\
\caption{Evolution of the D$_2$CS/H$_2$CS ratio (colour bar) as a function of time and temperature for a CR ionisation rate $\xi$=1.3$\times$10$^{-17}$ s$^{-1}$, a hydrogen number density $n_{\mathrm{H}}$=10$^5$, 10$^6$, and 10$^7$ cm$^{-3}$, and an initial sulphur abundance $S$$^+$=1.5$\times$10$^{-5}$, 1.5$\times$10$^{-6}$, and 1.5$\times$10$^{-7}$.}
\label{figure:D2CS/H2CS_evolution}
\end{figure*}

In particular, we calculated the evolution of HDCS/H$_2$CS and D$_2$CS/H$_2$CS as a function of time (considering a time range of 10$^3$-10$^7$ yr) and the core temperature (ranging from 7 K to 20 K). We also considered a visual extinction $A$$_{\mathrm{V}}$=15 mag, a UV field strength $\chi$$_{\mathrm{UV}}$=1 (in Draine units), different volume densities (10$^5$, 10$^6$, and 10$^7$ cm$^{-3}$), and also different initial sulphur abundances (1.5$\times$10$^{-5}$, 1.5$\times$10$^{-6}$, and 1.5$\times$10$^{-7}$; see Table \ref{table:abundances_Nautilus}). We considered different initial sulphur abundances
since the value of elemental sulphur abundance is uncertain. In particular, while the observed gaseous sulphur accounts for its total cosmic abundance in the primeval diffuse clouds and HII  regions \citep{Neufeld2015, Goicoechea2021}, there is an unexpected paucity of S-bearing molecules within dense molecular clouds \citep[see e.g.][]{Agundez2013}. In starless cores (i.e. the cold and dense cores where stars are formed), the sulphur atoms locked within the observed molecules constitute only $<$1$\%$ of the expected amount \citep[see e.g.][]{Vastel2018}). One could think that most of the sulphur is locked on the icy mantles that cover dust grains but, surprisingly, a similar trend is encountered within the solid phase. Nowadays, solid OCS and tentatively solid SO$_2$, are the unique sulphur compounds observed, while only upper limits to the solid H$_2$S abundance have been derived \citep{Palumbo1995, Palumbo1997, Boogert1997, Jimenez-Escobar2011}. According to these measurements, the abundances of the icy species could account for $<$5$\%$ of the total sulphur. It has been suggested that this so-called depleted sulphur may be locked as refractory material \citep[in particular as S$_8$,][]{Shingledecker2020, Cazaux2022}, claiming sulphur depletion of more than two orders of magnitude. Intense observational and theoretical work has been carried out in the last decade to constrain the value of sulphur depletion, and there is increasing evidence for a moderate factor of 1-20 of sulphur depletion in cold and warm clouds \citep{Esplugues2013, Esplugues2014, Fuente2019, Navarro-Almaida2020, Navarro-Almaida2021, Bulut2021, Codella2021}. For that, we consider in our models the un-depleted case (1.5$\times$10$^{-5}$), as well as sulphur depletions of a factor of 10 and 100 (Table \ref{table:abundances_Nautilus}).

Figures \ref{figure:HDCS/H2CS_evolution} and \ref{figure:D2CS/H2CS_evolution} show the evolution of HDCS/H$_2$CS and D$_2$CS/H$_2$CS, respectively. For the case of HDCS/H$_2$CS, when the number density is $n_{\mathrm{H}}$=10$^5$ cm$^{-3}$, its maximum reached value at any time is $\sim$0.2. However, for higher densities, HDCS/H$_2$CS increases up to $\sim$0.5 at $T_{\mathrm{gas}}$$<$12 K. Nevertheless, this maximum value is reached with a difference of about 5 million years depending on if the considered density is $n_{\mathrm{H}}$=10$^6$ cm$^{-3}$ or 10$^7$ cm$^{-3}$ (the smaller the density, the longer the evolution time). Apart from density, we also observe in Fig. \ref{figure:HDCS/H2CS_evolution} a variation in the deuterated ratio with sulphur fractional abundance in such a way that the lower the initial $X$(S), the higher the HDCS/H$_2$CS ratio, especially at low temperatures. Similar trend is also observed in Fig. \ref{figure:D2CS/H2CS_evolution} for D$_2$CS/HDCS. 
In particular, we find that decreasing the sulphur initial fractional abundance from 1.5$\times$10$^{-5}$ \citep[which corresponds to the Solar value,][]{Asplund2009} to 1.5$\times$10$^{-7}$, leads to an increase by a factor of up to $\sim$1.5 and $\sim$2.5 in the maximum values of the HDCS/H$_2$CS and D$_2$CS/H$_2$CS ratios, respectively.

From Figs. \ref{figure:HDCS/H2CS_evolution} and \ref{figure:D2CS/H2CS_evolution}, we also obtain a strong dependence of HDCS/H$_2$CS and D$_2$CS/H$_2$CS with time, since their values progressively increase with the evolution of the starless cores. These ratios, therefore, can be used as powerful tools to derive the evolutionary stage of starless cores. However, they cannot be used to derive the temperature of the starless cores since their evolution presents a double peak in two ranges of temperature ($T_{\mathrm{gas}}$$\sim$7-11 K and $\sim$15-19 K).

\subsection{Cosmic-ray impact on deuterium fractionation}
\label{Cosmic rays deuterium fractionation}

\begin{figure*}[h!]
\centering
\hspace{0.5cm}
\includegraphics[scale=0.5, angle=0]{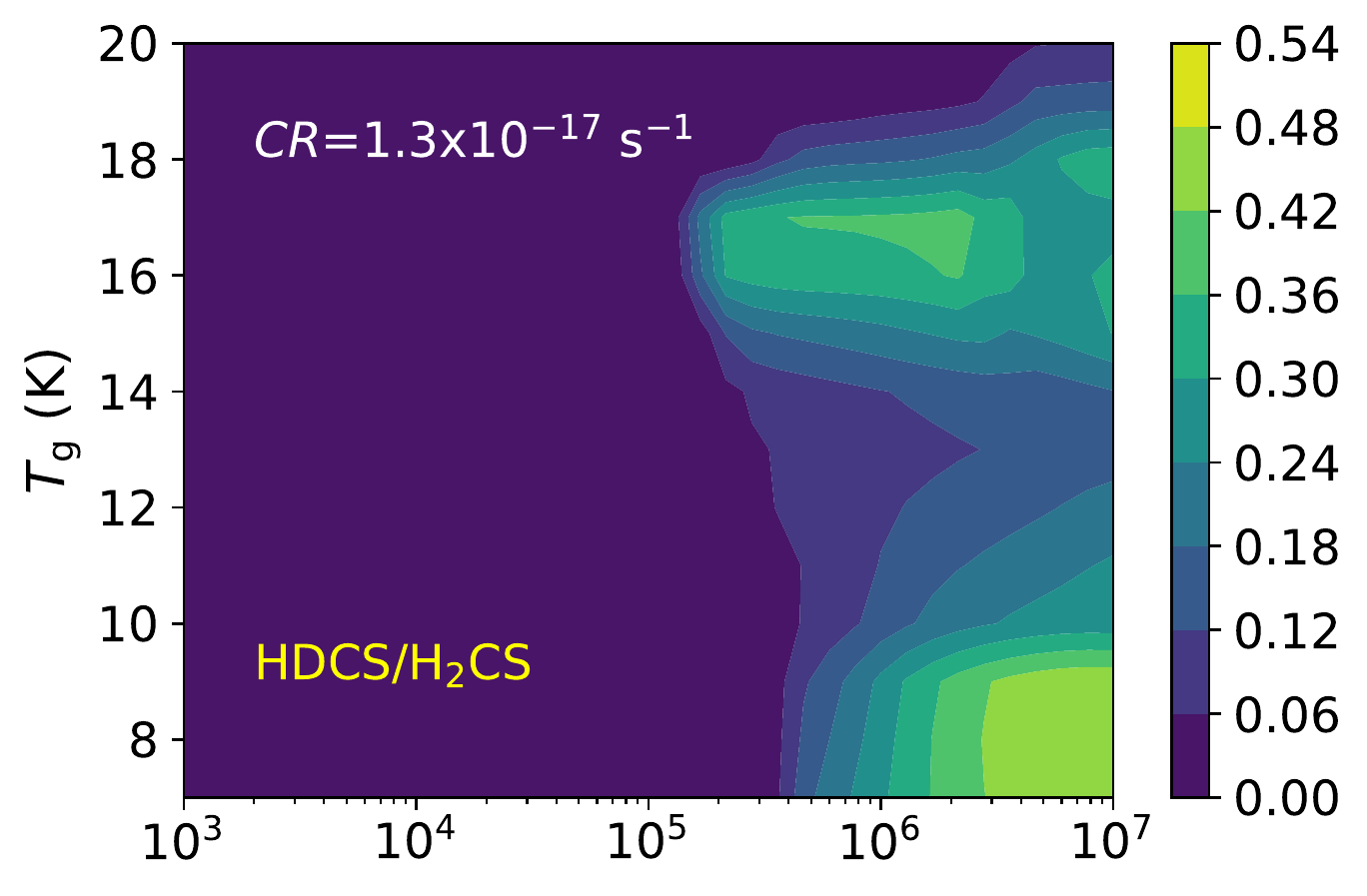}  
\hspace{0.8cm}
\includegraphics[scale=0.5, angle=0]{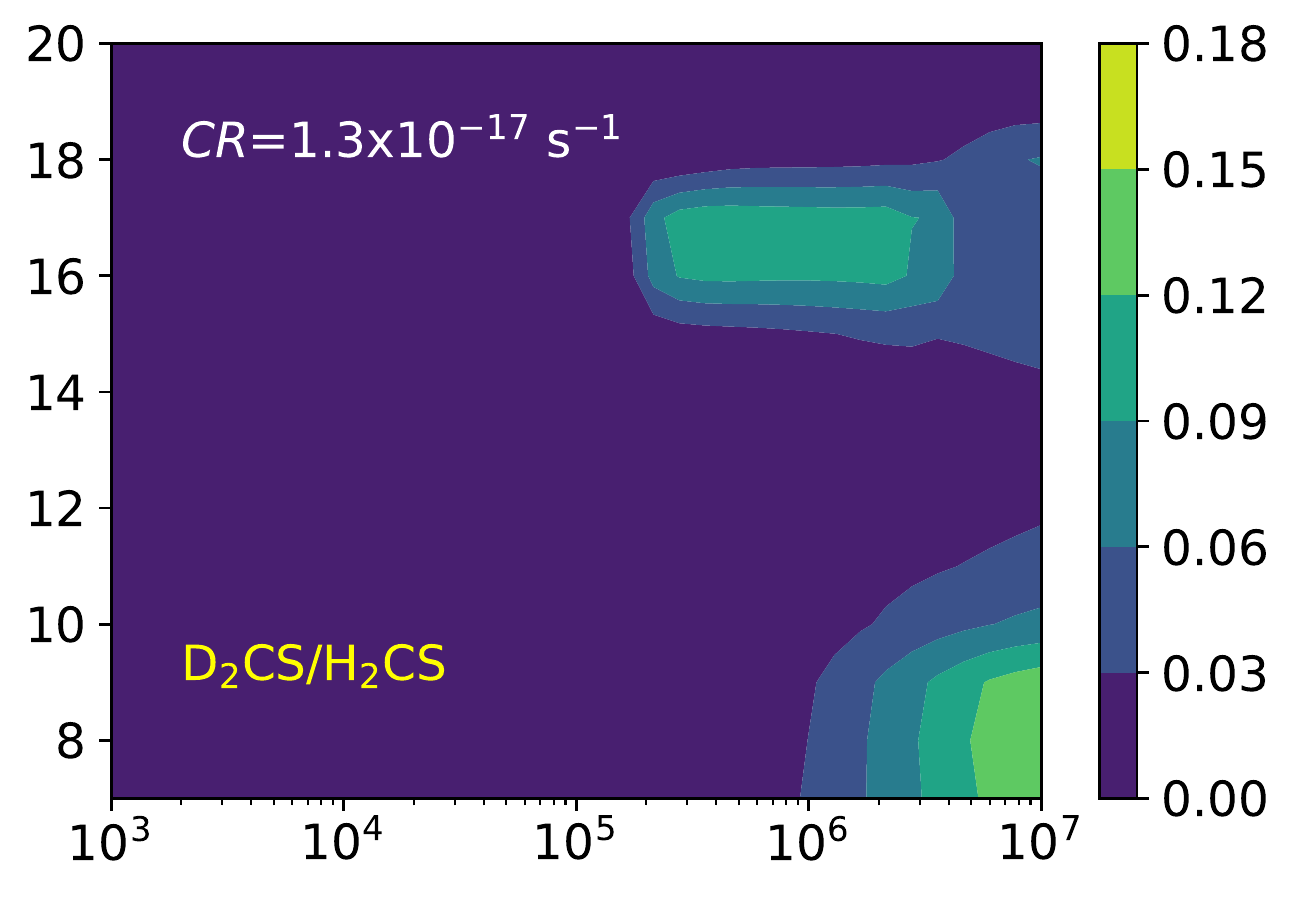}  
\hspace{0.5cm}

\includegraphics[scale=0.5, angle=0]{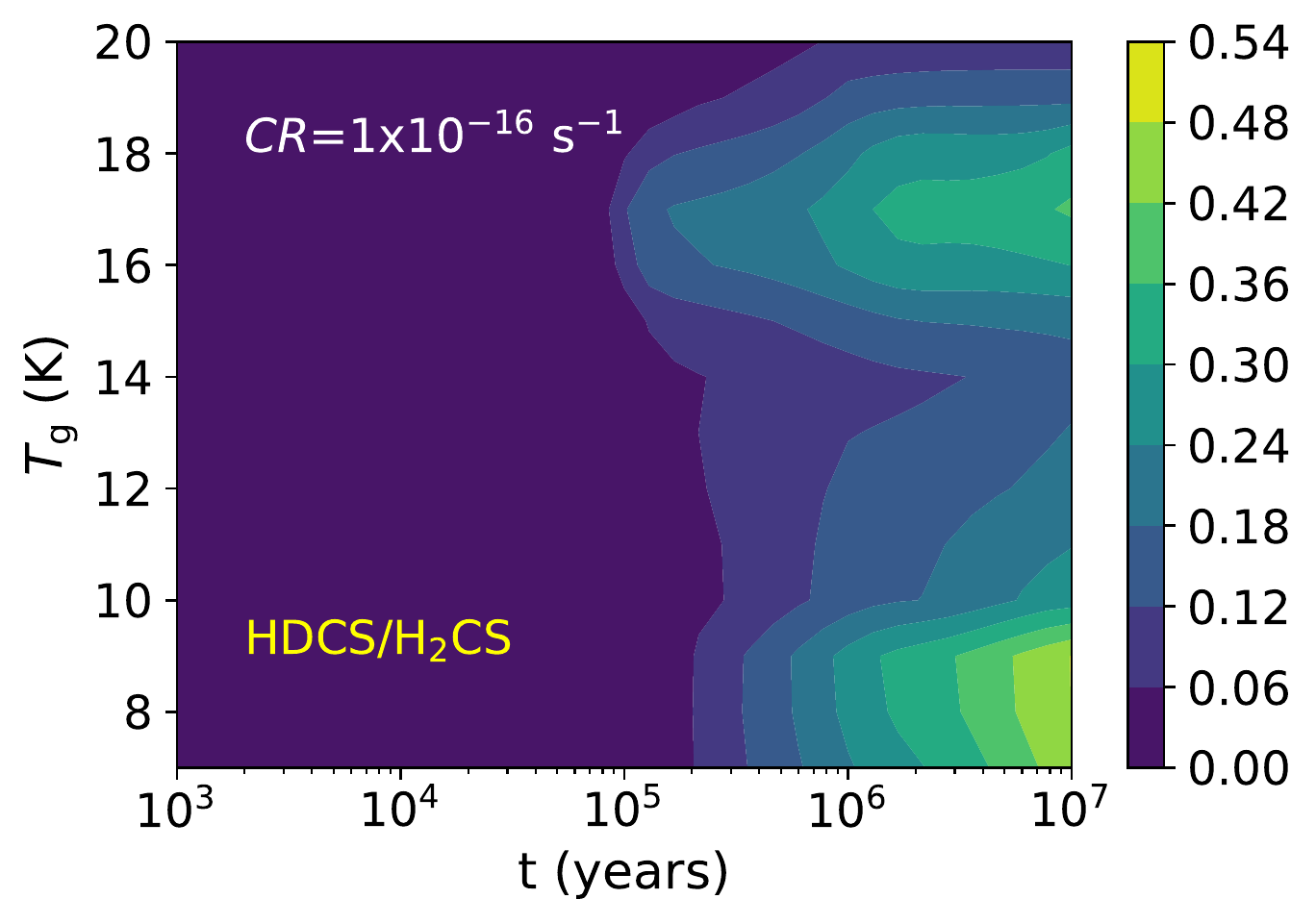}  
\hspace{0.8cm}
\includegraphics[scale=0.5, angle=0]{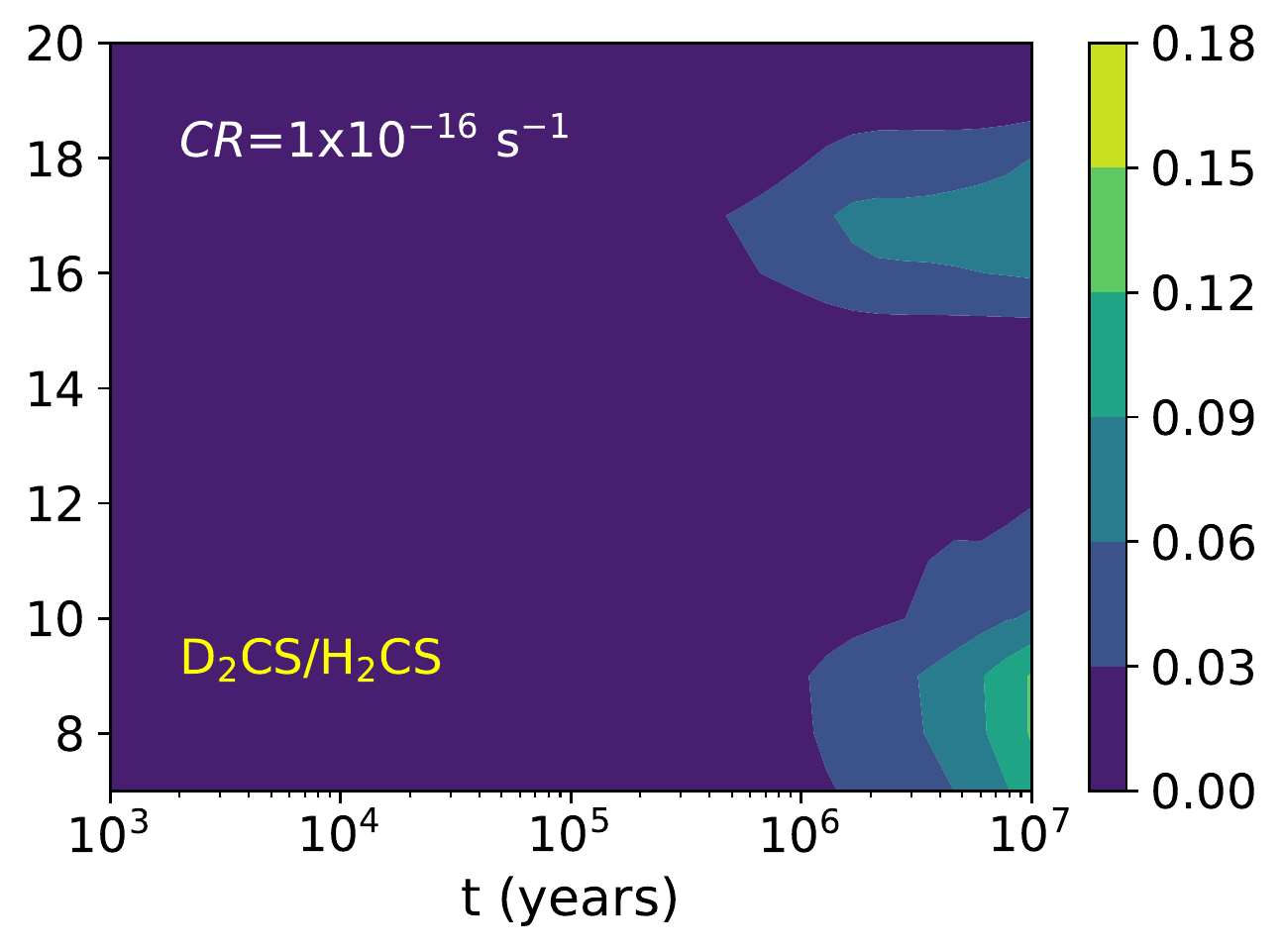}  
\hspace{-0.5cm}
\\
\caption{Evolution of the HDCS/H$_2$CS (left) and D$_2$CS ratio (right) as a function of time and temperature for an initial sulphur fractional abundance $S$$^+$=1.5$\times$10$^{-7}$, a hydrogen number density $n_{\mathrm{H}}$=10$^6$ cm$^{-3}$, and a CR ionisation rate $\zeta$=1$\times$10$^{-16}$ (bottom row) and 1.3$\times$10$^{-17}$ s$^{-1}$ (top row).}
\label{figure:HDCS/H2CS_D2CS/H2CS_evolution_CRhigh}
\end{figure*}

Another important parameter regulating deuterium fractionation in the cold phases of the ISM is the ubiquitous presence of CRs, since in the denser regions where UV photons are efficiently absorbed and cannot penetrate, CRs become the main ionising agent of the gas. In fact, its interaction with H$_2$ produces H$^{+}_{2}$, which rapidly reacts with another hydrogen molecule leading to H$^{+}_{3}$. This ion is the starting point for reactions between charged and neutral species, producing several other key molecules, such as HCO$^+$ and H$_2$D$^+$ \citep{Dalgarno1984}. In spite of its role in deuterium chemistry, the CR ionisation rate is still uncertain and spreads over a range of values. While \citet{van der Tak2000} determined a CR ionisation rate of $\sim$3$\times$10$^{-17}$ s$^{-1}$ in dense clouds, \citet{Indriolo2012} obtained a range (1.7-10.6)$\times$10$^{-16}$ s$^{-1}$ of values considering a sample of diffuse clouds, and \citet{Neufeld2017} derived a CR ionisation rate of the order of a few 10$^{-16}$ s$^{-1}$ in the Galactic disc. A very recent study of the pre-stellar core L\,1544 shows that observations are better reproduced with a CR ionisation rate $\zeta$=3$\times$10$^{-17}$ s$^{-1}$ than with $\zeta$=1.3$\times$10$^{-17}$ s$^{-1}$ \citep{Redaelli2021}. 

In the same line, \citep{Fuente2019} derived a CR ionisation rate of $\zeta$=(0.5-1.8)$\times$10$^{-16}$ s$^{-1}$ in the translucent phase of TMC1, probing that high CR ionisation rate is found in the outer regions of molecular clouds. A lower value of $\zeta$=1.3$\times$10$^{-17}$ s$^{-1}$ was derived by \citet{Navarro-Almaida2021} in the dense cores TMC1 (C) and TMC1 (CP), in qualitative agreement with the expected behaviour of $\zeta$ decreasing with visual extinction along the molecular cloud \citep{Padovani2013, Ivlev2018, Ivlev 2021}.

\begin{table*}
\caption{Classification of the core sample according to the evolutionary stage.}
\begin{center}
\begin{tabular}{l|ll|ll}
\hline 
\hline
        & \multicolumn{2}{c}{$t$<1 My}  &  \multicolumn{2}{c}{$t$$\sim$1-5 My}  \\ 
\hline
Region  & Core         & HDCS/H$_2$CS   &  Core       & HDCS/H$_2$CS \\
\hline
Taurus  & B\,213-C7-1  & 0.2$\pm$0.1    & B\,213-C1-1 & 0.5$\pm$0.2       \\
        & B\,213-C16-1 & 0.14$\pm$0.06  & B\,213-C2-1 & 0.3$\pm$0.1       \\
        & B\,213-C10-1 & 0.13$\pm$0.06  & B\,213-C6-1 & 0.3$\pm$0.1       \\
        & B\,213-C5-1  & 0.12$\pm$0.05  &             &                   \\
\hline
Perseus & 79-C1-1      & 0.2$\pm$0.1    & 1333-C6-1   & 0.8$\pm$0.4       \\
        & 1333-C3-14   & 0.15$\pm$0.07  & 1333-C7-1   & 0.5$\pm$0.2      \\
        & 1333-C3-1    & 0.10$\pm$0.05  & 1333-C4-1   & 0.3$\pm$0.1       \\
        & -            & -              & L\,1448     & 0.3$\pm$0.1      \\
\hline
Orion   & ORI-C2-3     & 0.10$\pm$0.04  &     -       &       -           \\

\hline 
\end{tabular}
\end{center}
\label{table:evolutionary_stage}
\end{table*}

In Fig. \ref{figure:HDCS/H2CS_D2CS/H2CS_evolution_CRhigh}, we show the impact of varying  the CR ionisation rate (from $\zeta$=10$^{-16}$ s$^{-1}$ to $\zeta$=1.3$\times$10$^{-17}$ s$^{-1}$, which is considered as the standard CR rate for  dense cores) on the deuterium fractionation for the HDCS/H$_2$CS and D$_2$CS/H$_2$CS ratios along 10$^7$ yr. We obtain that, for both models with different $\zeta$, the evolutionary trend is similar for each deuterium fractionation ratio, with them progressively increasing with time and showing two peaks at the range temperatures previously mentioned. Nevertheless, for HDCS/H$_2$CS, we also observe that, although the maximum reached value is the same ($\sim$0.54) independently on the considered $\zeta$, in the model with the higher CR ionisation rate the maximum HDCS/H$_2$CS ratio is reached about 3 million years later than in the model with the low CR ionisation rate. For D$_2$CS/H$_2$CS, we also obtain lower values for the higher $\zeta$ in each evolution time with respect to the model with the low CR ionisation rate. These results indicate that the main effect on the deuterium fractionation when increasing by one order of magnitude the CR ionisation rate is to slow down its evolution by a few million years.

\subsection{Comparison with observations}

In the Taurus complex, we find the highest HDCS/H$_2$CS ratios ($\sim$0.3-0.5, Table   \ref{table:deuterated_ratios}) in the starless cores located in the north of the B\,213  filament (C1, C2, and C6). Models shown in Fig. \ref{figure:HDCS/H2CS_evolution} reproduce  these values at $t$$\gtrsim$5$\times$10$^5$ yr, densities $n_{\mathrm{H}}$$>$10$^5$  cm$^{-3}$, and with a high sulphur initial depletion. In fact, the highest deuterium ratio  (HDCS/H$_2$CS=0.5) from the starless core B\,213-C1-1 is only reproduced when the sulphur  fractional abundance is as low as 1.5$\times$10$^{-7}$ and for a long evolution time of 1-5 Myr. 
Regarding the double deuterated ratio, the starless cores C1, C2, and C6 also present the highest values ($\sim$0.2-0.3) obtained observationally. These D$_2$CS/H$_2$CS ratios are reproduced by a model with high density, $t$$\gtrsim$10$^6$ yr, and also with a low sulphur initial fractional abundance ($\leq$1.5$\times$10$^{-6}$; see Fig. \ref{figure:D2CS/H2CS_evolution}). 
By contrast, the cores C10 and C16 located in the south part of the B\,213 filament present low HDCS/H$_2$CS and D$_2$CS/H$_2$CS ratios ($\lesssim$0.14). These values are reproduced at earlier times ($t$$<$10$^6$ yr) than those for the cores C1, C2, and C6, indicating that cores located in the south part of B\,213 are less chemically evolved than the ones located in the north of the filament. In any case, we should also consider the uncertainties associated with the D/H ratios of these cores (except for the core C1 with the highest D/H even considering its uncertainty; Table \ref{table:deuterated_ratios}), which makes it difficult to clearly determine the evolutionary stages of the cores since some of these uncertainties represent up to the 50$\%$ of the D/H value. Nevertheless, comparing with results from other authors, we find that our results are in agreement with the ones found by \citet{Hacar2013} through observations of CO and N$_2$H$^+$. In particular, they found that the region harbouring the cores C1, C2, and C6 is weak in C$^{18}$O and has a number of N$_2$H$^+$-bright dense cores, which suggests that some CO depletion has already taken place, and that this region must therefore be more chemically evolved than other regions located more to the south, such as the one hosting C7, 
which is more bright in C$^{18}$O. In fact, we obtain a low deuterium ratio in C7 (0.2$\pm$0.1). Table \ref{table:evolutionary_stage} summarises the evolutionary stages of our core sample obtained by comparing the observations and the model results.

In Perseus, the starless core NGC\,1333-C7 presents HDCS/H$_2$CS and D$_2$CS/H$_2$CS ratios as high as 0.5 and 0.2, respectively. As we observe in Figs. \ref{figure:HDCS/H2CS_evolution} and \ref{figure:D2CS/H2CS_evolution}, we reproduce these values for a model with a density $>$10$^5$ cm$^{-3}$, a $T_{\mathrm{K}}$$<$12 K, a sulphur fractional abundance of at least 1.5$\times$10$^{-6}$, and an evolutionary time between 1 and 5 Myr. This is a slightly older age than that derived (0.5-1.5 Myr) by \citet{Lada1996} and \citet{Wilking2004} for NGC\,1333. 
For the cores NGC\,1333-C4 and L\,1448-1, located in the south-western part of the Perseus complex like NGC\,1333-C7, we also obtain high ratios of HDCS/H$_2$CS (0.3) and, in particular, of D$_2$CS/H$_2$CS (0.3 and 0.2, respectively). We reproduce them for a chemical age of $\sim$2 Myr for L\,1448-1, and of $\sim$5 Myr for NGC\,1333-C4 (Fig.  \ref{figure:D2CS/H2CS_evolution}).  
On the other hand, the core 79-C1-1, which is located farthest north-east of the Perseus molecular cloud, presents low deuterium ratios (0.2 and 0.16 for HDCS and D$_2$CS, respectively). For the core IC\,348 (also situated in the north-east region), we do not detect emission of HDCS nor D$_2$CS. All this suggests that, in general terms, the north-eastern part of Perseus is in an earlier chemical evolutionary stage ($<$1 Myr) than the south-western part ($\geq$1-5 Myr; see Table \ref{table:evolutionary_stage}).  In any case, Perseus observations are reproduced considering a sulphur initial fractional abundance $\leq$1.5$\times$10$^{-6}$ as in the case of Taurus.  

\begin{figure*}
\centering
\hspace{-0.5cm}
\includegraphics[scale=0.4, angle=0]{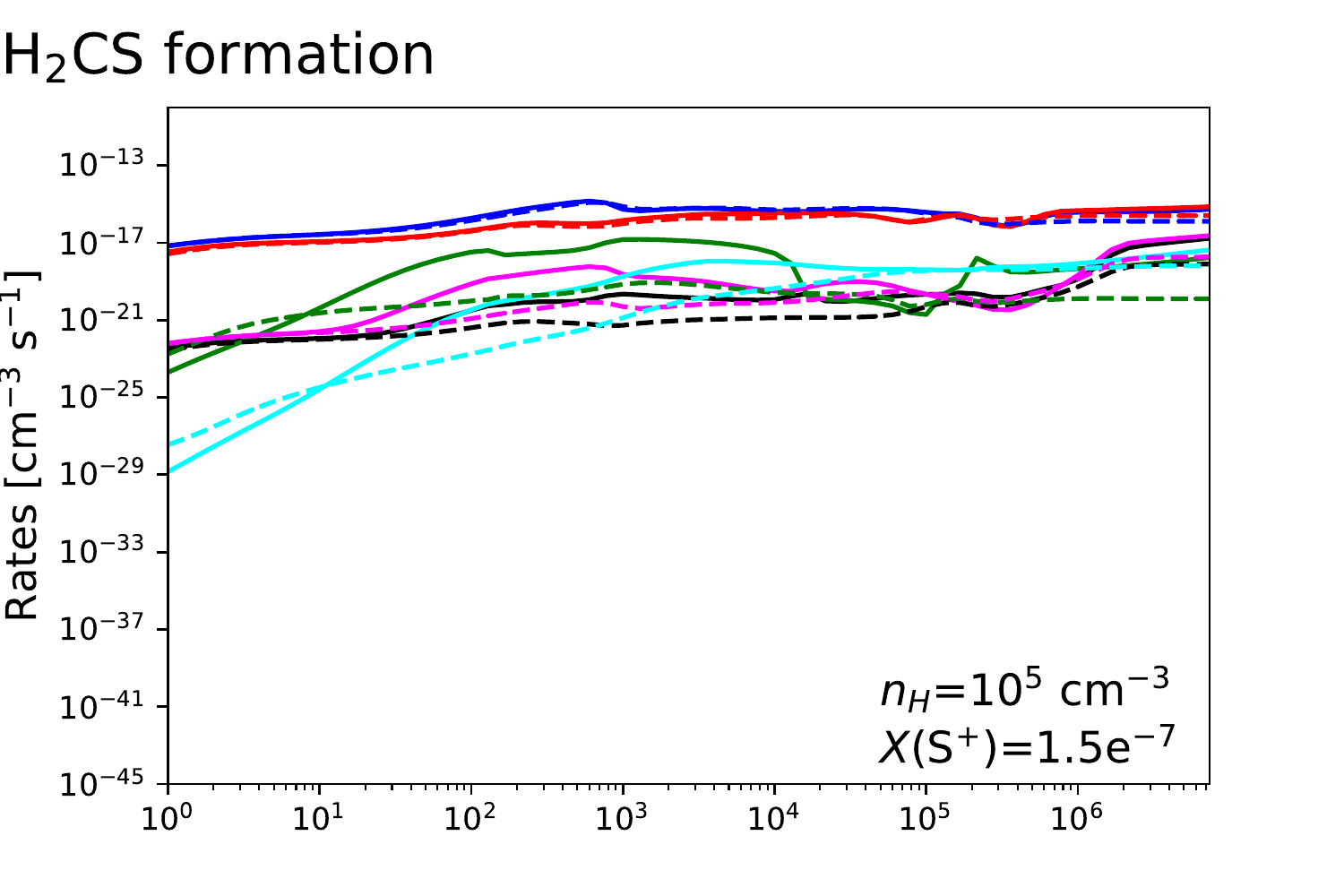}  
\hspace{-0.5cm}
\includegraphics[scale=0.4, angle=0]{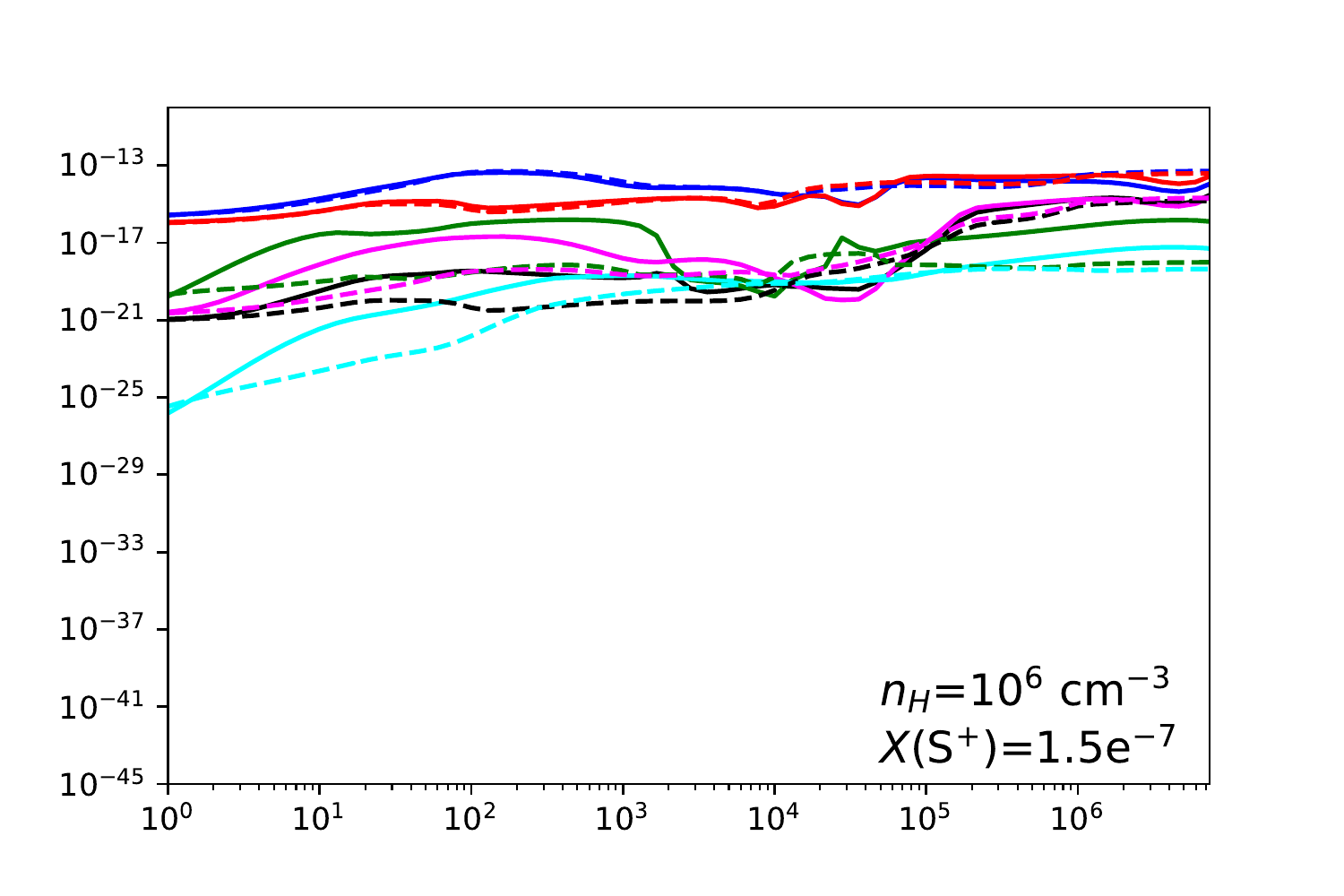} 
\hspace{-0.5cm}
\includegraphics[scale=0.4, angle=0]{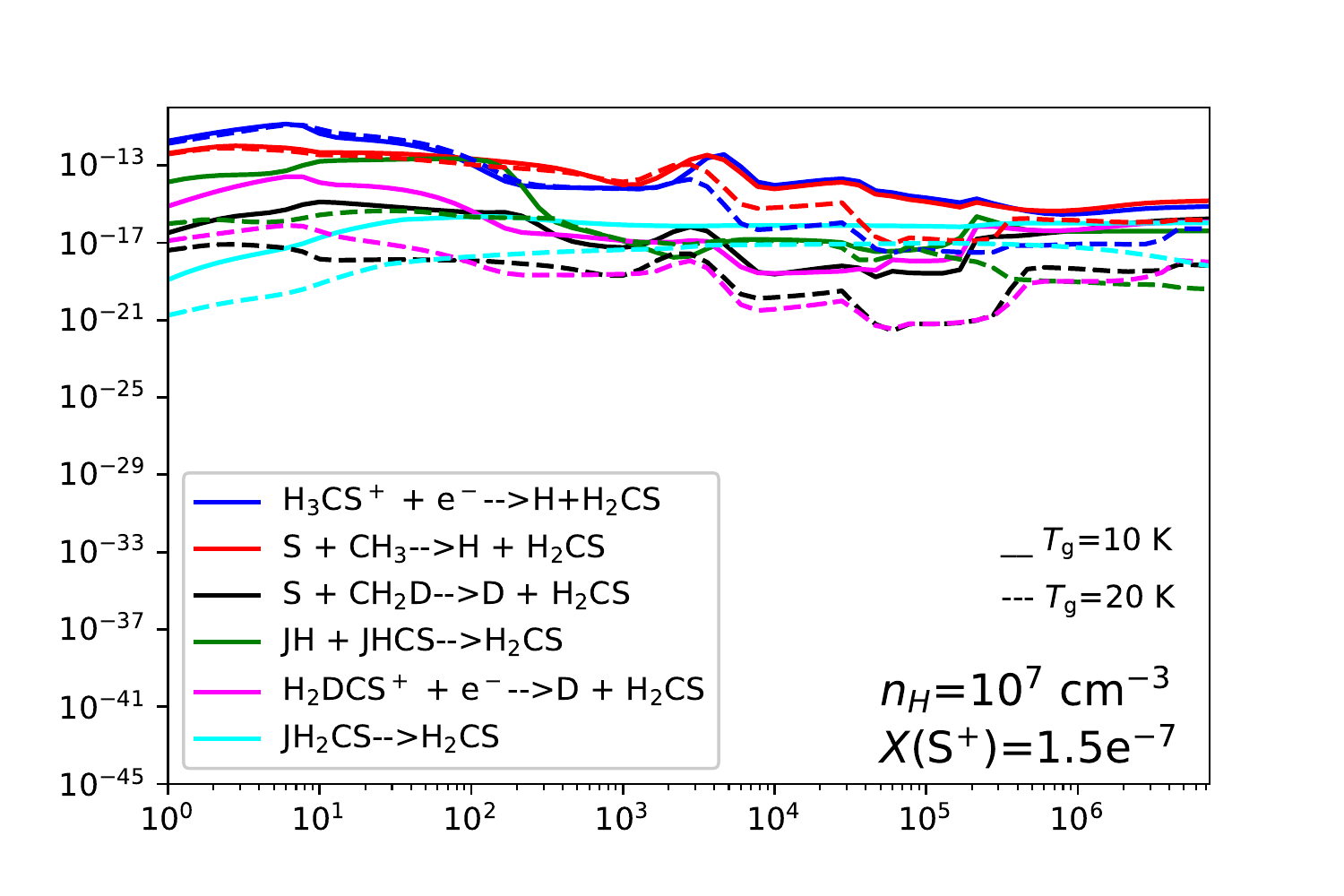}  
\hspace{-0.5cm}

\vspace{0.0cm}
\includegraphics[scale=0.4, angle=0]{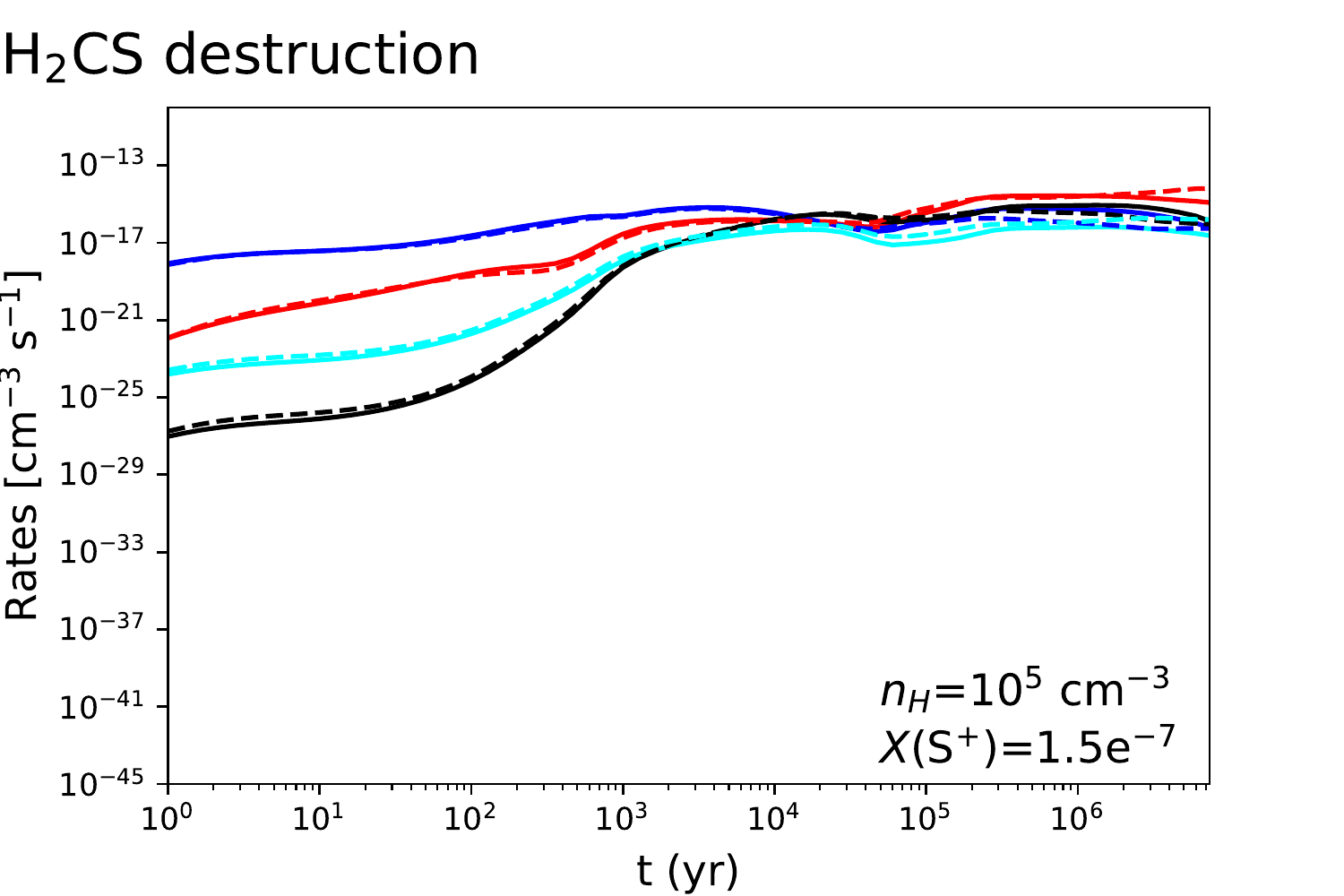}  
\hspace{-0.5cm}
\includegraphics[scale=0.4, angle=0]{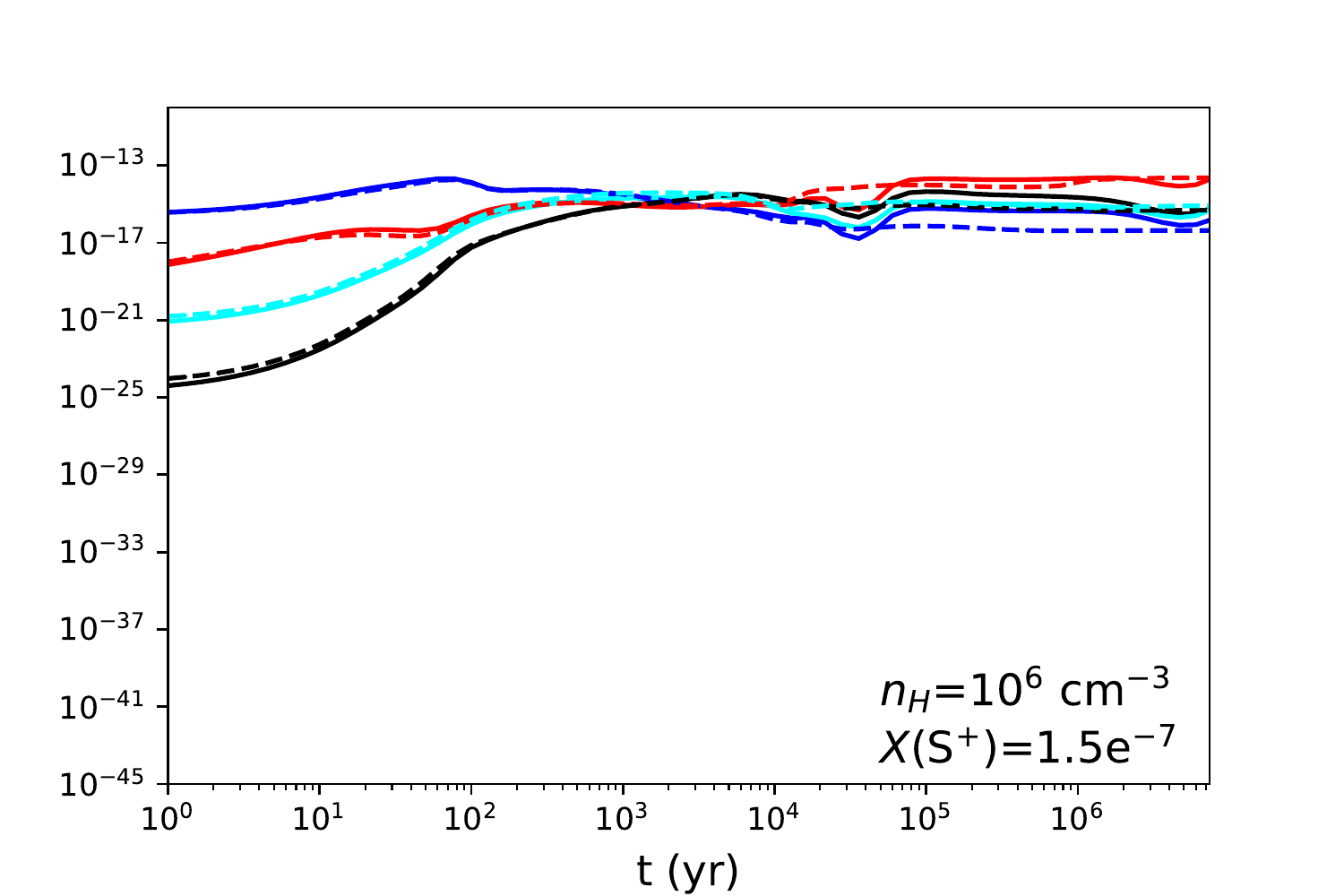}  
\hspace{-0.5cm}
\includegraphics[scale=0.4, angle=0]{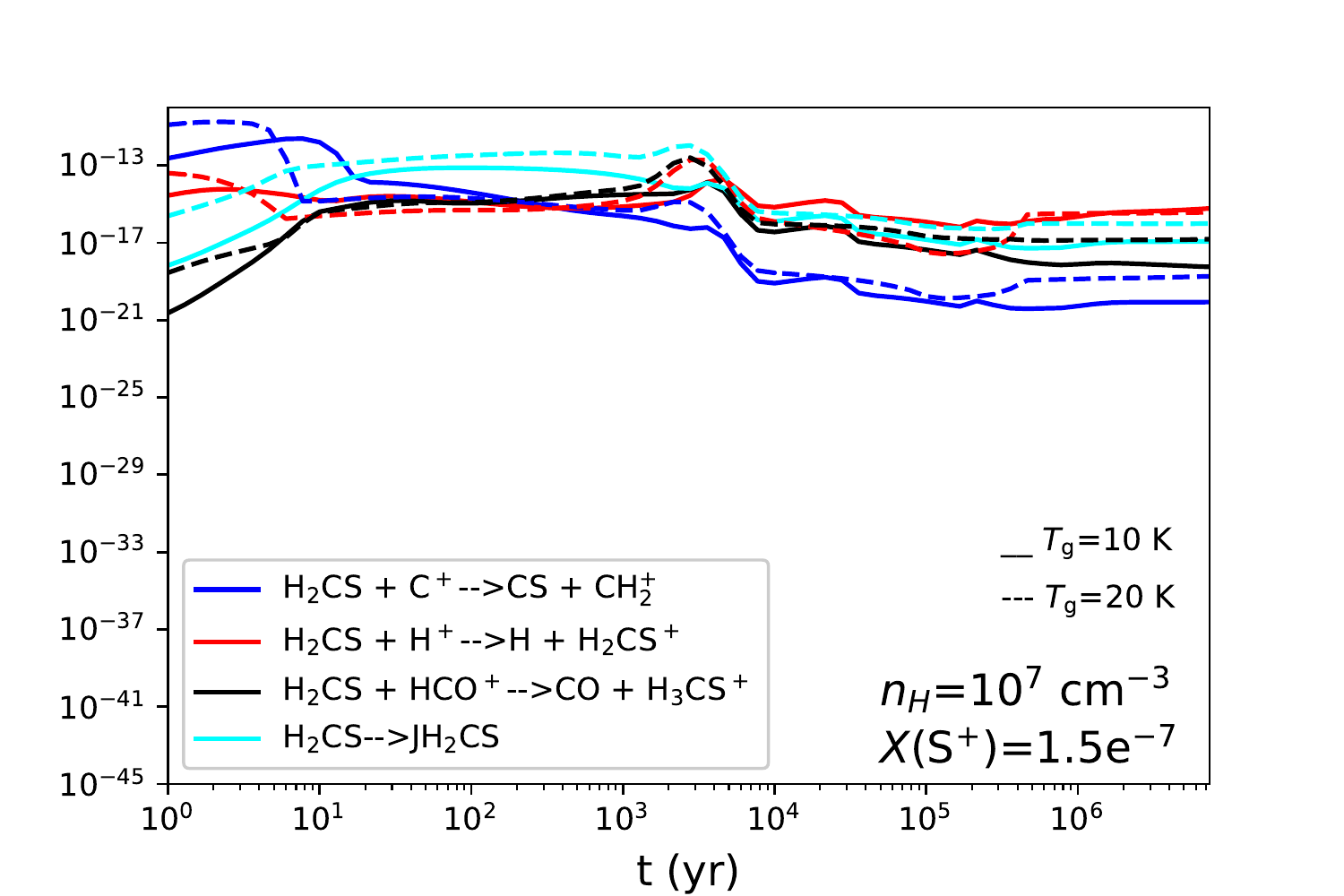}  
\hspace{-0.5cm}
\\
\caption{Main chemical reaction rates forming (top) and destroying (bottom) H$_2$CS.}
\label{figure:H2CS_rates}
\end{figure*}

Regarding Orion, we detect H$_2$CS in the three cores of the sample ($N$=1.9-3.6$\times$10$^{12}$ cm$^{-2}$, Table \ref{table:column_densities_Radex}), but we  only detect HDCS in ORI-C2-3. This is probably due to the more quiescent properties of OMC-4 (where ORI-C2 is located) compared to the other two regions harbouring the cores ORI-C1 and ORI-C3, as mentioned in Sect. \ref{Deuterated_thioformaldehyde}. The deuterium fractionation in ORI-C2-3 is as low as 0.1, which, for a high-density ($n_{\mathrm{H}}$$\geq$10$^6$ cm$^{-3}$) model characteristic of the Orion region, is reproduced at $t$$\lesssim$0.5 Myr.

\subsection{Thioformaldehyde formation and destruction}

Out of more than 200 molecules detected in space, about 10\% contain sulphur. In spite of the use of astronomical observations coupled with laboratory experiments and astrochemical modelling to study interstellar sulphur chemistry, results have been unsuccessful in predicting fractional abundances of species, such as carbonyl monosulfide (CS), falling short by up to two orders of magnitude compared to astronomical observations \citep[e.g.][]{Esplugues2014, Laas2019}. 
In this way, the analysis of formation mechanisms of (organo)sulphur molecules in different astrochemical environments is key to shed light on the origin and evolution of sulphur in our Galaxy.

\begin{figure*}
\centering
\hspace{-0.5cm}
\includegraphics[scale=0.4, angle=0]{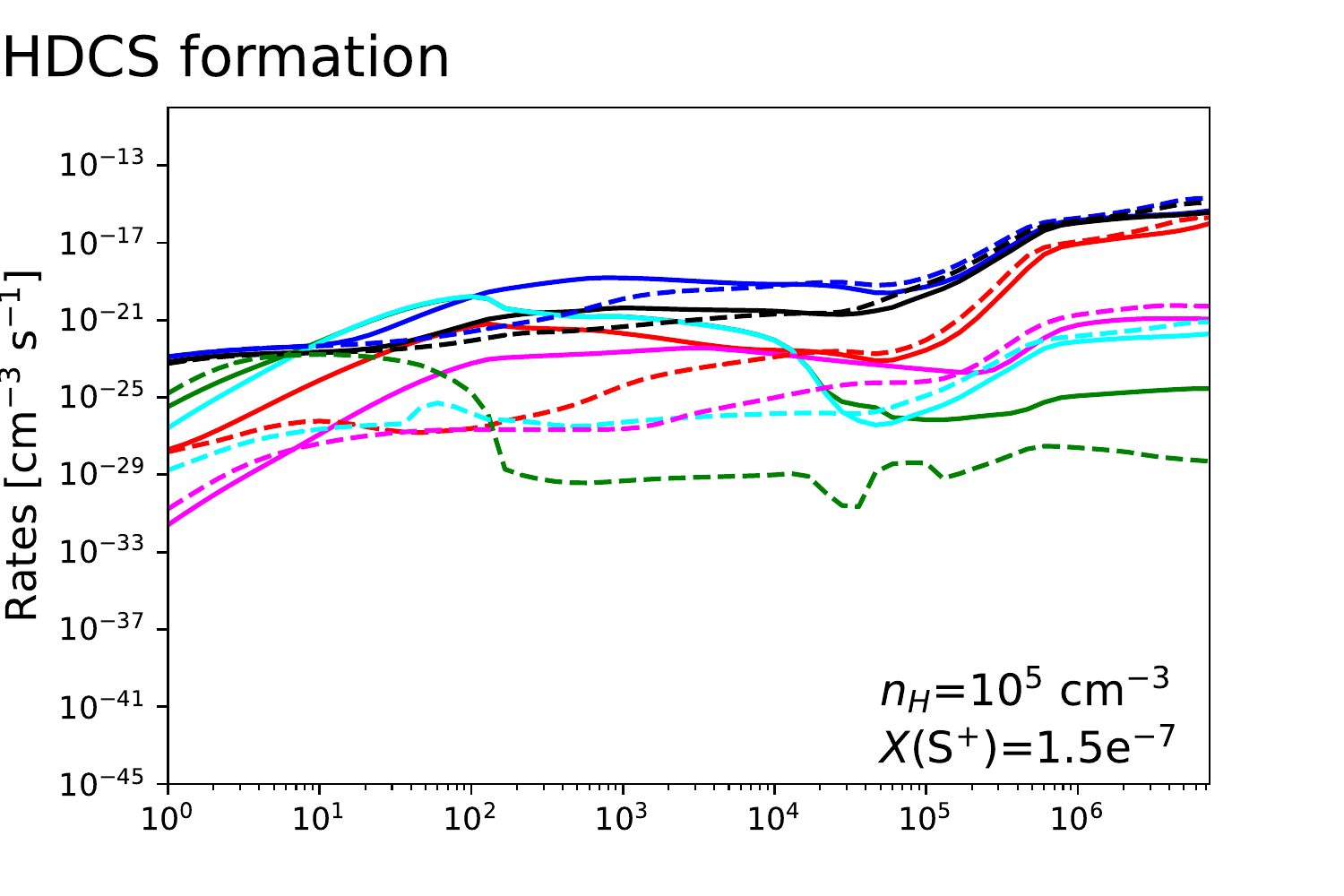}  
\hspace{-0.5cm}
\includegraphics[scale=0.4, angle=0]{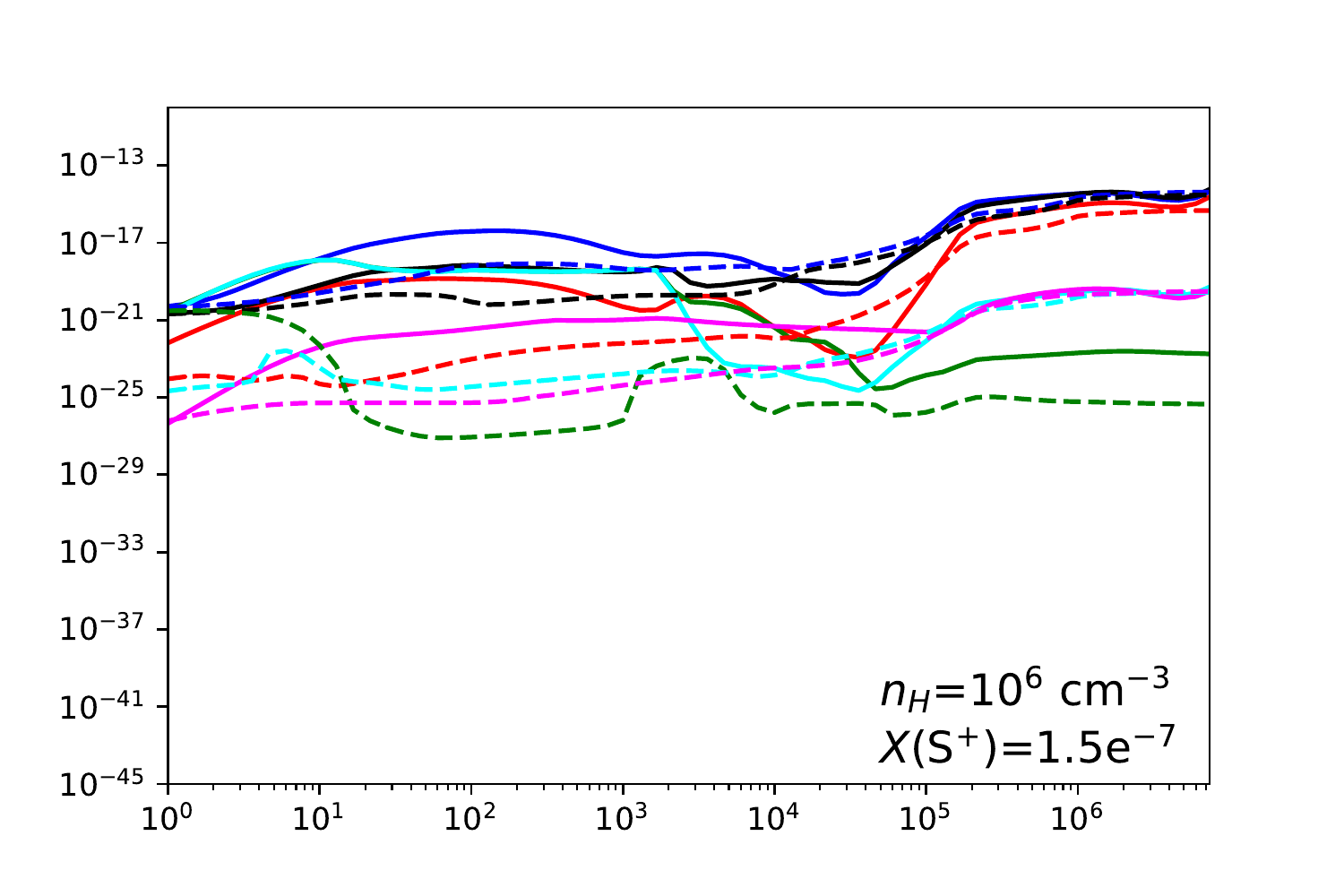} 
\hspace{-0.5cm}
\includegraphics[scale=0.4, angle=0]{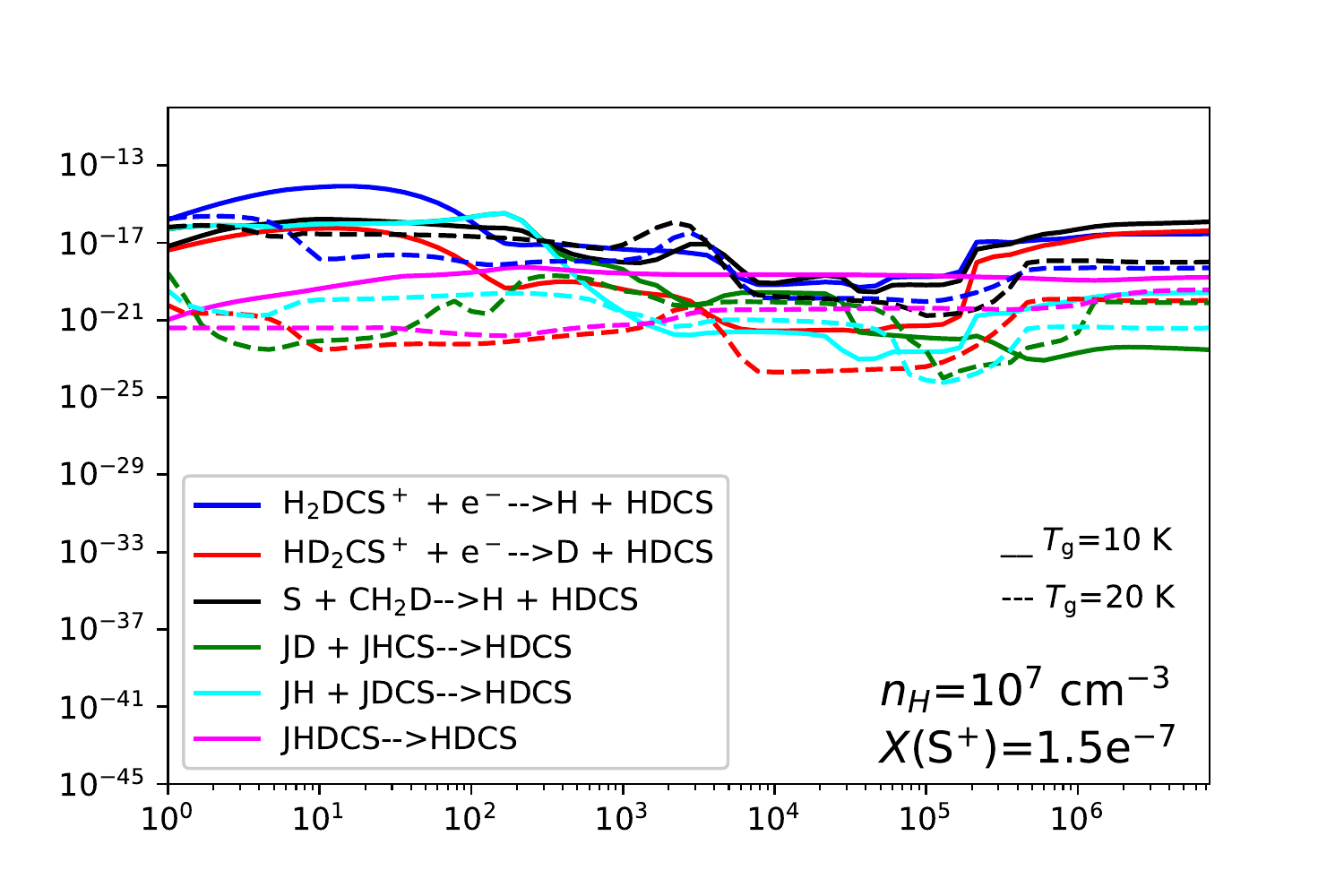}  
\hspace{-0.5cm}

\vspace{0.0cm}
\includegraphics[scale=0.4, angle=0]{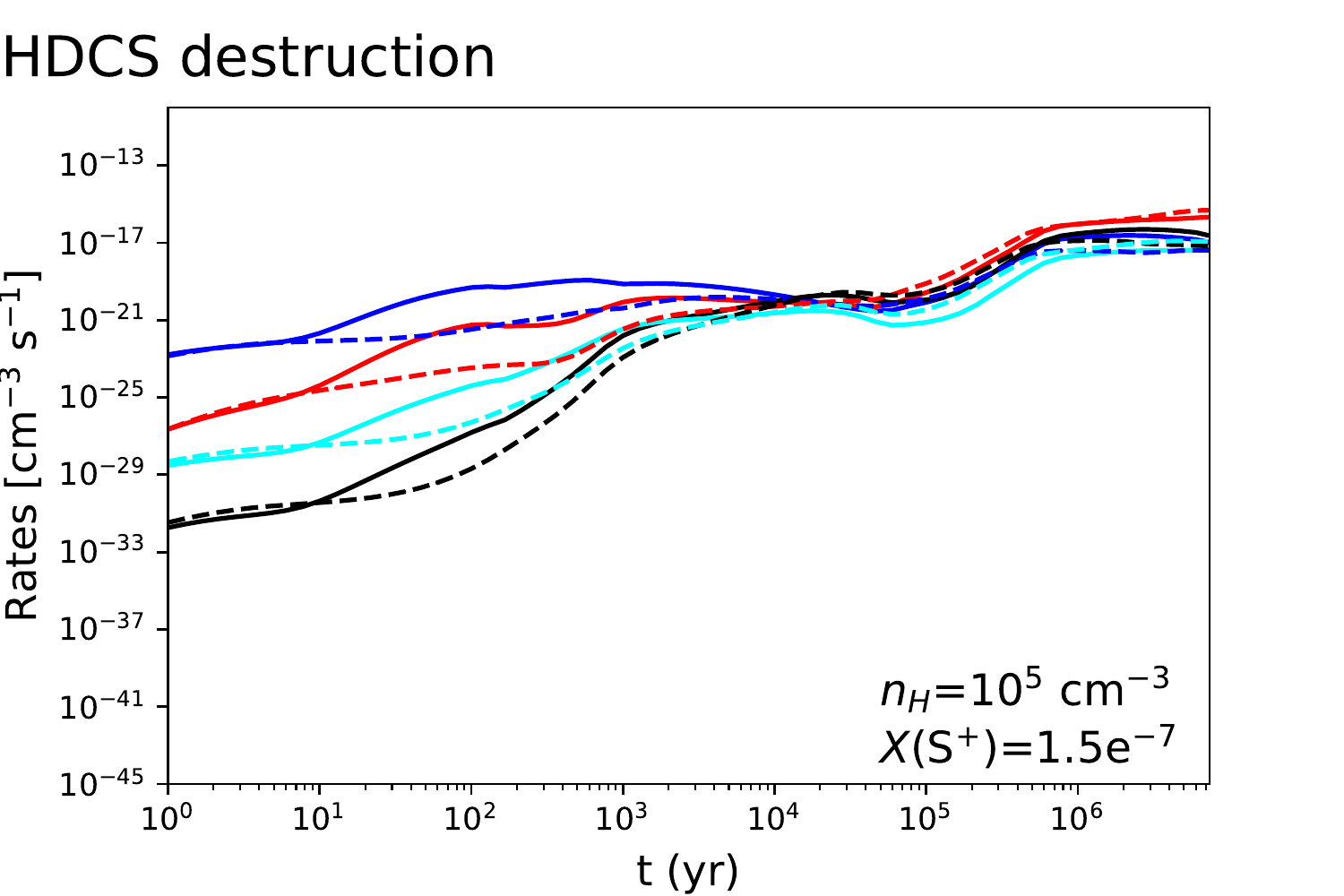}  
\hspace{-0.5cm}
\includegraphics[scale=0.4, angle=0]{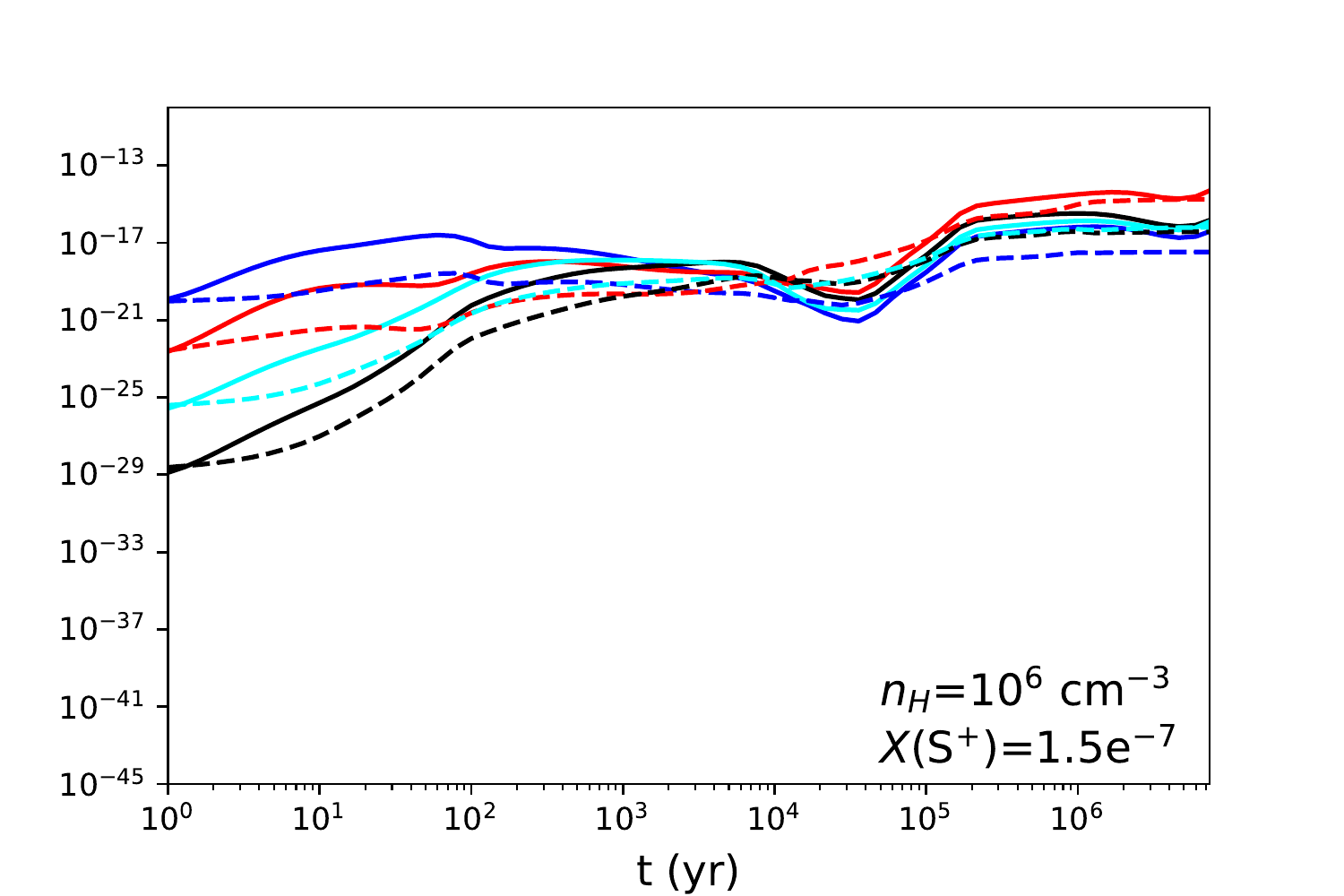}  
\hspace{-0.5cm}
\includegraphics[scale=0.4, angle=0]{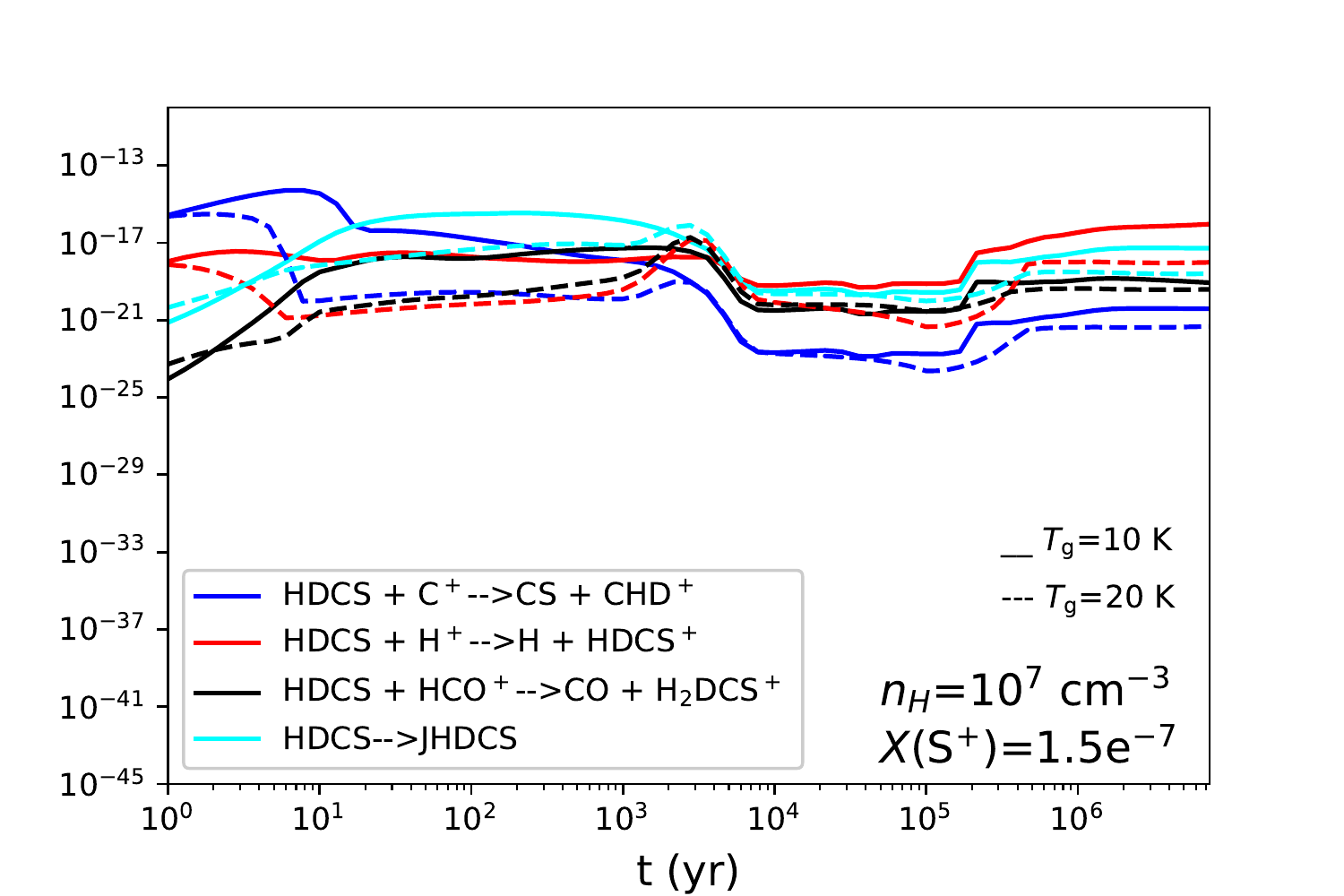}  
\hspace{-0.5cm}
\\
\caption{Main chemical reaction rates forming (top) and destroying (bottom) HDCS.}
\label{figure:HDCS_rates}
\end{figure*}

\begin{figure*}
\centering
\hspace{-0.5cm}
\includegraphics[scale=0.4, angle=0]{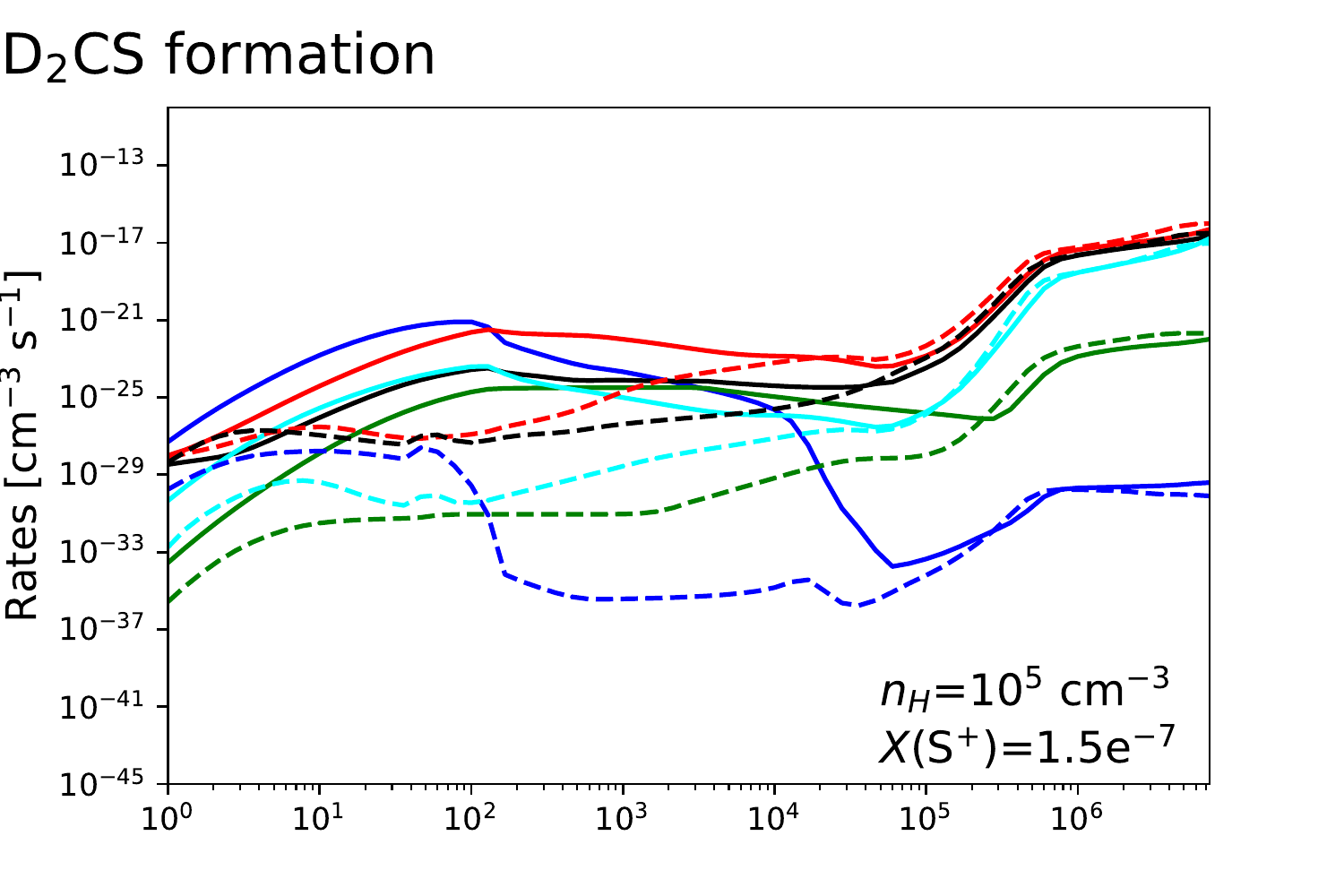}  
\hspace{-0.5cm}
\includegraphics[scale=0.4, angle=0]{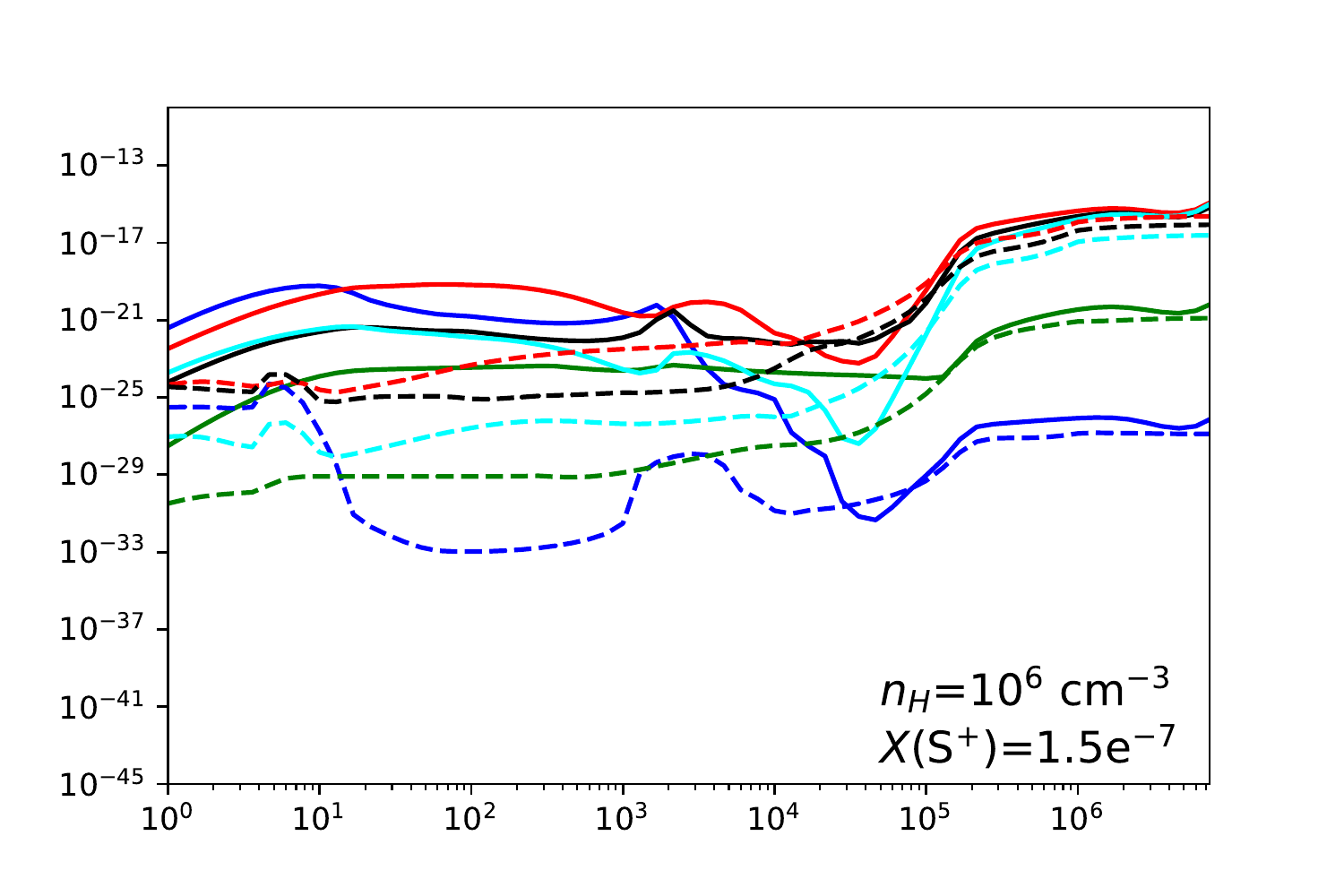} 
\hspace{-0.5cm}
\includegraphics[scale=0.4, angle=0]{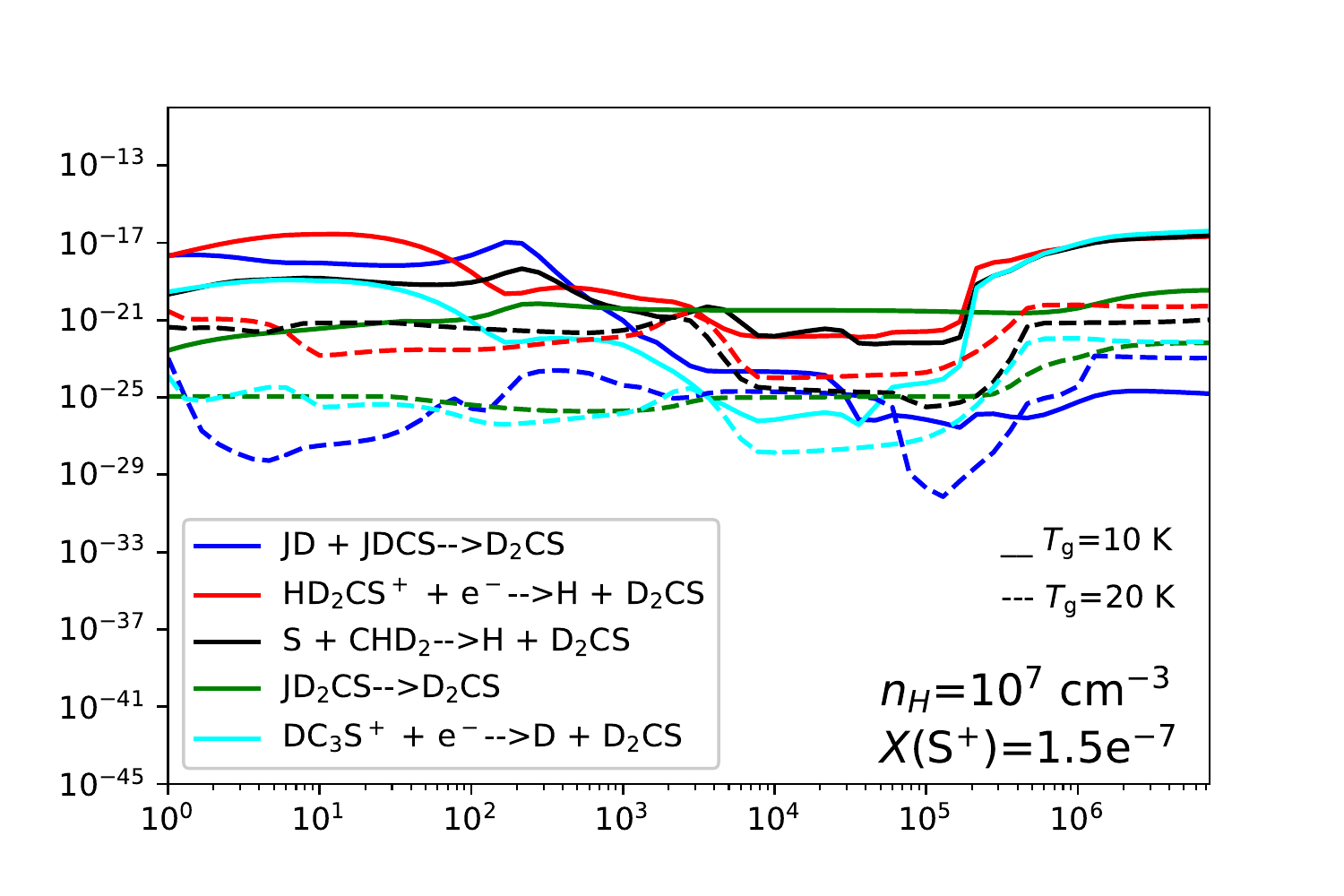}  
\hspace{-0.5cm}

\vspace{0.0cm}
\includegraphics[scale=0.4, angle=0]{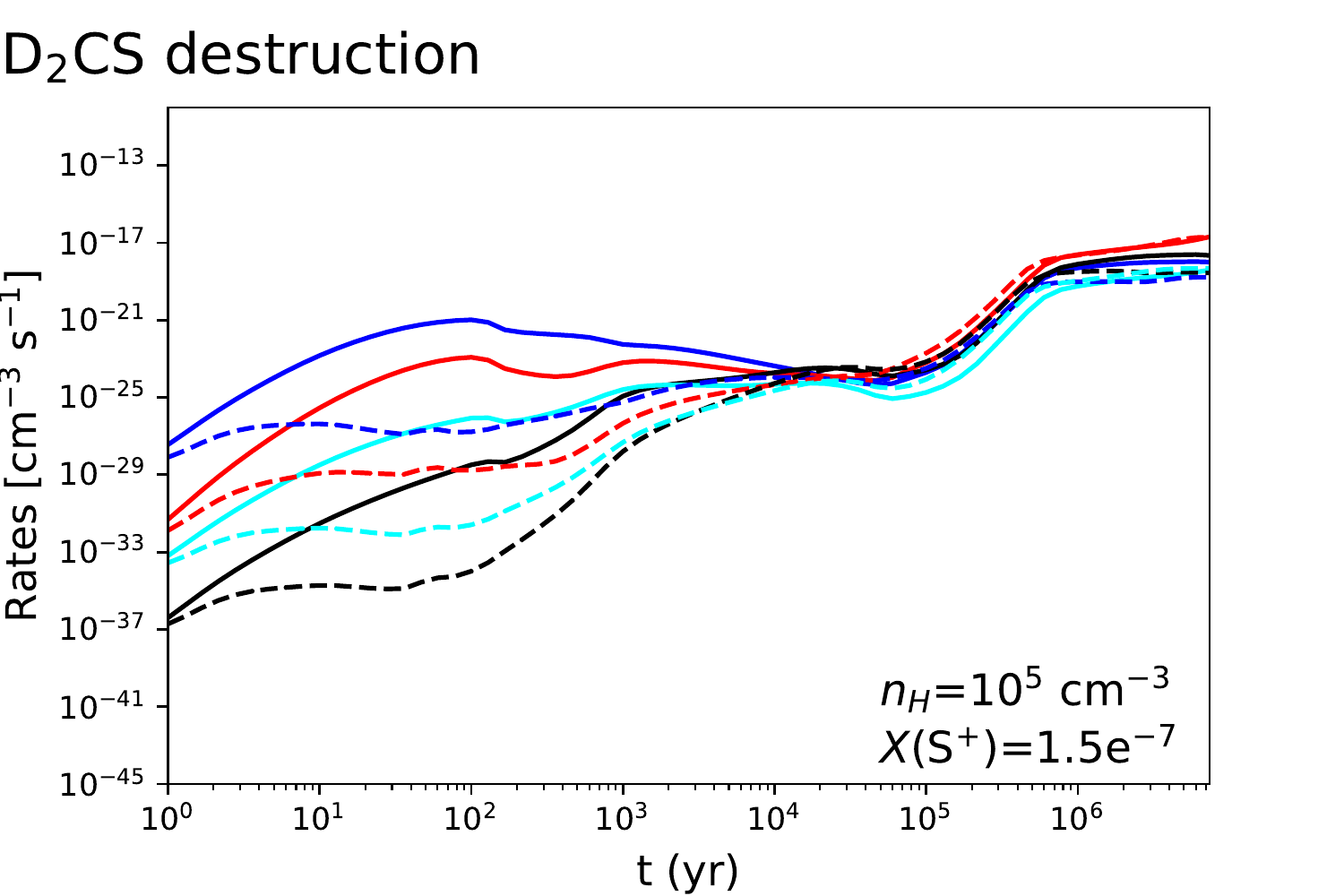}  
\hspace{-0.5cm}
\includegraphics[scale=0.4, angle=0]{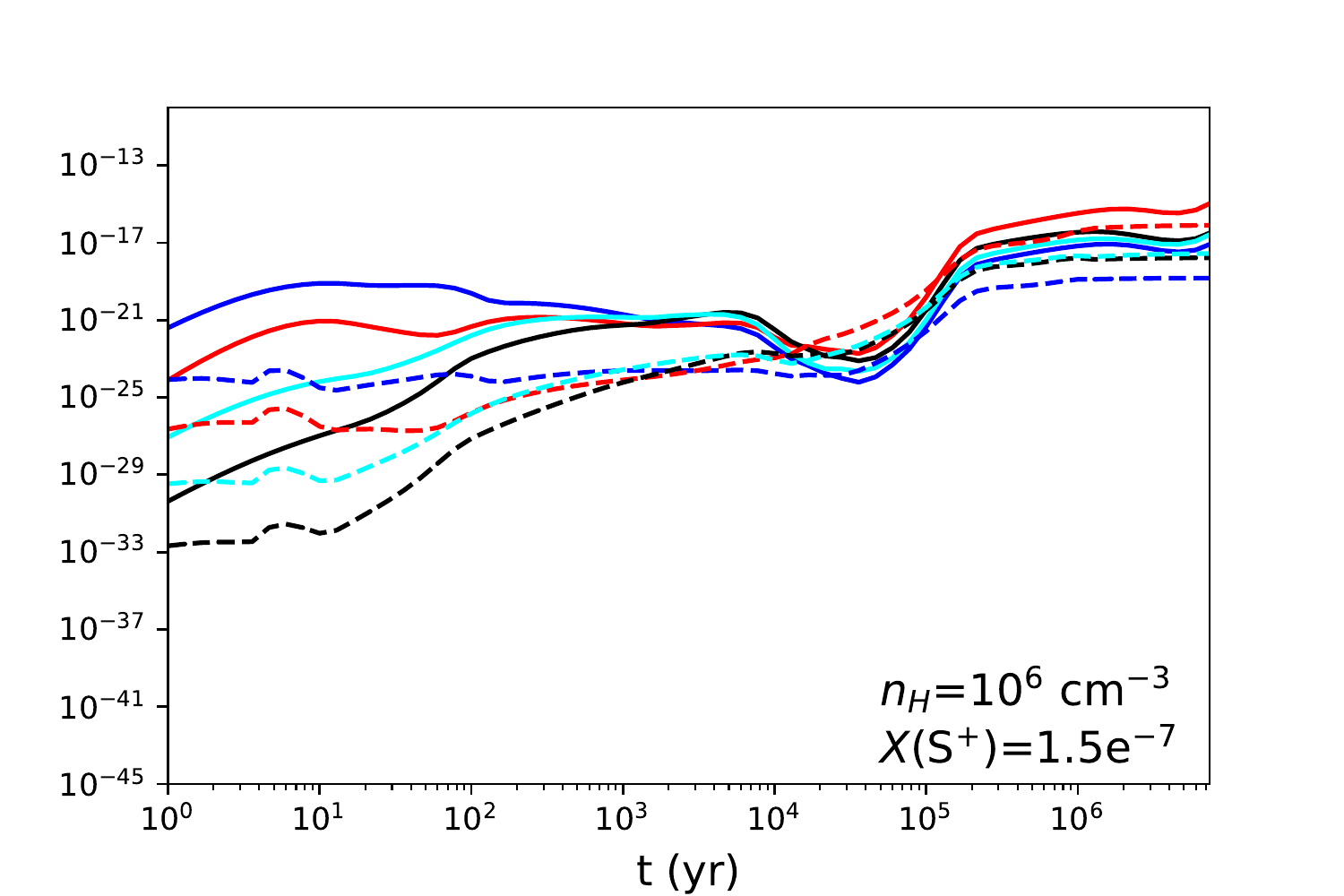}  
\hspace{-0.5cm}
\includegraphics[scale=0.4, angle=0]{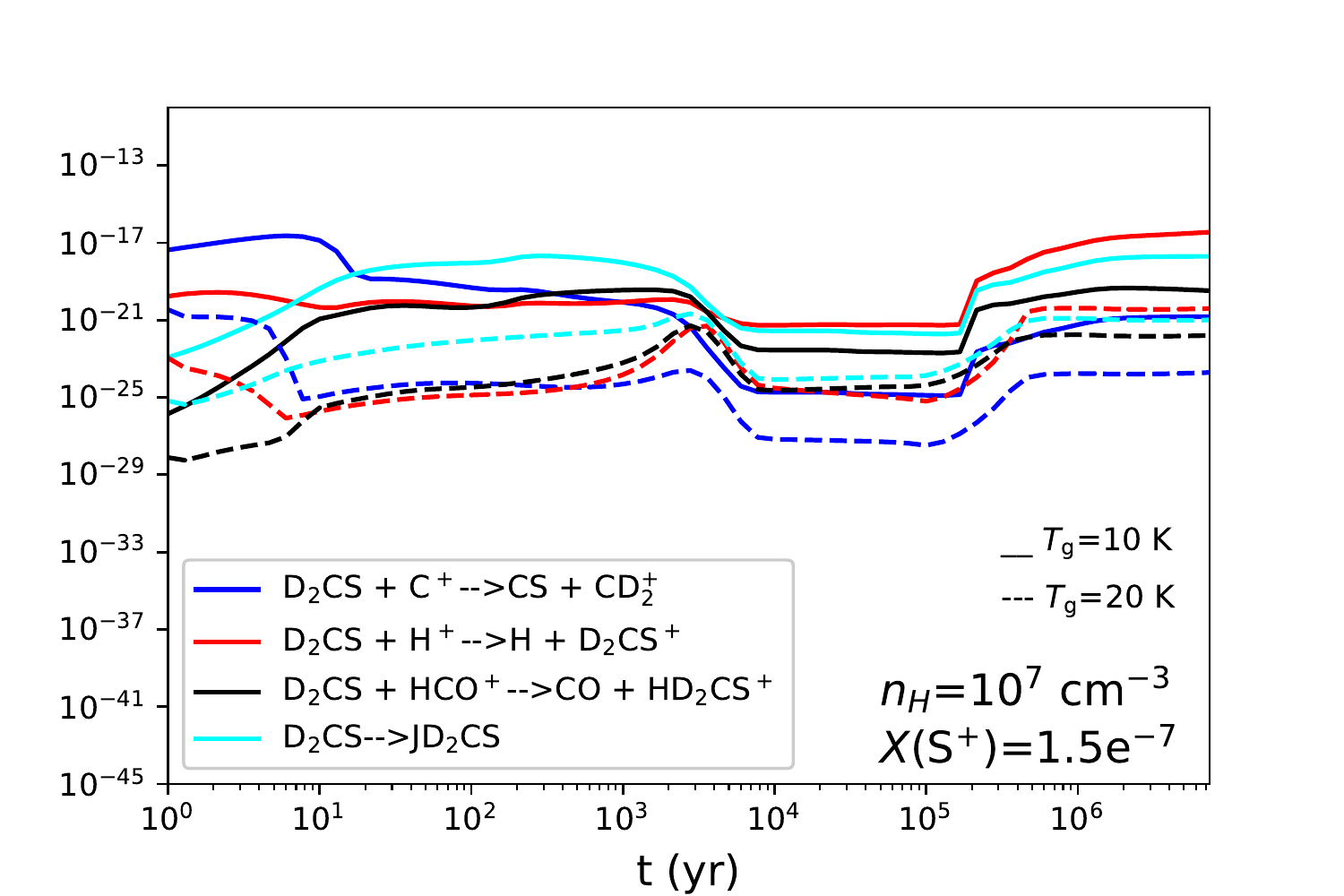}  
\hspace{-0.5cm}
\\
\caption{Main chemical reaction rates forming (top) and destroying (bottom) D$_2$CS.}
\label{figure:D2CS_rates}
\end{figure*}

Figures \ref{figure:H2CS_rates}-\ref{figure:D2CS_rates} show chemical rates for the main chemical reactions forming and destroying H$_2$CS, HDCS, and D$_2$CS calculated with the Nautilus code, considering a sulphur fractional abundance of 1.5$\times$10$^{-7}$, core temperatures of 10 K and 20 K, and densities ranging from 10$^{5}$ to 10$^{7}$ cm$^{-3}$. In the case of H$_2$CS (Fig. \ref{figure:H2CS_rates}), the main gas-phase routes (i.e. the reactions with the highest reaction rates) forming this species at any evolutionary time are

\begin{equation}
\mathrm{S + CH_3 \rightarrow H + H_2CS} 
\end{equation}

\noindent and

\begin{equation}
\mathrm{H_3CS^+ + e^- \rightarrow H + H_2CS} 
\end{equation}

\noindent in agreement with \citet{Prasad1982} and \citet{Laas2019}. Nevertheless, we note that while the ion H$_3$CS$^+$ is mainly formed through the reaction between CH$_4$ and S$^+$ at $t$$\lesssim$10$^4$ yr, for longer times H$_3$CS$^+$ is formed in turn through H$_2$CS.
In the case of a very high-density (10$^{7}$ cm$^{-3}$) starless core (right top panel) and $T$$_{\mathrm{g}}$=20 K (dashed lines), the surface reactions

\begin{equation}
\mathrm{JH + JHCS \rightarrow H_2CS} 
\end{equation}

\noindent and 

\begin{equation}
\mathrm{JH_2CS \rightarrow H_2CS} 
\end{equation}

\noindent also become dominant at 5$\times$10$^4$$\lesssim$$t$$\lesssim$5$\times$10$^5$ yr. These chemical and thermal desorption reactions barely influence instead the formation of H$_2$CS when the density of the core is $\leq$10$^6$ cm$^{-3}$. 
Regarding destruction, H$_2$CS is mainly destroyed through its reaction with C$^+$ during the early timescales ($t$$\lesssim$10$^4$, $\lesssim$10$^3$, and $\lesssim$10$^2$ yr for $n_{\mathrm{H}}$=10$^5$, 10$^6$, and 10$^7$ cm$^{-3}$, respectively). For longer times, H$_2$CS depletion becomes one of the main destruction routes, especially at very high densities, together with the reaction between H$_2$CS and H$^+$.

Formation and destruction reaction rates for HDCS are shown in Fig. \ref{figure:HDCS_rates}. HDCS is mainly formed in all the considered cases by the reactions

\begin{equation}
\mathrm{H_2DCS^+ + e^- \rightarrow H + HDCS} 
\end{equation}

\noindent and 

\begin{equation}
\mathrm{S + CH_2D \rightarrow H + HDCS}. 
\end{equation}

\noindent Nevertheless, in the low temperature case (10 K), the chemical desorption

\begin{equation}
\mathrm{JH + JDCS \rightarrow HDCS} 
\end{equation}

\noindent also plays an important role in forming HDCS at $t$$<$10$^3$ yr. Regarding HDCS destruction, its reaction with C$^+$ represents the main destroying mechanism for $t$$\lesssim$10$^4$ yr with $n$$\leq$10$^6$ cm$^{-3}$. At longer times, HDCS is mainly destroyed by H$^+$. In a very high-density core (right bottom panel), depletion of HDCS becomes also important removing this deuterated species from the gas phase.

With respect to the double deuterated thioformaldehyde, for an early time ($t$$\lesssim$10$^3$ yr) and a cold core ($T$$_{\mathrm{g}}$=10 K), D$_2$CS is efficiently formed both on dust grains (through chemical desorption from the reaction between JD and JDCS) and through the gas-phase reaction

\begin{equation}
\mathrm{HD_2CS^+ + e^- \rightarrow H + D_2CS}, 
\end{equation}

\noindent while for a warmer core ($T$$_{\mathrm{g}}$=20 K) the chemical desorption becomes less efficient in forming D$_2$CS, in favour of the neutral–neutral gas-phase reaction

\begin{equation}
\mathrm{S + CHD_2 \rightarrow H + D_2CS}. 
\end{equation}

\noindent At late timescales ($>$10$^5$ yr), D$_2$CS is mainly formed in the gas phase through the reactions

\begin{equation}
\mathrm{HD_2CS^+ + e^- \rightarrow H + D_2CS}, 
\end{equation}

\begin{equation}
\mathrm{S + CHD_2 \rightarrow H + D_2CS}, 
\end{equation}

\noindent and

\begin{equation}
\mathrm{DC_3S^+ + e^- \rightarrow D + D_2CS}. 
\end{equation}

\noindent By contrast, surface reactions are irrelevant forming D$_2$CS at long times ($>$10$^5$ yr), independently on the temperature and density of the starless core. Regarding D$_2$CS destruction, as in the case of HDCS and H$_2$CS, the ions C$^+$ and H$^+$ are the main destroyers of D$_2$CS. D$_2$CS depletion has also an important role, especially in cores with very high density, where it becomes the main D$_2$CS destroying mechanism during a period of $\sim$10$^4$ years. 

All these results show that (deuterated) thioformaldehyde is mainly formed through gas-phase reactions, where the ions H$_3$CS$^+$, H$_2$DCS$^+$, and HD$_2$CS$^+$ (formed in turn from H$_2$D$^+$ and D$_2$H$^+$), as well as the neutral precursors CH$_3$, CH$_2$D, and CHD$_2$, have a key role in the formation of H$_2$CS, HDCS, and D$_2$CS, respectively.

\section{Summary and conclusions}
\label{section:summary}

Within the IRAM 30m Large Program GEMS, we have carried out a comprehensive observational and theoretical study of thioformaldehyde and its deuterated counterparts in a sample of starless cores in filaments of the nearby star-forming regions Taurus, Perseus, and Orion. These regions have different degrees of star formation activity, and therefore different physical conditions, providing a possibility to explore the effect of environment on gas chemistry. 

Using the molecular excitation and radiative transfer code RADEX, we have obtained column densities and fractional abundances for H$_2$CS, HDCS, and D$_2$CS in each core of the sample, also providing upper limits for cases where these species were not detected. We have also derived deuterium fractionation ratios for both single and double deuterated thioformaldehyde. The obtained ratios are up to four orders of magnitude higher than the cosmic D/H ratio. This result is similar to the ones obtained in other starless cores, such as L\,183, and it is due to the low temperatures of these regions, as well as the large fraction of neutral heavy species that freeze-out onto dust grains, favouring the formation of deuterated species. We have also found that the HDCS/H$_2$CS ratio is similar to that found in regions characterised by higher temperatures, such as Class 0 objects, high-mass star-forming clouds, and PDRs. In these warm regions, the deuterium fractionation is driven by a chemistry different from that of cold regions, with D atoms being transferred to molecules by CH$_2$D$^+$ instead of H$_2$D$^+$. 

Since the deuterium fractionation ratio can be considered as an evolutionary tracer, we have also used it to derive the chemical evolution stage of the starless core sample through the Nautilus gas-grain chemical code, which computes the evolution of chemical abundances as a function of time. The comparison of model results with observations reveals that the north region of the B\,213 filament in Taurus, harbouring cores such as C1, C2, and C6, is more chemically evolved ($t$$\gtrsim$1 Myr) than other regions located more to the south of the filament ($t$$<$1 Myr), where we find the cores C7, C10, and C16. In Perseus, the results indicate that the north-eastern part presents an earlier chemical evolutionary stage ($t$$<$1 Myr) than the south-western part ($t$$\geq$1-5 Myr). In Orion, for the only core where deuterated thioformaldehyde is detected, we have obtained a chemical age of $t$$\lesssim$0.5 Myr.  

The theoretical study of deuterium fractionation ($D$$_{\mathrm{frac}}$) evolution has not only allowed us to analyse how the ratios HDCS/H$_2$CS and D$_2$CS/H$_2$CS vary with time, but also with other parameters, such as density, temperature, initial sulphur depletion, and the CR ionisation rate. Our results confirm a strong dependence of $D$$_{\mathrm{frac}}$ with time since it progressively increases with the evolution of the starless cores. This makes HDCS/H$_2$CS and D$_2$CS/H$_2$CS powerful tools for deriving the chemical evolution of these regions. However, they cannot be used to derive the temperature of the starless cores since their value presents a double peak in two temperature ranges ($T$$_{\mathrm{gas}}$=$\sim$7-11 K and $T$$_{\mathrm{gas}}$=$\sim$15-19 K).  
On the other hand, our results also show that both deuterium fractionation ratios increase with the density of the region and with the degree of sulphur depletion. In particular, we find that decreasing the sulphur initial abundance from 1.5$\times$10$^{-5}$ (the solar elemental sulphur fractional abundance) to 1.5$\times$10$^{-7}$ leads to an increase by a factor of up to $\sim$1.5 and $\sim$2.5 in the maximum values of the HDCS/H$_2$CS and D$_2$CS/H$_2$CS ratios, respectively. In any case, the comparison with the observations of the starless cores of our sample reveals that they are reproduced when the initial sulphur fractional abundance is as low as $\leq$1.5$\times$10$^{-6}$ (i.e. at least one order of magnitude lower than the solar elemental sulphur abundance), in agreement with previous GEMS results \citep[e.g.][]{Bulut2021, Navarro-Almaida2021}. Regarding the CR ionisation rate, we find that its increase by one order of magnitude slows down the deuterium fraction evolution by a few million years.  

Finally, we have also studied the main formation and destruction mechanisms of H$_2$CS, HDCS, and D$_2$CS using the Nautilus code. We have obtained that they are mainly formed through gas-phase reactions (double-replacement reactions and neutral–neutral displacement reactions). By contrast, grain chemistry (depletion) has an important role in the destruction of these species, especially at $t$$>$10$^4$ yr.

We finally stress that this work has allowed the sulphur chemistry in cold regions to be studied in detail through thioformaldehyde observations. We have also demonstrated the power of using its deuterated counterparts for the analysis of the properties and chemical evolution of starless cores.

\begin{acknowledgements}

We thank the Spanish MICINN for funding support from PID2019-106235GB-I00. L.M. acknowledges the financial support of DAE and DST-SERB research grants (SRG/2021/002116 and MTR/2021/000864) of the Government of India.
  
\end{acknowledgements}

\begin{appendix}

\onecolumn

\section{Tables and figures}
\label{}

\clearpage

\begin{table}
\caption{Observational parameters.}             
\begin{tabular}{llll}     
\hline\hline       
Telescope & Observed frequency range (GHx) & rms (mK) & HPBW($\arcsec$)  \\ 
\hline 
IRAM 30m  & 85.0-86.8      & 10-20     & 29 \\
          & 88.3-91.5      & 10-20     & 27  \\
          & 92.0-94.8      & 10-20     & 26 \\
          & 95.3-97.0      & 10-20     & 24 \\
          & 97.6-99.4      & 10-20     & 23 \\
          & 101.0-103.0    & 10-20     & 21  \\
          & 105.4-107.2    & 10-20     & 20 \\
          & 108.7-110.5    & 10-20     & 19 \\ 
          & 133.7-135.5    & 10-20     & 18  \\
          & 137.0-138.8    & 10-20     & 17 \\
          & 143.2-145.0    & 10-20     & 16  \\
          & 146.4-148.3    & 10-20     & 16  \\
          & 158.8-160.7    & 20-30     & 15  \\
          & 162.1-163.9    & 20-30     & 15  \\
          & 167.2-169.0    & 20-30     & 14  \\
          & 170.5-172.3    & 20-30     & 14  \\
\hline 
\label{table:observational_parameters}                 
\end{tabular}
\end{table}

\begin{table*}
\caption{H$_{2}$CS parameters from Gaussian fits.}
\begin{center}
\begin{tabular}{lllll||llll}
\hline 
\hline
Core & Transition & \multicolumn{3}{c}{o-H$_{2}$CS} & Transition & \multicolumn{3}{c}{p-H$_{2}$CS}  \\ 
& $J_{K,k}-J'_{K',k'}$  & $V$$_{\mathrm{LSR}}$ (km s$^{-1}$) & $\Delta$$V$ (km s$^{-1}$) & $T$$_{\mathrm{MB}}$ (K) & $J_{K,k}-J'_{K',k'}$ &
$V$$_{\mathrm{LSR}}$ (km s$^{-1}$) & $\Delta$$V$ (km s$^{-1}$) & $T$$_{\mathrm{MB}}$ (K) \\
\hline
B\,213-C1-1 & 3$_{1}$$_{,}$$_{3}$-2$_{1}$$_{,}$$_{2}$   & 5.89$\pm$0.01 & 0.38$\pm$0.08  & 0.5$\pm$0.1 & 4$_{0}$$_{,}$$_{4}$-3$_{0}$$_{,}$$_{3}$               & 6.02$\pm$0.01 & 0.27$\pm$0.05             & 0.37$\pm$0.07   \\
& 4$_{1}$$_{,}$$_{4}$-3$_{1}$$_{,}$$_{3}$             & 6.23$\pm$0.01   & 0.27$\pm$0.05  & 0.37$\pm$0.07 & 5$_{0}$$_{,}$$_{5}$-4$_{0}$$_{,}$$_{4}$             & 5.63$\pm$0.03   & 0.35$\pm$0.07           & 0.08$\pm$0.02    \\

B\,213-C2-1 & 3$_{1}$$_{,}$$_{3}$-2$_{1}$$_{,}$$_{2}$   & 6.98$\pm$0.01 & 0.42$\pm$0.08  & 0.15$\pm$0.03 & 4$_{0}$$_{,}$$_{4}$-3$_{0}$$_{,}$$_{3}$               & 7.11$\pm$0.04   & 0.44$\pm$0.09           & 0.06$\pm$0.01    \\
& 4$_{1}$$_{,}$$_{4}$-3$_{1}$$_{,}$$_{3}$             & 7.28$\pm$0.03   & 0.35$\pm$0.07  & 0.07$\pm$0.01 & 5$_{0}$$_{,}$$_{5}$-4$_{0}$$_{,}$$_{4}$               & 7.02$\pm$0.05   & 0.28$\pm$0.06           & 0.04$\pm$0.01    \\

B\,213-C5-1 & 3$_{1}$$_{,}$$_{3}$-2$_{1}$$_{,}$$_{2}$   & 6.32$\pm$0.01 & 0.6$\pm$0.1  & 0.27$\pm$0.05 & 4$_{0}$$_{,}$$_{4}$-3$_{0}$$_{,}$$_{3}$               & 6.42$\pm$0.03   & 0.7$\pm$0.1           & 0.08$\pm$0.02  \\
& 4$_{1}$$_{,}$$_{4}$-3$_{1}$$_{,}$$_{3}$             & 6.60$\pm$0.02   & 0.5$\pm$0.1  & 0.10$\pm$0.02 & 5$_{0}$$_{,}$$_{5}$-4$_{0}$$_{,}$$_{4}$               & 6.94$\pm$0.05   & $<$0.25$\pm$0.05          & $<$0.07$\pm$0.01   \\

B\,213-C6-1 & 3$_{1}$$_{,}$$_{3}$-2$_{1}$$_{,}$$_{2}$   & 6.90$\pm$0.01 & 0.36$\pm$0.07 & 0.30$\pm$0.06 & 4$_{0}$$_{,}$$_{4}$-3$_{0}$$_{,}$$_{3}$               & 6.96$\pm$0.02   & 0.31$\pm$0.06           & 0.09$\pm$0.02     \\
& 4$_{1}$$_{,}$$_{4}$-3$_{1}$$_{,}$$_{3}$             & 6.72$\pm$0.02   & 0.34$\pm$0.06  & 0.13$\pm$0.03 & 5$_{0}$$_{,}$$_{5}$-4$_{0}$$_{,}$$_{4}$               & 6.78$\pm$0.05   & $<$0.21$\pm$0.04  & $<$0.05$\pm$0.01  \\

B\,213-C7-1 & 3$_{1}$$_{,}$$_{3}$-2$_{1}$$_{,}$$_{2}$   & 6.80$\pm$0.01 & 0.39$\pm$0.08  & 0.45$\pm$0.09 & 4$_{0}$$_{,}$$_{4}$-3$_{0}$$_{,}$$_{3}$               & 6.83$\pm$0.01   & 0.25$\pm$0.05           & 0.23$\pm$0.05       \\
& 4$_{1}$$_{,}$$_{4}$-3$_{1}$$_{,}$$_{3}$             & 6.60$\pm$0.01   & 0.31$\pm$0.06  & 0.26$\pm$0.06 & 5$_{0}$$_{,}$$_{5}$-4$_{0}$$_{,}$$_{4}$               & 7.14$\pm$0.09   & $<$0.35$\pm$0.07  & $<$0.04$\pm$0.01   \\

B\,213-C10-1 & 3$_{1}$$_{,}$$_{3}$-2$_{1}$$_{,}$$_{2}$  & 6.74$\pm$0.02   & 0.45$\pm$0.09  & 0.13$\pm$0.03 & 4$_{0}$$_{,}$$_{4}$-3$_{0}$$_{,}$$_{3}$             & 6.72$\pm$0.03   & 0.23$\pm$0.05  & 0.05$\pm$0.01   \\
& 4$_{1}$$_{,}$$_{4}$-3$_{1}$$_{,}$$_{3}$              & 6.01$\pm$0.03   & 0.21$\pm$0.04  & 0.07$\pm$0.01 &  5$_{0}$$_{,}$$_{5}$-4$_{0}$$_{,}$$_{4}$             & 6.02$\pm$0.04   & $<$0.18$\pm$0.04      & $<$0.06$\pm$0.01   \\

B\,213-C12-1 & 3$_{1}$$_{,}$$_{3}$-2$_{1}$$_{,}$$_{2}$  & 6.64$\pm$0.02   & 0.42$\pm$0.08  & 0.09$\pm$0.02 & 4$_{0}$$_{,}$$_{4}$-3$_{0}$$_{,}$$_{3}$             & 6.65$\pm$0.04   & 0.24$\pm$0.05  & 0.06$\pm$0.01    \\
& 4$_{1}$$_{,}$$_{4}$-3$_{1}$$_{,}$$_{3}$              & 5.82$\pm$0.04    & 0.20$\pm$0.04  & 0.06$\pm$0.01 &   5$_{0}$$_{,}$$_{5}$-4$_{0}$$_{,}$$_{4}$             & 6.58$\pm$0.02   & $<$0.12$\pm$0.02  & $<$0.07$\pm$0.01   \\

B\,213-C16-1 & 3$_{1}$$_{,}$$_{3}$-2$_{1}$$_{,}$$_{2}$  & 6.64$\pm$0.01 & 0.41$\pm$0.08  & 0.6$\pm$0.1 & 4$_{0}$$_{,}$$_{4}$-3$_{0}$$_{,}$$_{3}$               & 6.77$\pm$0.01   & 0.39$\pm$0.08  & 0.22$\pm$0.04    \\
& 4$_{1}$$_{,}$$_{4}$-3$_{1}$$_{,}$$_{3}$               & 6.72$\pm$0.01 & 0.30$\pm$0.06  & 0.33$\pm$0.06 &  5$_{0}$$_{,}$$_{5}$-4$_{0}$$_{,}$$_{4}$            & 6.75$\pm$0.04   & $<$0.12$\pm$0.02  & $<$0.04$\pm$0.01   \\  
\hline
L\,1448-1 & 3$_{1}$$_{,}$$_{3}$-2$_{1}$$_{,}$$_{2}$     & 4.30$\pm$0.01 & 0.6$\pm$0.1  & 0.7$\pm$0.1 & 4$_{0}$$_{,}$$_{4}$-3$_{0}$$_{,}$$_{3}$               & 4.44$\pm$0.01 & 0.7$\pm$0.1  & 0.31$\pm$0.06    \\
& 4$_{1}$$_{,}$$_{4}$-3$_{1}$$_{,}$$_{3}$               & 4.42$\pm$0.01 & 0.6$\pm$0.1  & 0.35$\pm$0.07 & 5$_{0}$$_{,}$$_{5}$-4$_{0}$$_{,}$$_{4}$           & 4.79$\pm$0.03   & 0.41$\pm$0.08  & 0.25$\pm$0.05    \\ 
\hline
1333-C3-1 & 3$_{1}$$_{,}$$_{3}$-2$_{1}$$_{,}$$_{2}$   & 8.49$\pm$0.01 & 0.7$\pm$0.1  & 0.24$\pm$0.05 & 4$_{0}$$_{,}$$_{4}$-3$_{0}$$_{,}$$_{3}$                 & 8.65$\pm$0.01   & 0.42$\pm$0.08  & 0.11$\pm$0.02   \\
& 4$_{1}$$_{,}$$_{4}$-3$_{1}$$_{,}$$_{3}$             & 8.87$\pm$0.01   & 0.6$\pm$0.2  & 0.22$\pm$0.04 &  5$_{0}$$_{,}$$_{5}$-4$_{0}$$_{,}$$_{4}$               & 8.58$\pm$0.04   & 0.32$\pm$0.06  & 0.11$\pm$0.02    \\ 

1333-C4-1 & 3$_{1}$$_{,}$$_{3}$-2$_{1}$$_{,}$$_{2}$   & 7.67$\pm$0.01 & 1.0$\pm$0.2  & 0.7$\pm$0.1 & 4$_{0}$$_{,}$$_{4}$-3$_{0}$$_{,}$$_{3}$                 & 7.73$\pm$0.01   & 0.7$\pm$0.1   & 0.45$\pm$0.09    \\
& 4$_{1}$$_{,}$$_{4}$-3$_{1}$$_{,}$$_{3}$             & 7.79$\pm$0.08  & 0.9$\pm$0.2  & 0.6$\pm$0.1 &  5$_{0}$$_{,}$$_{5}$-4$_{0}$$_{,}$$_{4}$               & 7.83$\pm$0.05   & 0.6$\pm$0.1  & 0.17$\pm$0.03    \\

1333-C5-1 & 3$_{1}$$_{,}$$_{3}$-2$_{1}$$_{,}$$_{2}$   & 7.63$\pm$0.03   & 0.8$\pm$0.2  & 0.09$\pm$0.02 & 4$_{0}$$_{,}$$_{4}$-3$_{0}$$_{,}$$_{3}$               & 7.62$\pm$0.02   & 0.18$\pm$0.04  & 0.08$\pm$0.02    \\
& 4$_{1}$$_{,}$$_{4}$-3$_{1}$$_{,}$$_{3}$             & 7.00$\pm$0.04   & 0.45$\pm$0.09  & 0.11$\pm$0.02 &  5$_{0}$$_{,}$$_{5}$-4$_{0}$$_{,}$$_{4}$              & 6.49$\pm$0.05   & $<$0.10$\pm$0.02  & $<$0.11$\pm$0.02    \\

1333-C6-1 & 3$_{1}$$_{,}$$_{3}$-2$_{1}$$_{,}$$_{2}$   & 7.42$\pm$0.04   & 0.36$\pm$0.07  & 0.08$\pm$0.02 & 4$_{0}$$_{,}$$_{4}$-3$_{0}$$_{,}$$_{3}$               & 7.36$\pm$0.05   & $<$0.17$\pm$0.03  & $<$0.05$\pm$0.01  \\
& 4$_{1}$$_{,}$$_{4}$-3$_{1}$$_{,}$$_{3}$             & 7.06$\pm$0.04   & 0.31$\pm$0.06  & 0.08$\pm$0.02 & 5$_{0}$$_{,}$$_{5}$-4$_{0}$$_{,}$$_{4}$               & 7.50$\pm$0.09   & $<$0.15$\pm$0.03  & $<$0.5$\pm$0.01    \\

1333-C3-14 & 3$_{1}$$_{,}$$_{3}$-2$_{1}$$_{,}$$_{2}$  & 8.21$\pm$0.01 & 0.6$\pm$0.1  & 0.5$\pm$0.1 & 4$_{0}$$_{,}$$_{4}$-3$_{0}$$_{,}$$_{3}$                 & 8.28$\pm$0.01   & 0.6$\pm$0.1  & 0.23$\pm$0.05    \\
& 4$_{1}$$_{,}$$_{4}$-3$_{1}$$_{,}$$_{3}$             & 8.26$\pm$0.01 & 0.5$\pm$0.1  & 0.32$\pm$0.06 & 5$_{0}$$_{,}$$_{5}$-4$_{0}$$_{,}$$_{4}$                 & 8.31$\pm$0.03   & 0.30$\pm$0.06  & 0.17$\pm$0.03    \\
 
1333-C7-1 & 3$_{1}$$_{,}$$_{3}$-2$_{1}$$_{,}$$_{2}$   & 7.56$\pm$0.02   & 0.8$\pm$0.2  & 0.17$\pm$0.03 & 4$_{0}$$_{,}$$_{4}$-3$_{0}$$_{,}$$_{3}$               & 7.66$\pm$0.02   & 0.25$\pm$0.05  & 0.13$\pm$0.03     \\
& 4$_{1}$$_{,}$$_{4}$-3$_{1}$$_{,}$$_{3}$             & 7.04$\pm$0.02   & 0.33$\pm$0.07  & 0.16$\pm$0.03 & 5$_{0}$$_{,}$$_{5}$-4$_{0}$$_{,}$$_{4}$               & 7.59$\pm$0.06   & $<$0.10$\pm$0.02  & $<$0.06$\pm$ 0.01  \\
\hline
79-C1-1 & 3$_{1}$$_{,}$$_{3}$-2$_{1}$$_{,}$$_{2}$     & 10.33$\pm$0.01 & 0.6$\pm$0.1  & 0.36$\pm$0.07 & 4$_{0}$$_{,}$$_{4}$-3$_{0}$$_{,}$$_{3}$                & 10.06$\pm$0.01   & 0.46$\pm$0.09   & 0.19$\pm$0.04    \\
& 4$_{1}$$_{,}$$_{4}$-3$_{1}$$_{,}$$_{3}$             & 10.20$\pm$0.01 & 0.5$\pm$0.1  & 0.23$\pm$0.05 & 5$_{0}$$_{,}$$_{5}$-4$_{0}$$_{,}$$_{4}$             & 10.45$\pm$0.03   & 0.28$\pm$0.06  & 0.09$\pm$0.02    \\
\hline
IC\,348-1 & 3$_{1}$$_{,}$$_{3}$-2$_{1}$$_{,}$$_{2}$   & 8.94$\pm$0.02   & 0.7$\pm$0.1  & 0.11$\pm$0.02 & 4$_{0}$$_{,}$$_{4}$-3$_{0}$$_{,}$$_{3}$             & 8.94$\pm$0.02   & 0.43$\pm$0.09  & 0.09$\pm$0.02     \\
& 4$_{1}$$_{,}$$_{4}$-3$_{1}$$_{,}$$_{3}$             & 8.65$\pm$0.02   & 0.33$\pm$0.07  & 0.11$\pm$0.02 &   5$_{0}$$_{,}$$_{5}$-4$_{0}$$_{,}$$_{4}$             & 7.91$\pm$0.04   & $<$0.13$\pm$0.03  & $<$0.06$\pm$0.01  \\

IC\,348-10 & 3$_{1}$$_{,}$$_{3}$-2$_{1}$$_{,}$$_{2}$  & 8.46$\pm$0.02   & 0.6$\pm$0.1  & 0.13$\pm$0.03 & 4$_{0}$$_{,}$$_{4}$-3$_{0}$$_{,}$$_{3}$             & 8.46$\pm$0.02   & 0.29$\pm$0.06  & 0.08$\pm$0.02     \\
& 4$_{1}$$_{,}$$_{4}$-3$_{1}$$_{,}$$_{3}$             & 8.15$\pm$0.02   & 0.36$\pm$0.07  & 0.11$\pm$0.02 &   5$_{0}$$_{,}$$_{5}$-4$_{0}$$_{,}$$_{4}$             & 8.70$\pm$0.04   & $<$0.10$\pm$0.02  & $<$0.05$\pm$0.01  \\
\hline
ORI-C1-2 & 3$_{1}$$_{,}$$_{3}$-2$_{1}$$_{,}$$_{2}$    & 10.96$\pm$0.02  & 0.8$\pm$0.2  & 0.36$\pm$0.07 & 4$_{0}$$_{,}$$_{4}$-3$_{0}$$_{,}$$_{3}$               & 11.05$\pm$0.07  & 0.6$\pm$0.1  & 0.15$\pm$0.03    \\
& 4$_{1}$$_{,}$$_{4}$-3$_{1}$$_{,}$$_{3}$             & 11.10$\pm$0.03  & 0.7$\pm$0.1  & 0.27$\pm$0.05 & 5$_{0}$$_{,}$$_{5}$-4$_{0}$$_{,}$$_{4}$               & 11.31$\pm$0.05  & 0.5$\pm$0.1  & 0.09$\pm$0.02    \\

ORI-C2-3 & 3$_{1}$$_{,}$$_{3}$-2$_{1}$$_{,}$$_{2}$    & 8.16$\pm$0.01   & 0.41$\pm$0.08  & 0.24$\pm$0.05 & 4$_{0}$$_{,}$$_{4}$-3$_{0}$$_{,}$$_{3}$               & 8.13$\pm$0.02   & 0.29$\pm$0.06  & 0.15$\pm$ 0.03    \\
& 4$_{1}$$_{,}$$_{4}$-3$_{1}$$_{,}$$_{3}$             & 7.30$\pm$0.03   & 0.40$\pm$0.08  & 0.13$\pm$0.03 & 5$_{0}$$_{,}$$_{5}$-4$_{0}$$_{,}$$_{4}$               & 8.54$\pm$0.03   & $<$0.22$\pm$0.04  & $<$0.12$\pm$0.02  \\

ORI-C3-1 & 3$_{1}$$_{,}$$_{3}$-2$_{1}$$_{,}$$_{2}$    & 10.63$\pm$0.02   & 1.0$\pm$0.2  & 0.18$\pm$0.04 & 4$_{0}$$_{,}$$_{4}$-3$_{0}$$_{,}$$_{3}$              & 10.53$\pm$0.02  & 0.33$\pm$0.07  & 0.14$\pm$0.03     \\
& 4$_{1}$$_{,}$$_{4}$-3$_{1}$$_{,}$$_{3}$             & 10.40$\pm$0.02   & 0.7$\pm$0.1  & 0.23$\pm$0.05 & 5$_{0}$$_{,}$$_{5}$-4$_{0}$$_{,}$$_{4}$              & 11.4$\pm$0.05   & $<$0.19$\pm$0.04  & $<$0.18$\pm$0.04   \\
\hline 
\end{tabular}
\end{center}
\label{table:H2CS_parameters_gaussian}
\end{table*}

\begin{table*}
\caption{HDCS parameters from Gaussian fits.}
\begin{center}
\begin{tabular}{lllll}
\hline 
\hline
Core & Transition & \multicolumn{3}{c}{HDCS}   \\ 
& $J_{K,k}-J'_{K',k'}$ & $V$$_{\mathrm{LSR}}$ (km s$^{-1}$) & $\Delta$$V$ (km s$^{-1}$) & $T$$_{\mathrm{MB}}$ (K) \\
\hline
B\,213-C1-1 & 3$_{1}$$_{,}$$_{3}$-2$_{1}$$_{,}$$_{2}$   & 5.88$\pm$0.01 & 0.6$\pm$0.2  & 0.06$\pm$0.02   \\
& 3$_{0}$$_{,}$$_{3}$-2$_{0}$$_{,}$$_{2}$             & 5.94$\pm$0.01 & 0.37$\pm$0.09  & 0.29$\pm$0.07    \\
B\,213-C2-1 & 3$_{1}$$_{,}$$_{3}$-2$_{1}$$_{,}$$_{2}$   & 7.07$\pm$0.04   & 0.4$\pm$0.1  & 0.04$\pm$0.01  \\
& 3$_{0}$$_{,}$$_{3}$-2$_{0}$$_{,}$$_{2}$             & 7.07$\pm$0.01 & 0.4$\pm$0.1  & 0.10$\pm$0.02   \\
B\,213-C5-1 & 3$_{1}$$_{,}$$_{3}$-2$_{1}$$_{,}$$_{2}$   & 6.33$\pm$0.05   & 0.36$\pm$0.09  & 0.04$\pm$0.01  \\
& 3$_{0}$$_{,}$$_{3}$-2$_{0}$$_{,}$$_{2}$             & 6.40$\pm$0.02   & 0.5$\pm$0.1  & 0.11$\pm$0.03   \\
B\,213-C6-1 & 3$_{0}$$_{,}$$_{3}$-2$_{0}$$_{,}$$_{2}$   & 6.98$\pm$0.01 & 0.36$\pm$0.09  & 0.19$\pm$0.05   \\
B\,213-C7-1 & 3$_{1}$$_{,}$$_{3}$-2$_{1}$$_{,}$$_{2}$   & 6.90$\pm$0.03   & 0.5$\pm$0.1  & 0.07$\pm$0.02    \\
& 3$_{0}$$_{,}$$_{3}$-2$_{0}$$_{,}$$_{2}$             & 6.86$\pm$0.01 & 0.36$\pm$0.09  & 0.18$\pm$0.05   \\
B\,213-C10-1 & 3$_{0}$$_{,}$$_{3}$-2$_{0}$$_{,}$$_{2}$  & 6.68$\pm$0.03   & 0.30$\pm$0.08  & 0.04$\pm$0.01   \\
B\,213-C16-1 & 3$_{1}$$_{,}$$_{3}$-2$_{1}$$_{,}$$_{2}$  & 6.74$\pm$0.03   & 0.32$\pm$0.08  & 0.04$\pm$0.01    \\
& 3$_{0}$$_{,}$$_{3}$-2$_{0}$$_{,}$$_{2}$             & 6.72$\pm$0.01   & 0.4$\pm$0.1  & 0.15$\pm$0.04   \\  
\hline
L\,1448-1 & 3$_{1}$$_{,}$$_{3}$-2$_{1}$$_{,}$$_{2}$     & 4.51$\pm$0.02   & 0.5$\pm$0.1  & 0.12$\pm$0.04    \\
& 3$_{0}$$_{,}$$_{3}$-2$_{0}$$_{,}$$_{2}$             & 4.46$\pm$0.01 & 0.5$\pm$0.1  & 0.4$\pm$0.1   \\ 
\hline
1333-C3-1 & 3$_{0}$$_{,}$$_{3}$-2$_{0}$$_{,}$$_{2}$   & 8.57$\pm$0.06   & 0.4$\pm$0.1  & 0.05$\pm$0.01    \\
1333-C4-1 & 3$_{1}$$_{,}$$_{3}$-2$_{1}$$_{,}$$_{2}$   & 7.79$\pm$0.04   & 1.2$\pm$0.3  & 0.09$\pm$0.02    \\
& 3$_{0}$$_{,}$$_{3}$-2$_{0}$$_{,}$$_{2}$             & 7.67$\pm$0.02   & 1.2$\pm$0.3  & 0.18$\pm$0.05   \\
1333-C6-1 & 3$_{0}$$_{,}$$_{3}$-2$_{0}$$_{,}$$_{2}$   & 7.09$\pm$0.04   & 0.5$\pm$0.1  & 0.05$\pm$0.01  \\
1333-C3-14 & 3$_{0}$$_{,}$$_{3}$-2$_{0}$$_{,}$$_{2}$  & 8.23$\pm$0.01   & 0.6$\pm$0.1  & 0.16$\pm$0.04    \\
1333-C7-1 & 3$_{0}$$_{,}$$_{3}$-2$_{0}$$_{,}$$_{2}$   & 7.58$\pm$0.02   & 0.6$\pm$0.1  & 0.13$\pm$0.03    \\
\hline
79-C1-1 & 3$_{0}$$_{,}$$_{3}$-2$_{0}$$_{,}$$_{2}$     & 10.43$\pm$0.01  & 0.6$\pm$0.1  & 0.16$\pm$0.04  \\
\hline
ORI-C2-3 & 3$_{0}$$_{,}$$_{3}$-2$_{0}$$_{,}$$_{2}$    & 8.28$\pm$0.05   & 0.5$\pm$0.1  & 0.04$\pm$0.01     \\
\hline 
\end{tabular}
\end{center}
\label{table:HDCS_parameters_gaussian}
\end{table*}

\begin{table*}
\caption{D$_{2}$CS parameters from Gaussian fits.}
\begin{center}
\begin{tabular}{lllll||llll}
\hline 
\hline
Core & Transition & \multicolumn{3}{c}{o-D$_{2}$CS} & Transition & \multicolumn{3}{c}{p-D$_{2}$CS}  \\ 
& $J_{K,k}-J'_{K',k'}$ & $V$$_{\mathrm{LSR}}$ (km s$^{-1}$) & $\Delta$$V$ (km s$^{-1}$) & $T$$_{\mathrm{MB}}$ (K) & $J_{K,k}-J'_{K',k'}$ &
$V$$_{\mathrm{LSR}}$ (km s$^{-1}$) & $\Delta$$V$ (km s$^{-1}$) & $T$$_{\mathrm{MB}}$ (K) \\
\hline
B\,213-C1-1 & 3$_{0}$$_{,}$$_{3}$-2$_{0}$$_{,}$$_{2}$   & 5.95$\pm$0.01   & 0.31$\pm$0.08  & 0.21$\pm$0.05 & 5$_{1}$$_{,}$$_{5}$-4$_{1}$$_{,}$$_{4}$             & 5.78$\pm$0.04   & 0.30$\pm$0.09          & 0.05$\pm$0.01   \\
B\,213-C2-1 & 3$_{0}$$_{,}$$_{3}$-2$_{0}$$_{,}$$_{2}$  & 7.09$\pm$0.02    & 0.32$\pm$0.08  & 0.08$\pm$0.02 & 5$_{1}$$_{,}$$_{5}$-4$_{1}$$_{,}$$_{4}$              & 7.05$\pm$0.03   & $<$0.18$\pm$0.05          & $<$0.04$\pm$0.01 \\
B\,213-C5-1 & 3$_{0}$$_{,}$$_{3}$-2$_{0}$$_{,}$$_{2}$   & 6.43$\pm$0.04   & 0.36$\pm$0.09  & 0.05$\pm$0.01 & 5$_{1}$$_{,}$$_{5}$-4$_{1}$$_{,}$$_{4}$              & 6.37$\pm$0.09   & $<$0.20$\pm$0.06   & $<$0.02$\pm$0.01    \\
B\,213-C6-1 & 3$_{0}$$_{,}$$_{3}$-2$_{0}$$_{,}$$_{2}$   & 7.02$\pm$0.02   & 0.36$\pm$0.08  & 0.14$\pm$0.04 & 5$_{1}$$_{,}$$_{5}$-4$_{1}$$_{,}$$_{4}$              & 6.99$\pm$0.03   & $<$0.14$\pm$0.04          & $<$0.05$\pm$0.01     \\
B\,213-C7-1 & 3$_{0}$$_{,}$$_{3}$-2$_{0}$$_{,}$$_{2}$   & 6.93$\pm$0.02   & 0.30$\pm$0.08  & 0.10$\pm$0.03 & 5$_{1}$$_{,}$$_{5}$-4$_{1}$$_{,}$$_{4}$              & 6.84$\pm$0.04   & $<$0.17$\pm$0.05          & $<$0.05$\pm$0.01        \\
B\,213-C12-1 & 3$_{0}$$_{,}$$_{3}$-2$_{0}$$_{,}$$_{2}$  & 6.77$\pm$0.06   & 0.5$\pm$0.1  & 0.04$\pm$0.01 & 5$_{1}$$_{,}$$_{5}$-4$_{1}$$_{,}$$_{4}$              & 6.60$\pm$0.05   & $<$0.15$\pm$0.05   & $<$0.03$\pm$0.01      \\
B\,213-C16-1 & 3$_{0}$$_{,}$$_{3}$-2$_{0}$$_{,}$$_{2}$  & 6.79$\pm$0.04   & 0.4$\pm$0.1  & 0.07$\pm$0.02 & 5$_{1}$$_{,}$$_{5}$-4$_{1}$$_{,}$$_{4}$              & 6.97$\pm$0.07   & $<$0.18$\pm$0.05 & $<$0.02$\pm$0.01    \\
\hline
L\,1448-1 & 3$_{0}$$_{,}$$_{3}$-2$_{0}$$_{,}$$_{2}$     & 4.54$\pm$0.01 & 0.4$\pm$0.1  & 0.30$\pm$0.07 & 5$_{1}$$_{,}$$_{5}$-4$_{1}$$_{,}$$_{4}$                & 4.51$\pm$0.03   & 0.30$\pm$0.09  & 0.05$\pm$0.02    \\
\hline
1333-C4-1 & 3$_{0}$$_{,}$$_{3}$-2$_{0}$$_{,}$$_{2}$     & 7.76$\pm$0.01   & 1.0$\pm$0.3  & 0.21$\pm$0.05 & 5$_{1}$$_{,}$$_{5}$-4$_{1}$$_{,}$$_{4}$              & 7.37$\pm$0.06   & $<$0.25$\pm$0.08  & $<$0.04$\pm$0.01    \\
1333-C3-14 & 3$_{0}$$_{,}$$_{3}$-2$_{0}$$_{,}$$_{2}$    & 8.29$\pm$0.04   & 0.5$\pm$0.1  & 0.12$\pm$0.03 & 5$_{1}$$_{,}$$_{5}$-4$_{1}$$_{,}$$_{4}$              & 8.00$\pm$0.04   & $<$0.23$\pm$0.07  & $<$0.04$\pm$0.01    \\
1333-C7-1 & 3$_{0}$$_{,}$$_{3}$-2$_{0}$$_{,}$$_{2}$     & 7.73$\pm$0.03   & 0.34$\pm$0.09  & 0.07$\pm$0.02 & 5$_{1}$$_{,}$$_{5}$-4$_{1}$$_{,}$$_{4}$              & 7.84$\pm$0.04   & $<$0.22$\pm$0.07  & $<$0.06$\pm$0.02      \\
\hline
79-C1-1 & 3$_{0}$$_{,}$$_{3}$-2$_{0}$$_{,}$$_{2}$       & 10.46$\pm$0.02  & 0.4$\pm$0.1  & 0.12$\pm$0.03 & 5$_{1}$$_{,}$$_{5}$-4$_{1}$$_{,}$$_{4}$             &  10.24$\pm$0.06  & $<$0.4$\pm$0.1 & $<$0.03$\pm$0.01    \\
\hline    
\end{tabular}
\end{center}
\label{table:D2CS_parameters_gaussian}
\end{table*}

\begin{landscape}
\begin{table}
\centering
\caption{Column densities, $N$, for H$_{2}$CS, HDCS, and D$_{2}$CS and the OPR obtained from H$_2$CS.}
\begin{tabular}{l|llllllllll}
\hline 
\hline
Region & Core & $N$(o-H$_{2}$CS) & $N$(p-H$_{2}$CS )& OPR(H$_2$CS) & $N$(H$_{2}$CS) & $N$(HDCS) &  $N$(o-D$_{2}$CS) & $N$(p-D$_{2}$CS) & OPR(D$_2$CS)& $N$(D$_{2}$CS) \\ 
       &      & $\times$10$^{12}$ cm$^{-2}$ & $\times$10$^{12}$ cm$^{-2}$ &  & $\times$10$^{12}$ cm$^{-2}$ & $\times$10$^{12}$ cm$^{-2}$ &  $\times$10$^{12}$ cm$^{-2}$ & $\times$10$^{12}$ cm$^{-2}$ & & $\times$10$^{12}$ cm$^{-2}$     \\
\hline
\hline 
& B\,213-C1-1           & 2.9$\pm$0.6   & 1.0$\pm$0.2        & 3$\pm$1        & 3.9$\pm$0.8        & 1.8$\pm$0.5       & 0.7$\pm$0.2             & 0.4$\pm$0.1            & -           & 1.1$\pm$0.3 \\ 
& B\,213-C2-1           & 1.6$\pm$0.3   & 0.9$\pm$0.2        & 1.8$\pm$0.7    & 2.5$\pm$0.5        & 0.8$\pm$0.2       & 0.4$\pm$0.1             & 0.19$\pm$0.05$^*$      & 2$\pm$1     & 0.6$\pm$0.2$^*$ \\ 
& B\,213-C5-1           & 5$\pm$1       & 2.2$\pm$0.4        & 2.2$\pm$0.9    & 7$\pm$1            & 0.8$\pm$0.2       & 0.25$\pm$0.06           & 0.12$\pm$0.03$^*$      & -           & 0.37$\pm$0.09$^*$ \\
{Taurus} & B\,213-C6-1  & 3.0$\pm$0.6   & 1.3$\pm$0.3        & 2.3$\pm$0.9    & 4.3$\pm$0.9        & 1.2$\pm$0.3       & 0.8$\pm$0.2             & 0.38$\pm$0.09$^*$      & -           & 1.1$\pm$0.3$^*$ \\   
& B\,213-C7-1           & 2.8$\pm$0.6   & 1.6$\pm$0.3        & 1.8$\pm$0.7    & 4.4$\pm$0.9        & 1.0$\pm$0.3       & 0.35$\pm$0.09           & 0.18$\pm$0.04$^*$      & -           & 0.5$\pm$0.1$^*$ \\  
& B\,213-C10-1          & 1.0$\pm$0.2   & 0.6$\pm$0.1        & 1.6$\pm$0.6    & 1.6$\pm$0.3        & 0.21$\pm$0.05     & $<$0.18$\pm$0.05        & $<$0.09$\pm$0.02$^*$   & -           & $<$0.27$\pm$0.07$^*$ \\ 
& B\,213-C12-1          & 0.42$\pm$0.08 & 0.23$\pm$0.05      & 1.8$\pm$0.7    & 0.7$\pm$0.1        & $<$0.22$\pm$0.05  & 0.20$\pm$0.05           & 0.10$\pm$0.03$^*$      & -           & 0.30$\pm$0.08$^*$ \\ 
& B\,213-C16-1          & 3.9$\pm$0.8   & 1.7$\pm$0.3        & 2.3$\pm$0.9    & 6$\pm$1            & 0.8$\pm$0.2       & 0.4$\pm$0.1             & 0.21$\pm$0.05$^*$      & -           & 0.6$\pm$0.2$^*$ \\
\hline\hline
& L\,1448-1             & 7$\pm$1       & 4.0$\pm$0.8        & 1.8$\pm$0.7    & 11$\pm$2           & 3.5$\pm$0.9        & 1.8$\pm$0.4            & 0.7$\pm$0.2            &  2$\pm$1    & 2.6$\pm$0.7 \\ 
& 1333-C3-1             & 2.0$\pm$0.4   & 0.5$\pm$0.1        & 4$\pm$2        & 2.5$\pm$0.5        & 0.26$\pm$0.06      & $<$0.04$\pm$0.01       & $<$0.02$\pm$0.01$^*$   & -           & $<$0.07$\pm$0.02$^*$ \\  
& 1333-C4-1             & 8$\pm$2       & 3.0$\pm$0.6        & 3$\pm$1        & 11$\pm$3           & 3.4$\pm$0.9        & 2.4$\pm$0.6            & 1.2$\pm$0.3$^*$        & -           & 3.6$\pm$0.9$^*$  \\ 
& 1333-C5-1             & 0.8$\pm$0.2   & 0.16$\pm$0.03      & 5$\pm$2        & 1.0$\pm$0.2        & $<$0.07$\pm$0.02   & $<$0.10$\pm$0.03       & $<$0.05$\pm$0.01$^*$   & -           & $<$0.15$\pm$0.04$^*$ \\ 
{Perseus} & 1333-C6-1   & 0.34$\pm$0.07 & 0.14$\pm$0.03$^{**}$  & -           & 0.5$\pm$0.1$^{**}$ & 0.4$\pm$0.1        & $<$0.027$\pm$0.007     & $<$0.014$\pm$0.003$^*$ & -           & $<$0.04$\pm$0.01$^*$ \\
& 1333-C3-14            & 5$\pm$1       & 2.1$\pm$0.4        & 2.4$\pm$0.9    & 7$\pm$1            & 1.1$\pm$0.3        & 0.7$\pm$0.2            & 0.33$\pm$0.08$^*$      & -           & 1.0$\pm$0.3$^*$ \\ 
& 1333-C7-1             & 1.3$\pm$0.3   & 0.6$\pm$0.1        & 2.1$\pm$0.8    & 1.9$\pm$0.4        & 1.0$\pm$0.2        & 0.26$\pm$0.07          & 0.13$\pm$0.03$^*$      & -           & 0.4$\pm$0.1$^*$ \\  
& 79-C1-1               & 3.4$\pm$0.7   & 1.9$\pm$0.4        & 1.8$\pm$0.7    & 5$\pm$1            & 1.2$\pm$0.3        & 0.6$\pm$0.1            & 0.28$\pm$0.07$^*$      & -           & 0.9$\pm$0.2$^*$ \\  
& IC\,348-1             & 0.8$\pm$0.2   & 0.38$\pm$0.08      & 2.1$\pm$0.8    & 1.2$\pm$0.3        & $<$0.06$\pm$0.01   & $<$0.14$\pm$0.03       & $<$0.07$\pm$0.02$^*$   & -           & $<$0.21$\pm$0.05$^*$ \\ 
& IC\,348-10            & 0.8$\pm$0.2   & 0.35$\pm$0.07      & 2.3$\pm$0.9    & 1.2$\pm$0.3        & $<$0.04$\pm$0.01   & $<$0.09$\pm$0.02       & $<$0.05$\pm$0.01$^*$   & -           & $<$0.14$\pm$0.03$^*$ \\ 
\hline\hline
& ORI-C1-2              & 3.6$\pm$0.7   & 1.3$\pm$0.3        & 3$\pm$1        & 5$\pm$1            & $<$0.11$\pm$0.03   & $<$0.04$\pm$0.01       & $<$0.021$\pm$0.005$^*$ & -           & $<$0.06$\pm$0.02$^*$ \\
{Orion} & ORI-C2-3      & 2.0$\pm$0.4   & 1.8$\pm$0.4        & 1.2$\pm$0.5    & 3.8$\pm$0.8        & 0.37$\pm$0.09      & $<$0.06$\pm$0.02       & $<$0.032$\pm$0.008$^*$ & -           & $<$0.10$\pm$0.02$^*$ \\ 
& ORI-C3-1              & 1.9$\pm$0.4   & 0.6$\pm$0.1        & 3$\pm$1        & 2.5$\pm$0.5        & $<$0.16$\pm$0.04   & $<$0.08$\pm$0.02       & $<$0.04$\pm$0.01$^*$   & -           & $<$0.12$\pm$0.03$^*$ \\ 
\hline
\end{tabular}

$^{**}$Assuming an average OPR(H$_2$CS)=2.4$\pm$0.9 obtained from all the cores of the sample, except 1333-C6-1, where we only detect o-H$_2$CS \\  
$^*$Assuming an average OPR(D$_2$CS)=2.0$\pm$1.0 obtained from the cores B\,213-C1-1 and L\,1448-1, where we detect both o- and p-D$_2$CS \\
\label{table:column_densities_Radex}
\end{table}
\end{landscape}


\begin{figure*}
\centering
\includegraphics[scale=0.65, angle=0]{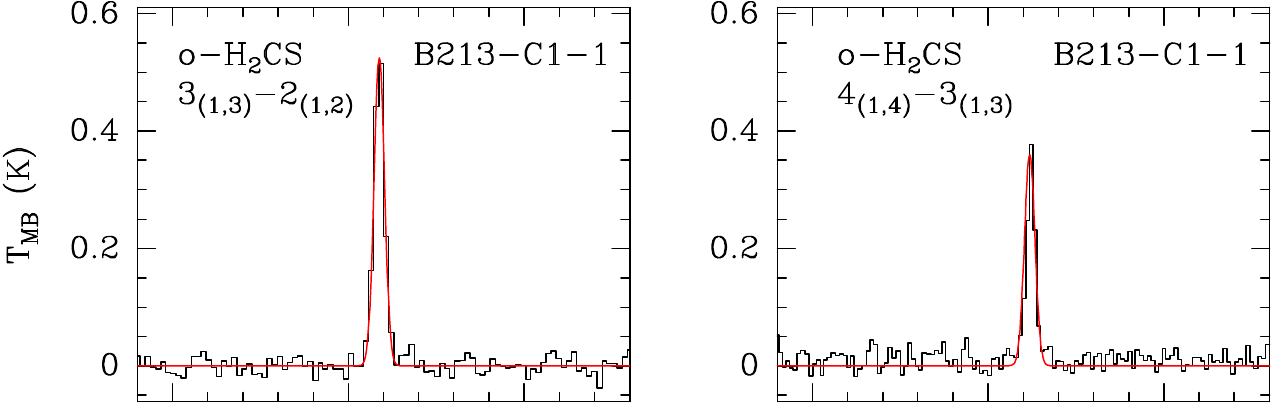}  
\hspace{1cm}
\includegraphics[scale=0.65, angle=0]{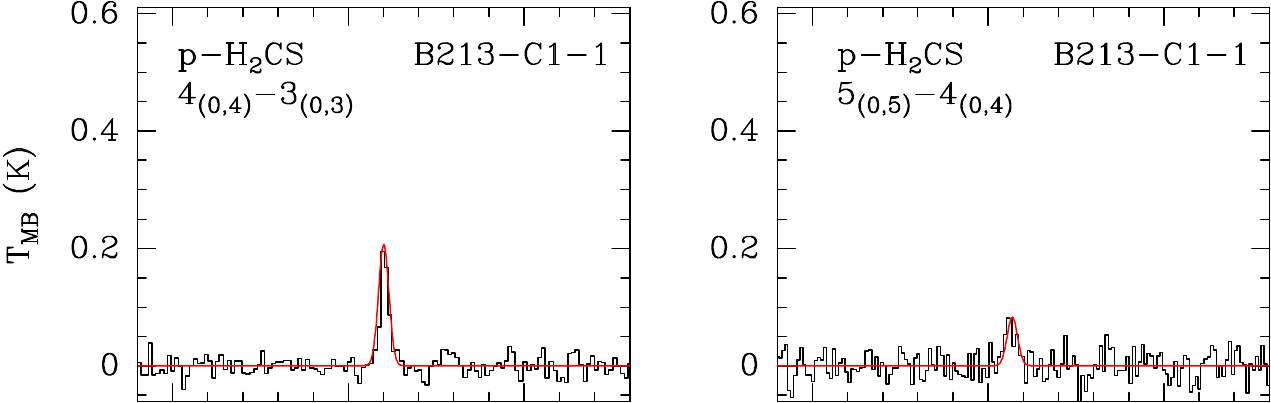} 

\vspace{0.1cm}

\includegraphics[scale=0.65, angle=0]{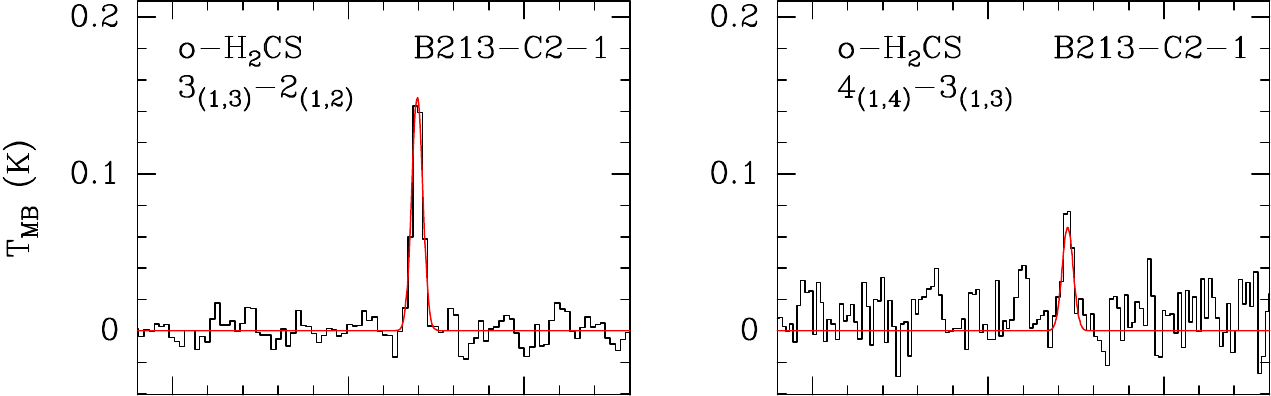}  
\hspace{1cm}
\includegraphics[scale=0.65, angle=0]{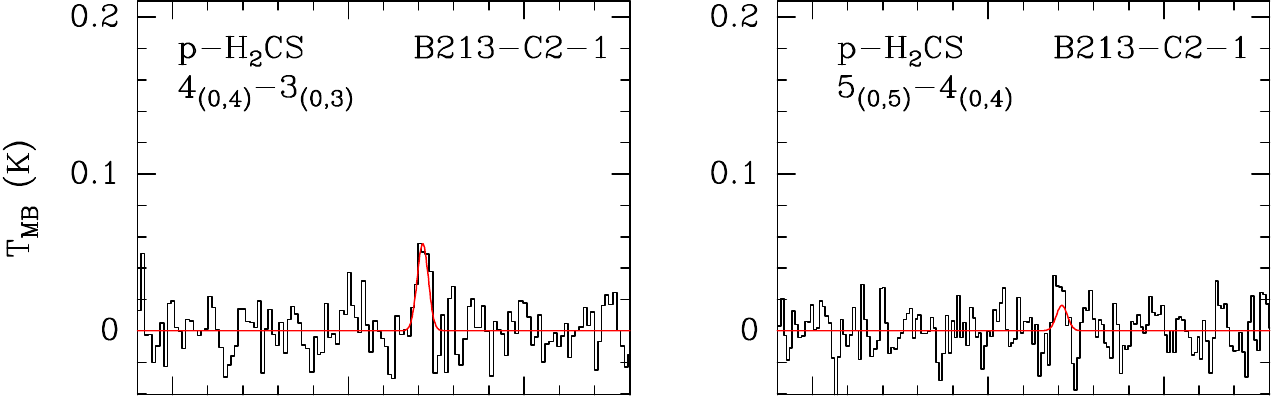} 

\vspace{0.1cm}

\includegraphics[scale=0.65, angle=0]{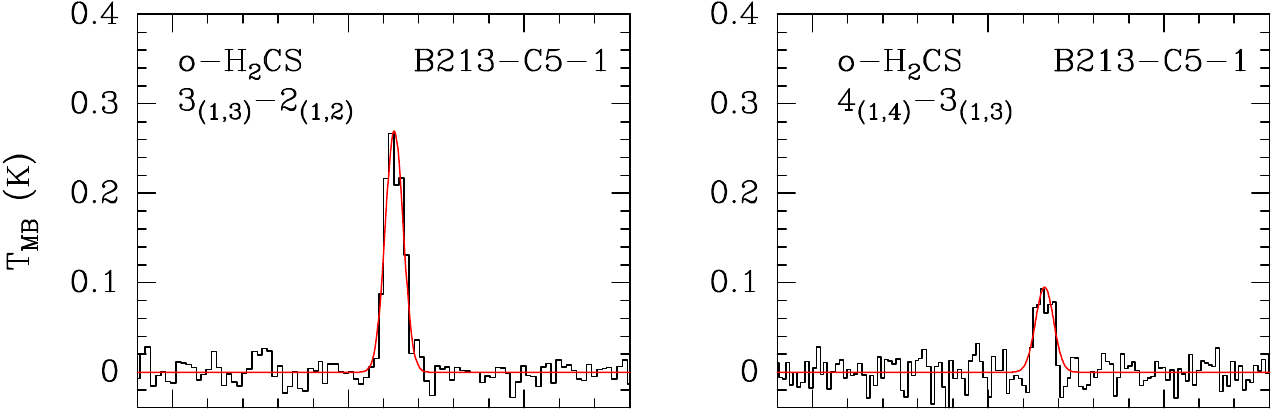}  
\hspace{1cm}
\includegraphics[scale=0.65, angle=0]{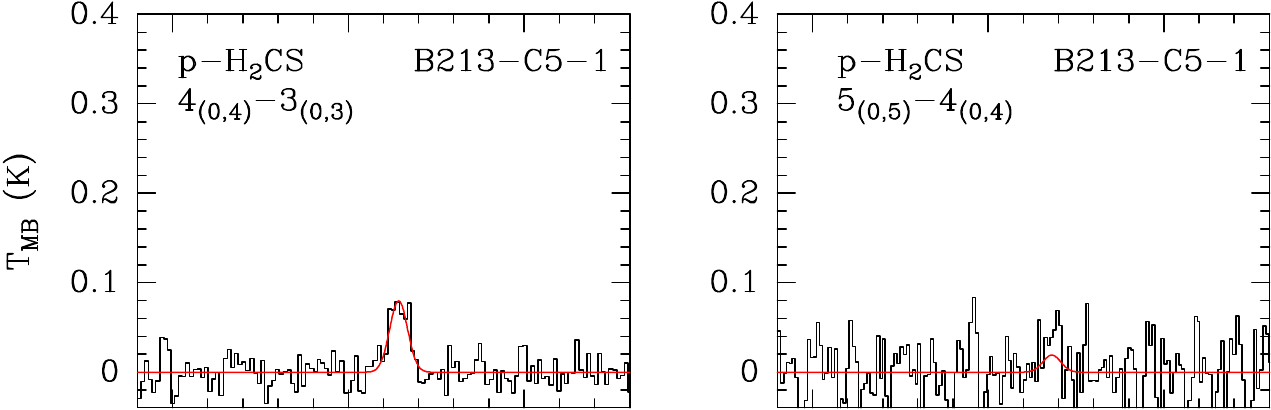} 

\vspace{0.1cm}

\includegraphics[scale=0.65, angle=0]{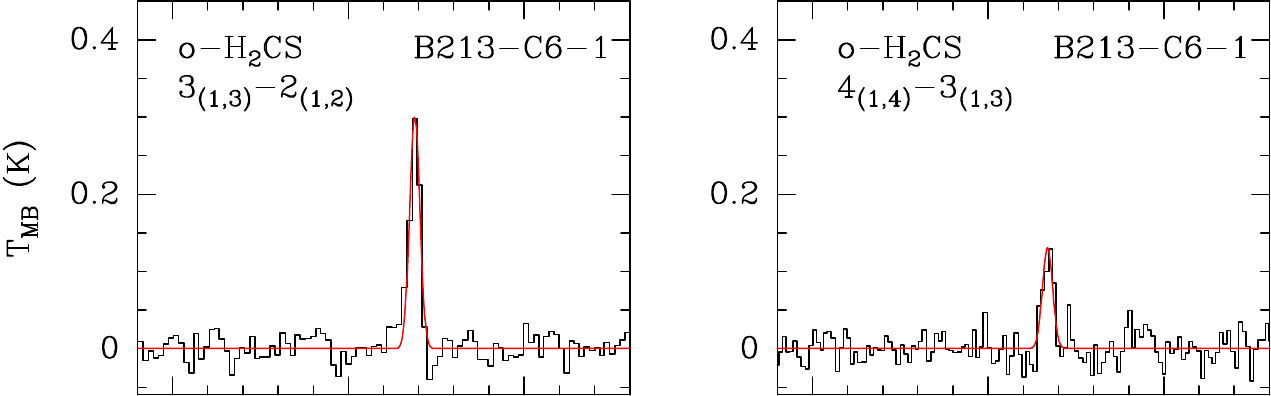}  
\hspace{1cm}
\includegraphics[scale=0.65, angle=0]{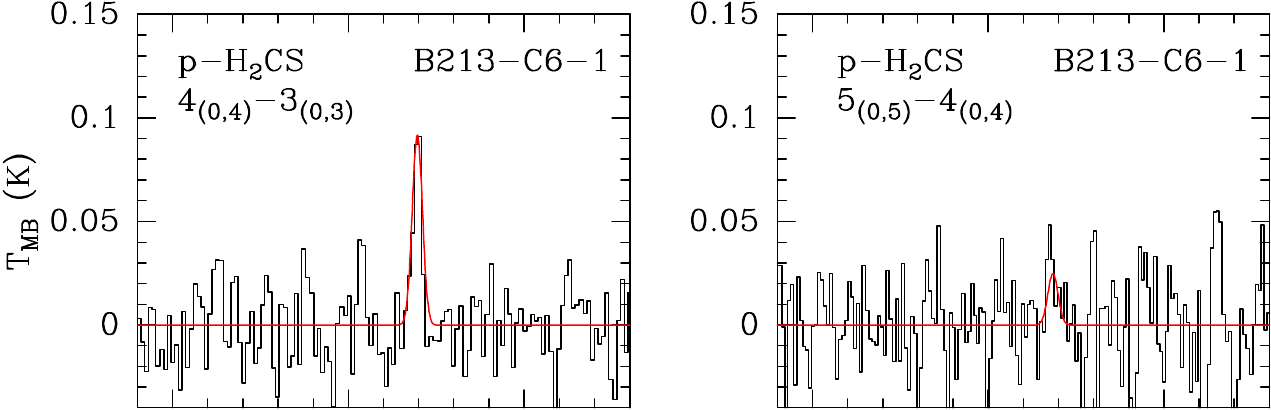} 

\vspace{0.1cm}

\includegraphics[scale=0.65, angle=0]{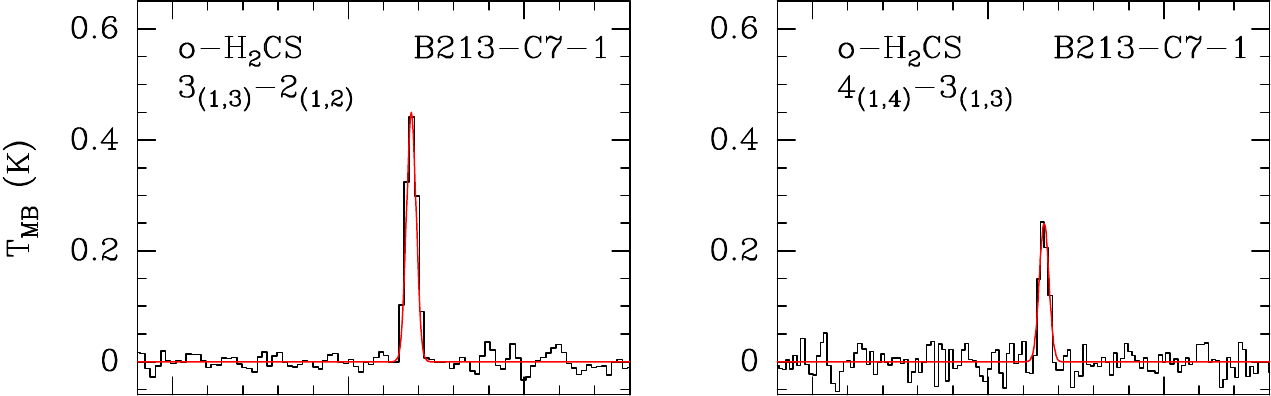}  
\hspace{1cm}
\includegraphics[scale=0.65, angle=0]{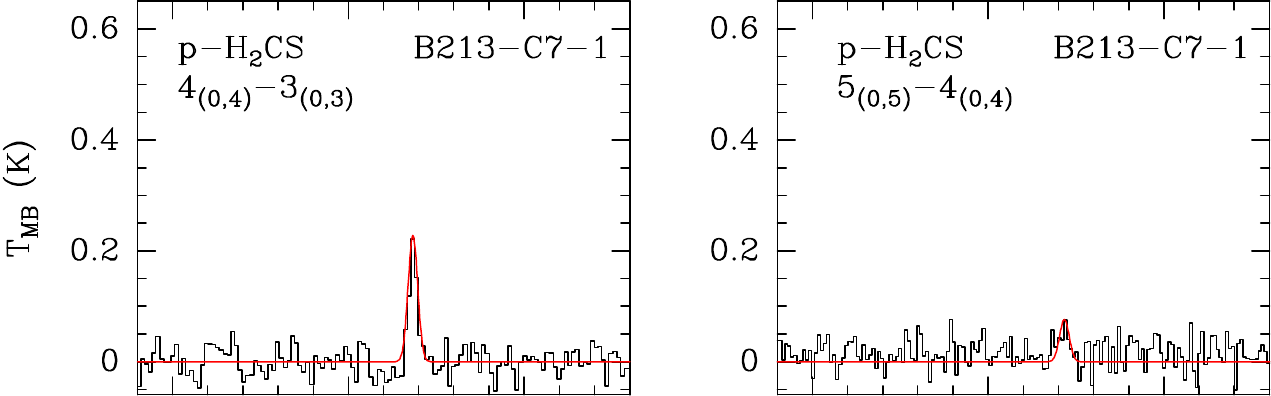} 

\vspace{0.1cm}

\includegraphics[scale=0.65, angle=0]{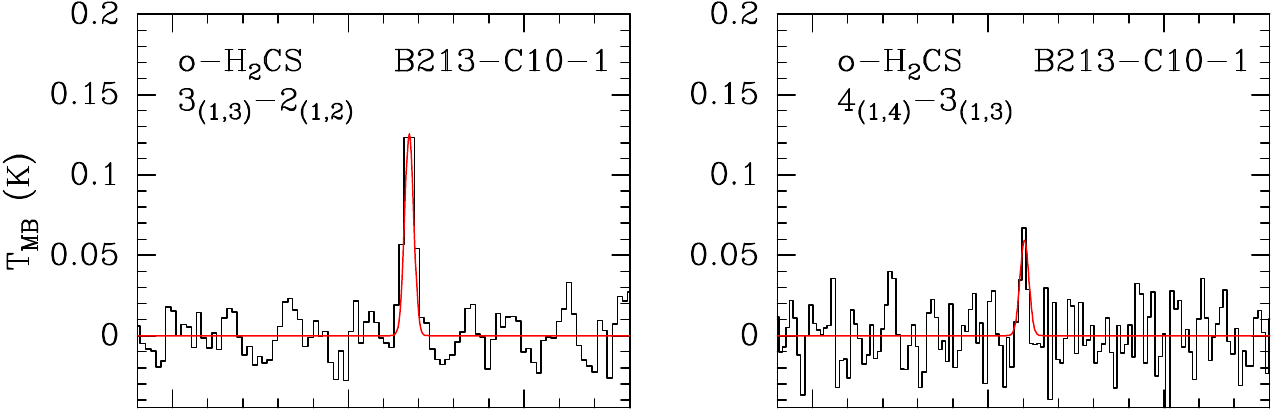}  
\hspace{1cm}
\includegraphics[scale=0.65, angle=0]{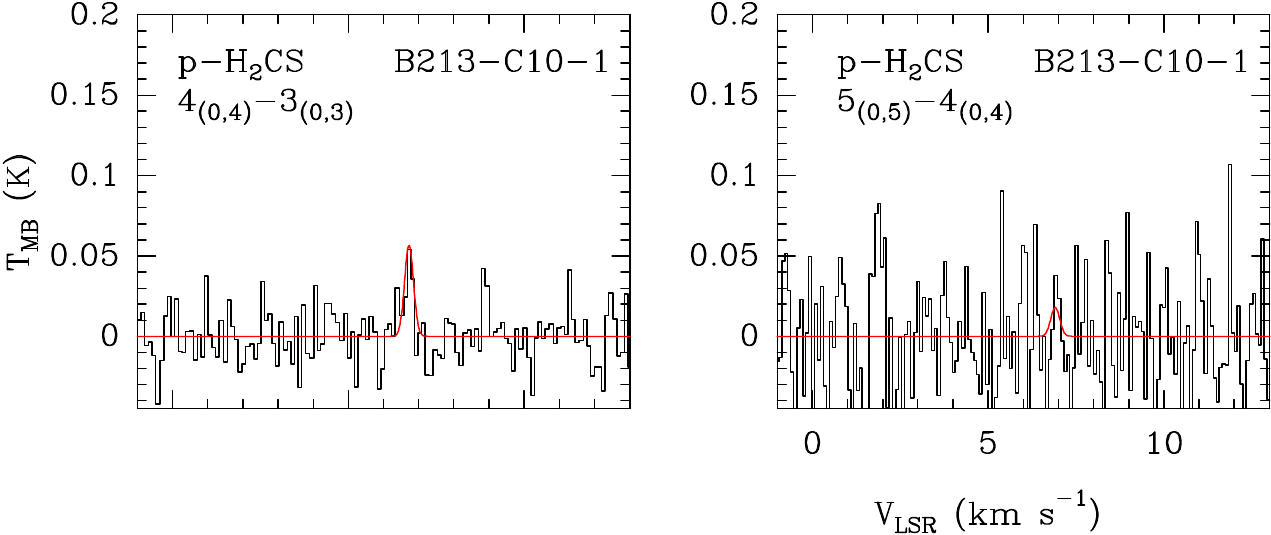} 

\vspace{0.1cm}

\hspace{-4.3cm}
\includegraphics[scale=0.65, angle=0]{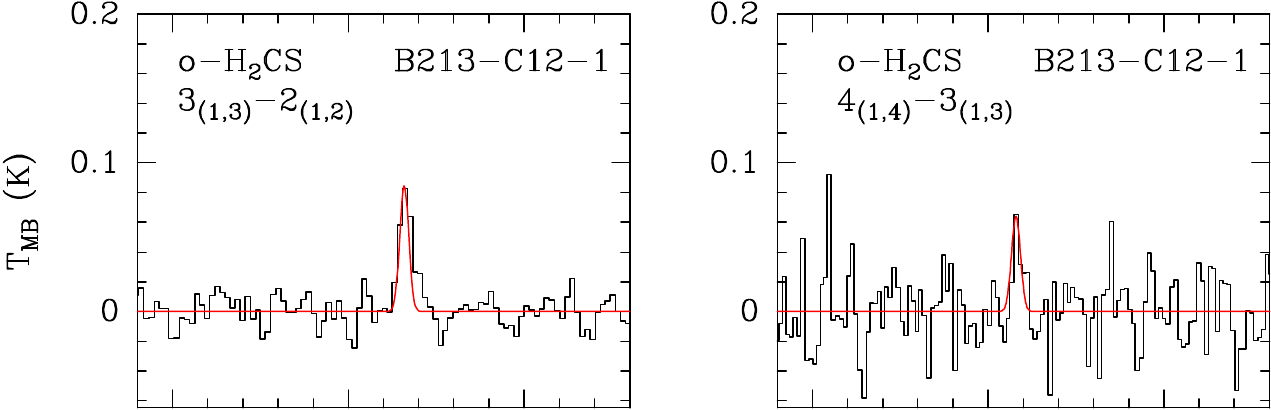}  
\hspace{1.0cm}
\includegraphics[scale=0.65, angle=0]{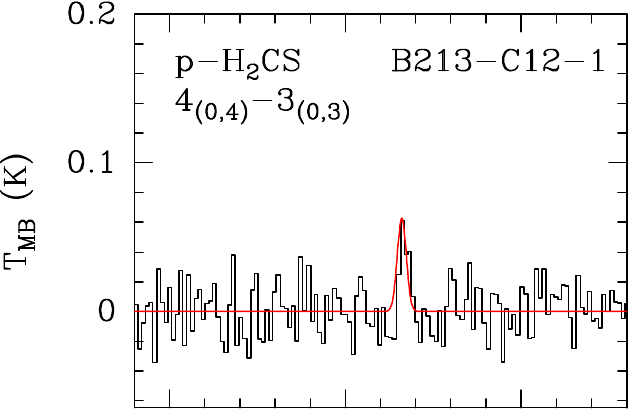} 

\vspace{0.1cm}

\hspace{-4.3cm}
\includegraphics[scale=0.65, angle=0]{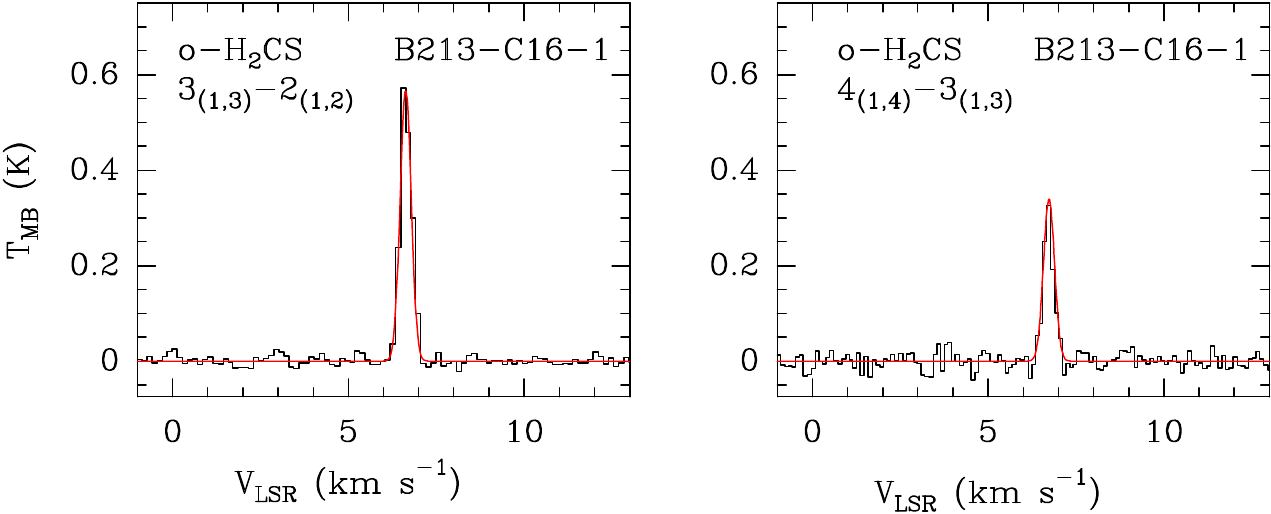}  
\hspace{1.0cm}
\includegraphics[scale=0.65, angle=0]{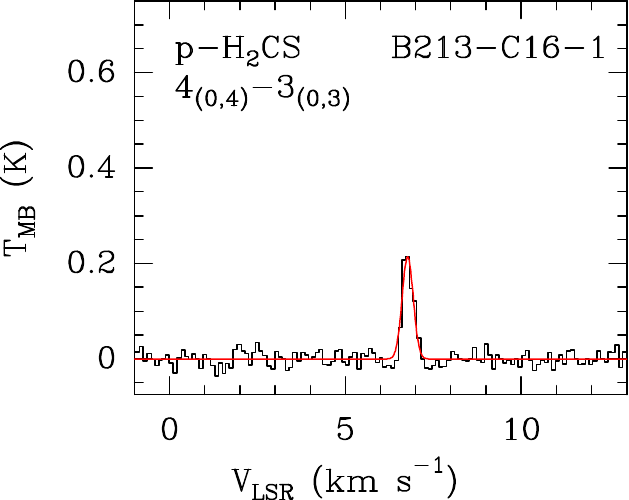} 
\hspace{4cm}
\\
\caption{Observed lines of o-H$_{2}$CS and p-H$_{2}$CS in B\,213-C1, C2, C5, C6, C7, C10, C12, and C16 (black), and the best fit (red).}
\label{figure:H2CS_B213}
\end{figure*}

\begin{figure*}
\centering
\includegraphics[scale=0.65, angle=0]{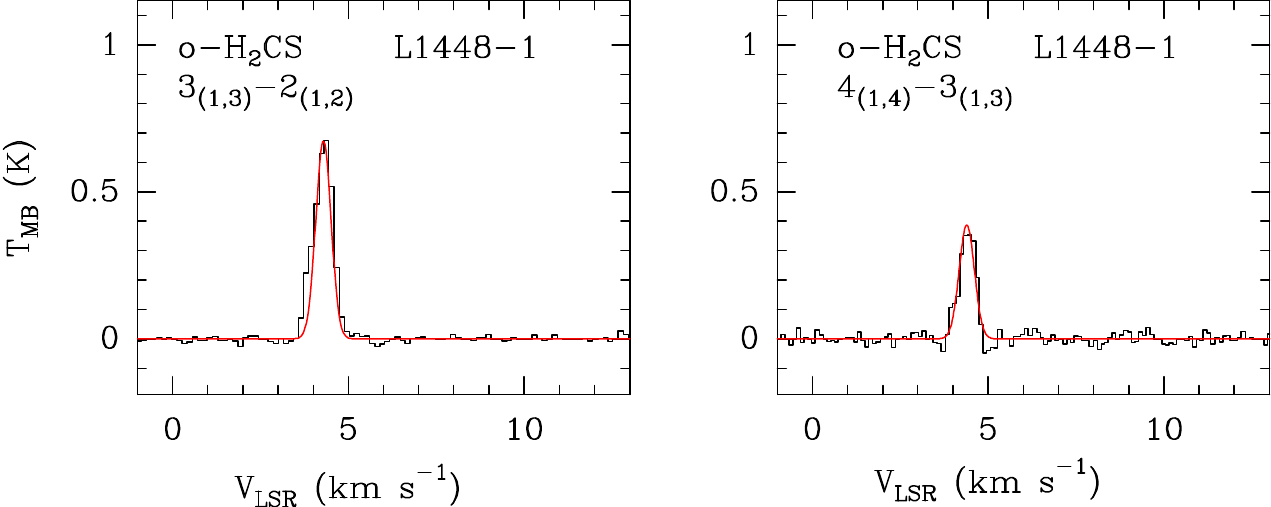}  
\hspace{1cm}
\includegraphics[scale=0.65, angle=0]{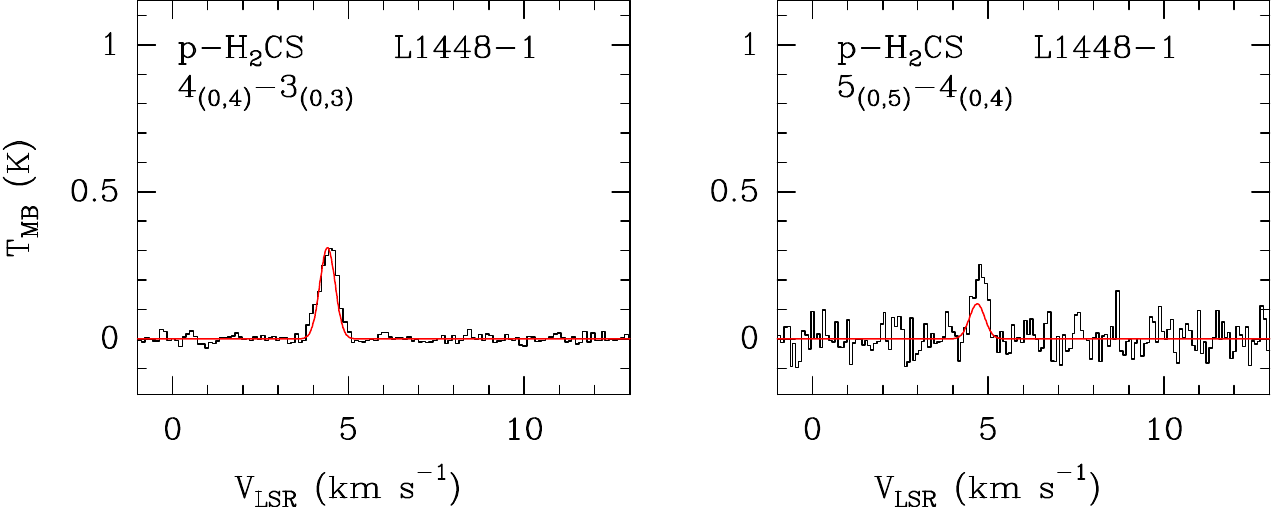} 
\hspace{4cm}
\\
\caption{Observed lines of o-H$_{2}$CS and p-H$_{2}$CS in L\,1448 (black), and the best fit (red).}
\label{figure:H2CS_L1448}
\end{figure*}

\begin{figure*}
\centering
\includegraphics[scale=0.65, angle=0]{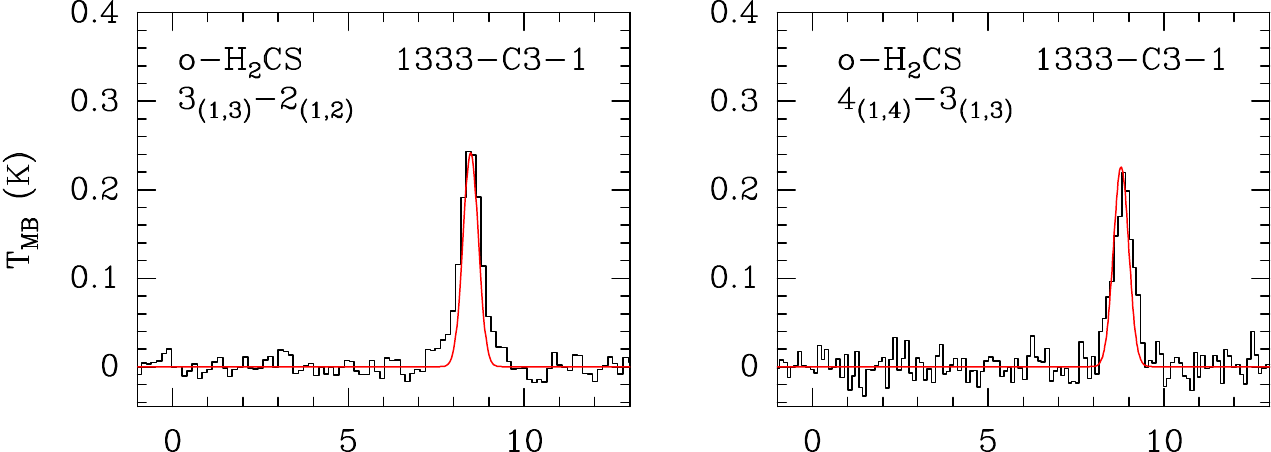}  
\hspace{1cm}
\includegraphics[scale=0.65, angle=0]{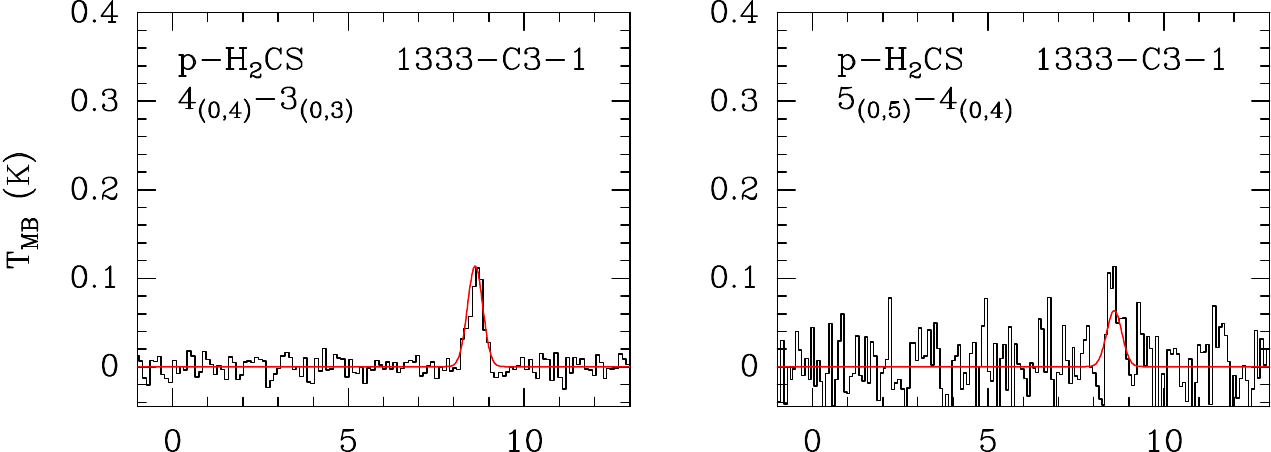} 
\vspace{0.1cm}


\includegraphics[scale=0.65, angle=0]{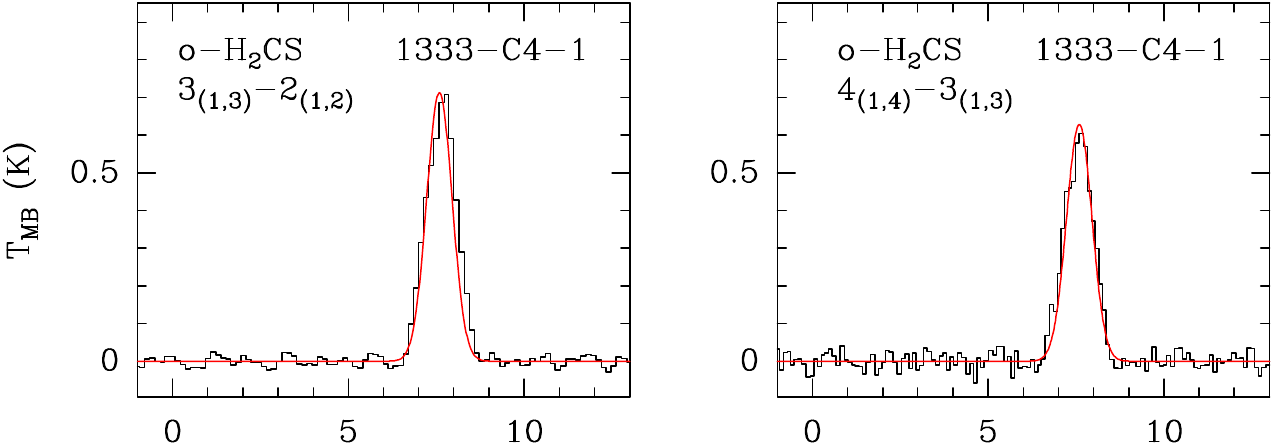}  
\hspace{1cm}
\includegraphics[scale=0.65, angle=0]{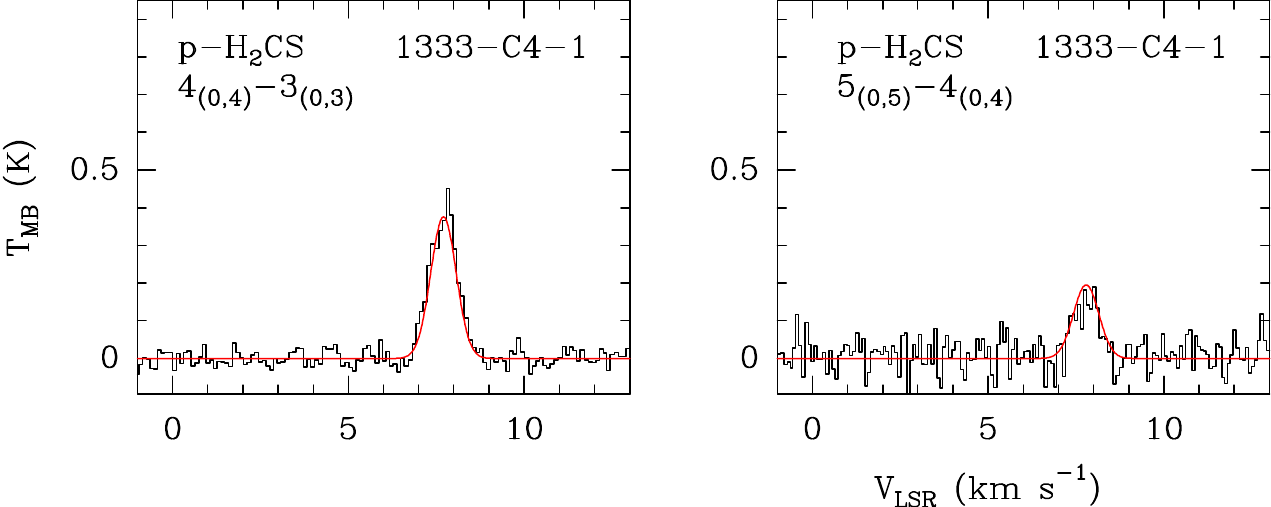} 

\vspace{-0.2cm}

\hspace{-4.3cm}
\includegraphics[scale=0.65, angle=0]{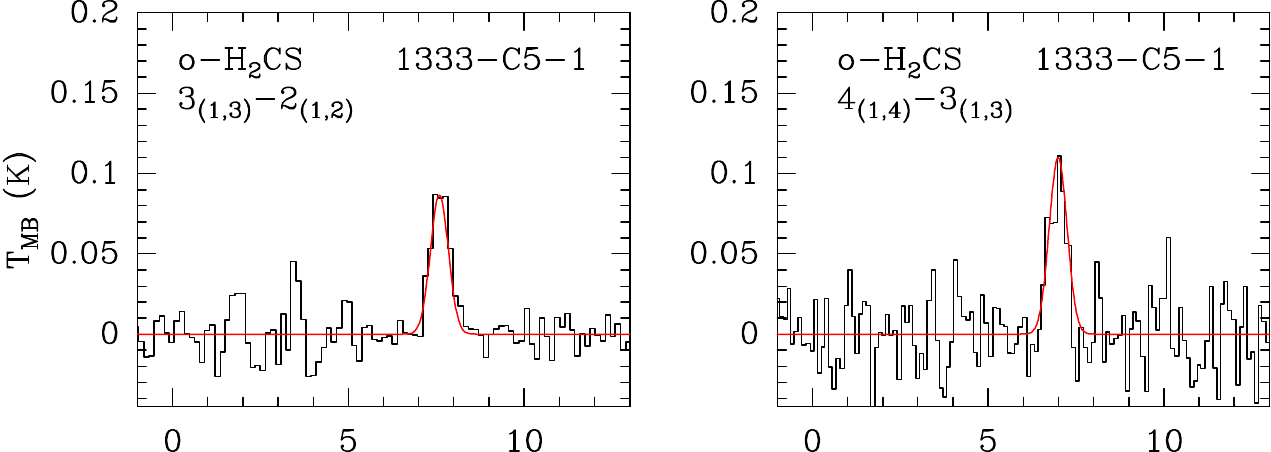}  
\hspace{1cm}
\includegraphics[scale=0.65, angle=0]{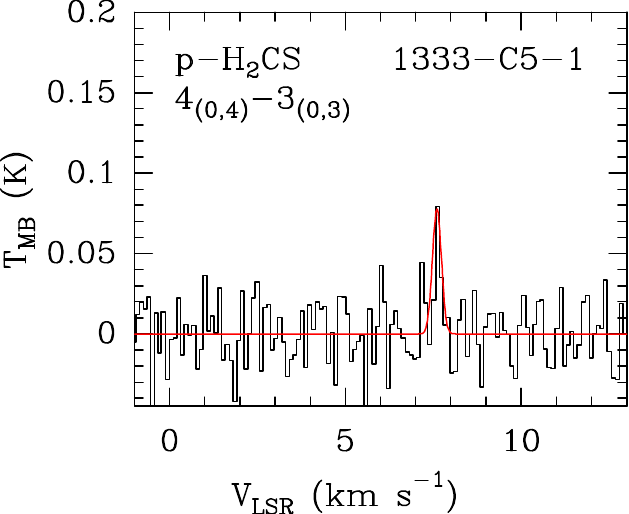} 

\vspace{0.3cm}

\hspace{-8.5cm}
\includegraphics[scale=0.65, angle=0]{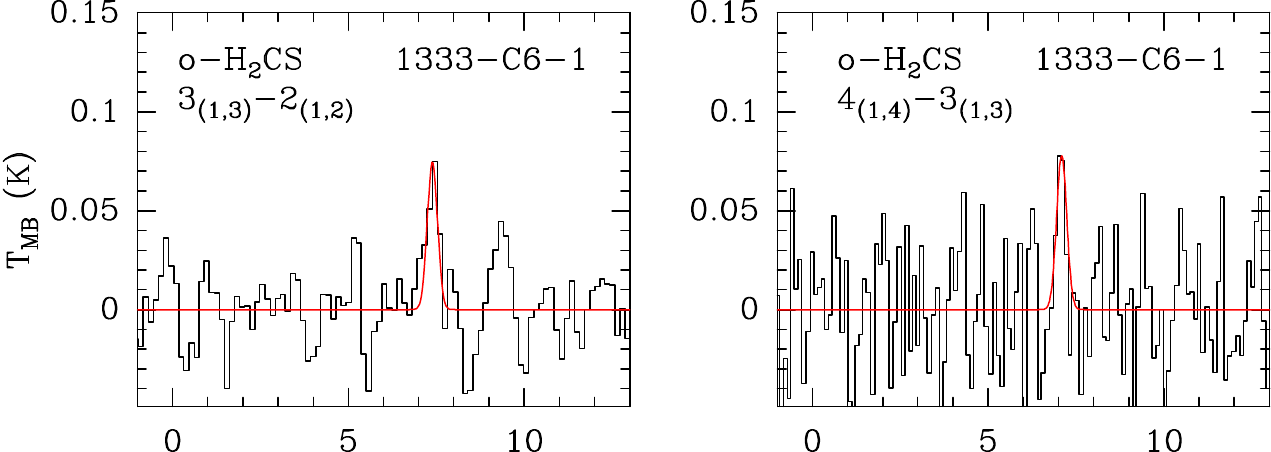}  
\hspace{1cm} 

\vspace{-0.1cm}

\includegraphics[scale=0.65, angle=0]{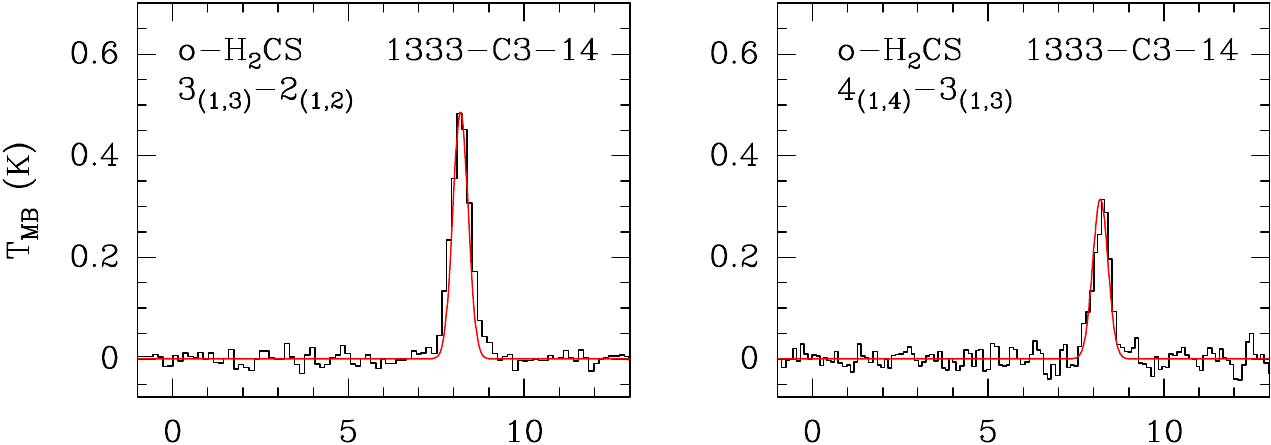}  
\hspace{1cm}
\includegraphics[scale=0.65, angle=0]{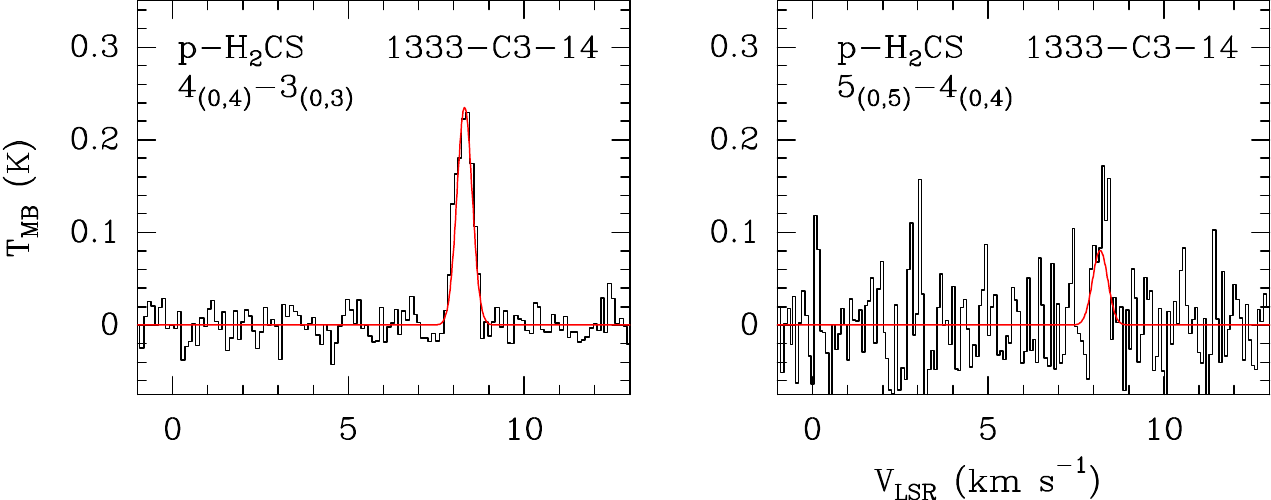} 

\vspace{0.0cm}

\hspace{-4.3cm}
\includegraphics[scale=0.65, angle=0]{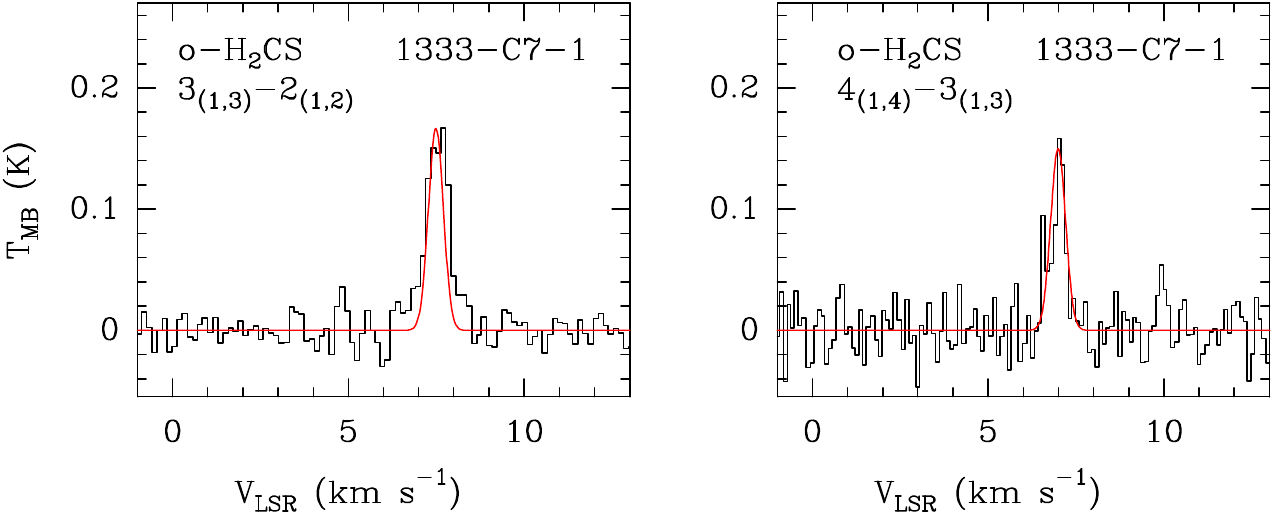}  
\hspace{1cm}
\includegraphics[scale=0.65, angle=0]{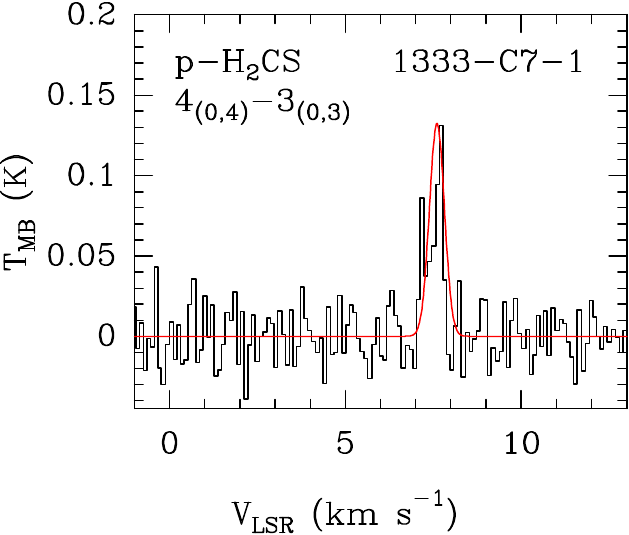} 
\hspace{4cm}
\\
\caption{Observed lines of o-H$_{2}$CS and p-H$_{2}$CS in the core sample of NGC\,1333 (black) and the best fit (red).}
\label{figure:H2CS_NGC1333}
\end{figure*}

\begin{figure*}
\centering
\includegraphics[scale=0.65, angle=0]{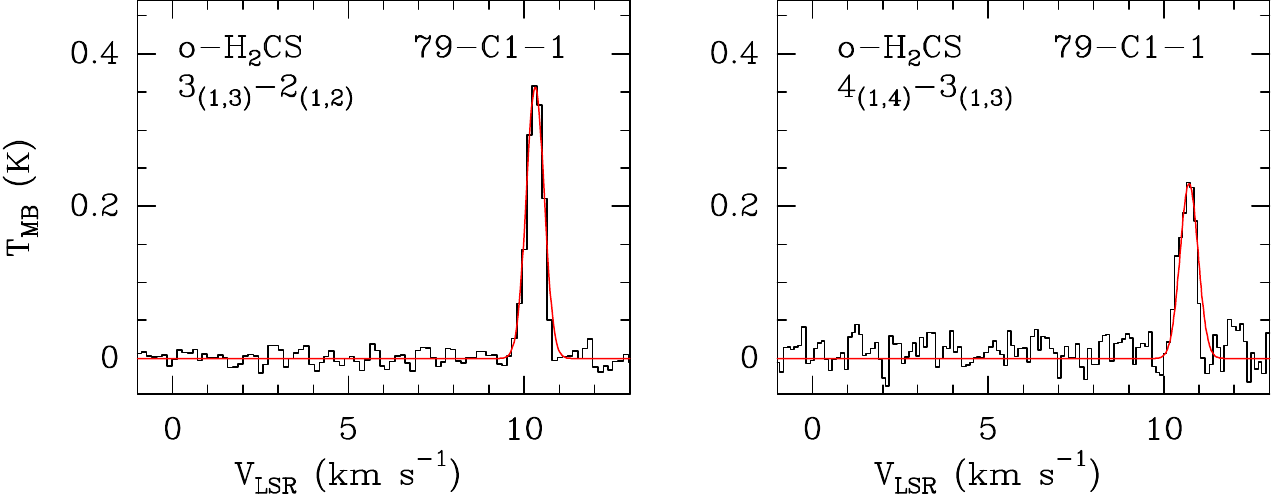}  
\hspace{1cm}
\includegraphics[scale=0.65, angle=0]{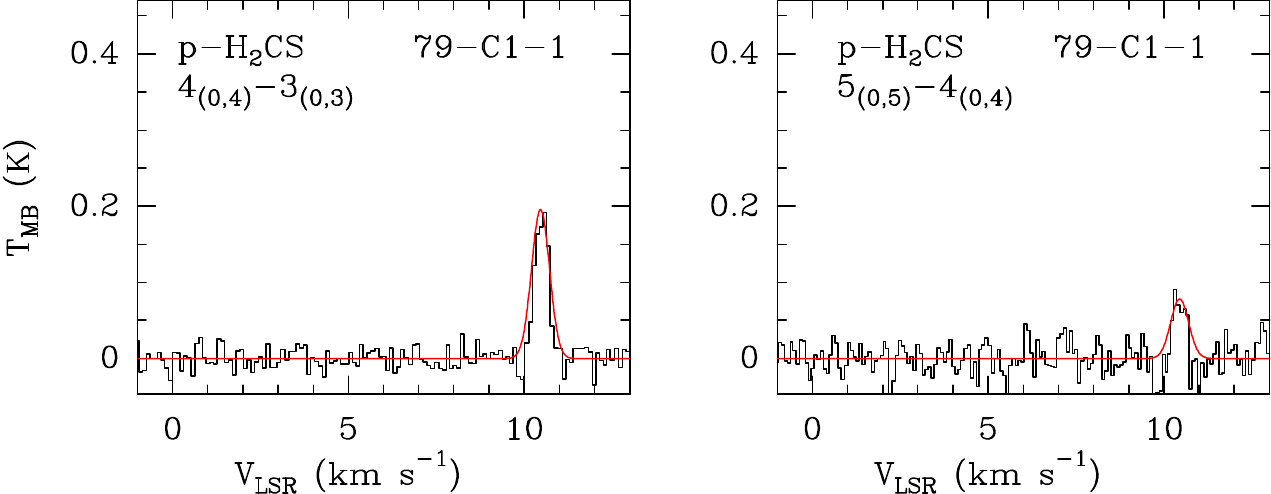} 
\hspace{4cm}
\\
\caption{Observed lines of o-H$_{2}$CS and p-H$_{2}$CS in Barnard\,5 (black) and the best fit (red).}
\label{figure:H2CS_79-C1}
\end{figure*}

\begin{figure*}
\centering
\hspace{-4.3cm}
\includegraphics[scale=0.65, angle=0]{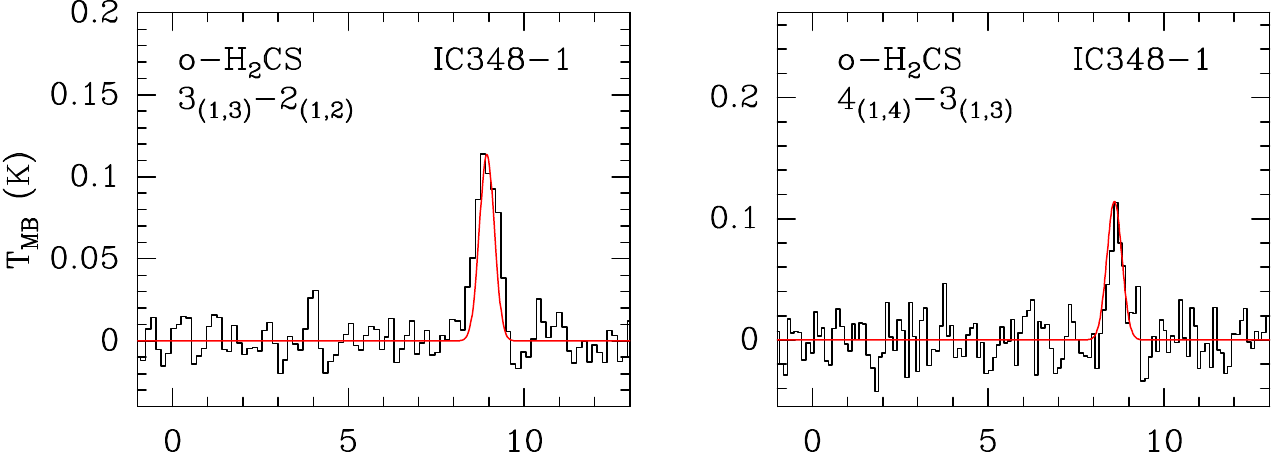}  
\hspace{1cm}
\includegraphics[scale=0.65, angle=0]{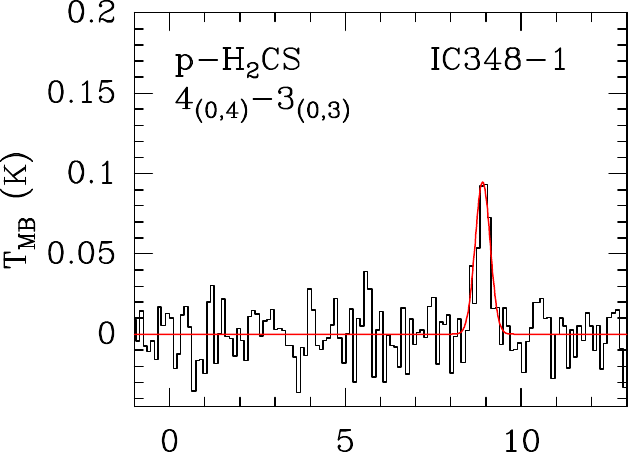} 

\vspace{0.2cm}

\hspace{-4.3cm}
\includegraphics[scale=0.65, angle=0]{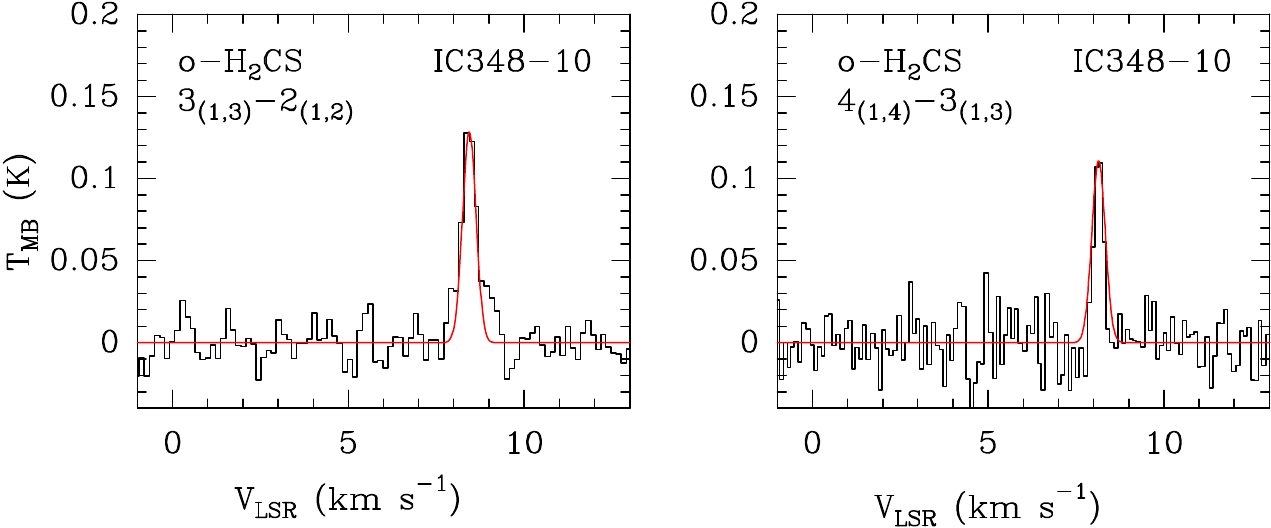}  
\hspace{1cm}
\includegraphics[scale=0.65, angle=0]{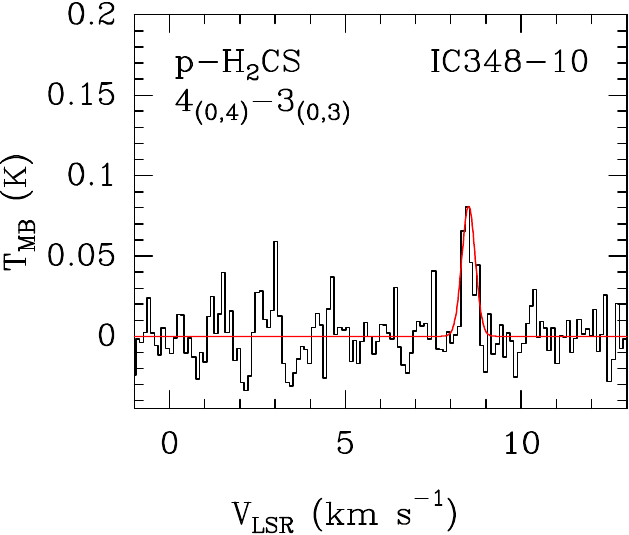} 
\hspace{4cm}
\\
\caption{Observed lines of o-H$_{2}$CS and p-H$_{2}$CS in the core sample of IC\,348 (black) and the best fit (red).}
\label{figure:H2CS_IC348}
\end{figure*}

\begin{figure*}
\centering
\includegraphics[scale=0.65, angle=0]{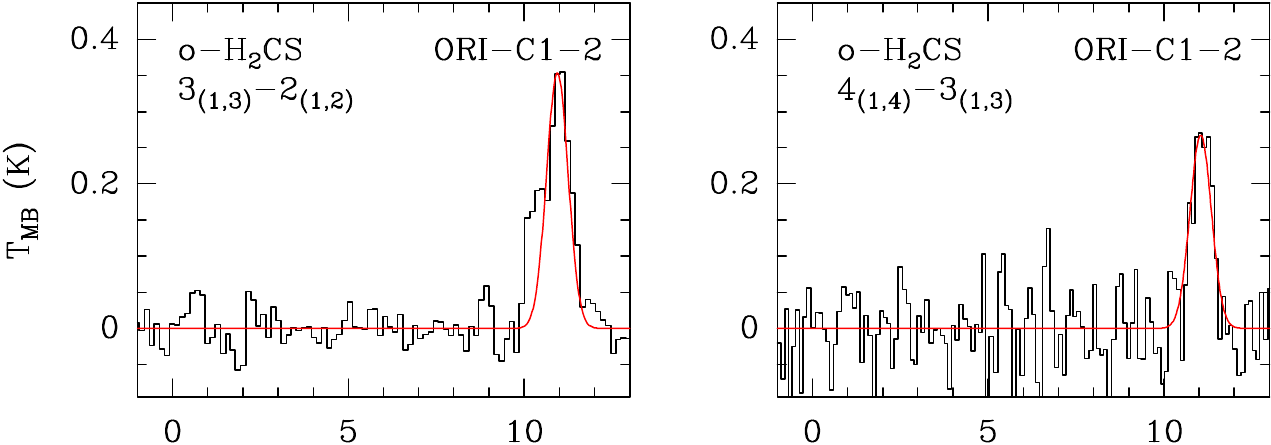}  
\hspace{1cm}
\includegraphics[scale=0.65, angle=0]{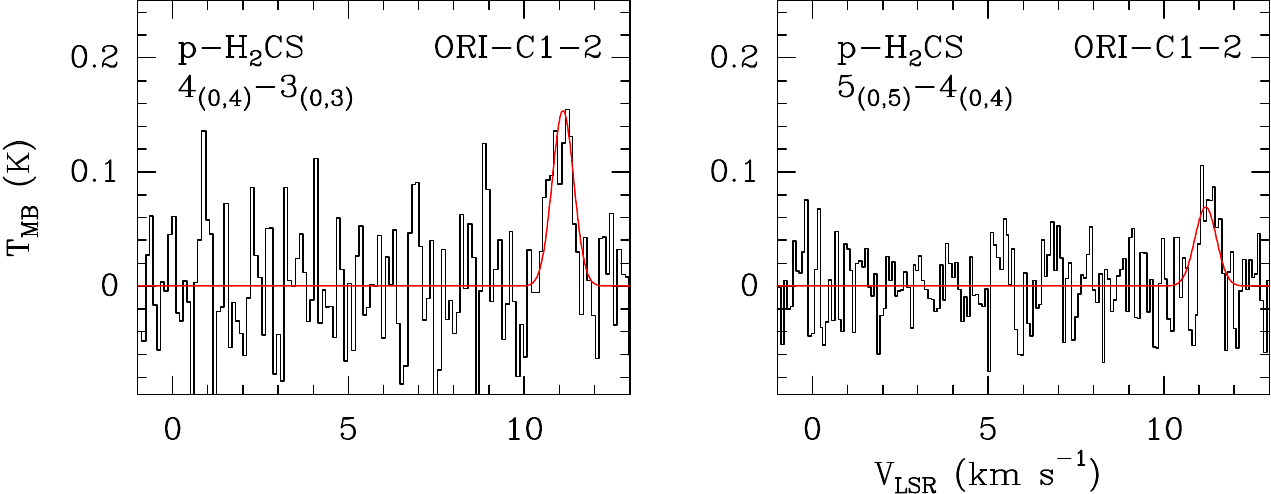} 

\vspace{0.3cm}

\hspace{-4.3cm}
\includegraphics[scale=0.65, angle=0]{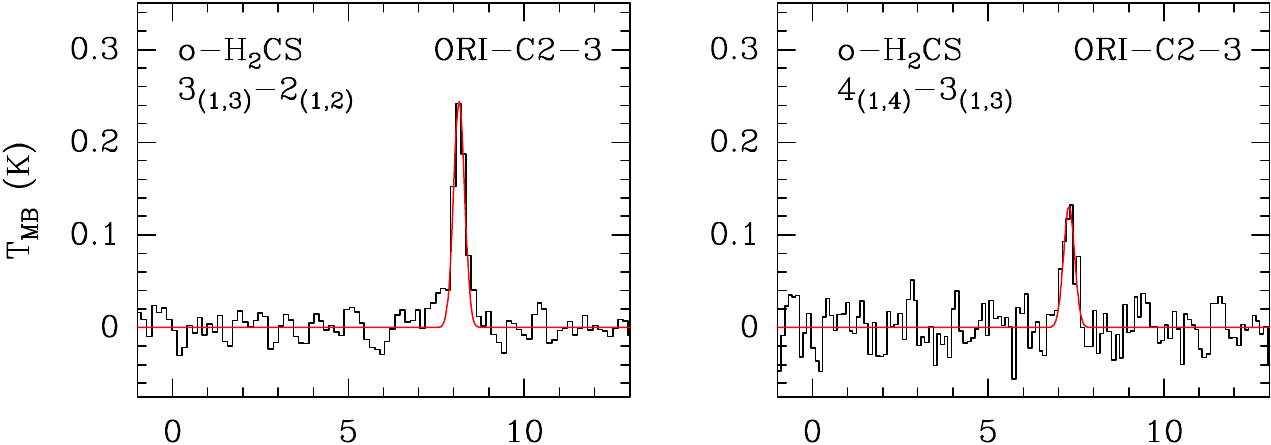}  
\hspace{1cm}
\includegraphics[scale=0.65, angle=0]{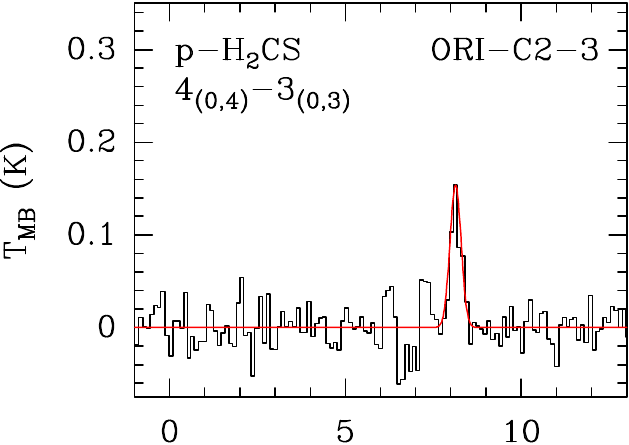} 

\vspace{0.4cm}

\hspace{-4.3cm}
\includegraphics[scale=0.65, angle=0]{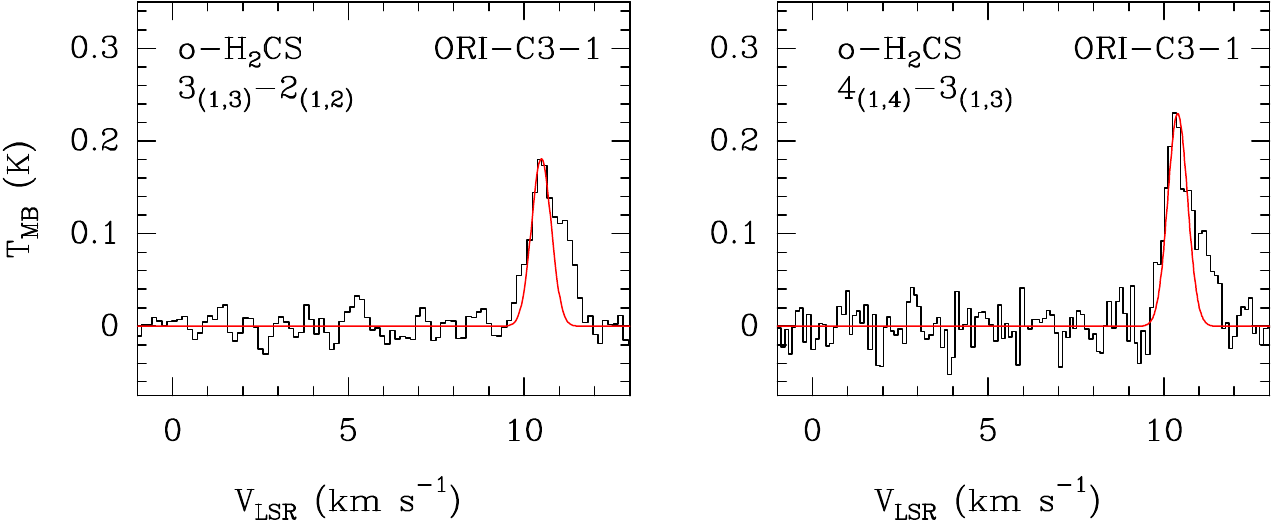}  
\hspace{1cm}
\includegraphics[scale=0.65, angle=0]{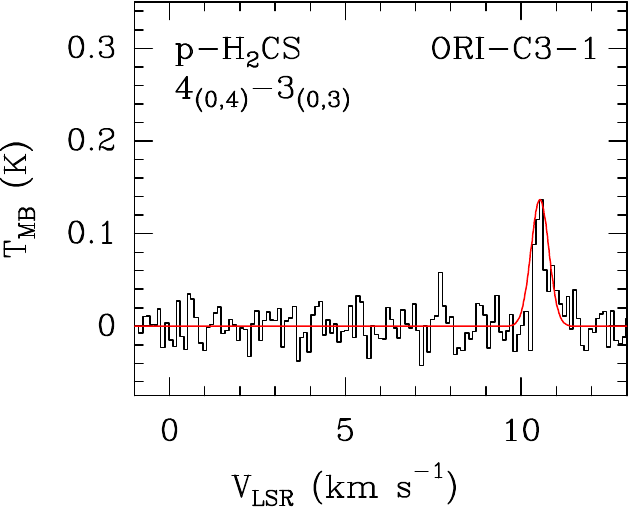} 
\\
\caption{Observed lines of o-H$_{2}$CS and p-H$_{2}$CS in the core sample of Orion\,A (black) and the best fit (red).}
\label{figure:H2CS_ORI}
\end{figure*}


\begin{figure*}
\centering
\includegraphics[scale=0.65, angle=0]{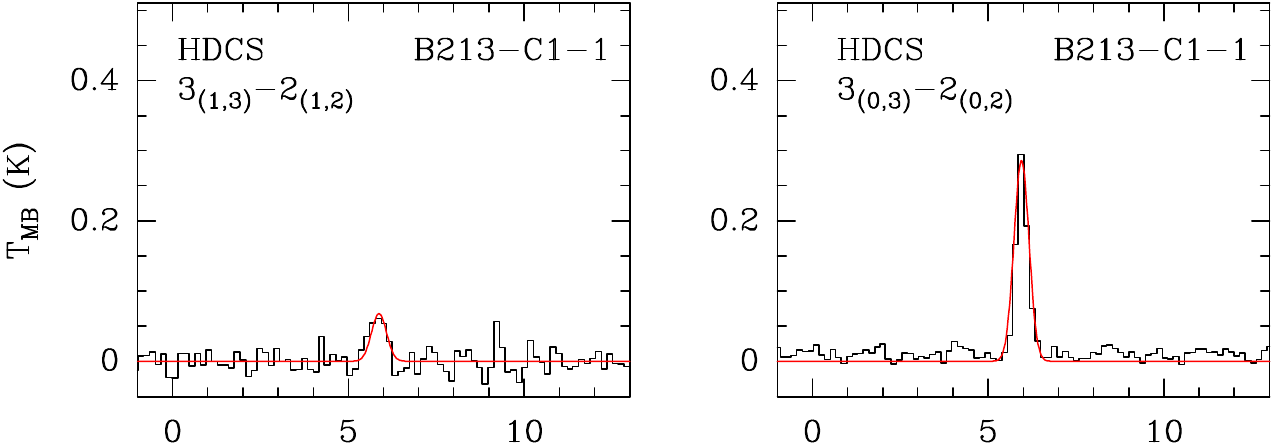}
\hspace{1cm}  
\includegraphics[scale=0.65, angle=0]{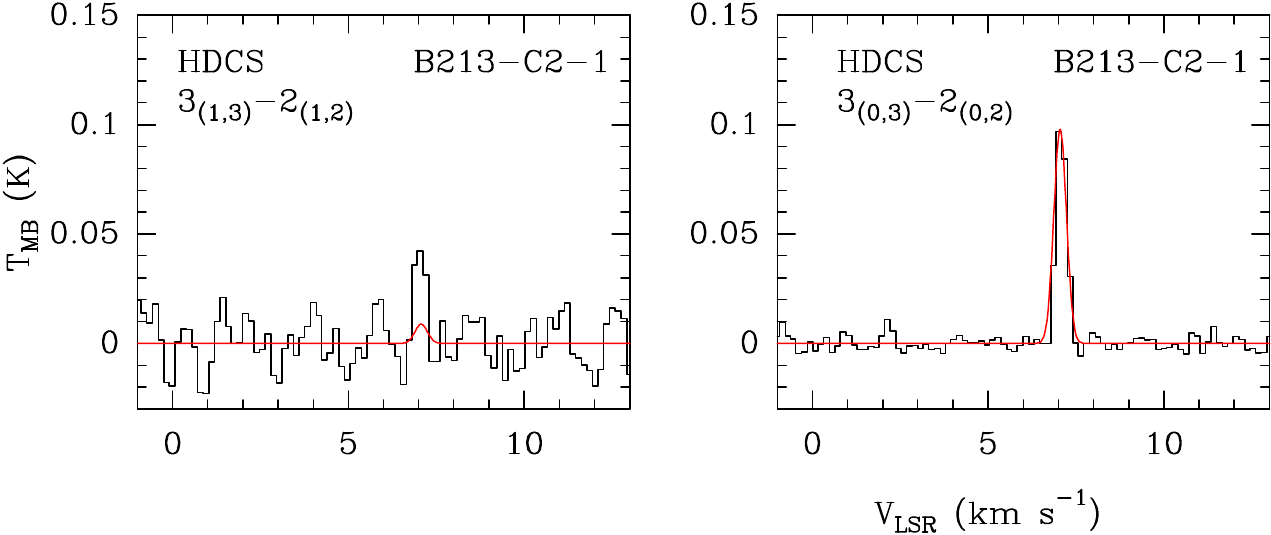} 

\vspace{0.1cm}

\hspace{-4.3cm}
\includegraphics[scale=0.65, angle=0]{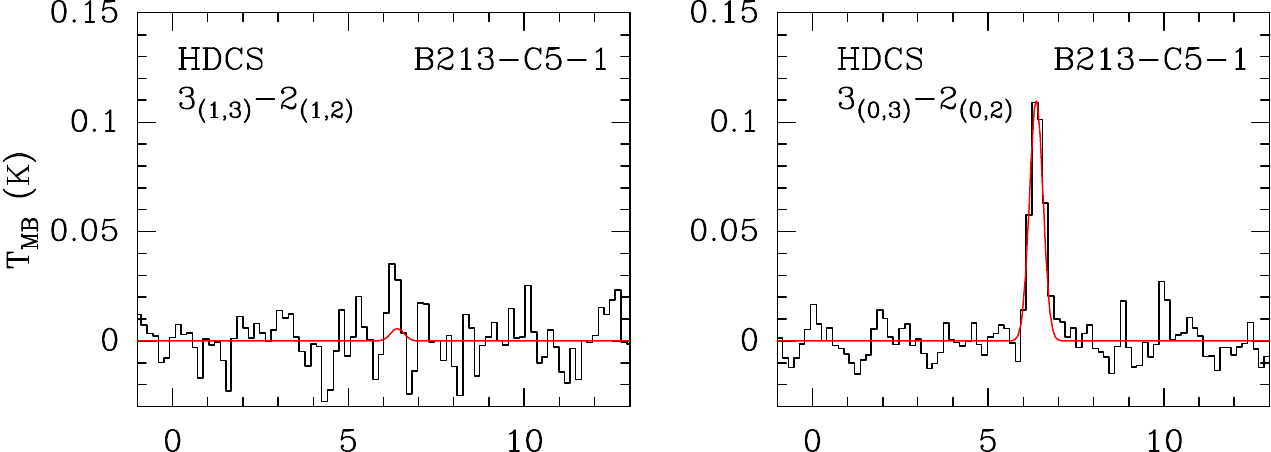}  
\hspace{1cm}  
\includegraphics[scale=0.65, angle=0]{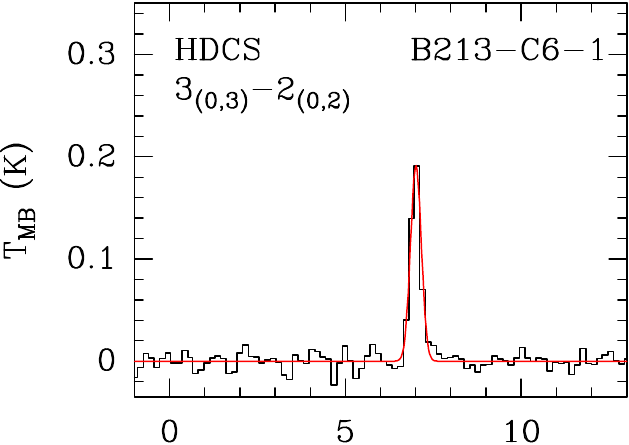} 

\vspace{0.2cm}

\hspace{-4.3cm}
\includegraphics[scale=0.65, angle=0]{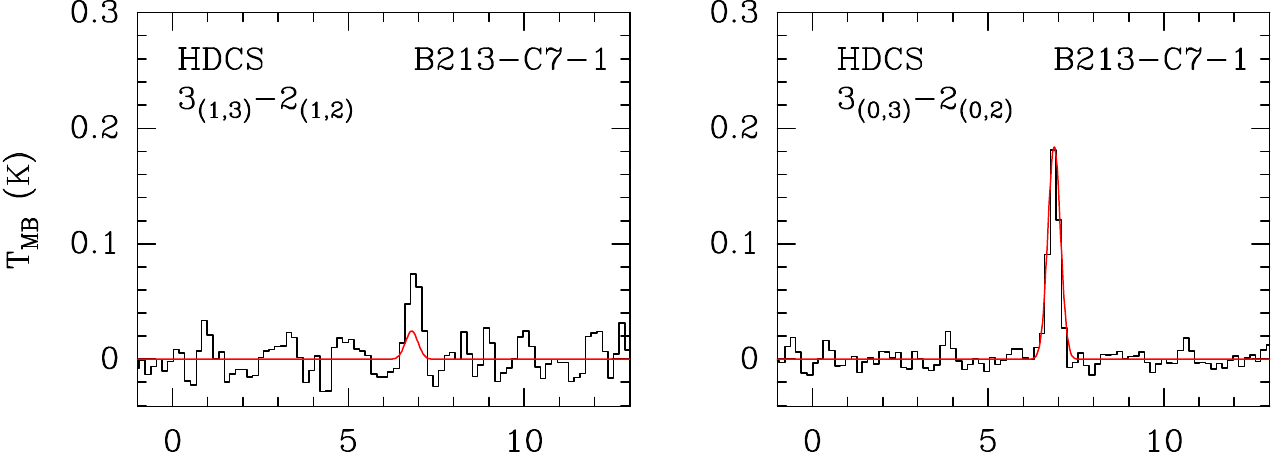}  
\hspace{1cm}  
\includegraphics[scale=0.65, angle=0]{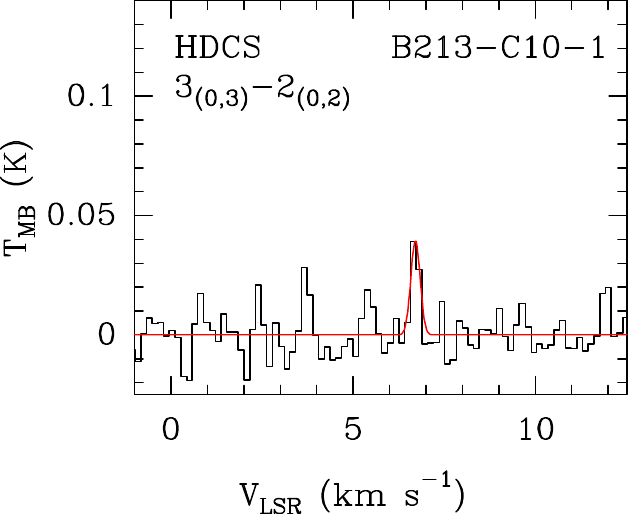} 

\vspace{0.2cm}

\hspace{-9.5cm}
\includegraphics[scale=0.65, angle=0]{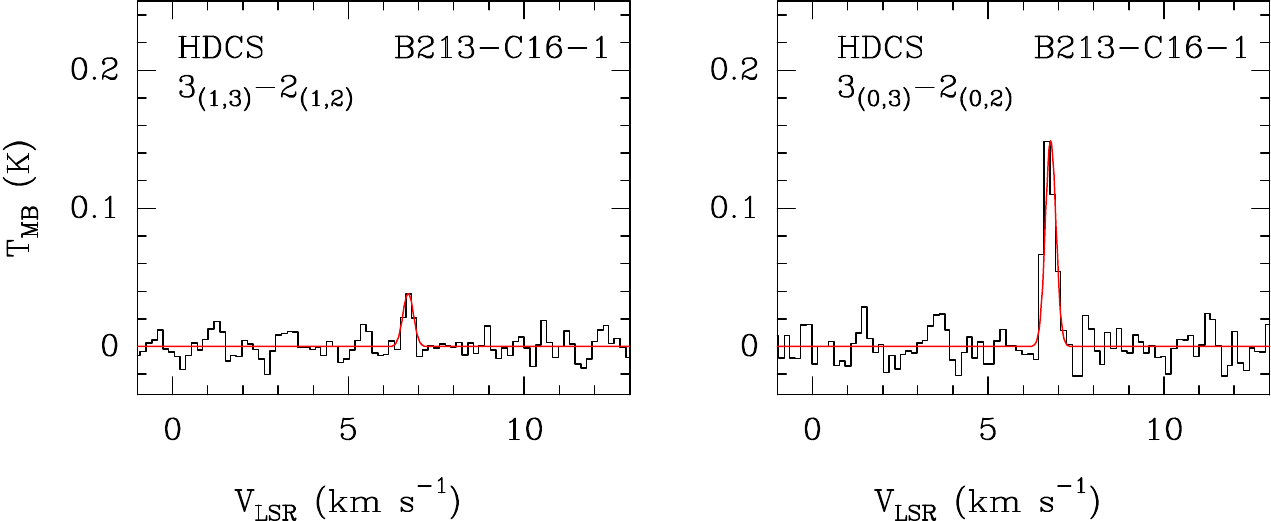}  
\hspace{4cm}
\\
\caption{Observed lines of HDCS in B\,213-C1, C2, C5, C6, C7, C10, and C16 (black) and the best fit (red).}
\label{figure:HDCS_B213}
\end{figure*}

\begin{figure}
\centering

\hspace{-9.4cm}
\includegraphics[scale=0.65, angle=0]{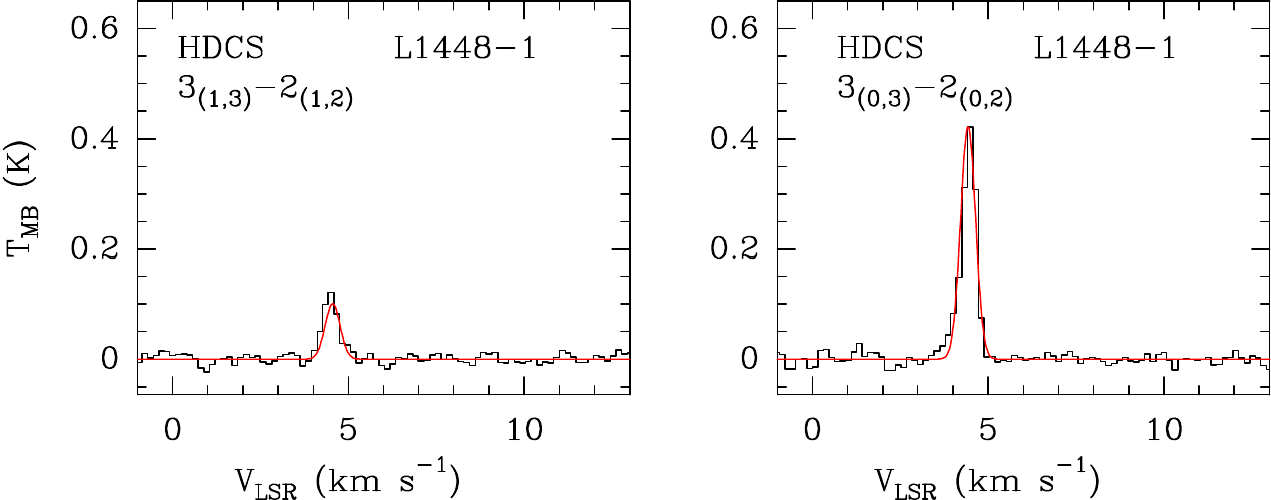} 
\hspace{4cm}
\\
\caption{Observed lines of HDCS in L1448 (black) and the best fit (red).}
\label{figure:HDCS_L1448}
\end{figure}

\begin{figure}
\centering
\hspace{4cm}
\hspace{-9.4cm}
\includegraphics[scale=0.65, angle=0]{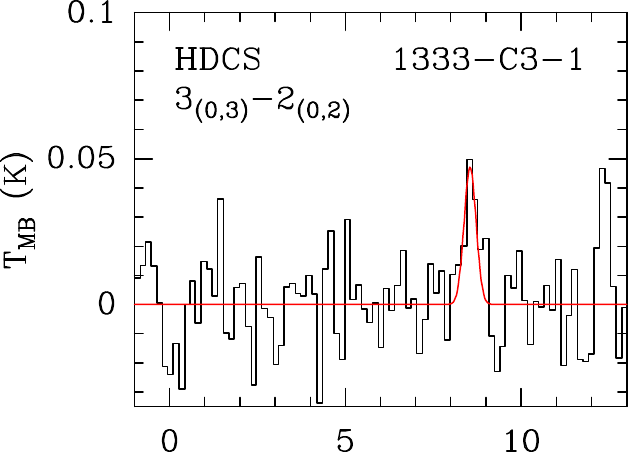}  

\vspace{0.2cm}

\hspace{-9.6cm}
\includegraphics[scale=0.65, angle=0]{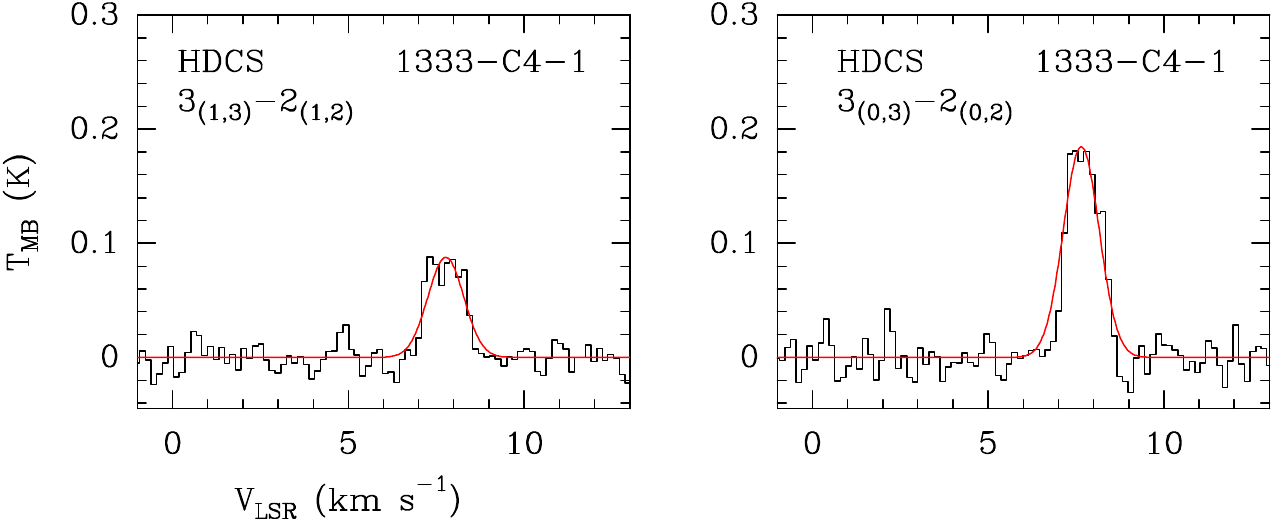}  

\vspace{0.1cm}

\hspace{-8.4cm}
\hspace{4cm}
\includegraphics[scale=0.65, angle=0]{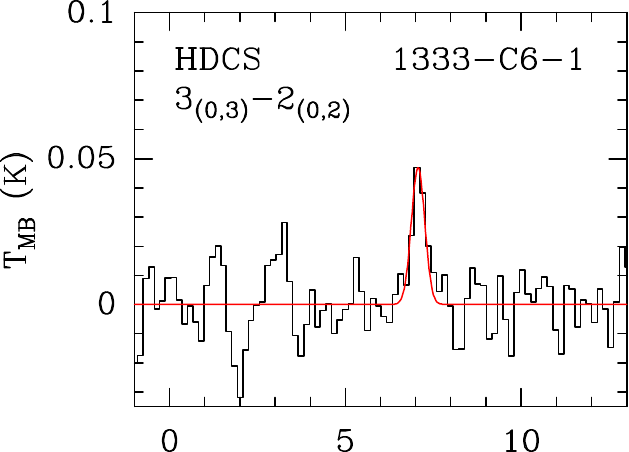}  
\hspace{1cm} 

\vspace{0.2cm}

\hspace{-9.4cm}
\hspace{4cm}
\includegraphics[scale=0.65, angle=0]{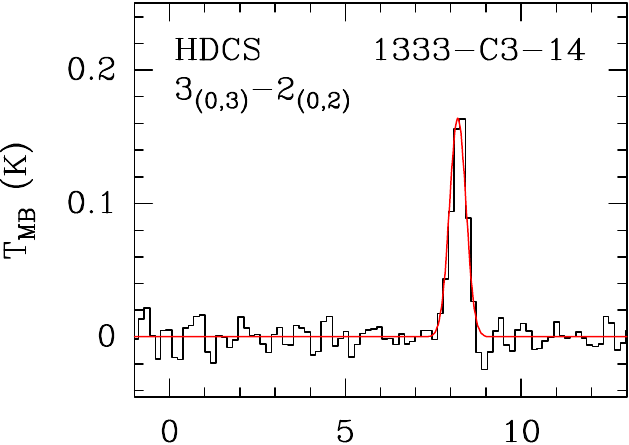}  

\vspace{0.2cm}

\hspace{-9.4cm}
\hspace{4cm}
\includegraphics[scale=0.65, angle=0]{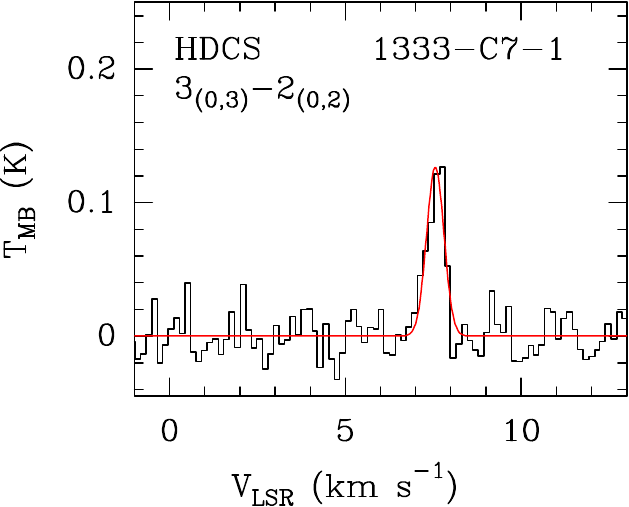}  
\hspace{4cm}
\\
\caption{Observed lines of HDCS in the core sample of NGC\,1333 (black) and the best fit (red).}
\label{figure:HDCS_NGC1333}
\end{figure}

\begin{figure}
\centering
\hspace{4cm}

\hspace{-5.3cm}
\includegraphics[scale=0.65, angle=0]{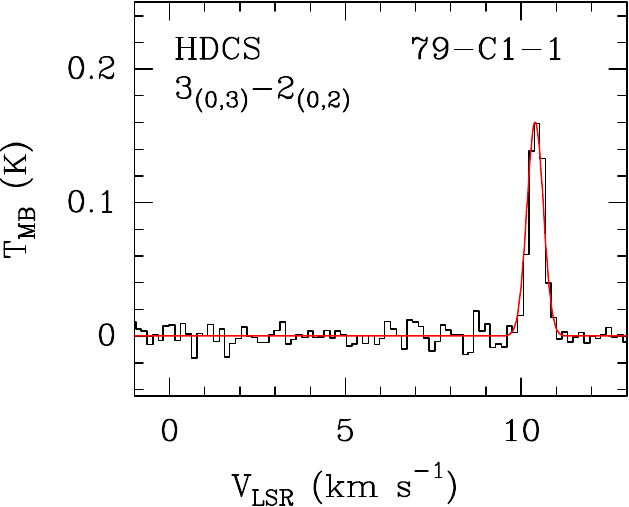}  
\hspace{4cm}
\\
\caption{Observed line of HDCS in Barnard\,5 (black) and the best fit (red).}
\label{figure:HDCS_79-C1}
\end{figure}

\begin{figure}
\centering
\hspace{4cm}
\hspace{-9.1cm}
\includegraphics[scale=0.65, angle=0]{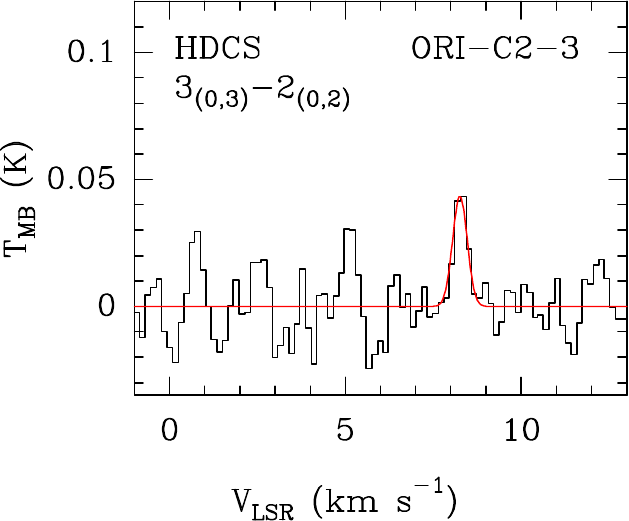}  
\hspace{4cm}
\\
\caption{Observed line of HDCS in Orion\,A (black) and the best fit (red).}
\label{figure:HDCS_ORI}
\end{figure}


\begin{figure}
\centering
\hspace{-8.5cm}
\includegraphics[scale=0.65, angle=0]{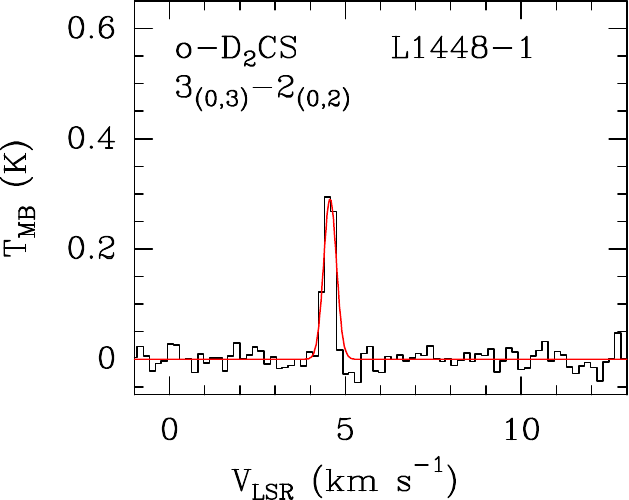}  
\hspace{0.3cm}
\includegraphics[scale=0.65, angle=0]{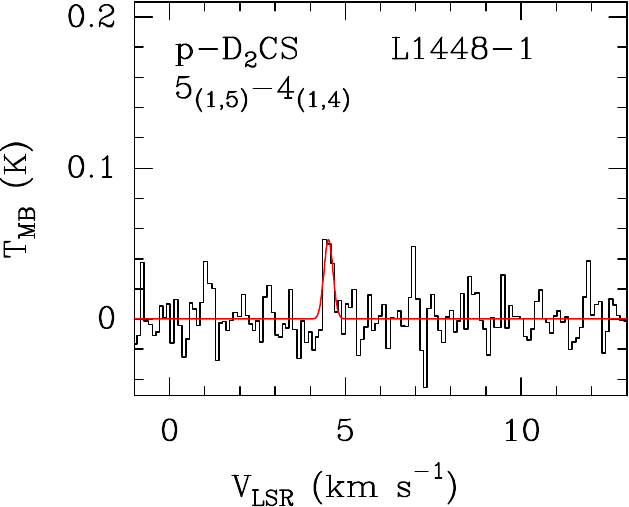} 
\hspace{4cm}
\\
\caption{Observed lines of o-D$_{2}$CS and p-D$_{2}$CS in L1448 (black) and the best fit (red).}
\label{figure:D2CS_L1448}
\end{figure}

\begin{figure}
\centering
\hspace{-4.3cm}
\hspace{-8.5cm}
\includegraphics[scale=0.65, angle=0]{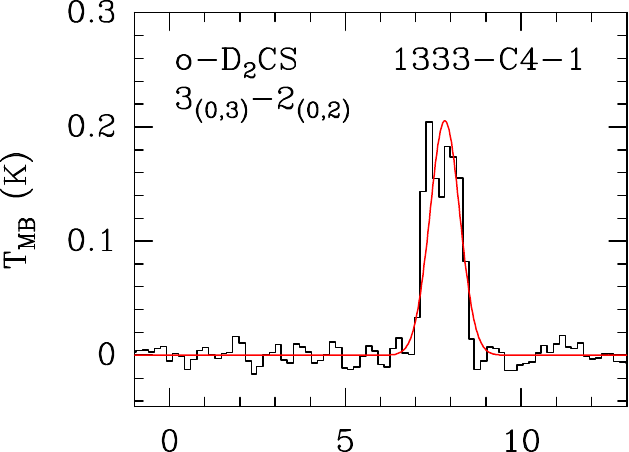}  

\vspace{0.2cm}

\hspace{-12.8cm}
\includegraphics[scale=0.65, angle=0]{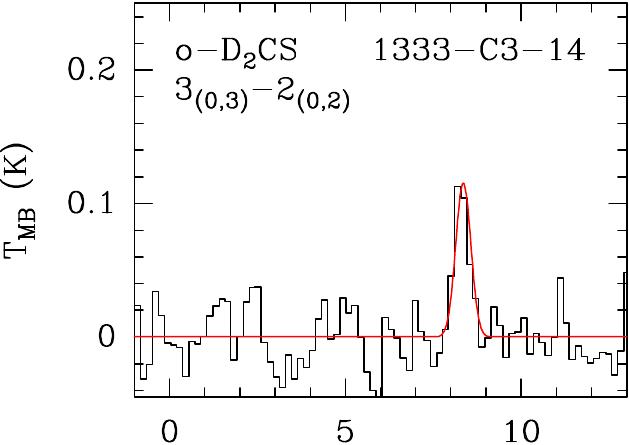}  

\vspace{0.2cm}

\hspace{-12.8cm}
\includegraphics[scale=0.65, angle=0]{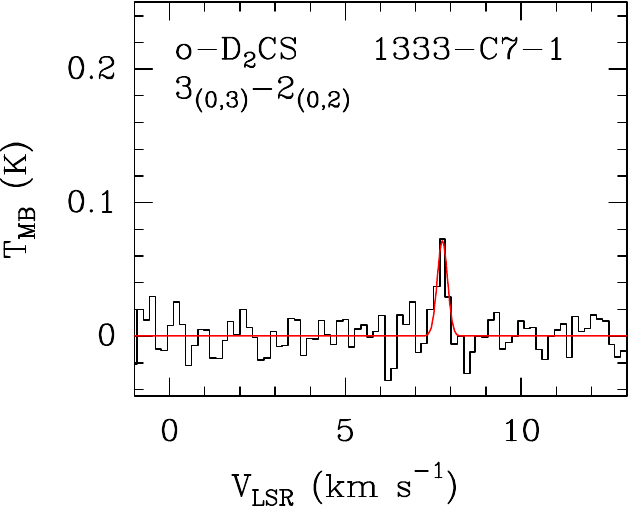}  
\hspace{4cm}
\\
\caption{Observed lines of o-D$_{2}$CS in the core sample of NGC\,1333 (black) and the best fit (red).}
\label{figure:D2CS_NGC1333}
\end{figure}

\begin{figure}
\centering

\hspace{-8.5cm}
\includegraphics[scale=0.65, angle=0]{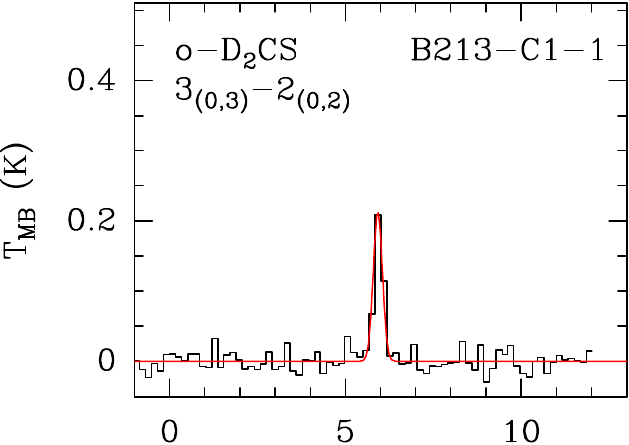}  
\hspace{0.3cm}
\includegraphics[scale=0.65, angle=0]{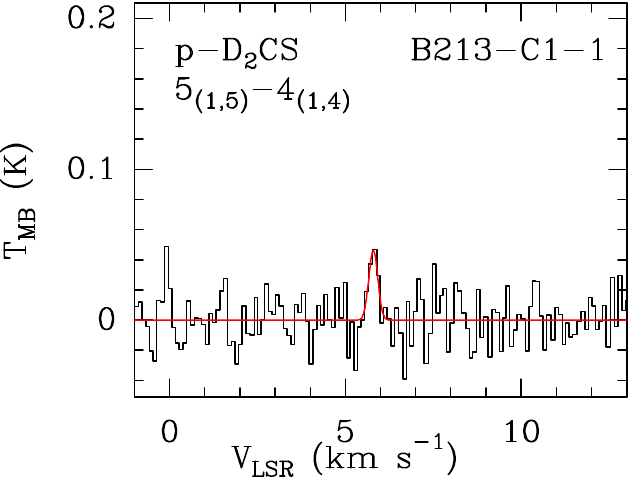} 

\vspace{0.2cm}

\hspace{-12.8cm}
\includegraphics[scale=0.65, angle=0]{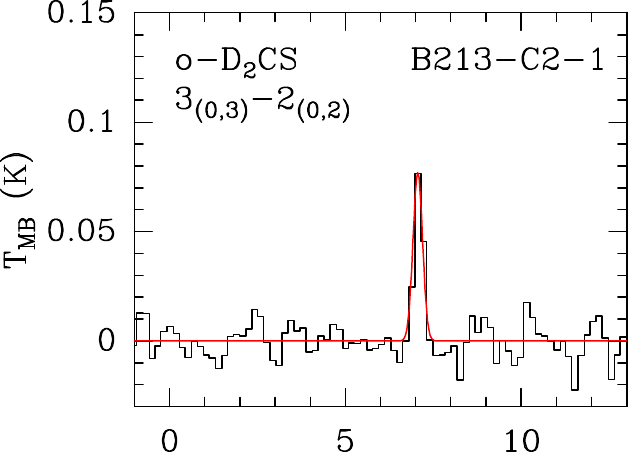}  

\vspace{0.2cm}

\hspace{-12.8cm}
\includegraphics[scale=0.65, angle=0]{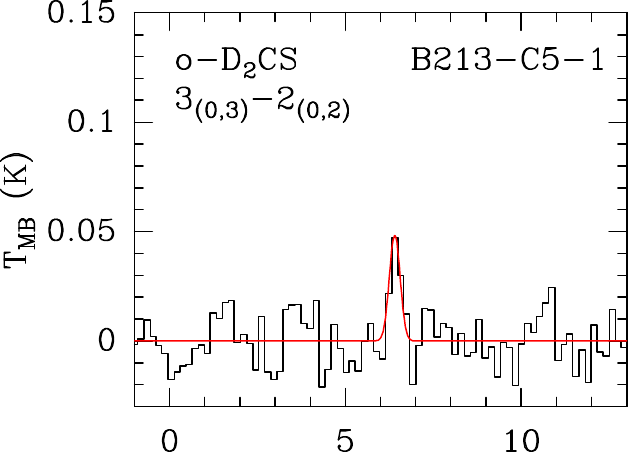}  

\vspace{0.2cm}

\hspace{-12.8cm}
\includegraphics[scale=0.65, angle=0]{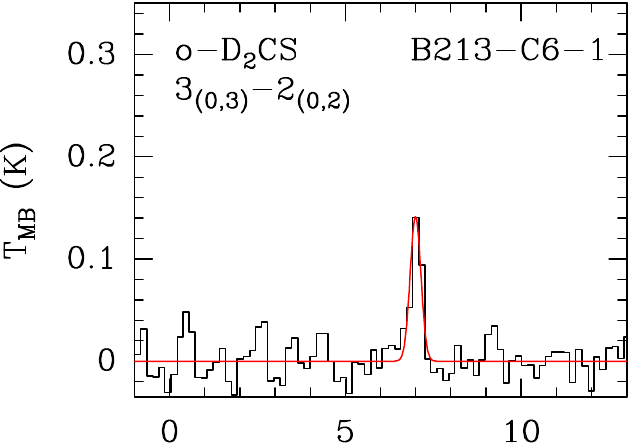}  

\vspace{0.2cm}

\hspace{-12.8cm}
\includegraphics[scale=0.65, angle=0]{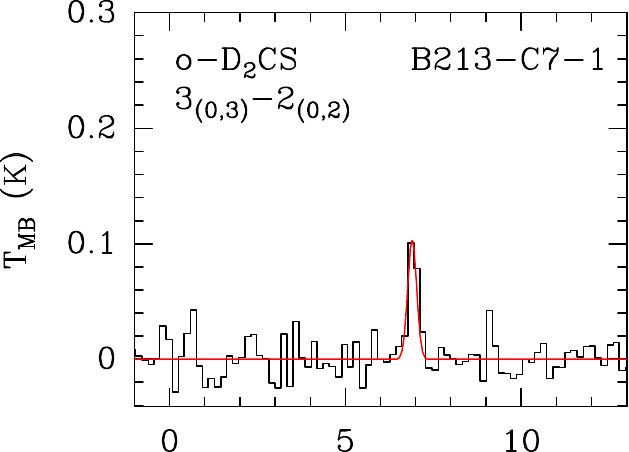}  

\vspace{0.2cm}

\hspace{-12.8cm}
\includegraphics[scale=0.65, angle=0]{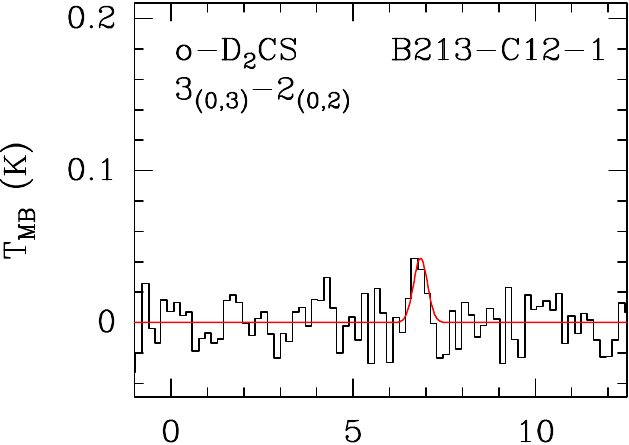}  

\vspace{0.2cm}

\hspace{-12.8cm}
\includegraphics[scale=0.65, angle=0]{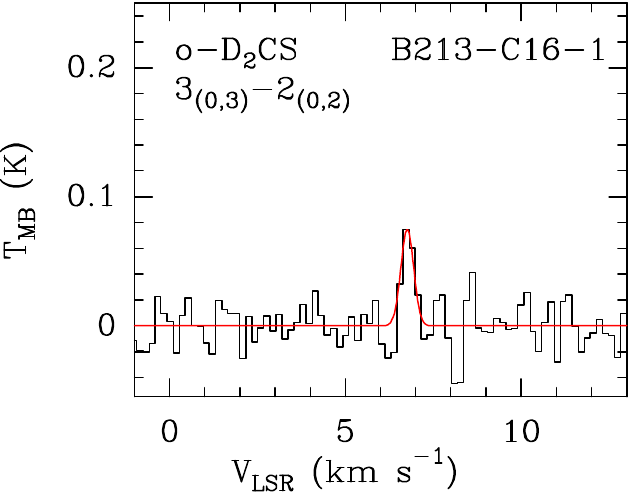}  
\hspace{4cm}
\\
\caption{Observed lines of o-D$_{2}$CS and p-D$_{2}$CS in B\,213-C1, C2, C5, C6, C7, C12, and C16 (black) and the best fit (red).}
\label{figure:D2CS_B213}
\end{figure}

\begin{figure*}
\centering
\includegraphics[scale=0.55, angle=0]{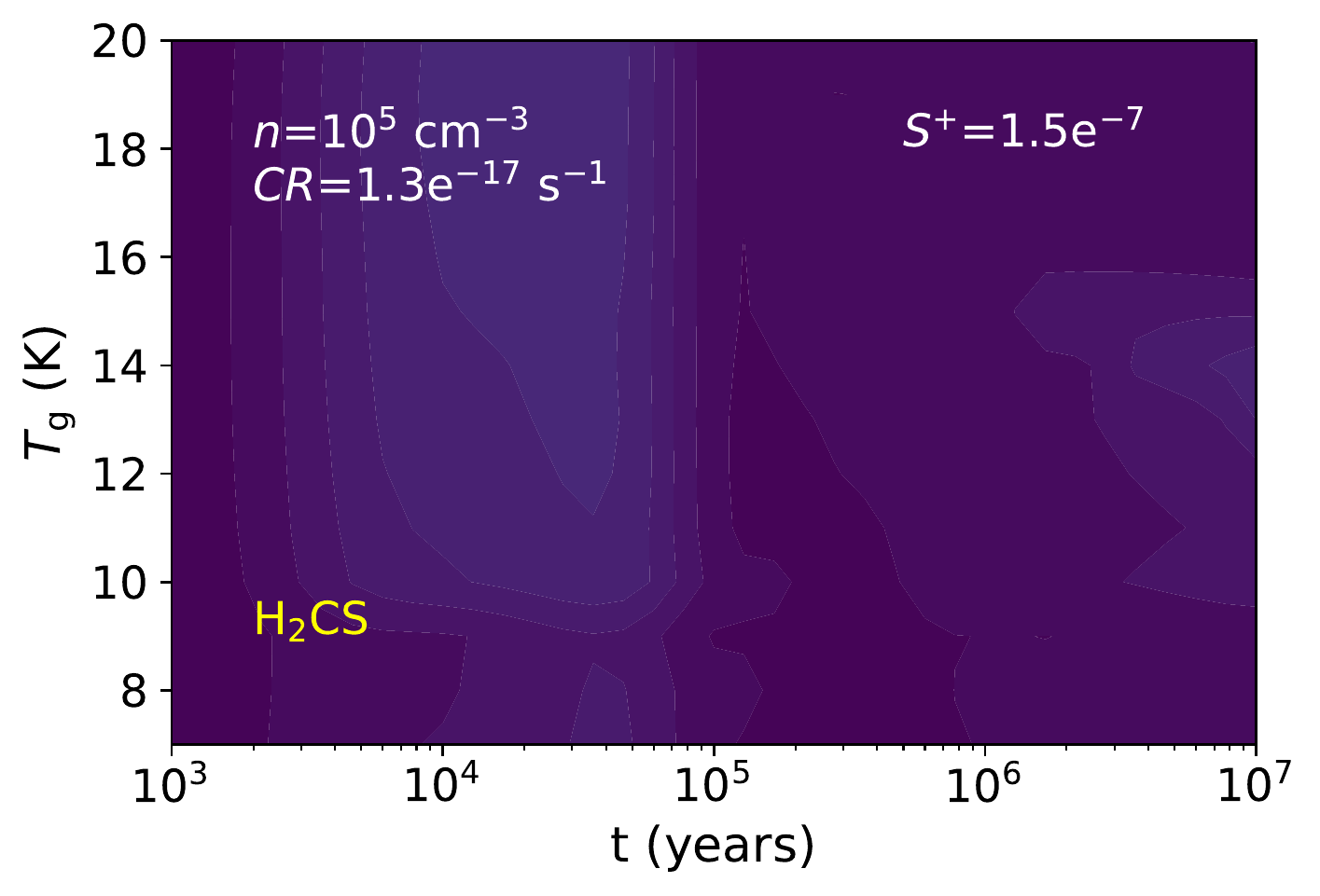}  
\hspace{-0.2cm}
\includegraphics[scale=0.55, angle=0]{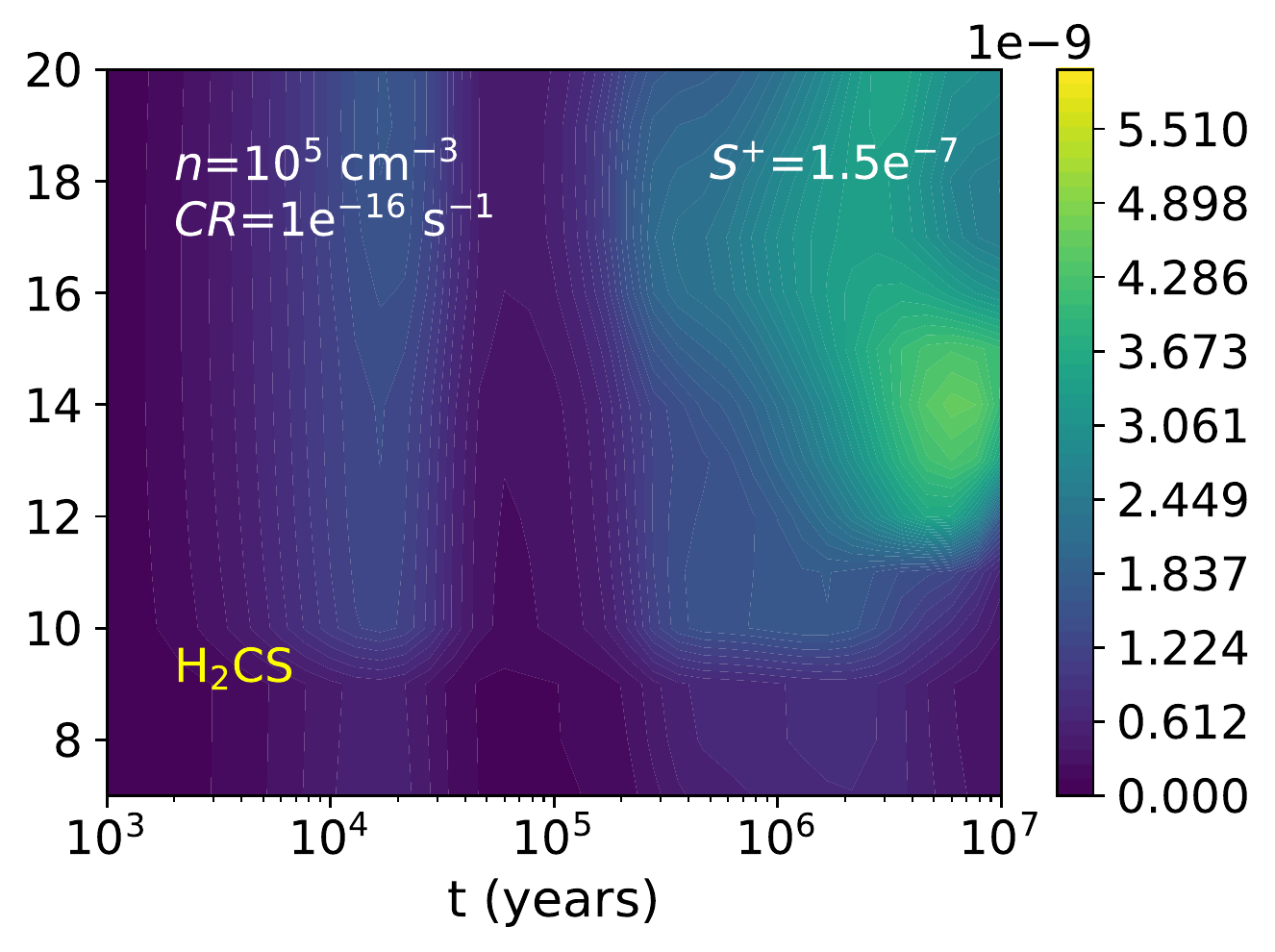}

\hspace{-0.5cm}
\\
\caption{Evolution of the H$_2$CS fractional abundance as a function of time for an initial sulphur abundance $S$$^+$=1.5$\times$10$^{-7}$, a hydrogen number density $n_{\mathrm{H}}$=10$^5$ cm$^{-3}$, and a CR ionisation rate $\zeta$=1.3$\times$10$^{-17}$ s$^{-1}$ (left) and $\zeta$=1.3$\times$10$^{-16}$ s$^{-1}$ (right).}
\label{figure:H2CS_evolution_vs_CR}
\end{figure*}

\pagebreak
\end{appendix}

\bibliography{biblio}

\end{document}